\documentclass[11pt]{article}

\usepackage[a4paper, margin=2.8cm]{geometry}
\usepackage{amsmath, amssymb, amsthm}
\usepackage{booktabs}
\usepackage[normalem]{ulem}
\usepackage{float}
\usepackage{multicol}
\usepackage{caption}
\usepackage[most]{tcolorbox}

\newtcolorbox[auto counter]{example}[1][]{
  enhanced,
  colback=gray!8, colframe=gray!50, boxrule=0.4pt, arc=1.5pt,
  colbacktitle=gray!22, coltitle=black!60,
  fonttitle=\small\scshape, title={example~\thetcbcounter},
  toptitle=1.5pt, bottomtitle=1.5pt,
  left=7pt, right=7pt, top=5pt, bottom=5pt,
  before upper={\setlength{\parindent}{15pt}}, #1}
\usepackage{graphicx}
\usepackage{colortbl}
\usepackage{tikz}

\newcommand{\pbar}[2][1]{%
  \begin{tikzpicture}[baseline=-0.55ex]
    \useasboundingbox (0,-0.17) rectangle (1.3,0.17);
    \fill[blue!25] (0,-0.15) rectangle ({#2*1.3*#1},0.15);
  \end{tikzpicture}}

\newcommand{\hm}[1]{\cellcolor{blue!#1}}

\newcommand{\pbarO}[2][1]{%
  \begin{tikzpicture}[baseline=-0.55ex]
    \useasboundingbox (-1.3,-0.17) rectangle (0,0.17);
    \fill[blue!40] ({-#2*1.3*#1},-0.15) rectangle (0,0.15);
  \end{tikzpicture}}
\newcommand{\pbarP}[1]{%
  \begin{tikzpicture}[baseline=-0.55ex]
    \useasboundingbox (0,-0.17) rectangle (1.3,0.17);
    \pgfmathsetmacro{\pblen}{max(0, 1.3*(1 + ln(#1)/2.302585/5))}%
    \fill[orange!70] (0,-0.15) rectangle (\pblen,0.15);
  \end{tikzpicture}}
\newcommand{\pbarB}[2][1]{%
  \begin{tikzpicture}[baseline=-0.55ex]
    \useasboundingbox (0,-0.17) rectangle (1.3,0.17);
    \fill[blue!40] (0,-0.15) rectangle ({#2*1.3*#1},0.15);
  \end{tikzpicture}}
\newcommand{\wild}{\mathord{\text{\normalsize$\ast$}}}
\usepackage{pgfplots}
\pgfplotsset{compat=1.18}
\usetikzlibrary{svg.path}
\usetikzlibrary{arrows.meta}
\usetikzlibrary{tikzmark}
\usetikzlibrary{calc}
\usetikzlibrary{matrix, fit}
\usetikzlibrary{decorations.pathreplacing}

\newcommand{\fslot}[6]{%
  \draw[fill=blue!#2, draw=black!60, line width=0.3pt]
    (#1,0) rectangle ({#1+2.1*#3},#4);
  \draw[fill=blue!#2, draw=black!60, line width=0.3pt]
    ({#1+2.1*#3},0) rectangle ({#1+2.1},#5);
  \draw[dashed, black!70, line width=0.4pt] (#1,#6) -- ({#1+2.1},#6);
  \node[above, font=\scriptsize] at ({#1+1.05},{max(#4,#5)}) {$#6$};
  \pgfmathsetmacro{\yza}{ifthenelse(#4<0.2,-0.09,0.1)}
  \pgfmathsetmacro{\yzb}{ifthenelse(#5<0.2,-0.09,0.1)}
  \node[fill=white, inner sep=0.8pt, font=\tiny] at ({#1+1.05*#3},\yza) {$0$};
  \node[fill=white, inner sep=0.8pt, font=\tiny] at ({#1+1.05+1.05*#3},\yzb) {$1$};
}
\newcommand{\fbrace}[2]{%
  \draw[decorate, decoration={brace, mirror, amplitude=3pt}, black!70]
    (#1,-0.19) -- ({#1+2.1},-0.19);
  \node[below, font=\scriptsize] at ({#1+1.05},-0.28) {$\mathbf{#2}$};
}
\newcommand{\ledgerbars}[1]{%
\begin{center}
\begin{tikzpicture}[x=1cm, y=1cm]
  \fill[gray!10] (-0.35,-1.85) rectangle (0.40,1.3);
  \fill[gray!28] (0.40,-1.85) rectangle (2.85,1.3);
  \fill[gray!10] (2.85,-1.85) rectangle (6.40,1.3);
  \fill[gray!28] (6.40,-1.85) rectangle (8.85,1.3);
  \fill[gray!10] (8.85,-1.85) rectangle (9.60,1.3);
  \draw[->] (-0.6,-1.15) -- (-0.6,1.3);
  \foreach \y/\t in {-1/-1, 0/0, 1/+1}
    \draw (-0.66,\y) -- (-0.54,\y) node[left, font=\scriptsize] {$\t$};
  \draw[black!60, line width=0.4pt] (-0.6,0) -- (9.7,0);
  \foreach \x/\v/\l/\c in {#1} {
    \draw[fill=\c, draw=black!60, line width=0.3pt]
      ({\x-0.21},0) rectangle ({\x+0.21},\v);
    \node[rotate=90, anchor=east, font=\tiny] at (\x,-1.22) {\l};
  }
\end{tikzpicture}
\end{center}}

\newcommand{\faxis}{%
  \draw[->] (-0.45,0) -- (-0.45,1.5);
  \node[above, font=\scriptsize] at (-0.45,1.5) {bits};
  \node[rotate=90, font=\scriptsize] at (-0.85,0.75) {$\Delta H(M)$};
  \foreach \y in {0,1}
    \draw (-0.51,\y) -- (-0.39,\y) node[left, font=\scriptsize] {$\y$};
  \draw[black!60, line width=0.4pt] (-0.45,0) -- (10.15,0);
}
\usepackage[colorlinks=true, linkcolor=blue!50!black, citecolor=blue!50!black, urlcolor=blue!50!black]{hyperref}
\newlength{\dgcw}\newlength{\dgcl}\newlength{\dgcr}
\newcommand{\dgc}[3][\mathbf{0}]{%
  \settowidth{\dgcw}{$#1$}%
  \ifdim\dgcw<0.62cm\setlength{\dgcw}{0.62cm}\fi
  \setlength{\dgcl}{-\arraycolsep}%
  \setlength{\dgcr}{\dgcw}\addtolength{\dgcr}{\arraycolsep}%
  \begin{tikzpicture}[baseline=0pt]
    \useasboundingbox (0,-0.27) rectangle (\dgcw,0.43);
    \coordinate (dtl) at (\dgcl,0.43);
    \coordinate (dbr) at (\dgcr,-0.27);
    \coordinate (dbl) at (\dgcl,-0.26);
    \coordinate (dtr) at (\dgcr,0.42);
    \draw[line width=0.4pt] (dtl) -- (dbr);
    \node[anchor=south west, inner sep=0.6pt]
      at ($(dbl)!0.2!($(dtl)!(dbl)!(dbr)$)$) {$#2$};
    \node[anchor=north east, inner sep=0.6pt]
      at ($(dtr)!0.2!($(dtl)!(dtr)!(dbr)$)$) {$#3$};
  \end{tikzpicture}}

\let\stdsection\section
\renewcommand{\section}{\clearpage\stdsection}

\newcommand{\Mtab}[4]{%
  \begin{array}{c|cc}
    \dgc{\mathbf a}{\mathbf q} & \mathbf{1} & \mathbf{0} \\
    \hline
    \mathbf{0} & #1 & #2 \\[2pt]
    \mathbf{1} & #3 & #4
  \end{array}}

\newcommand{\MtabIV}[8]{%
  \begin{array}{c|cccc}
    \dgc{\mathbf a}{\mathbf q}
      & \mathbf{11} & \mathbf{10} & \mathbf{01} & \mathbf{00} \\
    \hline
    \mathbf{0} & #1 & #2 & #3 & #4 \\[2pt]
    \mathbf{1} & #5 & #6 & #7 & #8
  \end{array}}

\definecolor{cok}{HTML}{1BAF7A}
\definecolor{cbad}{HTML}{D62D2D}
\definecolor{cexp}{HTML}{555555}
\definecolor{cpre}{HTML}{2A78D6}
\definecolor{clim}{HTML}{EB6834}
\definecolor{cn1}{HTML}{A8CCEF}
\definecolor{cn2}{HTML}{72A9DF}
\definecolor{cn3}{HTML}{367FC4}
\definecolor{cn4}{HTML}{174F91}
\definecolor{cf1}{HTML}{1BAF7A}
\definecolor{cf2}{HTML}{2A78D6}
\definecolor{cf3}{HTML}{EB6834}
\definecolor{cf4}{HTML}{EDA100}
\definecolor{cf5}{HTML}{E87BA4}

\newif\ifdraftdata
\draftdatafalse
\newcommand{\dataplot}[4]{%
  \ifdraftdata\else\addplot[#1] table[x=#2, y=#3] {#4};\fi}
\newcommand{\datalegend}[1]{\ifdraftdata\else\legend{#1}\fi}

\title{Leveraged Learning: entropy cleared per bit received}
\author{Daniel Chernowitz}
\date{\today}

\begin{document}

\maketitle

In 1948 Claude Shannon introduced the world to information entropy \cite{shannon48}.
He quantified the amount of information in a signal. What if we wanted to quantify learning in the same spirit, as the increased understanding of a problem?
This work is an attempt to follow that thread. It models the problem as a boolean map, and the understanding as the aggregate ability to predict that map.
And learning, as the speed at which that understanding increases, specifically as information is supplied to the learner. Some interesting relationships can be formed between the two.

This essay rests on the following tenets:
\begin{enumerate}
  \item Learning is the incremental ability to answer questions correctly that follow a pattern.
  \item The number of questions
reviewed is not the right measure of a learner's investment. What
counts is the information received from asking them. Not all
questions are created equal: about some, the learner already knows
something a priori, and their answers count for \emph{less}.
\item Conversely, with instincts about the world, one can infer \emph{more} than the knowledge of one answer from one question.
It may settle confusion about any number of yet unseen questions. 
\item This ability to extrapolate from prior information is codified here as \textbf{leverage}.
\end{enumerate}
\begin{equation}
  \text{Leverage of answered question} = \frac{\text{Expected predictive gain}}{\text{Expected surprisal received}}.
\end{equation}

The expected surprisal of a yes--no answer is its entropy: at most $1$ bit. Its
expected predictive gain (uncertainty cleared up) is also an entropy, and can exceed $1$ bit through deductions about
other questions. We model successful learning as a high ratio.

\stdsection*{Summary of results}

\begin{enumerate}
\item \textbf{Setup.} Let there be $Q$ questions $q$, which can each beget $A$ answers $a$,
for a total of $N = A^Q$ maps. The true map is $\psi$, and each candidate map $\phi_j$
has prior belief $p_j$. Answers $a=\psi(q)$ fed one-by-one eliminate maps. We study
the surprisal of each answer, vs the remaining entropy over all $Q$ questions.

\item \textbf{The answer matrix.} What the prior belief \emph{predicts} at any point in the 
learning process is the answer matrix, one column per question, holding the total weight
of the maps that would answer $a$ to $q$. Its column entropies added
measure the total remaining predictive uncertainty under separate logarithmic scoring,
\[
  M_{a,q} \;=\; \sum_{j\,:\,\phi_j(q) = a} p_j,
  \qquad
  H(M) \;=\; \sum_q H\big(M_{\cdot,q}\big).
\]

\item \textbf{Leverage.} Asking $q$ and hearing $a$ costs the surprisal
$s = -\log_2 M_{a,q}$ and, by culling inconsistent maps $\phi_j$, settles some of $H(M)$. 
The two are not equal and the columns of questions never asked move with every update. 
Let us have seen $\ell$ questions. The ratio of total reduced uncertainty to surprisal is \textbf{leverage},
\[
  L_\ell \;=\;
  \frac{H\big(M^{(0)}\big) - H\big(M^{(\ell)}\big)}
       {\sum_{k \le \ell} s_k} .
\]
$L = 1$ is what we expect from a uniform prior, one bit destroyed per bit paid.
An individual run can exceed this baseline through luck. But if the excess is \emph{expected}: that is due to deduction from good prior instinct.

\item \textbf{The finite average.} Average over the truth $\psi \sim p_j$ and uniformly 
over the order in which questions arrive. 
The whole trajectory is then generated by a
single sequence, the mean entropy $G_\ell$ of the answers to $\ell$
questions,
\[
  \mathcal L_\ell
  \;=\; \frac{Q\,G_1 - (Q-\ell)\big(G_{\ell+1} - G_\ell\big)}{G_\ell} .
\]
Here $\mathcal L_\ell$ divides expected predictive gain by expected
surprisal. Anything above $1$ in this $\mathcal L_\ell$ is deduction.
Received surprisal averages to $G_\ell$, and what remains is $Q-\ell$ copies of one
\textbf{increment}. This is exact at every size and for every prior, with no
assumption about where the belief came from. The sequence is a combinatoric function 
of the prior, summing over the $\binom Q\ell$ sets $\mathcal S$ of
$\ell$ questions and the $A^\ell$ answer patterns $y$ they admit,
\[
  G_\ell \;=\; -\binom{Q}{\ell}^{-1} \sum_{|\mathcal S| = \ell}\;
  \sum_{y} P_{\mathcal S,y}\log_2 P_{\mathcal S,y},
  \qquad
  P_{\mathcal S,y} \;=\; \sum_{j\,:\,\phi_j(\mathcal S) = y} p_j .
\]

\item \textbf{The thermodynamic limit.} We send $n \to \infty$ at a fixed asked fraction
$t = \ell/Q$. The increments become a profile $\gamma(t)$, the received
entropy its integral, and the limit is obtained,
\[
  L(t) \;=\; \frac{\eta_0 - (1-t)\,\gamma(t)}{g(t)},
  \qquad
  g(t) = \int_0^t \gamma(x)\,dx .
\]
Bounded and nonincreasing is all that is asked of $\gamma$.

\item \textbf{An Occam prior's macroscopic limit.} Write a boolean map as a polynomial
over $\{0,1\}^n$ and group maps by their degree. Allocate prior weight to shells of varying degrees by a CDF
$F(x)$, dividing the $p_j$ uniformly over the maps inside it. A map's weights decrease strictly with degree, 
making this genuinely Occam, and its macroscopic profile is closed form,
\[
  \gamma(t) \;=\; 1 - F(t),
  \qquad
  L_F(t) \;=\; \frac{t + (1-t)F(t)}{\int_0^t \big(1-F(x)\big)\,dx} .
\]
which allows us to model a simplicity bias that permits an explicit TDL.
\end{enumerate}

\clearpage
\tableofcontents

\section{A thought experiment}\label{sec:thought}

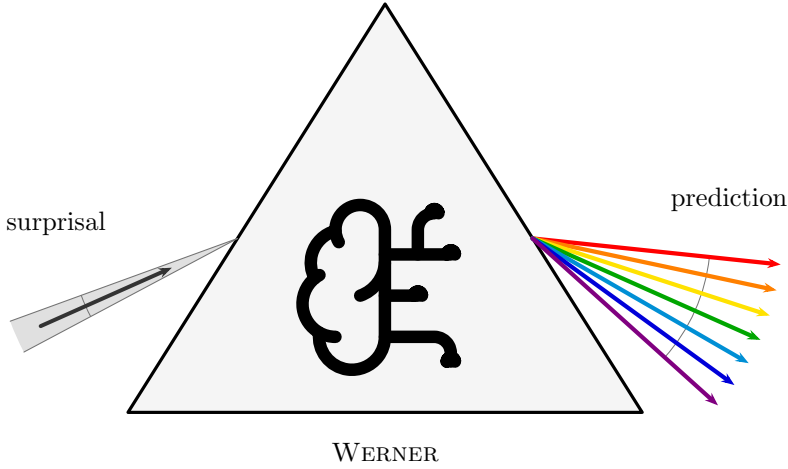
\begin{figure}[h]
\centering
\begin{tikzpicture}[line cap=round, line join=round]
  \fill[black!4] (0,3.2) -- (-3.4,-2.2) -- (3.4,-2.2) -- cycle;
  \draw[very thick] (0,3.2) -- (-3.4,-2.2) -- (3.4,-2.2) -- cycle;

  \begin{scope}[shift={(-1.31cm,0.76cm)}, xscale=3.1, yscale=-3.1,
                line width=4.7pt]
    \draw svg {M 12 5 a 3 3 0 1 0 -5.997 0.125 a 4 4 0 0 0 -2.526 5.77
               a 4 4 0 0 0 0.556 6.588 A 4 4 0 1 0 12 18 Z};
    \draw svg {M 9 13 a 4.5 4.5 0 0 0 3 -4};
    \draw svg {M 6.003 5.125 A 3 3 0 0 0 6.401 6.5};
    \draw svg {M 3.477 10.896 a 4 4 0 0 1 0.585 -0.396};
    \draw svg {M 6 18 a 4 4 0 0 1 -1.967 -0.516};
    \draw svg {M 12 13 h 4};
    \draw svg {M 12 18 h 6 a 2 2 0 0 1 2 2 v 1};
    \draw svg {M 12 8 h 8};
    \draw svg {M 16 8 V 5 a 2 2 0 0 1 2 -2};
    \draw svg {M 15.5 13 a 0.5 0.5 0 1 0 1 0 a 0.5 0.5 0 1 0 -1 0};
    \draw svg {M 17.5 3 a 0.5 0.5 0 1 0 1 0 a 0.5 0.5 0 1 0 -1 0};
    \draw svg {M 19.5 21 a 0.5 0.5 0 1 0 1 0 a 0.5 0.5 0 1 0 -1 0};
    \draw svg {M 19.5 8 a 0.5 0.5 0 1 0 1 0 a 0.5 0.5 0 1 0 -1 0};
  \end{scope}

  \fill[black!12] (-1.951,0.1) -- ++(200:3.2) arc (200:208:3.2) -- cycle;
  \draw[black!55, line width=0.4pt] (-1.951,0.1) -- ++(200:3.2);
  \draw[black!55, line width=0.4pt] (-1.951,0.1) -- ++(208:3.2);
  \draw[black!55, line width=0.35pt]
    (-1.951,0.1) ++(200:2.2) arc (200:208:2.2);
  \draw[-{Stealth[length=5pt]}, black!80, line width=1.5pt]
    (-1.951,0.1) ++(204:2.85) -- ++(24:1.9);
  \node[font=\small] at (-4.35,0.30) {surprisal};

  \draw[black!55, line width=0.35pt]
    (1.951,0.1) ++(-42:2.35) arc (-42:-6:2.35);
  \foreach \ang/\col in {-6/red, -12/orange, -18/yellow!85!orange,
    -24/green!65!black, -30/cyan!75!blue, -36/blue!85!black,
    -42/violet}
    \draw[\col, line width=1.6pt, -{Stealth[length=5pt]}]
      (1.951,0.1) -- ++(\ang:3.3);
  \node[font=\small] at (4.55,0.62) {prediction};

  \node[font=\small] at (0,-2.72) {\textsc{Werner}};
\end{tikzpicture}
\caption{Werner the learner.}
\label{fig:werner}
\end{figure}

Let us start with a small thought experiment, by introducing
\emph{Werner the learner} (Figure~\ref{fig:werner}).

In this work, we model learning as the increased ability to give the
correct answer to questions, as more of the ground truth is revealed.
Learning a hidden function by asking for its values is the query model
of learning theory \cite{angluin88}.
It is Werner's job to learn a pattern. The most basic shape of a
pattern is a \textbf{binary map}, and the most basic binary map is a map
from one bit to one bit.

The setup of the experiment:
\begin{itemize}
  \item a \textbf{question} $q \in \mathcal Q = \{0,1\}$;
  \item an \textbf{answer} $a \in \mathcal A = \{0,1\}$;
  \item the \textbf{truth} $\psi: \mathcal Q \to \mathcal A$, the one fixed map nature
    uses to answer questions. $\psi$ is unknown to Werner. By definition $\psi(q) = a$.
\end{itemize}
There are exactly four candidate maps $\phi_j$ that $\psi$ could be
(Table~\ref{tab:fourmaps}): each of the two inputs can be sent to
either of the two outputs, independently. We index them by reading
the truth table $\phi_j(1)\,\phi_j(0)$ as a binary numeral: that
number \emph{is} $j$.

\begin{table}[H]
\centering
\begin{tabular}{c cc l}
\toprule
$j$ & $\phi_j(1)$ & $\phi_j(0)$ & name \\
\midrule
0 & 0 & 0 & constant $0$ \\
1 & 0 & 1 & NOT ($\lnot q$) \\
2 & 1 & 0 & identity ($q$) \\
3 & 1 & 1 & constant $1$ \\
\bottomrule
\end{tabular}
\caption{The four binary maps from one bit to one bit.}
\label{tab:fourmaps}
\end{table}

\subsection{An unintelligent learner}\label{sec:unintelligent}

Now, if Werner is unintelligent, he has no preconceived notion of
which maps are more useful, more likely, or somehow more logical than
others. In practice, we would have to tell Werner the answer $a$ to every single question $q$,
before he has learned the whole pattern. 

We model this with a \textbf{uniform prior}: writing
$p_j = \Pr(\psi = \phi_j)$ for Werner's belief, every candidate map
carries the same weight.

\begin{example}
For the four maps at hand:
\begin{equation}
  p_j = \tfrac{1}{4}, \qquad j = 0, 1, 2, 3.
\end{equation}
\end{example}

What does this belief predict? For each question $q$ separately, we
can tabulate the probability of each possible answer $a$: it is the
total belief sum of the maps that would answer $a$ to $q$,
\begin{equation}
  M_{a,q} \;=\; \sum_{j\,:\,\phi_j(q) = a} p_j .
\end{equation}
Arranged with one row per answer ($a = 0, 1$ from the top) and one
column per question ($q = 1, 0$ from the left, descending like the
digits of the numeral convention), these numbers form the matrix
$M$. Each column is a probability distribution over answers, so $M$
is column-stochastic. Werner reads it as follows: find the column of the
question at hand. It's the marginal distribution over answers, according to his belief.

\begin{example}
Under the uniform prior, two of the four maps
answer $0$ and two answer $1$, at either question:
\begin{equation}
  M \;=\;
  \Mtab{M_{0,1}}{M_{0,0}}{M_{1,1}}{M_{1,0}}
  \;=\;
  \Mtab{p_0 + p_1}{p_0 + p_2}{p_2 + p_3}{p_1 + p_3}
  \;=\;
  \Mtab{\tfrac12}{\tfrac12}{\tfrac12}{\tfrac12}\;.
\end{equation}
Every column is a fair coin: the unintelligent Werner predicts
nothing about any answer before it is revealed.

Werner gets to predict the answer to the question $q = 0$. By
his own belief it is a coin toss: $M_{0,0} = M_{1,0} = \tfrac12$.
Let us say the answer comes back $a = 1$. The \textbf{surprisal} of
this answer, to Werner, is $1$ bit, because the probability he
assigned to it was one half.
\end{example}

In general, when an event that a belief holds
with probability $P$ actually occurs, its surprisal is
\begin{equation}
  s \;=\; -\log_2 P \quad \text{bits},
\end{equation}
here $s = -\log_2 M_{a,q}$ for answer $a$ arriving at question $q$.
Surprisal measures how much the received information deviates from
what was already expected: an answer held certain ($P = 1$) carries
$0$ bits: nothing new \cite{baldiitti10}. A coin toss carries exactly $1$ bit; and
the surprisal grows without bound as the answer gets more unexpected,
$P \to 0$. It is additive: the surprisal of two independent events
occurring is the sum of their surprisals, just as their probabilities
multiply.

Moreover, Werner can use a process of elimination (an extreme form
of Bayes' rule) to update his prior into a posterior. Semantically:
every candidate map that would have answered the question wrongly is
now \emph{ruled out}, not merely disfavored. They are inconsistent with what's been learned.
The survivors are what concept learning calls the \emph{version space}
\cite{mitchell82}.

\begin{example}
The truth $\psi$ answered $1$ at $q = 0$, and a map with
$\phi_j(0) = 0$ cannot be $\psi$. This kills the constant-$0$ map
($j = 0$) and the identity ($j = 2$). The two survivors, NOT
($j = 1$) and constant $1$ ($j = 3$), both predicted the observed
answer, so this round of evidence does not distinguish between them
at all (Table~\ref{tab:elimination}).

\begin{center}
\begin{tabular}{c cc l}
\toprule
$j$ & $\phi_j(1)$ & $\phi_j(0)$ & name \\
\midrule
\tikzmark{strikeA}0 & 0 & $\mathbf{0}$ & constant $0$\tikzmark{strikeAend} \\
1 & 0 & 1 & NOT ($\lnot q$) \\
\tikzmark{strikeB}2 & 1 & $\mathbf{0}$ & identity ($q$)\tikzmark{strikeBend} \\
3 & 1 & 1 & constant $1$ \\
\bottomrule
\end{tabular}
\begin{tikzpicture}[overlay, remember picture]
  \draw[line width=0.5pt]
    ([shift={(-2pt,0.55ex)}]pic cs:strikeA) --
    ([shift={(2pt,0.55ex)}]pic cs:strikeAend);
  \draw[line width=0.5pt]
    ([shift={(-2pt,0.55ex)}]pic cs:strikeB) --
    ([shift={(2pt,0.55ex)}]pic cs:strikeBend);
\end{tikzpicture}
\captionof{table}{The surviving candidate maps after the answer $\psi(0) = 1$.}
\label{tab:elimination}
\end{center}
\end{example}

The surviving ratio of believabilities must be preserved, and the
only update that does so is to renormalize the surviving weights to
sum to one. (This is Bayes' rule with a $0/1$ likelihood:
$\Pr(\psi = \phi_j \mid a) \propto \Pr(a \mid \phi_j)\, p_j$, and
$P(a \mid \phi_j)$ is $1$ for a map that answers $a$ and $0$ for a
map that does not.)

\begin{example}
The posterior is
\begin{equation}
  p' \;=\; \big(0,\ \tfrac12,\ 0,\ \tfrac12\big).
\end{equation}
Constructing the new answer matrix from $p'$ by the same recipe as
before:
\begin{equation}
  M' \;=\;
  \Mtab{p'_0 + p'_1}{p'_0 + p'_2}{p'_2 + p'_3}{p'_1 + p'_3}
  \;=\;
  \Mtab{\tfrac12}{0}{\tfrac12}{1}\;.
\end{equation}
Indeed, Werner knows the answer to $q = 0$: that column has
collapsed onto $a = 1$. He is still as agnostic as possible
about the other answer: the $q = 1$ column remains a fair coin. He
gained $1$ bit of surprisal information, and cleaned up his table to
the tune of $1$ bit: the $q = 0$ column went from a full bit of
answer-uncertainty to none, while the $q = 1$ column is unchanged.

After the next question, $q = 1$, he learns $a = 1$, again a coin
toss beforehand ($M'_{1,1} = \tfrac12$), so again worth $1$ bit of
surprisal. Elimination now rules out NOT ($\phi_1(1) = 0$), leaving a
single survivor:
\begin{equation}
  p'' \;=\; \big(0,\ 0,\ 0,\ 1\big),
  \qquad
  M'' \;=\; \Mtab{0}{0}{1}{1}\;.
\end{equation}
So this is the constant map: $\psi = \phi_3$. Werner is done. Every
column is one-hot, and he will be able to answer all questions
correctly from now on.
\end{example}

\pagebreak
\subsection{Intelligent priors}\label{sec:intelligent}

Now, let us assume that Werner knows \emph{something} about the world
before he starts. Eliminating hypotheses under a prior that favors some
of them over others is Bayesian concept learning \cite{tenenbaum01}.

\begin{example}
Let's restart, with a different $\psi$. Moreover, let Werner know the map is not constant. His belief then
spreads evenly over the two remaining candidates, NOT ($j = 1$) and
the identity ($j = 2$):
\begin{equation}
  p \;=\; \big(0,\ \tfrac12,\ \tfrac12,\ 0\big).
\end{equation}
Is he any better at predicting $\psi(0)$? Constructing $M$ by the
usual recipe:
\begin{equation}
  M \;=\;
  \Mtab{p_0 + p_1}{p_0 + p_2}{p_2 + p_3}{p_1 + p_3}
  \;=\;
  \Mtab{\tfrac12}{\tfrac12}{\tfrac12}{\tfrac12}\;.
\end{equation}
He is still just as bad as before: every column a fair coin.
\end{example}

So what has all this veteran wisdom about the nature of patterns in
the world bought him? His increased ability for \emph{inference}.

\begin{example}
Ask $q = 0$ as the first question, and let the answer again come back $a = 1$.
Again a coin toss beforehand, so again worth exactly $1$ bit of
surprisal. Elimination rules out the identity ($\phi_2(0) = 0$), and
the lone survivor is NOT:
\begin{equation}
  p' \;=\; \big(0,\ 1,\ 0,\ 0\big),
  \qquad
  M' \;=\; \Mtab{1}{0}{0}{1}\;.
\end{equation}
This time \emph{both} columns collapse at once. Werner received the
same single bit of surprisal as before, but cleaned up $2$ bits of
table: the answer at $q = 0$ he observed, and the answer at $q = 1$
he \emph{deduced}, without ever asking. Preempting the constants
means $\phi(1) \neq \phi(0)$, so his prior held the two answers
perfectly anticorrelated, and settling one settles the other. He is
done after one question instead of two.
\end{example}

We say the information from the first answer was \textbf{leveraged}.
We will make this more precise in the upcoming chapters.

\subsubsection{A one-parameter world}\label{sec:oneparam}

For more structure, let us relax the claim that there are \emph{no}
constant maps. Instead, we create a one-parameter world where the
prior probability that the map is constant is $p$, spread evenly
within each class (Table~\ref{tab:oneparam}):

\begin{table}[H]
\centering
\begin{tabular}{c cc l c}
\toprule
$j$ & $\phi_j(1)$ & $\phi_j(0)$ & name & $p_j$ \\
\midrule
0 & 0 & 0 & constant $0$ & $p/2$ \\
1 & 0 & 1 & NOT ($\lnot q$) & $(1-p)/2$ \\
2 & 1 & 0 & identity ($q$) & $(1-p)/2$ \\
3 & 1 & 1 & constant $1$ & $p/2$ \\
\bottomrule
\end{tabular}
\caption{The one-parameter prior: constant with probability $p$.}
\label{tab:oneparam}
\end{table}

\noindent
In vector form, and with the answer matrix constructed by the usual
recipe:
\begin{equation}
  \big(p_0,\ p_1,\ p_2,\ p_3\big)
  \;=\; \tfrac12\big(p,\ 1-p,\ 1-p,\ p\big),
  \qquad
  M \;=\; \Mtab{\tfrac12}{\tfrac12}{\tfrac12}{\tfrac12}\;,
\end{equation}
fair coins in every column, for every value of $p$.

How much is there to learn? We measure that by the predictive uncertainty that remains in total. Suppose Werner
predicts each answer separately and is scored by logarithmic loss:
when $a=\psi(q)$ is revealed, his loss is $-\log_2 M_{a,q}$.
Under his own belief, the expected loss on question $q$ is the entropy
of its column. Adding the scores over the questions (extensivity) therefore gives
the table entropy, 
\begin{equation}\label{eq:HM}
  H(M) \;:=\; \sum_{q} H\big(M_{\cdot,q}\big),
  \qquad
  H\big(M_{\cdot,q}\big) \;:=\; -\sum_{a} M_{a,q}\,\log_2 M_{a,q}.
\end{equation}
This is an operational choice: $H(M)$ is the aggregate Bayes
prediction risk under logarithmic loss~\cite{covthomas}. A prediction
risk is the expected loss of a stated prediction under a fixed scoring
rule, and it is a Bayes risk when the prediction stated is the one that
minimizes that expectation. Under logarithmic loss the minimizer is
Werner's own column $M_{\cdot,q}$, and the value it attains is that
column's entropy, which is why the two coincide here. It scores
answers separately, even when they are correlated.
One observation can improve many separate predictions at once.
 Identifying the
whole map is a different task, whose uncertainty is its joint entropy. 
This may never happen. There is a practical argument to measuring the full uncertainty as it stands.
In the present example, both columns are fair coins carrying $1$ bit each, so the table
starts at its maximum: $H(M) = 2$ bits. It stays there for every $p$:
the parameter tilts the belief over the four maps without moving a
single column of $M$. Whatever $p$ buys Werner, then, it is not a
sharper forecast of either answer on its own; it can only be a link
between the two.

Now let us ask a question. Again we could use $q = 0$; by the
symmetry of the prior it actually does not matter for this argument.
And we are now going to talk about \emph{expected} values, instead
of realized ones: averages over the possible maps, denoted
$\langle\,\cdot\,\rangle$. We must assume our prior is \emph{a
priori} correct: if we had better assumptions, we would have put
them in the prior. In other words, we model nature to draw 
the truth from the very distribution
Werner believes. Therefore, the average over $\psi$ simply zips with the
weights: $\langle X \rangle = \sum_j p_j\, X(\psi{=}\phi_j)$.

How much surprisal should Werner expect from asking a question $q$?
The answer comes back $a$ exactly when the truth answers $a$, which
under the zipped average happens with probability
$\sum_{j:\phi_j(q)=a} p_j = M_{a,q}$ - the very probability Werner
assigned to it - and when it does, it costs him $-\log_2 M_{a,q}$
of surprisal. Grouping the average over maps by the answer each map
gives:
\begin{equation}\label{eq:surprisal-entropy}
  \langle s_q \rangle
  \;=\; \sum_{j} p_j \,\big({-}\log_2 M_{\phi_j(q),\,q}\big)
  \;=\; \sum_{a} \underbrace{\sum_{j\,:\,\phi_j(q)=a} p_j}_{M_{a,q}}
        \big({-}\log_2 M_{a,q}\big)
  \;=\; {-}\sum_{a} M_{a,q}\,\big(\log_2 M_{a,q}\big).
\end{equation}
The right-hand side is exactly the column entropy
$H(M_{\cdot,q})$ of equation~\eqref{eq:HM}: the expected
surprisal of a question is the entropy of its column. 

Each answer occurs with
exactly the probability Werner assigned to it, so his average
surprisal is his own uncertainty about that answer. This identity will do a great deal of work in what follows.

For a two-answer column, this entropy is a function of a single
number, the probability $x$ of one of the two answers: the
binary entropy function \cite{covthomas}:
\begin{equation}\label{eq:binary-entropy}
  h_2(x) \;:=\; -x\log_2 x \;-\; (1-x)\log_2(1-x),
\end{equation}
zero at $x \in \{0, 1\}$ (a settled answer) and maximal, $1$ bit, at
the fair coin $x = \tfrac12$. For the first question here the column
is a fair coin, so
\begin{equation}
  \langle s_1 \rangle \;=\; H\big(M_{\cdot,0}\big)
  \;=\; h\big(\tfrac12\big) \;=\; 1 \text{ bit},
\end{equation}
whatever the value of $p$.

\begin{figure}[H]
\centering
\begin{tikzpicture}[
  declare function={ hh(\x) = (-\x*ln(\x) - (1-\x)*ln(1-\x))/ln(2); }]
\begin{axis}[
  width=0.48\textwidth, height=6.2cm,
  xlabel={$p$},
  ylabel={bits},
  xmin=0, xmax=1, ymin=0, ymax=1.05,
  grid=major, grid style={gray!25},
  legend style={at={(0.5,0.04)}, anchor=south, draw=gray!60,
                fill=white, font=\small},
  legend cell align=left]
  \addplot[blue, thick, smooth, domain=0.001:0.999, samples=201]
    { hh(x) };
  \addlegendentry{$h_2(p)$}
\end{axis}
\begin{axis}[
  xshift=0.52\textwidth,
  width=0.48\textwidth, height=6.2cm,
  xlabel={$p$},
  ylabel={leverage},
  xmin=0, xmax=1, ymin=0.95, ymax=2.05,
  grid=major, grid style={gray!25},
  legend style={at={(0.5,0.96)}, anchor=north, draw=gray!60,
                fill=white, font=\small},
  legend cell align=left]
  \addplot[red!80!black, thick, smooth, domain=0.001:0.999, samples=201]
    { 2 - hh(x) };
  \addlegendentry{$\mathcal L_1 = 2 - h_2(p)$}
  \addplot[green!55!black, thick, smooth, domain=0.001:0.999, samples=201]
    { 2/(1 + hh(x)) };
  \addlegendentry{$\mathcal L_2 = 2/(1+h_2(p))$}
\end{axis}
\end{tikzpicture}
\caption{The one-parameter world as a function of the constant-map
probability $p$. Left: the table entropy $h_2(p)$ remaining after one
question, in bits. Right: the one-question leverage $\mathcal L_1$ and the
cumulative leverage $\mathcal L_2$ after both questions. At $p = \tfrac12$
(the uniform prior) nothing is leveraged; toward the
deterministic-class extremes the first answer doubles its value.}
\label{fig:oneparam}
\end{figure}
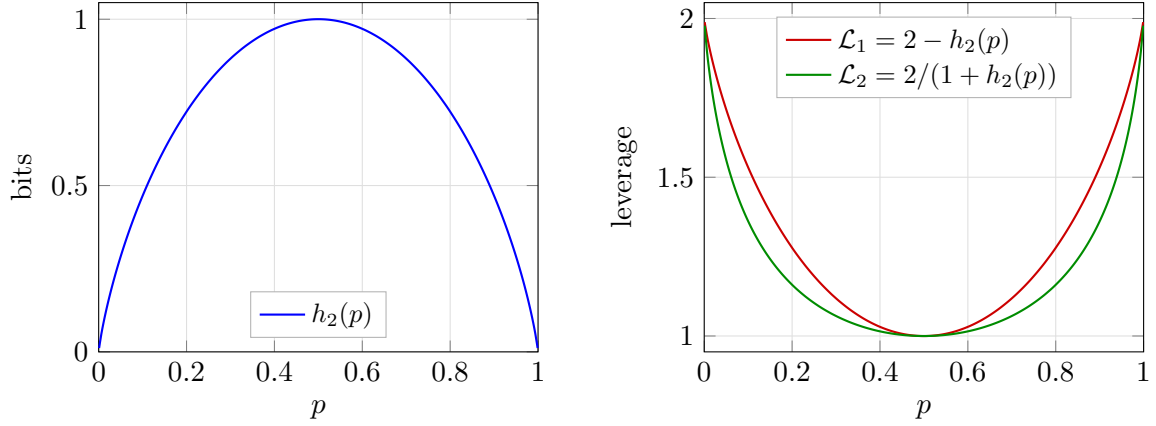

And the other column is now cleaned up. Breaking the update down over
the possible maps: with probability $\tfrac12$ the answer is $1$
(truth is NOT or constant $1$), with probability $\tfrac12$ it is $0$
(truth is FALSE or the identity), and
\begin{equation}
  M' \;=\;
  \underbrace{\Mtab{1-p}{0}{p}{1}}_{a_1 = 1,\ \text{prob } 1/2}
  \quad\text{or}\quad
  \underbrace{\Mtab{p}{1}{1-p}{0}}_{a_1 = 0,\ \text{prob } 1/2}\;.
\end{equation}
Either way, the asked column is one-hot and the unasked column is a
coin of bias $p$: what survives on $q = 1$ is precisely the
class uncertainty ``constant or not''. The total
entropy of the table is now
\begin{equation}
  H(M') \;=\; 0 + h_2(p) \;=\; h_2(p),
\end{equation}
a reduction of $2 - h_2(p)$ bits, purchased with a single bit of
surprisal. Define the \textbf{leverage} after one question as the
ratio of expected table entropy cleared to expected surprisal received:
\begin{equation}
  \mathcal L_1 \;:=\; \frac{2 - h_2(p)}{1} \;=\; 2 - h_2(p) \;\in\; [1, 2].
\end{equation}
The expected surprisal on the second question is, once more, the
entropy of its column,
\begin{equation}
  \langle s_2 \rangle \;=\; h_2(p) \;\leq\; 1 \text{ bit},
\end{equation}
after which the table is empty. Cumulatively over the two questions,
$1 + h_2(p)$ expected bits of received information reduced the table by the
full $2$ bits it started with, so the \textbf{cumulative leverage}
after two questions is
\begin{equation}
  \mathcal L_2 \;:=\; \frac{2}{1 + h_2(p)} \;\in\; [1, 2].
\end{equation}
Our knowledge about the world has allowed us to be more certain
after our inference. Set
$p = \tfrac12$, and the prior is the uniform one of
Section~\ref{sec:unintelligent}: $h_2 = 1$, $\mathcal L_1 = \mathcal L_2 = 1$, and nothing is
leveraged at all. As $p$ skews further from $\tfrac12$, the first question unlocks more. The
second question carries less and less expected information, and Werner's
potential for learning from it goes down with it. The column it could clean up is nearly settled already. In the limit $p = 0$, we recover the non-constant
world from the start of Section~\ref{sec:intelligent}. There is no
surprisal and no improvement to $M$ left in the second question at
all: it contributes nothing to denominator or numerator, and $\mathcal L_1 = \mathcal L_2$. Figure~\ref{fig:oneparam} traces all three curves
between these extremes.

In general, in a larger world, if there are $d$ questions to disagree on between the two maps, the reader can convince themselves that the limit as $p \to 1$ is $\mathcal L_1=d$. 

\section{The general case}\label{sec:general}

Now let us give the definitions in full generality. Let questions be bit
strings of length $n$, and answers bit strings of length $m$:
\begin{equation}
  q \in \mathcal Q = \{0,1\}^n \equiv \mathbb{F}_2^{\otimes n},
  \qquad
  a \in \mathcal A = \{0,1\}^m \equiv \mathbb{F}_2^{\otimes m},
\end{equation}
where $\mathbb{F}_2$ is the binary field. We write the sizes of those two sets without bars,
\begin{equation}
  Q \;:=\; |\mathcal Q| \;=\; 2^n,
  \qquad
  A \;:=\; |\mathcal A| \;=\; 2^m,
  \label{eq:sizes}
\end{equation}
so there are $Q$ distinct questions and $A$ possible answers to
each. As before, we read bit strings as the numbers they spell, using
the two interchangeably: $q \in \{0, \ldots, Q{-}1\}$ and
$a \in \{0, \ldots, A{-}1\}$. The truth
$\psi: \mathcal Q \to \mathcal A$ is one of
the candidate maps $\phi_j$; since each of the $Q$ inputs can be
sent to any of the $A$ outputs independently, there are
\begin{equation}
  N \;=\; A^{Q} \;=\; 2^{mQ}
\end{equation}
of them, and the index $j$ is again the truth table read as a
numeral, now with $Q$ digits in base $A$:
$j = \sum_q \phi_j(q)\,A^{q}$, running from the constant-$0$ map
($j = 0$) to the constant-$(A{-}1)$ map ($j = N{-}1$). Werner's
belief is a distribution over all of them, $\sum_{j=0}^{N-1} p_j = 1$.

\begin{example}
The example we will use to guide the general case is $n = 2$,
$m = 1$, which has $N = 2^{1 \cdot 4} = 16$ maps: every boolean
function of the two input bits $b_1, b_0$, with $q$ represented as $b_1 b_0$, or read as
an ordinary binary number, $q = 2 b_1 + b_0$. Each $\phi_j$ in
Table~\ref{tab:sixteen} is shown as its truth table, printed in
descending question order $(11, 10, 01, 00)$ so that the answer to
question $q$ occupies the digit of weight $2^q$. The printed
string is exactly $j$ in binary (e.g.\ AND, which answers $1$ only to
$q = 11$, reads $1000$, so it is $j = 8$). Alongside we conjure an arbitrary prior $p_j$ for
illustration.

\begin{center}
\begin{tabular}{c |cccc| l c c}
\toprule
 & \multicolumn{4}{|c|}{$\phi_j(q)$} & & & \\
$j$ & $11$ & $10$ & $01$ & $00$ & name & $p_j$ & \\
\midrule
0  & 0 & 0 & 0 & 0 & FALSE (constant $0$) & 0.394 & \pbar[2]{0.394} \\
1  & 0 & 0 & 0 & 1 & $\lnot(b_1 \vee b_0)$ (NOR) & 0.010 & \pbar[2]{0.010} \\
2  & 0 & 0 & 1 & 0 & $\lnot b_1 \wedge b_0$ & 0.010 & \pbar[2]{0.010} \\
3  & 0 & 0 & 1 & 1 & $\lnot b_1$ & 0.010 & \pbar[2]{0.010} \\
4  & 0 & 1 & 0 & 0 & $b_1 \wedge \lnot b_0$ & 0.131 & \pbar[2]{0.131} \\
5  & 0 & 1 & 0 & 1 & $\lnot b_0$ & 0.010 & \pbar[2]{0.010} \\
6  & 0 & 1 & 1 & 0 & $b_1 \oplus b_0$ (XOR) & 0.010 & \pbar[2]{0.010} \\
7  & 0 & 1 & 1 & 1 & $\lnot(b_1 \wedge b_0)$ (NAND) & 0.010 & \pbar[2]{0.010} \\
8  & 1 & 0 & 0 & 0 & $b_1 \wedge b_0$ (AND) & 0.213 & \pbar[2]{0.213} \\
9  & 1 & 0 & 0 & 1 & $b_1 = b_0$ (XNOR) & 0.092 & \pbar[2]{0.092} \\
10 & 1 & 0 & 1 & 0 & $b_0$ (projection) & 0.040 & \pbar[2]{0.040} \\
11 & 1 & 0 & 1 & 1 & $b_1 \to b_0$ (IF THEN) & 0.010 & \pbar[2]{0.010} \\
12 & 1 & 1 & 0 & 0 & $b_1$ (projection) & 0.010 & \pbar[2]{0.010} \\
13 & 1 & 1 & 0 & 1 & $b_0 \to b_1$ (IF THEN) & 0.030 & \pbar[2]{0.030} \\
14 & 1 & 1 & 1 & 0 & $b_1 \vee b_0$ (OR) & 0.010 & \pbar[2]{0.010} \\
15 & 1 & 1 & 1 & 1 & TRUE (constant $1$) & 0.010 & \pbar[2]{0.010} \\
\bottomrule
\end{tabular}
\captionof{table}{The guiding example: all $16$ maps of the $n = 2$,
$m = 1$ hypothesis space, with a contrived prior
($\sum_j p_j = 1.000$).}
\label{tab:sixteen}
\end{center}
\end{example}

\subsection{The answer matrix and the update rule}

The answer matrix is defined by the same rule as before: the total
belief in the maps that would answer $a$ to $q$:
\begin{equation}\label{eq:M-general}
  M_{a,q} \;=\; \sum_{j\,:\,\phi_j(q) = a} p_j,
\end{equation}
now an $A \times Q$ matrix, still with each column a probability
distribution over the possible answers: column-stochastic.

It is worth taking a moment to describe the constitution of this
matrix. Because of the enumeration of $j$ (the index \emph{is} the
truth table) the answer matrix has a natural structure: it is the
contraction of the belief vector $p$ with a fixed tensor with $\{0,1\}$ entries $E$,
summed along the map index,
\begin{equation}\label{eq:E}
  M_{a,q} \;=\; \sum_{j=0}^{N-1} E_{a,q,j}\; p_j,
  \qquad
  E_{a,q,j} \;:=\; \delta\big(\phi_j(q),\, a\big)
  \;=\; \delta\Big(\big\lfloor j \cdot 2^{-qm} \big\rfloor \bmod A,\;
  a\Big).
\end{equation}
The second form is pure arithmetic on the index: by the
numeral convention, $\phi_j(q)$ is nothing but the $q$-th digit of
$j$ in base $A$. For $m = 1$ this reduces to: bit $q$ of $j$ equals $a$. 
So we never need to store $E$: we read membership of
$j$ in the entry $M_{a,q}$ directly off the digits of $j$. Just as well: $E$ is astronomically large for even moderate $m$ and $n$.
 The tensor
$E$ is moreover sparse and perfectly regular: each question-slice ($q$ fixed)
holds exactly $N$ nonzero entries: every map answers exactly one
$a$, and the maps divide evenly over the $A$ answer rows, $N/A$ apiece.
Within a slice the nonzero row follows the odometer pattern of a
numeral system: runs of $A^{q}$ consecutive $j$ share a digit,
cycling through all $A$ rows with period $A^{q+1}$. 

The odometer pattern gives every entry of $M$ the same compact
closed form: for $m = 1$, bit $q$ of $j$ equals $0$ exactly when
$j \bmod 2^{q+1} < 2^q$, and equals $1$ otherwise.

\begin{example}
Each entry of $M$ collects the eight maps from two to one bit of Table~\ref{tab:sixteen}
whose truth tables carry the right digit. Filling in the entries - note the period halves column by column. The outer columns
are written in their simplified forms, $j \bmod 16 < 8$ being simply
$j < 8$ and $j \bmod 2 < 1$ simply ``$j$ even''.
\begin{equation}\label{eq:M16}
\begin{aligned}
  M \;&=\;
  \MtabIV{\scriptstyle\sum\limits_{j < 8} p_j}
         {\scriptstyle\sum\limits_{j \bmod 8 < 4} p_j}
         {\scriptstyle\sum\limits_{j \bmod 4 < 2} p_j}
         {\scriptstyle\sum\limits_{j\ \mathrm{even}} p_j}
         {\scriptstyle\sum\limits_{j \geq 8} p_j}
         {\scriptstyle\sum\limits_{j \bmod 8 \geq 4} p_j}
         {\scriptstyle\sum\limits_{j \bmod 4 \geq 2} p_j}
         {\scriptstyle\sum\limits_{j\ \mathrm{odd}} p_j}
  \\[10pt]
  &=\;
  \MtabIV{0.585}{0.779}{0.890}{0.818}
         {0.415}{0.221}{0.110}{0.182}\;.
\end{aligned}
\end{equation}
\end{example}

The update rule also carries over verbatim. When question $q$ is
asked and the answer $a = \psi(q)$ comes back, the process of
elimination reads, in one formula,
\begin{equation}\label{eq:update}
  p'_j \;=\; \frac{\delta\big(\phi_j(q),\, a\big)\; p_j}{M_{a,q}}
  \;=\; \frac{E_{a,q,j}\; p_j}{M_{a,q}}.
\end{equation}
The Kronecker delta performs the elimination. Only maps that agree with $\psi(q)=a$ survive and the division renormalizes the survivors,
whose total prior mass is exactly $M_{a,q}$ by
equation~\eqref{eq:M-general}. Note that one and the same number
prices the answer and funds the update: the event arrives with the
probability $M_{a,q}$, costs Werner $-\log_2 M_{a,q}$ bits of
surprisal, and its belief mass becomes the normalization. Every update shrinks the support by $A^{-1}$, radically thinning, especially for large $m$. The survivors live on a block-wave in $j$-space. The map determines the offset and frequency.
The total set of survivors depends on the set of asked questions, those that fell through the slits, not the order: updating commutes. This is illustrated in the example below.

\begin{example}
The figure below makes the process visual for $n = m = 2$: a
schematic prior $\{p_j\}$ over $N = 4^4 = 256$ maps, drawn as vertical
lines with thickness representing weight. Each answer stops $3$ out of every $A=4$ surviving lines:
maps whose truth tables carry a wrong digit at the asked question. Renormalization inflates the weight of the rest, so the total probability is always unit. Four questions cut
$256 \to 64 \to 16 \to 4 \to 1$, and the last line standing is the
truth. Note the regularity of the cull: each answer keeps a single
residue class of $j$, so the cadence of the survivors is regular.

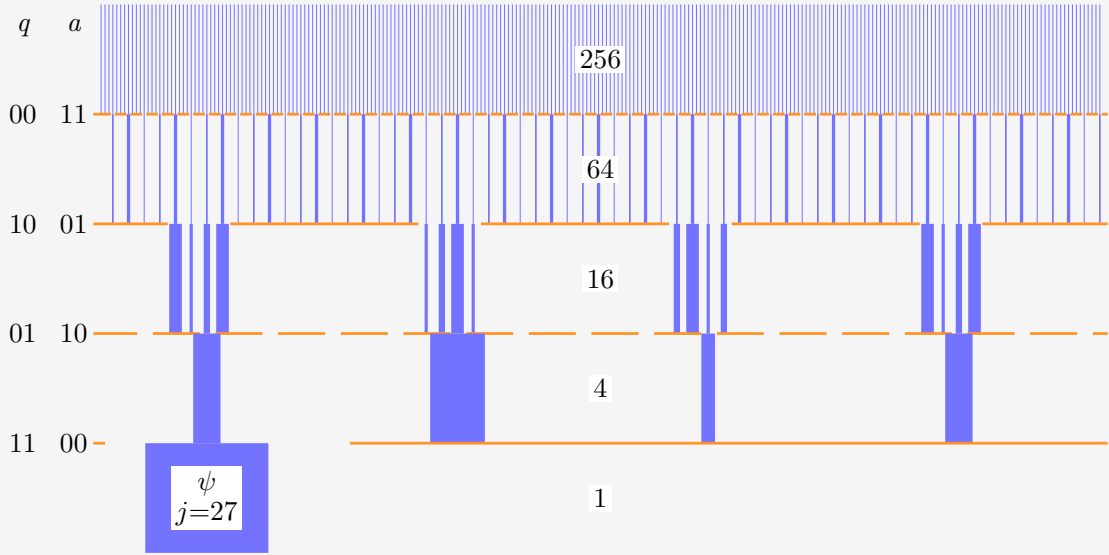
\begin{center}
\resizebox{\linewidth}{!}{%
\begin{tikzpicture}[x=0.05cm, y=1cm]
  \node[anchor=north, font=\small\itshape] at (-20,-0.05) {q};
  \node[anchor=north, font=\small\itshape] at (-7,-0.05) {a};
  \foreach \j in {0,...,255}
    \draw[blue!55, line width=0.1pt] (\j,0) -- (\j,-1.4);
  \draw[orange!85, line width=1pt] (-2,-1.4) -- (2.5,-1.4);
  \foreach \k in {0,...,62}
    \draw[orange!85, line width=1pt] (4*\k+3.5,-1.4) -- (4*\k+6.5,-1.4);
  \draw[orange!85, line width=1pt] (255.5,-1.4) -- (257,-1.4);
  \node[font=\small] at (-20,-1.4) {$00$};
  \node[font=\small] at (-7,-1.4) {$11$};
  \foreach \j in {3,15,...,255}
    \draw[blue!55, line width=0.6pt] (\j,-1.4) -- (\j,-2.8);
  \foreach \j in {7,19,...,247}
    \draw[blue!55, line width=1.2pt] (\j,-1.4) -- (\j,-2.8);
  \foreach \j in {11,23,...,251}
    \draw[blue!55, line width=0.3pt] (\j,-1.4) -- (\j,-2.8);
  \draw[orange!85, line width=1pt] (-2,-2.8) -- (17,-2.8);
  \foreach \s in {0,...,2}
    \draw[orange!85, line width=1pt] (64*\s+33,-2.8) -- (64*\s+81,-2.8);
  \draw[orange!85, line width=1pt] (225,-2.8) -- (257,-2.8);
  \node[font=\small] at (-20,-2.8) {$10$};
  \node[font=\small] at (-7,-2.8) {$01$};
  \foreach \j in {23,83,95,155,215}
    \draw[blue!55, line width=1.15pt] (\j,-2.8) -- (\j,-4.2);
  \foreach \j in {27,87,147,159,219}
    \draw[blue!55, line width=2.29pt] (\j,-2.8) -- (\j,-4.2);
  \foreach \j in {19,31,91,151,211,223}
    \draw[blue!55, line width=4.58pt] (\j,-2.8) -- (\j,-4.2);
  \draw[orange!85, line width=1pt] (-2,-4.2) -- (9.2,-4.2);
  \foreach \t in {0,...,14}
    \draw[orange!85, line width=1pt] (16*\t+13.2,-4.2) -- (16*\t+25.2,-4.2);
  \draw[orange!85, line width=1pt] (253.2,-4.2) -- (257,-4.2);
  \node[font=\small] at (-20,-4.2) {$01$};
  \node[font=\small] at (-7,-4.2) {$10$};
  \draw[blue!55, line width=9.93pt]  (27,-4.2)  -- (27,-5.6);
  \draw[blue!55, line width=19.87pt] (91,-4.2)  -- (91,-5.6);
  \draw[blue!55, line width=4.97pt]  (155,-4.2) -- (155,-5.6);
  \draw[blue!55, line width=9.93pt]  (219,-4.2) -- (219,-5.6);
  \draw[orange!85, line width=1pt] (-2,-5.6) -- (1,-5.6);
  \draw[orange!85, line width=1pt] (63.5,-5.6) -- (257,-5.6);
  \node[font=\small] at (-20,-5.6) {$11$};
  \node[font=\small] at (-7,-5.6) {$00$};
  \draw[blue!55, line width=44.7pt] (27,-5.6) -- (27,-7.0);
  \node[fill=white, inner sep=1.5pt, font=\small, align=center]
    at (27,-6.3) {$\psi$\\[-1pt] $j{=}27$};
  \node[fill=white, inner sep=1.5pt, font=\small] at (127.5,-0.7) {$256$};
  \node[fill=white, inner sep=1.5pt, font=\small] at (127.5,-2.1) {$64$};
  \node[fill=white, inner sep=1.5pt, font=\small] at (127.5,-3.5) {$16$};
  \node[fill=white, inner sep=1.5pt, font=\small] at (127.5,-4.9) {$4$};
  \node[fill=white, inner sep=1.5pt, font=\small] at (127.5,-6.3) {$1$};
\end{tikzpicture}}
\end{center}
The update rule as a cull, for $n = m = 2$. All $256$
candidate maps start as vertical lines under a schematic prior. The truth is
the bitwise negation $\psi(q) = \lnot q$ ($j = 27$, ``0123'' base 4), and the four
questions are asked in the order $00, 10, 01, 11$ (``0, 2, 1, 3'' base 4). Each answer terminates the
$3/4$ of surviving maps whose truth-table digit at the asked question
disagrees. Every survivor's thickness, proportional to its
weight, is rescaled by the renormalization factor of that update, preserving
their ratios. After four questions a single map remains. Interestingly, the first cull's survivors are evenly spaced because it's the ``smallest'' question. After the three smallest $q$, even spacing returns.
\end{example}

What an answer does to a learner admits more than one measure. Baldi and
Itti score such a datapoint by the relative entropy between the prior and the
posterior it induces, rather than by how unexpected the datum
was~\cite{baldiitti10}. Under the paradigm of this work, the elimination
of exact maps, the two coincide. Likelihood is either $0$ or $1$. The posterior is the prior restricted to
the survivors, each of them rescaled by the same $1/M_{a,q}$, so the
relative entropy of the posterior from the prior is $-\log_2 M_{a,q}$.
Surprisal is the measure of information received.

A more powerful treatment of the update rule, in terms of precomputed \emph{correlators} of questions, 
is explained in Appendix~\ref{app:correlators}.

\subsection{Entropy bookkeeping}\label{sec:bookkeeping}

The extensive predictive entropy of the table adds the total logarithmic
prediction risks of its columns, exactly as in equation~\eqref{eq:HM}:
\begin{equation}\label{eq:HM-general}
  H(M) \;=\; \sum_{q \in \mathcal Q} H\big(M_{\cdot,q}\big)
  \;=\; -\sum_{q \in \mathcal Q} \sum_{a \in \mathcal A} M_{a,q} \log_2 M_{a,q}
  \;\;\leq\;\; mQ,
\end{equation}
each column carrying at most $m$ bits (a uniform spread over its
$A$ answers). The ceiling $mQ$ is the size of the entire truth
table in bits, and it is attained precisely (but not exclusively) by the uniform prior:
maximal ignorance prices every answer at full cost.

The belief itself carries an entropy as well, and we give its gap to
the table a name:
\begin{equation}
  H(p) \;=\; -\sum_{j=0}^{N-1} p_j \log_2 p_j \leq mQ,
\end{equation}
\begin{equation}\label{eq:C}
  C \;:=\; H(M) - H(p) \;\geq\; 0.
\end{equation}
$H(p)$, Werner's uncertainty about the \emph{identity} of the truth,
obeys the same ceiling: a distribution over $N$ maps has entropy at
most $\log_2 N = mQ$, attained again by the uniform prior, and this time
by it alone. The information in the
prior is not directly useful for prediction, but it is good to track. 

Why is $C>0$? The answers jointly determine the map, so $H(p)$ is the
\emph{joint} entropy of all $Q$ answers, while $H(M)$ sums their
\emph{marginal} entropies and a joint entropy never exceeds the
sum of its marginals. Hence $H(p) \leq H(M)$, and the gap $C$, the
\textbf{total correlation} of the answers~\cite{watanabe60}, is nonnegative: knowledge
stored in correlations \emph{between} answers, invisible to every
single column of $M$.

Two independent things have now been named, and it is worth keeping
them apart. The first is how \textbf{unbiased} the table is: how close
each column of $M$ sits to an even spread over the answers, with
$H(M) = mQ$ when every column is flat. The second is how
\textbf{correlated} the answers are: the gap $C$, which vanishes exactly
when the prior is a product over the questions, so that no answer can
teach another. Neither pins the other down. The one-parameter world of
Subsection~\ref{sec:oneparam} keeps every column a fair coin at its
ceiling for every $p$, while its $C = 1 - h_2(p)$ runs the whole way
from $0$ to $1$ bit. That is why the ceiling on $H(M)$ is reached by
priors other than the uniform one, while the ceiling on $H(p)$ is not:
the uniform prior is the single point at which both are maximal.
Appendix~\ref{app:correlators} opens $C$ up: in the Walsh-Hadamard
basis it stops being a single number and becomes one coordinate per
set of questions, with the update rule acting as a two-term
convolution that moves weight down the ledger.

The same scoring interpretation holds through observation updates. Let
$\mathcal S$ be the questions already asked and $y=\psi(\mathcal S)$
the string of answers seen. The current table has
$M_{a',q'}=\Pr(\psi(q')=a'\mid\psi(\mathcal S)=y)$. Its expected
total logarithmic loss is
\begin{equation}\label{eq:HM-logloss}
  \mathbb E\!\left[
    \sum_{q'\in\mathcal Q}-\log_2 M_{\psi(q'),q'}
    \,\middle|\,\psi(\mathcal S)=y\right]
  =\sum_{q'\in\mathcal Q}H\big(\psi(q')\mid\psi(\mathcal S)=y\big)
  =H(M).
\end{equation}
Already answered columns contribute zero. Thus $H(M)$ measures the
remaining risk of separate predictions, while $H(p)$ measures the
remaining uncertainty about the map as a whole. 
Notably, \emph{in the independent case}, of which the ignorant case is an example,
\begin{enumerate}
  \item $H(M)$ is the total amount of information we must collect from questions to settle the map.
  \item $H(M) = H(p)$, so the entropy equals the risk, without correlations.
\end{enumerate}
This equality pins the scaling. How these quantities respond to an \emph{intelligent} prior is the main question of this work.

Their agreement in
the uniform-prior case is a consequence of independence; correlations
are what allow one answer to improve several predictions.

\subsubsection{What a question is worth}\label{sec:forecasts}

What are these ledgers worth to Werner before he asks anything? He can
compute two forecasts for every unasked question, in advance, from his
current belief: the expected
surprisal $\langle s \rangle$, which by
equation~\eqref{eq:surprisal-entropy} is the entropy of that
question's column, and the expected drop
$\langle \Delta H(M) \rangle$ of the \emph{whole} table if that
question is asked next. That second forecast is the numerator of the
leverage. It is an expected information gain in Lindley's sense
\cite{lindley56}, the kind of objective that information-based active
learning has used to choose its data \cite{mackay92}. What that
literature calls the \emph{expected predictive information gain}, or EPIG
\cite{bickfordsmith23}, is exactly this form: what an answer is expected
to reveal about the answers to other questions, rather than about which
map is true. Both forecasts are averages over the
possible $a$, weighted by their probability (according to $M$):
\begin{equation}
  \langle s \rangle(q)
  \;=\; \sum_{a} M_{a,q}\,\big({-}\log_2 M_{a,q}\big) = H(M_{\cdot,q}),
\end{equation}
\begin{equation}
  \langle \Delta H(M) \rangle(q)
  \;=\; \sum_{a} M_{a,q}\,\Big[H(M) - H\big(M^{(q,a)}\big)\Big],
\end{equation}
where $M^{(q,a)}$ is the table after the update
rule~\eqref{eq:update} processes answer $a$ to question $q$. We know the entropy of column $q$ will drop to zero, ditching $\langle s \rangle(q)$ entropy from $M$. But other columns will also respond.
The excess of the second over the first is
the expected transfer to unasked columns, and it can only be paid out
of correlations.

This transfer has a direct expression in mutual information. Work
at the current posterior, after the already-seen string
$y=\psi(\mathcal S)$, and consider asking $q\notin\mathcal S$ next,
with answer $a=\psi(q)$. For any other question $q'$, averaging its
entropy drop over the possible $a$ gives
\begin{equation}\label{eq:column-mutual-gain}
\begin{aligned}
 &H\big(\psi(q')\mid\psi(\mathcal S)=y\big)
 -\sum_a M_{a,q}\,
   H\big(\psi(q')\mid\psi(\mathcal S)=y,\psi(q)=a\big)\\
 &\hspace{2cm}=I\big(\psi(q');\psi(q)\mid\psi(\mathcal S)=y\big).
\end{aligned}
\end{equation}
The asked column loses its whole entropy, so summing gives
\begin{equation}\label{eq:mutual-gain}
  \langle\Delta H(M)\rangle(q)
  =H(M_{\cdot,q})+
   \sum_{q'\ne q}I\big(\psi(q');\psi(q)\mid\psi(\mathcal S)=y\big).
\end{equation}
The first term is the expected surprisal received; each term in the
sum is the expected improvement on another question. Their sum is
the expected withdrawal from the correlation store. This form also
makes the EPIG correspondence exact: since
$I\big(\psi(q);\psi(q) \mid \psi(\mathcal S) = y\big) = H(M_{\cdot,q})$,
the right-hand side of \eqref{eq:mutual-gain} is
$\sum_{q'} I\big(\psi(q');\psi(q) \mid \psi(\mathcal S) = y\big)$ over all
$Q$ questions, which is $Q$ times the EPIG of $q$ under a target
distribution uniform over the table. The surprisal
forecast needs $M$ alone, but the improvement forecast also needs
the pairwise joint laws of the asked answer and the other answers.
Two priors with the same $M$ can therefore give different forecasts
of improvement.

A forecast is not a promise. Playing the guiding example through all
four questions, worked out in Appendix~\ref{app:playing}, shows the
gap. Werner opens on $q = 11$, the question carrying the largest
forecast: he expects it to clear $1.104$ bits from the table, for
$0.979$ bits of surprisal. The answer that arrives is the less likely
one, $a = 1$ at probability $0.415$. It costs him $1.269$ bits and
clears only $0.748$ of table, worse on both counts than the forecast,
and short of breaking even. Later questions more than repay it.
Once every question has been asked every column is settled and $H(M)$
has reached zero, as it always must; so the run ends having cleared
the whole $2.925$ bits Werner started with, at a cost of $2.231$ bits
of surprisal.

That last pair of numbers is the thing to keep: what a run clears,
set against what it cost. It deserves a name, and a quantity that
tracks the ratio round by round.

\subsection{Leverage}

If the prior is good, meaning it actually models reality (the prevalence of maps),
we expect graphs such as the one in Example~\ref{ex:run21} of
Appendix~\ref{app:playing} to drop as questions come in.
The ratio of table entropy destroyed per bit of
surprisal received deserves a name.

The \textbf{leverage}. Run the protocol for $\ell$ rounds:
questions $q_1, \ldots, q_\ell$ are asked, the answers
$a_k = \psi(q_k)$ are each processed by the update
rule~\eqref{eq:update}, producing the tables
$M^{(1)}, \ldots, M^{(\ell)}$, and round $k$
costs the surprisal $s_k = -\log_2 M^{(k-1)}_{a_k,\,q_k}$. The
cumulative leverage after $\ell$ rounds is the table entropy destroyed
per bit of surprisal received:
\begin{equation}\label{eq:leverage}
  L_\ell \;:=\;
  \frac{H\big(M^{(0)}\big) - H\big(M^{(\ell)}\big)}
       {\sum_{k=1}^{\ell} s_k}\;.
\end{equation}
A leverage of $1$ is the uniform-prior baseline: every bit of table
entropy destroyed was paid for at face value, one bit of surprisal
each. On an individual run, leverage above $1$ can also come from
windfall, as the guiding example showed. The aggregate quantity
$\mathcal L_\ell$ defined in Chapter~\ref{sec:expectations} removes
that windfall in expectation: anything above $1$ in this
$\mathcal L_\ell$ is deduction.

\section{Expectations}\label{sec:expectations}

While the games of the previous chapters may be entertaining, they don't readily divulge any universal laws. When we average (integrate) out the question order, and the specific map drawn,
many more robust quantities appear. This is a stepping stone to our terminus: the thermodynamic limit, where we expect questions to flow in and learning to happen on a continuum, where such ensemble-average statements are
physically instructive and often quite descriptive of nature.

\subsection{Two exact laws}\label{sec:twolaws}

Until now $\langle\,\cdot\,\rangle$ has averaged over the truth alone,
for a question we had named. From here we average over which questions
were asked as well, and we double the brackets to say so.
So $\langle\!\langle\,\cdot\,\rangle\!\rangle$ denotes expectation over both

\begin{itemize}
  \item the truth $\psi \sim \{p_j\}$,
  \item uniformly over the $\binom Q\ell$ sets $\mathcal S$ of $\ell$ questions
    that could have been asked.
\end{itemize}
The size $\ell$ is left implicit: it is always the number of questions asked at
the point where the brackets stand. As this is usually clear from the object being averaged,
there is no confusion, and we can think of the $\langle\!\langle\,\cdot\,\rangle\!\rangle$ as a uniform average over the question order,
at some point where $\ell$ total questions have been answered.

Single brackets keep their earlier
meaning wherever they still appear, as in the correlators $\chi_{\mathcal S}$.
We take the ratio of
the expected entropy cleared from the table $M$ to the expected surprisal received:
\begin{equation}\label{eq:Lk-avg}
  \mathcal L_\ell \;:=\;
  \frac{H\big(M^{(0)}\big) - \big\langle\!\big\langle  H\big(M^{(\ell)}\big) \big\rangle\!\big\rangle}
       {\big\langle\!\big\langle \sum_{k=1}^{\ell} s_k \big\rangle\!\big\rangle}\;.
\end{equation}
The calligraphic $\mathcal L_\ell$ denotes this ratio of expectations,
and not the expectation of the individual-run ratio $L_\ell$. The two
are different quantities: averaging the ratio lets a lucky run with a
vanishing denominator dominate, while averaging numerator and
denominator separately does not. We already used this aggregate quantity in
Subsection~\ref{sec:oneparam}. There, the problem was actually symmetric in $q_0 \leftrightarrow q_1$. That is not the case in general.

As stated above, in order to progress, we will be interested in such averages. Both
are natural. The question order is uniform, without replacement:
Without control of the questions, Werner has no reason to expect one to arrive
before another (Appendix~\ref{app:strategies} expounds strategies when
he does control which to ask). The truth is drawn from the prior
itself: Werner scores his future against his own beliefs, which is the best he can do.

The bookkeeping, especially for questions asked, simplifies at once. Updating commutes, so after
$\ell$ rounds the state depends only on the asked \emph{set} $\mathcal S$,
a uniform random subset of size $\ell$. Let us say the answers
received form a vector 
\begin{equation}
  y = \{y_k = \psi(q_k)\},\quad \forall q_k \in \mathcal S. 
\end{equation}
Each surprisal is
paid at the posterior odds of its own round,
\begin{equation}
  s_k = -\log_2 M^{(k-1)}_{y_k, q_k},
\end{equation}
and the posterior column is,
by construction of the update rule, exactly the conditional
probability of the next answer given everything seen so far:
\begin{equation}
M^{(k-1)}_{y_k, q_k} = \Pr(\psi(q_k) = y_k \mid \psi(q_1) = y_1,
\ldots, \psi(q_{k-1}) = y_{k-1}). 
\end{equation}
Take $P_{\mathcal S, y}$ to be the mass of
the maps that answer the pattern $y$ on a set $\mathcal S$. 
The sums of the
surprisals display the chain rule of probability
\begin{equation}
\sum_{k=1}^{\ell} s_k = -\log_2 \left[\prod_{k=1}^{\ell}  P_{\mathcal S_1, y}\,
        \frac{P_{\mathcal S_2, y}}{P_{\mathcal S_1, y}}\,
        \frac{P_{\mathcal S_3, y}}{P_{\mathcal S_2, y}}
        \,\cdots\,
        \frac{P_{\mathcal S_\ell, y}}{P_{\mathcal S_{\ell-1}, y}}\right],
\end{equation}
where we use the shorthand that $\mathcal S_k$ contains the first $k$ questions, and is answered by the first $k$ elements of pattern $y$.
Each conditioning cancels against the next term's outcome, and $\mathcal S_\ell = \mathcal S$. Thus we arrive at the expression with no path dependence:
\begin{equation}\label{eq:telescope}
  \sum_{k=1}^{\ell} s_k \;=\; -\log_2 P_{\mathcal S, y},
  \qquad
  P_{\mathcal S, y} \;=\; \Pr\big(\psi(\mathcal S) = y\big)
  \;=\; \sum_{j \,:\, \phi_j(q_k) = y_k \;\forall q_k \in
  \mathcal S} p_j .
\end{equation}
Averaging
\eqref{eq:telescope} over $\psi$ gives the entropy of the block
distribution $P_{\mathcal S,\cdot}$: the distribution over such sets. Everything is generated by
one sequence, the \textbf{mean block entropy} at size $k$:
\begin{equation}\label{eq:Gk}
  G_k \;:=\; \binom{Q}{k}^{-1} \sum_{|\mathcal S| = k}
  H(\psi(\mathcal S))
  \;=\; -\binom{Q}{k}^{-1} \sum_{|\mathcal S| = k}\;
  \sum_{y} P_{\mathcal S,y} \log_2 P_{\mathcal S,y},
\end{equation}
the inner sum running over the $A^k$ answer patterns on the block;
nothing here is specific to $m = 1$. Appendix~\ref{app:tensor-marginals}
arranges the elimination rules as a tensor and reads every $G_k$ off one
sweep of it, rather than
rebuilding each block distribution from scratch. The endpoints are
\begin{equation}\label{eq:Gk-ends}
  G_0 \;=\; 0,
  \qquad
  G_{Q} \;=\; H(p).
\end{equation}
$G_1$ is by definition the mean column entropy of the first question, so the prior table entropy is
\begin{equation}\label{eq:HM0}
  H\big(M^{(0)}\big) \;=\; Q\, G_1 .
\end{equation}

\begin{example}
What a block entropy is, on the guiding space ($n = 2$, $m = 1$).
Take the block $\mathcal S = \{11, 10\}$, two questions. An
answer pattern is $y = (y_{11}, y_{10})$, with four options, and by
the numeral convention the maps answering each pattern are four
consecutive rows of Table~\ref{tab:sixteen}: the residue classes
are read off the two leading truth-table digits.

\begin{center}
\begin{tabular}{c c l c}
\toprule
$y$ & maps $j$ & names & $P_{\mathcal S,y}$ \\
\midrule
$(0,0)$ & $0$--$3$ & FALSE, NOR, $\lnot b_1 \wedge b_0$,
  $\lnot b_1$ & $0.424$ \\
$(0,1)$ & $4$--$7$ & $b_1 \wedge \lnot b_0$, $\lnot b_0$, XOR,
  NAND & $0.161$ \\
$(1,0)$ & $8$--$11$ & AND, XNOR, $b_0$, $b_1 \to b_0$ & $0.355$ \\
$(1,1)$ & $12$--$15$ & $b_1$, $b_0 \to b_1$, OR, TRUE & $0.060$ \\
\bottomrule
\end{tabular}
\end{center}

The four masses sum to one, and the block entropy is the Shannon
entropy of this four-outcome distribution,
$H(\psi(11,10)) = -\sum_y P_{\mathcal S,y}\log_2
P_{\mathcal S,y} = 1.723$ bits: Werner's joint uncertainty about
the two answers before either is asked. It is one of the six
size-two terms averaged into $G_2$ in the next example.
\end{example}

Two exact laws now follow. The first is
equation~\eqref{eq:telescope} averaged:
\begin{equation}\label{eq:law-received}
  \Big\langle\!\Big\langle \sum_{k=1}^{\ell} s_k \Big\rangle\!\Big\rangle \;=\; G_\ell.
\end{equation}
The second law describes the total column entropy of $M^{(\ell)}$, which is the answer matrix after $\ell = |\mathcal S|$ questions have been settled. 
Fix $\mathcal S$ and average the posterior table over the
answers. Each unasked $q'$ column's entropy becomes a conditional
entropy, 
\begin{equation}
H(\psi(q') \mid \psi(\mathcal S)) = H(\psi(\mathcal S \cup \{q'\})) - H(\psi(\mathcal S)),
\end{equation}
and the already answered columns contribute zero. Sum the $Q-\ell$
unasked columns, then average over which $\ell$ questions were asked:
\begin{equation}
\begin{aligned}
  \big\langle\!\big\langle H\big(M^{(\ell)}\big) \big\rangle\!\big\rangle
  &\;=\; \binom{Q}{\ell}^{-1} \sum_{|\mathcal S| = \ell}\;
     \sum_{q' \notin \mathcal S}
     \Big[ H\big(\psi(\mathcal S \cup \{q'\})\big)
           - H\big(\psi(\mathcal S)\big) \Big] \\[2pt]
  &\;=\; 
     (\ell{+}1)\binom{Q}{\ell}^{-1}\binom{Q}{\ell+1}\, G_{\ell+1}
           \;-\; (Q{-}\ell)\, G_\ell .
\end{aligned}
\label{eq:law-remaining-count}
\end{equation}
The first sum collects that way because every set of size $\ell + 1$
arises as $\mathcal S \cup \{q'\}$ in $\ell+1$ ways, one for each $q'$.
The second because $H(\psi(\mathcal S))$ is simply repeated for each of
the $Q-\ell$ unasked $q'$, and we invoke the definition of $G_k$. Substituting
$(\ell{+}1)\binom{Q}{\ell+1} = (Q{-}\ell)\binom{Q}{\ell}$ cancels the
prefactor and leaves
\begin{equation}\label{eq:law-remaining}
  \big\langle\!\big\langle H\big(M^{(\ell)}\big) \big\rangle\!\big\rangle
  \;=\; (Q - \ell)\,\big(G_{\ell+1} - G_\ell\big).
\end{equation}
The whole expected learning curve is the discrete derivative of
one monotone sequence.

Putting together \eqref{eq:HM0}, \eqref{eq:law-received}, and \eqref{eq:law-remaining}, equation~\eqref{eq:Lk-avg} becomes
\begin{equation}\label{eq:Lk-G}
  \mathcal L_\ell \;=\;
  \frac{Q\,G_1 - (Q - \ell)\big(G_{\ell+1} - G_\ell\big)}
       {G_\ell}\;,
  \qquad 1 \le \ell \le Q,
\end{equation}
the vanishing factor $Q - \ell$ retiring the undefined increment at
$\ell = Q$; at $\ell = 0$ nothing has been asked or paid and the
ratio is an empty $0/0$.
Some observations.
\begin{itemize}
  \item Expectation kills the windfall. Each step has
    $\langle\!\langle \Delta H(p) - s \rangle\!\rangle = 0$, so the expected deduced
    part is a pure tower withdrawal:
    $H(M^{(0)}) - \langle\!\langle H(M^{(\ell)}) \rangle\!\rangle - G_\ell
    = C_0 - \langle\!\langle C_\ell \rangle\!\rangle$, with
    $\langle\!\langle C_\ell \rangle\!\rangle =
    \langle\!\langle H(M^{(\ell)}) \rangle\!\rangle - \big(H(p) - G_\ell\big)$. It
    is nonnegative: anti-leverage is only ever bad luck.
  \item Deduction saturates one round early: substituting
    $G_{Q} = H(p)$ at $\ell = Q - 1$ already gives
    $\langle\!\langle C_{Q - 1} \rangle\!\rangle = 0$. The store is empty before
    the last question, which can only be observed, never
    leveraged.
  \item $\mathcal L_\ell \geq 1$ always, but it need not be
    monotone. The endpoint is forced:
    $\mathcal L_{Q} = H(M^{(0)}) / H(p) = 1 + C_0/H(p)$, the
    aggregate completion ratio.
\end{itemize}

There is a broader baseline than uniformity: any product prior over
the answers gives $\mathcal L_\ell=1$, whenever the denominator is
positive. Its columns may be biased, but observing one cannot improve
another, so expected predictive gain equals expected surprisal.

More generally, the finite aggregate leverage satisfies the sharp bounds
\begin{equation}\label{eq:leverage-bounds}
  1\leq\mathcal L_\ell\leq Q,\qquad G_\ell>0.
\end{equation}

\begin{example}
This example illustrates equation~\eqref{eq:law-remaining}: the
total column entropy of the table $M$ after $\ell$ questions, read
off a lattice. Take the guiding space, $\mathcal Q = \{11, 10, 01,
00\}$ with $Q = 4$ and $m = 1$, and abbreviate $\psi(11, 10)$ for
$\psi(\{11, 10\})$. Definition~\eqref{eq:Gk} averages
$H(\psi(\mathcal S))$ over the sets of each size, so the five block
entropies collect $1, 4, 6, 4, 1$ terms apiece:
\begin{align*}
  G_0 &= 0, \\
  G_1 &= \tfrac{1}{4}\big[\,
    H(\psi(11)) + H(\psi(10)) + H(\psi(01)) + H(\psi(00))
    \,\big], \\
  G_2 &= \tfrac{1}{6}\big[\,
    H(\psi(11,10)) + H(\psi(11,01)) + H(\psi(11,00)) \\[-2pt]
  &\qquad\quad {}+ H(\psi(10,01)) + H(\psi(10,00))
    + H(\psi(01,00)) \,\big], \\
  G_3 &= \tfrac{1}{4}\big[\,
    H(\psi(10,01,00)) + H(\psi(11,01,00)) \\[-2pt]
  &\qquad\quad {}+ H(\psi(11,10,00)) + H(\psi(11,10,01)) \,\big], \\
  G_4 &= H(\psi(11,10,01,00)) \;=\; H(p).
\end{align*}
Those sixteen sets are the nodes of a lattice, one rank per value of
$\ell$, and each $G_\ell$ is the average of the entropies on rank
$\ell$:
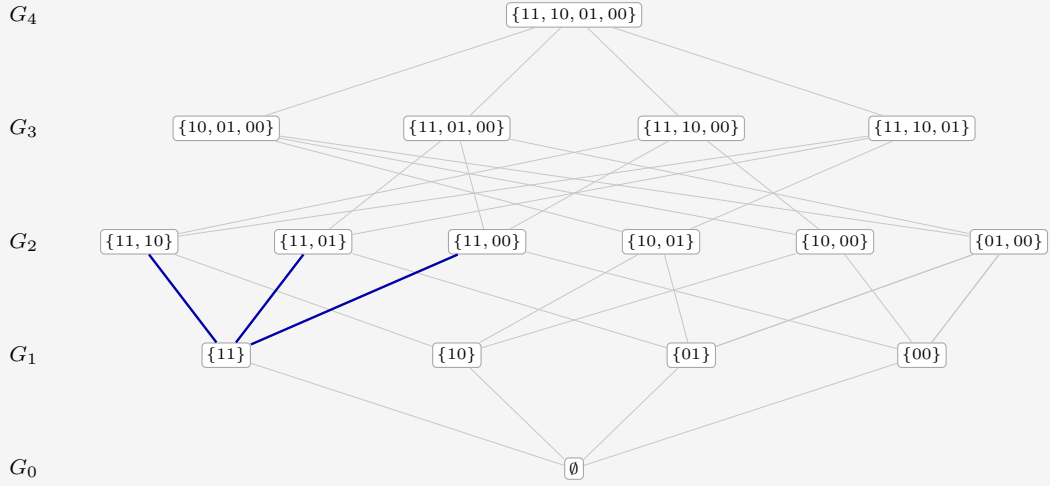
\begin{center}
\begin{tikzpicture}[
  nd/.style={draw=black!35, rounded corners=1.5pt, inner sep=1.6pt,
             fill=white, font=\tiny},
  ed/.style={draw=black!22, line width=0.3pt},
  up/.style={draw=blue!65!black, line width=0.9pt}]
  \node[nd] (r0) at (0,0) {$\emptyset$};
  \node[nd] (r4) at (0,6) {$\{11,10,01,00\}$};
  \node[nd] (a11) at (-4.6,1.5) {$\{11\}$};
  \node[nd] (a10) at (-1.55,1.5) {$\{10\}$};
  \node[nd] (a01) at (1.55,1.5) {$\{01\}$};
  \node[nd] (a00) at (4.6,1.5) {$\{00\}$};
  \node[nd] (b1) at (-5.75,3) {$\{11,10\}$};
  \node[nd] (b2) at (-3.45,3) {$\{11,01\}$};
  \node[nd] (b3) at (-1.15,3) {$\{11,00\}$};
  \node[nd] (b4) at (1.15,3) {$\{10,01\}$};
  \node[nd] (b5) at (3.45,3) {$\{10,00\}$};
  \node[nd] (b6) at (5.75,3) {$\{01,00\}$};
  \node[nd] (c1) at (-4.6,4.5) {$\{10,01,00\}$};
  \node[nd] (c2) at (-1.55,4.5) {$\{11,01,00\}$};
  \node[nd] (c3) at (1.55,4.5) {$\{11,10,00\}$};
  \node[nd] (c4) at (4.6,4.5) {$\{11,10,01\}$};
  \foreach \t in {a11,a10,a01,a00} \draw[ed] (r0) -- (\t);
  \foreach \s/\t in {a10/b1, a11/b2, a11/b3, a10/b4, a10/b5,
                     a01/b2, a01/b4, a01/b6, a00/b3, a00/b5, a00/b6}
    \draw[ed] (\s) -- (\t);
  \foreach \s/\t in {b1/c3, b1/c4, b2/c2, b2/c4, b3/c2, b3/c3,
                     b4/c1, b4/c4, b5/c1, b5/c3, b6/c1, b6/c2}
    \draw[ed] (\s) -- (\t);
  \foreach \s in {c1,c2,c3,c4} \draw[ed] (\s) -- (r4);
  \foreach \t in {b1,b2,b3} \draw[up] (a11) -- (\t);
  \foreach \s in {a01,a00} \draw[ed] (\s) -- (b6);
  \foreach \y/\g in {0/{$G_0$}, 1.5/{$G_1$}, 3/{$G_2$},
                     4.5/{$G_3$}, 6/{$G_4$}}
    \node[font=\scriptsize, anchor=east] at (-6.95,\y) {\g};
\end{tikzpicture}
\end{center}

Every edge up is the addition of one question, and it weighs the
conditional entropy of the answer it adds: the edge from
$\mathcal S$ up to $\mathcal S \cup \{q'\}$ represents
$H(\psi(q') \mid \psi(\mathcal S))$. Fix a single node. Its upward edges run to the
questions still unasked, and their weights are exactly the column
entropies of the posterior table. The
remaining table entropy of an asked set is the total marginal weight of the
up-edges leaving its node. There are $Q - \ell$ of them, drawn in
blue above for $\mathcal S = \{11\}$; summed and averaged over the
rank, they are exactly equation~\eqref{eq:law-remaining}'s $\big\langle\!\big\langle H\big(M^{(\ell)}\big)\big\rangle\!\big\rangle$, a component of
the leverage.
\end{example}

\subsection{The total predictive gain}\label{sec:predgain}

For a fixed asked set $\mathcal S$, averaging over its answer string
$y$ gives the total predictive gain
\[
  H(M^{(0)})-\mathbb E_y H(M^{(\mathcal S,y)})
  =\sum_{q\in\mathcal Q}I\big(\psi(q);\psi(\mathcal S)\big).
\]
Each term is at most $H(\psi(\mathcal S))$, proving the upper bound.
The terms with $q\in\mathcal S$ sum to
$\sum_{q\in\mathcal S}H(\psi(q))\geq H(\psi(\mathcal S))$,
and all other terms are nonnegative, proving the lower bound.
Averaging over the uniformly chosen sets and dividing by $G_\ell$
preserves both inequalities. Independence attains the lower bound;
a single nondegenerate random answer copied to every question attains
the upper bound. For the second, let $\nu$ be a discrete probability
distribution on the answer alphabet $\mathcal A$, with $\nu(a)$ the
weight it puts on the answer $a$, and let the prior draw one
$a \sim \nu$ and return that same $a$ to every question,
\begin{equation}
  p_j \;=\;
  \begin{cases}
    \nu(a), & \phi_j \equiv a \ \text{ for some } a \in \mathcal A,\\
    0, & \text{otherwise},
  \end{cases}
  \label{eq:cliqueQ}
\end{equation}
for any $\nu$ that is not a point mass. The support of $\nu$ fixes the
support of $p$: the maps of positive weight are exactly the constant
maps $\phi \equiv a$ with $\nu(a) > 0$, one apiece, and every
non-constant map has $p_j = 0$. Every nonempty block then carries
$H(\psi(\mathcal S)) = H(\nu)$, so $G_\ell = H(\nu)$ for all
$\ell \ge 1$, the increment vanishes, and $\mathcal L_\ell = Q$ at
every $\ell$. Werner knows the map is constant, so a single
observation collapses all the uncertainty at once, as in
Section~\ref{sec:intelligent}. It is the clique block law of
Appendix~\ref{app:blockpriors} taken at full size, $r = Q$.

Recall \eqref{eq:M-m} from the discussion on correlators,
Appendix~\ref{app:correlators}. The joint law of the answers to any set
$\mathcal S$ of questions is an inverse transform over the correlators
with support contained inside $\mathcal S$'s blocks. The mean block
entropy $G_{|\mathcal S|}$ does not depend on higher correlators
(involving more questions) than are in $\mathcal S$. Since the expected
leverage up to $\ell$ questions is assembled from $G_\ell$ and
$G_{\ell+1}$ alone, it depends only on correlators supported on at most
$\ell+1$ questions. Two priors agreeing in correlators to that order
are indistinguishable through round $\ell$. Deeper shells are
invisible.

There is also a monotonicity property hiding in $G$. Entropy is
\emph{submodular} \cite{fujishige78,covthomas}: a fresh answer is worth at most in a larger
context $\mathcal{S}$ what it was worth in a smaller one $\mathcal{V}$,
\begin{equation}
H(\psi(q) \mid \psi(\mathcal S)) \le H(\psi(q) \mid \psi(\mathcal V));\quad \forall \mathcal V \subseteq \mathcal S,
\end{equation}
because conditioning on more
answers can only explain away part of its novelty, never add to it.
Coupling a random $\ell$-set to a random $(\ell{+}1)$-set and
averaging, the increments $G_\ell - G_{\ell-1}$, which are the
expected surprisals of the $\ell$-th answer, decrease with $\ell$:
answers get cheaper as evidence accumulates. This only holds on uniformly random question order.
The $k$-subset average is also the setting of Han's inequality
\cite{han78}, that $G_k/k$ is nonincreasing; the concavity here implies
it.

\begin{example}
For the guiding prior of Table~\ref{tab:sixteen}, recall the block-entropy sequence is
$G_k = \big\langle\!\big\langle \sum_{i=1}^{k} s_i \big\rangle\!\big\rangle$, and the laws unfold
it into the expected run:
\begin{center}
\begin{tabular}{c c c c c}
\toprule
 & received & remaining & store & leverage \\
$\ell$ & $G_\ell$
  & $\big\langle\!\big\langle H\big(M^{(\ell)}\big) \big\rangle\!\big\rangle$
  & $\langle\!\langle C_\ell \rangle\!\rangle$
  & $\mathcal L_\ell$ \\
\midrule
1 & 0.731 & 2.113 & 0.137 & 1.110 \\
2 & 1.436 & 1.328 & 0.056 & 1.113 \\
3 & 2.100 & 0.608 & 0     & 1.104 \\
4 & 2.707 & 0     & 0     & 1.081 \\
\bottomrule
\end{tabular}
\end{center}
The aggregate leverage humps at $\ell = 2$ and lands on
$2.925/2.707 = 1.081$. The realized run of
Chapter~\ref{sec:general} scattered around this gentle
expectation: greedy questioning and a lucky truth delivered
$1.311$. The figure below stacks the expected
bits.

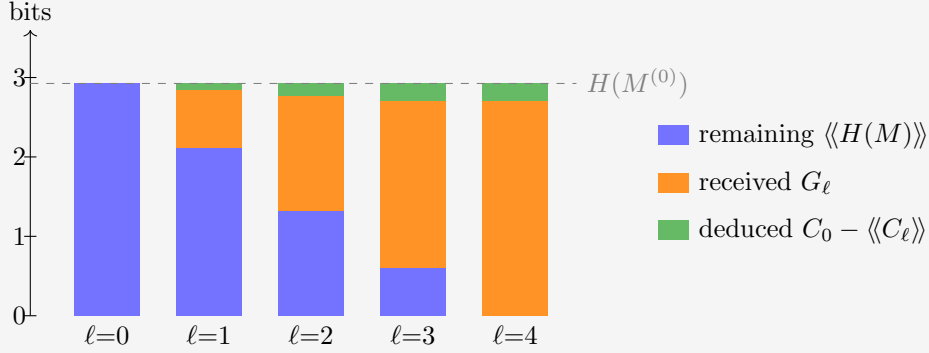
\begin{center}
\begin{tikzpicture}[x=1.35cm, y=1.05cm]
  \draw[->] (-0.75,0) -- (-0.75,3.6);
  \node[above, font=\small] at (-0.75,3.6) {bits};
  \foreach \y in {0,1,2,3}
    \draw (-0.81,\y) -- (-0.69,\y) node[left, font=\small] {$\y$};
  \fill[blue!55] (-0.32,0) rectangle (0.32,2.925);
  \fill[blue!55] (0.68,0) rectangle (1.32,2.113);
  \fill[orange!85] (0.68,2.113) rectangle (1.32,2.844);
  \fill[green!55!black!60] (0.68,2.844) rectangle (1.32,2.925);
  \fill[blue!55] (1.68,0) rectangle (2.32,1.328);
  \fill[orange!85] (1.68,1.328) rectangle (2.32,2.764);
  \fill[green!55!black!60] (1.68,2.764) rectangle (2.32,2.925);
  \fill[blue!55] (2.68,0) rectangle (3.32,0.608);
  \fill[orange!85] (2.68,0.608) rectangle (3.32,2.708);
  \fill[green!55!black!60] (2.68,2.708) rectangle (3.32,2.925);
  \fill[orange!85] (3.68,0) rectangle (4.32,2.707);
  \fill[green!55!black!60] (3.68,2.707) rectangle (4.32,2.925);
  \draw[dashed, gray] (-0.75,2.925) -- (4.6,2.925);
  \node[right, font=\small, gray] at (4.6,2.925) {$H(M^{(0)})$};
  \foreach \l in {0,...,4}
    \node[below, font=\small] at (\l,0) {$\ell{=}\l$};
  \fill[blue!55] (5.4,2.15) rectangle (5.7,2.4);
  \node[right, font=\small] at (5.7,2.27)
    {remaining $\langle\!\langle H(M) \rangle\!\rangle$};
  \fill[orange!85] (5.4,1.55) rectangle (5.7,1.8);
  \node[right, font=\small] at (5.7,1.67) {received $G_\ell$};
  \fill[green!55!black!60] (5.4,0.95) rectangle (5.7,1.2);
  \node[right, font=\small] at (5.7,1.07)
    {deduced $C_0 - \langle\!\langle C_\ell \rangle\!\rangle$};
\end{tikzpicture}
\captionsetup{skip=8pt}
\end{center}
Conservation is now exact at every $\ell$: remaining plus received
plus deduced reaches $H(M^{(0)})$, so the cake always tops out on the
dashed line, and no overshoot is possible. The top layer is the
expected withdrawal from the correlation store,
$C_0 - \langle\!\langle C_\ell \rangle\!\rangle$; it saturates at $C_0 = 0.218$
already at $\ell = 3$.
\end{example}

\section{The thermodynamic limit}\label{sec:tdl}

The earlier chapters describe learning on a finite question set. This
chapter is arguably the real-world payoff. Here we can draw conclusions about 
general learning, not just boolean maps. This chapter asks what remains when the number of input bits tends to
infinity. Sending a learning problem to a thermodynamic limit and reading
off its learning curve is an established program
\cite{seung92,watkin93,haussler94}; this chapter offers it a new
entry, one that keeps time by the fraction of the whole question table
already asked rather than by examples per parameter. The hope is to find traces of universality: that character of learning where the microscopic details are coarse-grained out. 

Recall definition~\ref{eq:sizes} where $Q = |\mathcal Q| = 2^n$, the limit $n \to \infty$ is
one of an exponentially growing table of possible questions. Let the measure of time be the
fraction of that table already queried,
\begin{equation}
  t \;=\; \frac{\ell}{Q} \;\in\; [0,1].
  \label{eq:askedfraction}
\end{equation}
The limit at fixed $t > 0$ is \textbf{macroscopic}: it records learning
that occupies a nonzero fraction of the run. Learning completed in
$O(1)$, in $O(\log Q)$, or more generally in $o(Q)$ questions is
compressed into a boundary layer at $t = 0$. That distinction
will determine the class of solutions that emerges in the TDL.

Throughout, $u := A^{-1} = 2^{-m}$ abbreviates the reciprocal answer
alphabet, and $N = A^{Q}$ counts the maps
$\mathcal Q \to \mathcal A$, enumerated $\phi_0, \ldots, \phi_{N-1}$
with prior weights $p_j = \Pr(\psi = \phi_j)$. One value of the truth
$\psi$ is drawn at the start of a run and supplies every answer in
it; posteriors are conditional distributions of that same random map,
and expected trajectories average over its prior law and the uniformly random $q$ order.

\subsection{A toy example: the spike prior}\label{sec:spike}

The most accessible example from micro- to macroscopics, as one calculus exercise,
is arguably the spike prior. It places a distinguished weight on one map and spreads
the remaining mass uniformly over every other map. The setup is also desirable due to the full
support: at every system size, Werner feels the effect of the dimensionality he inhabits.

Take the earmarked boolean
map to be the all-zero map, $\phi_0(q) = 0^m$ for every
$q \in \mathcal Q$, so that its answer to any $k$ questions is the
all-zero block $0^{mk}$. This is fully general by symmetry: the answer
alphabet may be relabeled separately at each question. We then adopt the convention
\begin{equation}
  p_0 = p,
  \qquad
  p_j = \omega := \frac{1-p}{N-1}
  \quad (j \neq 0).
  \label{eq:spikeprior}
\end{equation}
The plain $p = p_0$ is the mass of the all-zero map alone at start time. We will be interested in the behavior at large (but not necessarily unit) $p$.
The interpretation is that Werner is quite sure of his hypothesis, without ruling anything out.

The example is tractable because it is closed under Bayesian
conditioning. A confirming answer produces a sharper spike on a
smaller map space; a refuting answer culls (among others) the special map and
leaves equal weights on all survivors. The posterior therefore has a
two-state flow, in which the uniform family is the stable, absorbing
class: once a trajectory enters it, later conditioning keeps it
there. An interior spike is a source of one-way probability current
into that class, while conditional on avoiding refutation the spike sharpens.

\subsubsection{The two-state renormalization flow}

What happens when Werner handles the first question? He either gets a zero, or not.

The maps that produce $0^m$ are a proportion $u=A^{-1}$. One has probability $p$, and $Nu-1$ have probability $\omega$.
If we started in this prior with probability $P_0=1$ (full faith), Werner avoids refuting $\phi_0$ with probability
\begin{equation}
  P_1 = p + \left(Nu-1\right)\omega
\end{equation}
With probability $P_1$ we can still believe $\psi = \phi_0$.
In this case, what exactly is the shape of the posterior? The Bayesian consistency condition holds that 
unconditional probability cannot be created without measurement. At this point in the ansatz, we must have the same credence in $\phi_0$. 

At every step, there is a $p_0$, but as a member of an ever-shrinking cohort.
Let $r_1$ be the height of the peak $p_0$, after one question. Then we must have $r_1\cdot P_1 = p$.
The rest of the probability is spread again over the $Nu-1$ other surviving maps. Conversely, with probability $1-P_1$, we've refuted the zero map, and are in a uniform posterior corresponding to a nonzero answer to the question.

After repeating this process $k$ times, asking the set of questions $\mathcal{S}$, we can infer the probability that $\phi_0$ is still in the race. 
Let $P_k$ be the probability that Werner remains in the confirming scenario: he has only seen zeros.
A history remains on the spike branch exactly
when $\psi(\mathcal S) = 0^{mk}$, and exactly $N u^k$ maps
produce that confirming history, one of which is $\phi_0$. The probability of confirmation is then
\begin{equation}
  P_k \;:=\; \Pr\big(\psi(\mathcal S) = 0^{mk}\big)
  \;=\; p + (N u^k - 1)\,\omega .
  \label{eq:spikePk}
\end{equation}
On that event the posterior is another spike on those $N u^k$ maps,
\begin{equation}
  r_k \;=\; \Pr\big(\psi = \phi_0 \mid
  \psi(\mathcal S) = 0^{mk}\big) \;=\; \frac{p}{P_k},
  \qquad
  \omega_k \;=\; \frac{\omega}{P_k},
  \label{eq:spikeposterior}
\end{equation}
so $r_k$ is the time-indexed posterior height of the spike evaluated
on the confirming branch, with $r_0 = p$ the original
weight. It's pegged inversely to $P_k$ for all time, by the same Bayesian consistency:
\begin{equation}
  P_k\, r_k \;=\; p.
  \label{eq:spikeinvariant}
\end{equation}

If any answer is nonzero then $\phi_0$ is eliminated, every
compatible survivor had the same mass $\omega_k$, and the posterior is
uniform; subsequent conditioning on answers preserves uniformity. With an equivalence across system sizes, this induces a two state flow. Renormalization sharpens the spike. The two
macrostates and their directions are
\begin{equation}
  \left. \text{spike}\longrightarrow
  \begin{cases}
    \text{sharper spike}, & \text{with probability } P_{k+1}/P_k,\\
    \text{uniform}, & \text{with probability } 1 - P_{k+1}/P_k,
  \end{cases}
  \qquad \right|
  \text{uniform}\longrightarrow\text{uniform}.
  \label{eq:spikeflow}
\end{equation}
\begin{example}
The flow at $(n, m) = (3, 2)$, so $A = 4$ and $N = 4^8 = 65{,}536$,
started from $r_0 = p_0 = 0.3$. Each confirmed answer divides the
surviving map space by four and renormalizes the spike upward; each
refutation drops the belief into the uniform class, which nothing
ever leaves.

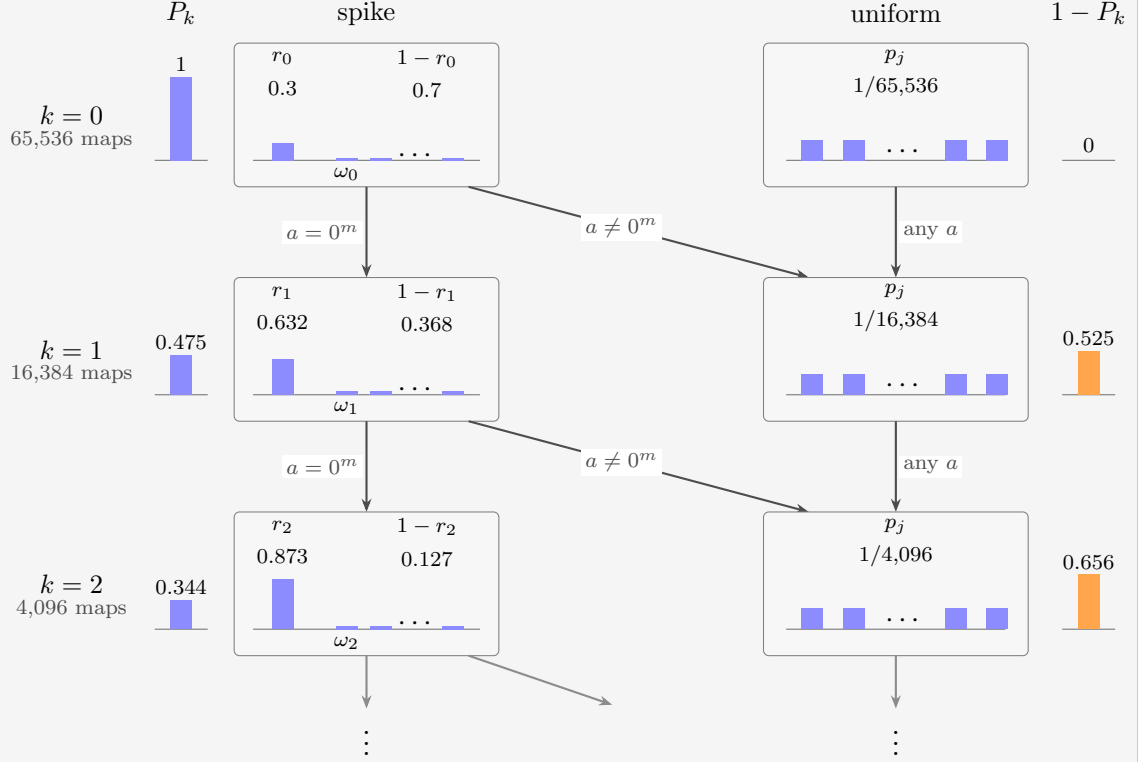
\begin{center}
\begin{tikzpicture}[x=1cm, y=1cm,
  state/.style={draw=black!50, rounded corners=2pt, fill=gray!6},
  flow/.style={-{Stealth[length=5pt]}, black!70, line width=0.8pt},
  alab/.style={font=\scriptsize, fill=white, inner sep=1.5pt}]

  \node[font=\small] at (-0.7,2.3) {$P_k$};
  \node[font=\small] at (1.75,2.3) {spike};
  \node[font=\small] at (8.75,2.3) {uniform};
  \node[font=\small] at (11.3,2.3) {$1 - P_k$};

  \node[font=\small] at (-2.15,0.95) {$k = 0$};
  \node[font=\scriptsize, black!70] at (-2.15,0.62) {$65{,}536$ maps};
  \draw[black!55] (-1.05,0.35) -- (-0.35,0.35);
  \fill[blue!45] (-0.84,0.35) rectangle (-0.56,1.45);
  \node[font=\scriptsize] at (-0.7,1.62) {$1$};

  \draw[state] (0,0) rectangle (3.5,1.9);
    \draw[black!55] (0.25,0.35) -- (3.25,0.35);
  \fill[blue!45] (0.5,0.35) rectangle (0.78,0.575);
  \node[font=\scriptsize, align=center] at (0.64,1.5) {$r_0$\\[3pt]$0.3$};
  \fill[blue!45] (1.35,0.35) rectangle (1.63,0.385);
  \fill[blue!45] (1.8,0.35) rectangle (2.08,0.385);
  \node at (2.42,0.42) {$\cdots$};
  \fill[blue!45] (2.75,0.35) rectangle (3.03,0.385);
  \node[font=\scriptsize] at (1.49,0.17) {$\omega_0$};
  \node[font=\scriptsize, align=center] at (2.55,1.5) {$1-r_0$\\[3pt]$0.7$};

  \draw[state] (7,0) rectangle (10.5,1.9);
    \draw[black!55] (7.3,0.35) -- (10.2,0.35);
  \foreach \xx in {7.5,8.05} \fill[blue!45]
    (\xx,0.35) rectangle (\xx+0.28,0.62);
  \node at (8.87,0.47) {$\cdots$};
  \node[font=\scriptsize, align=center] at (8.75,1.5) {$p_j$\\[3pt]$1/65{,}536$};
  \foreach \xx in {9.4,9.95} \fill[blue!45]
    (\xx,0.35) rectangle (\xx+0.28,0.62);

  \draw[black!55] (10.95,0.35) -- (11.65,0.35);
  \node[font=\scriptsize] at (11.3,0.55) {$0$};

  \draw[flow] (1.75,0) -- (1.75,-1.2)
    node[alab, midway, left=1pt] {$a = 0^m$};
  \draw[flow] (3.1,0) -- (7.6,-1.2)
    node[alab, pos=0.45] {$a \neq 0^m$};
  \draw[flow] (8.75,0) -- (8.75,-1.2)
    node[alab, midway, right=1pt] {any $a$};

  \node[font=\small] at (-2.15,-2.15) {$k = 1$};
  \node[font=\scriptsize, black!70] at (-2.15,-2.48) {$16{,}384$ maps};
  \draw[black!55] (-1.05,-2.75) -- (-0.35,-2.75);
  \fill[blue!45] (-0.84,-2.75) rectangle (-0.56,-2.23);
  \node[font=\scriptsize] at (-0.7,-2.06) {$0.475$};

  \draw[state] (0,-3.1) rectangle (3.5,-1.2);
    \draw[black!55] (0.25,-2.75) -- (3.25,-2.75);
  \fill[blue!45] (0.5,-2.75) rectangle (0.78,-2.276);
  \node[font=\scriptsize, align=center] at (0.64,-1.6) {$r_1$\\[3pt]$0.632$};
  \fill[blue!45] (1.35,-2.75) rectangle (1.63,-2.71);
  \fill[blue!45] (1.8,-2.75) rectangle (2.08,-2.71);
  \node at (2.42,-2.68) {$\cdots$};
  \fill[blue!45] (2.75,-2.75) rectangle (3.03,-2.71);
  \node[font=\scriptsize] at (1.49,-2.93) {$\omega_1$};
  \node[font=\scriptsize, align=center] at (2.55,-1.6) {$1-r_1$\\[3pt]$0.368$};

  \draw[state] (7,-3.1) rectangle (10.5,-1.2);
    \draw[black!55] (7.3,-2.75) -- (10.2,-2.75);
  \foreach \xx in {7.5,8.05} \fill[blue!45]
    (\xx,-2.75) rectangle (\xx+0.28,-2.48);
  \node at (8.87,-2.63) {$\cdots$};
  \node[font=\scriptsize, align=center] at (8.75,-1.6) {$p_j$\\[3pt]$1/16{,}384$};
  \foreach \xx in {9.4,9.95} \fill[blue!45]
    (\xx,-2.75) rectangle (\xx+0.28,-2.48);

  \draw[black!55] (10.95,-2.75) -- (11.65,-2.75);
  \fill[orange!70] (11.16,-2.75) rectangle (11.44,-2.17);
  \node[font=\scriptsize] at (11.3,-2.0) {$0.525$};

  \draw[flow] (1.75,-3.1) -- (1.75,-4.3)
    node[alab, midway, left=1pt] {$a = 0^m$};
  \draw[flow] (3.1,-3.1) -- (7.6,-4.3)
    node[alab, pos=0.45] {$a \neq 0^m$};
  \draw[flow] (8.75,-3.1) -- (8.75,-4.3)
    node[alab, midway, right=1pt] {any $a$};

  \node[font=\small] at (-2.15,-5.25) {$k = 2$};
  \node[font=\scriptsize, black!70] at (-2.15,-5.58) {$4{,}096$ maps};
  \draw[black!55] (-1.05,-5.85) -- (-0.35,-5.85);
  \fill[blue!45] (-0.84,-5.85) rectangle (-0.56,-5.47);
  \node[font=\scriptsize] at (-0.7,-5.3) {$0.344$};

  \draw[state] (0,-6.2) rectangle (3.5,-4.3);
    \draw[black!55] (0.25,-5.85) -- (3.25,-5.85);
  \fill[blue!45] (0.5,-5.85) rectangle (0.78,-5.195);
  \node[font=\scriptsize, align=center] at (0.64,-4.7) {$r_2$\\[3pt]$0.873$};
  \fill[blue!45] (1.35,-5.85) rectangle (1.63,-5.81);
  \fill[blue!45] (1.8,-5.85) rectangle (2.08,-5.81);
  \node at (2.42,-5.78) {$\cdots$};
  \fill[blue!45] (2.75,-5.85) rectangle (3.03,-5.81);
  \node[font=\scriptsize] at (1.49,-6.03) {$\omega_2$};
  \node[font=\scriptsize, align=center] at (2.55,-4.7) {$1-r_2$\\[3pt]$0.127$};

  \draw[state] (7,-6.2) rectangle (10.5,-4.3);
    \draw[black!55] (7.3,-5.85) -- (10.2,-5.85);
  \foreach \xx in {7.5,8.05} \fill[blue!45]
    (\xx,-5.85) rectangle (\xx+0.28,-5.58);
  \node at (8.87,-5.73) {$\cdots$};
  \node[font=\scriptsize, align=center] at (8.75,-4.7) {$p_j$\\[3pt]$1/4{,}096$};
  \foreach \xx in {9.4,9.95} \fill[blue!45]
    (\xx,-5.85) rectangle (\xx+0.28,-5.58);

  \draw[black!55] (10.95,-5.85) -- (11.65,-5.85);
  \fill[orange!70] (11.16,-5.85) rectangle (11.44,-5.13);
  \node[font=\scriptsize] at (11.3,-4.96) {$0.656$};

  \draw[flow, black!45] (1.75,-6.2) -- (1.75,-6.9);
  \draw[flow, black!45] (3.1,-6.2) -- (5.0,-6.85);
  \draw[flow, black!45] (8.75,-6.2) -- (8.75,-6.9);
  \node at (1.75,-7.25) {$\vdots$};
  \node at (8.75,-7.25) {$\vdots$};
\end{tikzpicture}
\end{center}

The two columns of boxes are the macrostates of \eqref{eq:spikeflow},
each arrow is labeled by the answer that drives its transition, and
each row's map space is one quarter of the row above. The outer columns track the branch
probabilities $P_k$ and $1 - P_k$, and the invariant $P_k r_k = p$ is
visible between the $P_k$ bars and the $r_k=p_0$ bars. The bars inside the boxes 
measure the $\{p_j\}$, on the left for the spike $\phi_0$ vs others. The spike climbs
exactly as fast as its branch loses probability,
$0.3 = 1 \times 0.3 = 0.475 \times 0.632 = 0.344 \times 0.873$.
\end{example}

For $p \in (0,1)$, we can combine equations \eqref{eq:spikeprior} and \eqref{eq:spikePk} to get the limiting odds on the surviving branch and, from equation \eqref{eq:spikeposterior}, the surviving posterior:

\begin{equation}
  P_k = p + (1-p)u^k + O(N^{-1}),
  \qquad
  r_k = \frac{p}{p + (1-p)u^k} + O(N^{-1}).
  \label{eq:spikelargeN}
\end{equation}
Each confirmation supplies approximately $m$ bits of evidence
for the spike, in the sense of log odds:
\begin{equation}
  \log_2\left(\frac{r_k}{1-r_k}\right) = \log_2\left(\frac{p}{(1-p)\cdot u^k}\right) = \log_2\frac{p}{1-p} + mk.
\end{equation}
After all $Q$ questions, the only way the spike can survive is by eliminating all alternative maps. This happens with 
$P_Q = p$ and hence $r_Q = 1$; the histories that confirm through the
complete table are precisely those with $\psi = \phi_0$. At the
uniform point $p = 1/N$ both posterior branches are uniform on their
surviving map spaces.

In anticipation of the TDL, we observe that the same $P_k$ has a second reading: it is the probability of the
single $k$-answer block $0^{mk}$. For $1 \le k \le Q$ every other answer block is realized by
$N u^k$ background maps and therefore carries the common probability
\begin{equation}
  U_k \;:=\; \frac{1 - P_k}{A^k - 1} \;=\; N u^k\,\omega,
  \qquad\text{so}\qquad
  \Pr(\psi(\mathcal S) = y)
  =\begin{cases}
    P_k, & y = 0^{mk},\\
    U_k, & y \neq 0^{mk},
  \end{cases}
  \label{eq:spikeUk}
\end{equation}
for any question set $\mathcal S$ of size $k$. The entropy of the above
distribution, coarse-grained to blocks of questions, is
\begin{equation}
  G_k \;=\; -P_k\log_2 P_k - (A^k-1)\,U_k\log_2 U_k,
  \label{eq:spikeGk}
\end{equation}
as there are $A^k-1$ terms where $y \neq 0^{mk}$.
Every choice of $k$ questions obeys the same law, so
\eqref{eq:spikeGk} is also the mean $k$-question block entropy of
Chapter~\ref{sec:expectations}. In particular $G_0 = 0$ and $H(M^{(0)}) = Q\,G_1$, with the exact prior entropy
\begin{equation}
  G_Q \;=\; h_2(p) + (1-p)\log_2(N-1),
  \label{eq:spikeHp}
\end{equation}
extensive in $Q$ as $N=A^Q$ and $P_Q=p$, at fixed $0 < p < 1$. Prior entropy vanishes at $p = 1$.

\subsubsection{The fixed-\texorpdfstring{$p$}{p} thermodynamic limit leverage}

The thermodynamic limit keeps the prior mass $0 < p < 1$ and asked fraction $t > 0$ fixed,
while sending $n \to \infty$. The guiding
principle is the exact finite law \eqref{eq:Lk-G} for $\mathcal L$: the aggregate
leverage is built from three ingredients, 

\begin{enumerate}
  \item the prior table entropy $H(M^{(0)})$
  \item the received entropy $G_\ell$
  \item the increment $G_{\ell+1} - G_\ell$.
\end{enumerate}

If we can derive each for the example, we are done. The block entropy \eqref{eq:spikeGk}
supplies every $G_k$ through the single counting formula
\eqref{eq:spikePk}. $H(M^{(0)}) = Q\,G_1$ holds for any prior:
$G_1$ is by definition the mean entropy of one column of $M$. Set
$\ell = \lfloor tQ\rfloor$ and take the terms of \eqref{eq:Lk-G} in
turn.

All three derivations run through one exact regrouping of
\eqref{eq:spikeGk}. The $A^k-1$ refuting blocks
\eqref{eq:spikeUk} share the mass $1-P_k$ evenly, so substituting
$U_k = (1-P_k)/(A^k-1)$ into \eqref{eq:spikeGk} and expanding the
logarithm regroups the block entropy into the binary entropy of the
confirm-or-refute split plus a uniform choice among the refuting
blocks:
\begin{equation}
  G_k \;=\; h_2(P_k) \;+\; (1-P_k)\log_2\!\big(A^k-1\big).
  \label{eq:spikeGsplit}
\end{equation}
At $k = 1$, by \eqref{eq:spikePk}
the spike answer $0^m$ carries
$P_1 = p + (Nu - 1)\,\omega \to p + (1-p)u$: the special map's whole
mass, plus the fraction $u$ of the background maps that happen to
agree with it at any one question. Substituting that limit,
\begin{equation}
\begin{aligned}
  \frac{H(M^{(0)})}{Q} \;=\; G_1
  \;&=\; h_2(P_1) + (1-P_1)\log_2(A-1) \\
  \;&\longrightarrow\;
  h_2\big(p + (1-p)u\big) + (1-p)(1-u)\log_2(A-1)
  \;=:\; \eta_0(p,m).
\end{aligned}
  \label{eq:hspike}
\end{equation}
An $\eta_0$-term features in the numerator of every macroscopic
leverage, the entropy of the first answer, as in \eqref{eq:master}. The spike produces this shape; 
in general $\eta_0$ depends on the prior and has no closed form.

The received entropy: Evaluate \eqref{eq:spikeGsplit} at
$k = \ell = \lfloor tQ\rfloor$. By \eqref{eq:spikelargeN}
$P_\ell \to p$, so the split entropy stays bounded while the
block-count logarithm grows linearly,
$\log_2(A^\ell-1) = \ell m + o(1)$:
\begin{equation}
  \frac{G_{\lfloor tQ\rfloor}}{Q}
  \;=\;
  \underbrace{\frac{h_2(P_\ell)}{Q}}_{\textstyle\to\; 0}
  \;+\;
  \underbrace{(1-P_\ell)\vphantom{\frac{h}{Q}}}
    _{\textstyle\to\; 1-p}
  \;\underbrace{\frac{\log_2\!\big(A^\ell-1\big)}{Q}}
    _{\textstyle\to\; t\,m}
  \;\longrightarrow\; t\,(1-p)\,m .
  \label{eq:spikeGlimit}
\end{equation}
This is the two-state flow watched at macroscopic time: for any
$t > 0$ the confirm-or-refute lottery has already resolved inside
the boundary layer at $t = 0$, where $r_k \to 1$ within $o(Q)$
questions. A surviving spike answers for free; only the refuted
class, probability $1-p$, keeps paying the full $m$ bits per answer.

The increment: Subtract consecutive copies of
\eqref{eq:spikeGsplit}: the bounded $h_2(P_k)$ terms cancel in the
limit, and the block-count logarithm grows by
$\log_2(A^{\ell+1}-1) - \log_2(A^\ell-1) \to \log_2 A = m$,
paid only by the refuted mass:
\begin{equation}
  G_{\ell+1} - G_\ell \;\longrightarrow\; (1-p)\,m,
  \qquad\text{so}\qquad
  \frac{(Q-\ell)\big(G_{\ell+1}-G_\ell\big)}{Q}
  \;\longrightarrow\; (1-t)(1-p)\,m .
  \label{eq:spikeincrement}
\end{equation}
Filling the three limits into \eqref{eq:Lk-G} term by term,
\[
  \mathcal L_\ell
  = \frac{\overbrace{H\big(M^{(0)}\big)}^{\textstyle\to\,
      Q\,\eta_0(p,m)}
    \;-\; \overbrace{(Q-\ell)\big(G_{\ell+1}-G_\ell\big)}
      ^{\textstyle\to\, Q\,(1-t)(1-p)m}}
    {\underbrace{G_\ell\vphantom{\big(}}_{\textstyle\to\,
      Q\,t(1-p)m}},
\]
and canceling the common factor $Q$ gives
\begin{equation}
  L_p(t)
  = \frac{\eta_0(p,m) - (1-t)(1-p)m}{t(1-p)m}
  = 1 + \frac{\eta_0(p,m) - (1-p)m}{t(1-p)m},
  \qquad 0 < t \le 1. 
  \label{eq:spikehyperbola}
\end{equation}
The $1/t$ term records information released in a vanishing initial
fraction of the run, divided by the $O(tQ)$ information received
afterwards.

\begin{figure}[H]
\centering
\begin{tikzpicture}[baseline=(current bounding box.north)]
\begin{semilogyaxis}[width=7.4cm, height=5.4cm,
  title={$m = 1$}, title style={font=\small},
  xlabel={$t = \ell/Q$},
  ylabel={$L_p$},
  xmin=0, xmax=1, ymin=0.9, ymax=400, domain=0.02:1, samples=120,
  tick label style={font=\scriptsize}, label style={font=\small},
  grid=major, grid style={black!12},
  legend style={font=\tiny, at={(0.97,0.97)}, anchor=north east},
  legend cell align=left]
\addplot[cf1, thick] {1+0.00100/x};
\addplot[cf2, thick] {1+0.10308/x};
\addplot[cf3, thick] {1+0.62256/x};
\addplot[cf4, thick] {1+1.86397/x};
\addplot[cf5, thick] {1+5.20406/x};
\addplot[black, dotted, forget plot] {1};
\legend{{$p=0.001$}, {$p=0.1$}, {$p=0.5$}, {$p=0.9$}, {$p=0.999$}}
\end{semilogyaxis}
\end{tikzpicture}\hfill
\begin{tikzpicture}[baseline=(current bounding box.north)]
\begin{semilogyaxis}[width=7.4cm, height=5.4cm,
  title={$m = 2$}, title style={font=\small},
  xlabel={$t = \ell/Q$},
  xmin=0, xmax=1, ymin=0.9, ymax=400, domain=0.02:1, samples=120,
  tick label style={font=\scriptsize}, label style={font=\small},
  grid=major, grid style={black!12}]
\addplot[cf1, thick] {1+0.00100/x};
\addplot[cf2, thick] {1+0.09977/x};
\addplot[cf3, thick] {1+0.54879/x};
\addplot[cf4, thick] {1+1.51592/x};
\addplot[cf5, thick] {1+4.02798/x};
\addplot[black, dotted] {1};
\end{semilogyaxis}
\end{tikzpicture}
\caption{The fixed-$p$ thermodynamic curve \eqref{eq:spikehyperbola},
on a logarithmic vertical scale. Every weight produces the same $1/t$
shape; the prior sets only the coefficient, which grows from
$0.001$ at $p = 0.001$ to above $5$ at $p = 0.999$.}
\label{fig:spikehyp}
\end{figure}
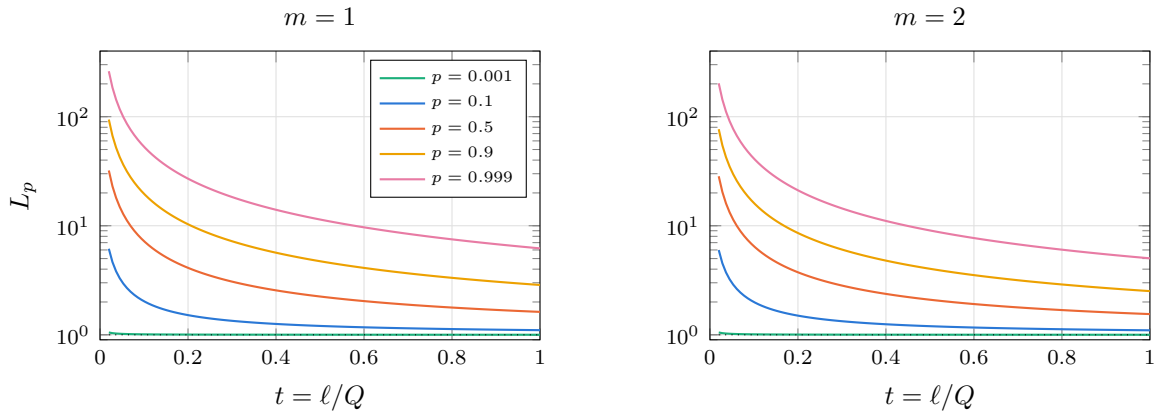

\begin{example}
The approach to the hyperbolic limit \eqref{eq:spikehyperbola}. 
Both panels evaluate the
exact finite law \eqref{eq:Lk-G} on the block entropies
\eqref{eq:spikeGsplit} at increasing finite $n$, and
set it against the thermodynamic curve in black.

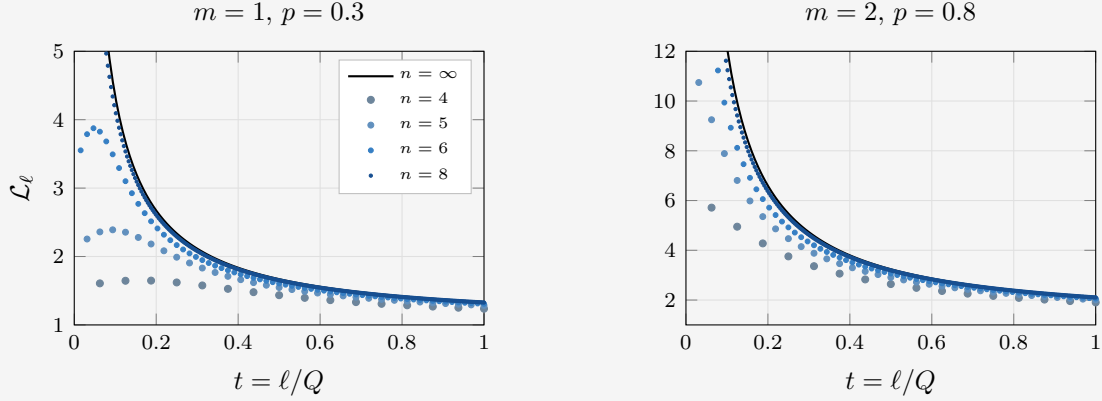
\begin{center}
\begin{tikzpicture}[baseline=(current bounding box.north)]
\begin{axis}[width=7.0cm, height=5.2cm,
  title={$m = 1$, $p = 0.3$}, title style={font=\small},
  xlabel={$t = \ell/Q$}, ylabel={$\mathcal L_{\ell}$},
  xmin=0, xmax=1, ymin=1, ymax=5, ytick={1,2,3,4,5},
  tick label style={font=\scriptsize}, label style={font=\small},
  grid=major, grid style={black!12},
  legend style={font=\tiny, at={(0.97,0.97)}, anchor=north east,
    draw=black!25, fill=white},
  legend cell align=left]
\addplot[black, thick, domain=0.02:1, samples=120]
  {1+0.33438/x};
\dataplot{cn1!65!black, only marks, mark=*, mark size=1.3}{t}{lev}{data/spikefin_m1p30_n4.dat}
\dataplot{cn2!85!black, only marks, mark=*, mark size=1.1}{t}{lev}{data/spikefin_m1p30_n5.dat}
\dataplot{cn3, only marks, mark=*, mark size=0.9}{t}{lev}{data/spikefin_m1p30_n6.dat}
\dataplot{cn4, only marks, mark=*, mark size=0.6}{t}{lev}{data/spikefin_m1p30_n8.dat}
\datalegend{{$n=\infty$}, {$n=4$}, {$n=5$}, {$n=6$}, {$n=8$}}
\end{axis}
\end{tikzpicture}\hfill
\begin{tikzpicture}[baseline=(current bounding box.north)]
\begin{axis}[width=7.0cm, height=5.2cm,
  title={$m = 2$, $p = 0.8$}, title style={font=\small},
  xlabel={$t = \ell/Q$},
  xmin=0, xmax=1, ymin=1, ymax=12, ytick={2,4,6,8,10,12},
  tick label style={font=\scriptsize}, label style={font=\small},
  grid=major, grid style={black!12}]
\addplot[black, thick, domain=0.02:1, samples=120]
  {1+1.11896/x};
\dataplot{cn1!65!black, only marks, mark=*, mark size=1.3}{t}{lev}{data/spikefin_m2p80_n4.dat}
\dataplot{cn2!85!black, only marks, mark=*, mark size=1.1}{t}{lev}{data/spikefin_m2p80_n5.dat}
\dataplot{cn3, only marks, mark=*, mark size=0.9}{t}{lev}{data/spikefin_m2p80_n6.dat}
\dataplot{cn4, only marks, mark=*, mark size=0.6}{t}{lev}{data/spikefin_m2p80_n8.dat}
\end{axis}
\end{tikzpicture}
\end{center}

Every finite curve approaches the limit from below, and the gap
closes last at the left edge: the hyperbola diverges as $t \to 0$,
while a run of $Q$ questions has its first point at $t = 1/Q$ with a
finite value there. The boundary layer is a limiting phenomenon.
\end{example}

\subsubsection{Exchangeable answers and de Finetti's theorem}
\label{sec:spikedefinetti}

The block law \eqref{eq:spikeUk} has a strong property. Whatever set $\mathcal S$
Werner asks, the all-zero pattern $y=0^{mk}$ carries probability $P_k$ and every other pattern
carries $U_k$. Both depend on $\mathcal S$ solely through its size $k$. 
Which questions were asked, and in which order, is not relevant. 
A prior with that property is called \textbf{exchangeable}, and
a classical theorem constrains such priors to be equivalent to a restrictive construction.

Write the answers in the order they arrive as a sequence of random
symbols,
\begin{equation}
  \xi_i \;:=\; \psi(q_i) \;\in\; \mathcal A,
  \qquad i = 1, \ldots, Q,
  \label{eq:xiseq}
\end{equation}
so that $\xi$ is the random object Werner actually observes (in this case $m$ bits each), one
coordinate per question-answer. The sequence is exchangeable when its joint law
is unchanged by permuting the indices: for every permutation $\sigma$
and every pattern $y$,
\begin{equation}
  \Pr\big(\xi_1 = y_1, \ldots, \xi_k = y_k\big)
  \;=\;
  \Pr\big(\xi_{\sigma(1)} = y_1, \ldots, \xi_{\sigma(k)} = y_k\big).
  \label{eq:exchangeable}
\end{equation}
Exchangeability is weaker than independence. Independent answers tell
each other nothing, whereas exchangeable ones may be strongly
dependent; all that is asked is that they be dependent
\emph{symmetrically}, so that no question occupies a special position.

De Finetti's theorem \cite{definetti37,hewittsavage} states that
symmetric dependence, for sequences of unbounded length, is equivalent to a \emph{mixture of coins}. 
Let $\nu$ be a distribution on the answer alphabet $\mathcal A$, an \textbf{answer law}, writing
$\nu(a) = \Pr(a)$ for the weight it puts on the answer $a$. Let $\Lambda$
be a distribution over such answer laws, the \textbf{directing
measure}. Then for an exchangeable sequence there is exactly one
$\Lambda$ with
\begin{equation}
  \Pr\big(\xi_1 = y_1, \ldots, \xi_k = y_k\big)
  \;=\; \int \prod_{i=1}^{k} \nu(y_i)\; d\Lambda(\nu) ,
  \label{eq:definetti}
\end{equation}
for every $k$ and every pattern. The RHS is executed as a
two-stage experiment. Nature first draws one answer law $\nu$ from
$\Lambda$, once and for all, and then answers every question
independently from that single $\nu$. Conditional on $\nu$ the answers
are independent; unconditionally they are correlated. The
correlation is mediated by Werner considering each $\nu$ possible a priori.
The theorem needs the sequence to extend to an infinite one, or
equivalently the family of finite laws to be consistent as $Q$ grows;
a finite exchangeable law taken in isolation need not be a mixture of
independent draws of $\nu$. Sampling without replacement is the standard
counterexample, and how far a finite exchangeable law can sit from every
mixture is quantified in \cite{diaconisfreedman80}. In the infinite set limit, sampling with and without replacement is 
equivalent.

In general $\Lambda$ may spread over any collection of answer laws
$\nu_1, \nu_2, \ldots$, finite, countable or a continuum, and the
richer that collection the more there is to learn about which one was
drawn.

Such a $\nu$ has appeared once already, at \eqref{eq:cliqueQ}, but it 
is used differently here. There, one draw from $\nu$
was copied as the answer to every question, which in the present language is
$\Lambda$ only sampling from such degenerate laws, carrying weight $\nu(a)$ on
the law that answers $a$ with certainty: perfect correlation of answers. Here, after drawing $\nu$ from $\Lambda$, each 
answer is independent, the opposite.

The spike prior is the simplest nontrivial such example: $\Lambda$ is a discrete
distribution on two points. Let $\nu_0$ be the spike,
and let $\nu_{>}$ be the uniform law, which allows answers $\neq 0^m$ as
well. These distributions on $\mathcal A$ have probabilities
\begin{equation}
  \nu_0(a) = \delta_{a,0^m},\quad\quad
  \nu_>(a) = u
\end{equation}
Then
\begin{equation}
  \Lambda \;=\;
  \Big(p - \frac{1-p}{N-1}\Big)\, \delta_{\nu_0}
  \;+\;
  (1-p)\,\frac{N}{N-1}\, \delta_{\nu_{>}},
  \label{eq:spikedefinetti}
\end{equation}
with $\delta_\nu$ the point mass at $\nu$, reproduces the spike
exactly. Substituting \eqref{eq:spikedefinetti} into
\eqref{eq:definetti}, the confirming pattern $0^{mk}$ receives full
weight from $\nu_0$ and $u^k$ from $\nu_{>}$, while every other
pattern receives nothing from $\nu_0$ and $u^k$ from $\nu_{>}$:
\begin{equation}
\begin{aligned}
  \Pr\big(\psi(\mathcal S) = 0^{mk}\big)
  &= \Big(p - \frac{1-p}{N-1}\Big) + (1-p)\,\frac{N u^k}{N-1}
  = p + (Nu^k - 1)\,\omega , \\[2pt]
  \Pr\big(\psi(\mathcal S) = y\big)
  &= (1-p)\,\frac{N u^k}{N-1} = N u^k \omega
  \qquad (y \neq 0^{mk}),
\end{aligned}
  \label{eq:spikedefinettitwo}
\end{equation}
which are \eqref{eq:spikePk} and \eqref{eq:spikeUk} exactly. The
prior on boolean maps comes back too. A map is answered by $\nu_0$ on all questions 
only if it is $\phi_0$, and by $\nu_{>}$ with weight $u^Q = 1/N$ whatever its
truth table holds. We can find the prior probability of $\phi_0$ by reconstructing the probability
of only receiving zero answers, through either $\nu$. Starting from \eqref{eq:definetti}, substituting \eqref{eq:spikedefinetti} at $k = Q$ returns
\begin{equation}
  \Pr\big(\xi_1 = 0^m, \ldots, \xi_Q = 0^m\big)
  \;=\; \Big(p - \frac{1-p}{N-1}\Big) \cdot 1^Q + (1-p)\,\frac{N}{N-1} \cdot u^Q = p,
  \label{eq:definettispecif}
\end{equation}
And the reader can convince themselves that the probabilities of every other map (where the first term vanishes) are equal.
$p_j = \omega$. Together, these reproduce \eqref{eq:spikeprior}. 

Why is the leverage a hyperbola? All the prior knows is which of the
two answer laws $\nu$ was drawn, and that is settled within $o(Q)$ questions. Afterwards, every remaining answer is an independent draw from a known law
and costs the same $\mathbb E_\Lambda[H(\nu)] \to (1-p)m$ bits of surprisal, so the
profile $\gamma$ is flat on $(0,1)$ and \eqref{eq:master} has no
freedom left but a $1/t$ curve. The coefficient can be read off as
well. Since
\begin{equation}
  \eta_0 - (1-p)m
  \;=\; H\big(\mathbb E_\Lambda\, \nu\big)
      - \mathbb E_\Lambda\big[H(\nu)\big]
  \;=\; I\big(\nu \,;\, \xi_1\big),
  \label{eq:spikejensen}
\end{equation}
the numerator of \eqref{eq:spikehyperbola} is the mutual information
between the drawn answer law and a single answer: the whole of the
spike's leverage is what one question reveals about which world Werner
is in, divided by what an answer costs him once he knows.

For the spike, $p_j = f(|j|)$ here in its simplest form: the weight of a map
depends on its truth table only through whether the Hamming weight $|j|$
vanishes. This generalizes, and both the exchangeability and its
consequence for the limit are recognizable from the prior alone. Any
$\{p_j\}$ whose weight depends on the truth table only through the
\emph{count} of each answer is exchangeable, and vice
versa. Under convergence of those counts in distribution to some
$\Lambda$ as $n \to \infty$, the profile $\gamma(t)$ is flat and the
leverage curve is a hyperbola, whether or not any finite prior along
the way is literally a mixture of independent draws.

We have explored a number of other characteristics of the spike
prior in Appendix~\ref{app:spike}: the branch-by-branch effect of
the first question, the extraction ratio, and various further
limits, among them the sharp prior $p \to 1^-$ and its
noncommutation with $n \to \infty$.

\subsection{Thermodynamic-limit calculus}\label{sec:tdlcalc}

The spike is solvable because all subsets of questions of the same size ($k$) have the
same entropy $G_k$. Averages are free. The general theory replaces that symmetry by an average
over subsets, which in general have distinct entropies.
Any symmetry of the prior reduces that average to a sum over orbits of
$k$-subsets, weighted by orbit size, and the spike is the extreme case:
invariant under every permutation of the questions, it has one orbit
per size and $G_k$ collapses to a single term. Less symmetrical priors
divide the number of terms by their order. Nonetheless, this is
instrumental in obtaining moderate $n$ results numerically. More
comments in Appendix~\ref{app:aig}, for instance.

\subsubsection{The finite sequence from which the limit is taken}

For each fixed question set $\mathcal S \subseteq \mathcal Q$ the vector
$\psi(\mathcal S)$ collects the answers to those questions, and its
entropy is taken under the prior law of $\psi$ with $\mathcal S$ held
fixed. The mean block entropy is that of
equation~\eqref{eq:Gk}, now carrying the system size explicitly:
\begin{equation}
  G_{n,k}
  \;:=\; \binom{Q}{k}^{-1}
  \sum_{\substack{
    |\mathcal S| = k}} H\big(\psi(\mathcal S)\big),
  \qquad 0 \le k \le Q,
  \label{eq:Gnk}
\end{equation}
so that $G_{n,0} = 0$ and $G_{n,Q} = H(\psi) = H(p^{(n)})$, where
$p^{(n)}$ is the prior law of the random map at size $n$, with
increments 
\begin{equation}
\gamma_{n,k} := G_{n,k+1} - G_{n,k}.
\label{eq:teleskippie}
\end{equation}
By the entropy chain rule, exactly as in the derivation of
equation~\eqref{eq:law-remaining}, the increment is the mean
conditional entropy of one fresh answer,
\begin{equation}
  \gamma_{n,k}
  \;=\; \mathbb E_{\mathcal S, q}
  \big[H\big(\psi(q)\mid\psi(\mathcal S)\big)\big],
  \qquad
  |\mathcal S| = k,
  \quad q \in \mathcal Q \setminus \mathcal S.
  \label{eq:gammank}
\end{equation}
The increments also connect directly to erasure-channel
\emph{extrinsic information transfer} (EXIT) functions~\cite{ashikhmin04,richardsonurbanke08}.
Three terms from coding theory are needed first. A \textbf{code} of
length $Q$ over $\mathcal A$ is a subset of $\mathcal A^{Q}$, its
members the \textbf{codewords}; a truth table is a string of exactly
that length, so any set of maps is a code, and the degree classes of
Chapter~\ref{sec:shapeprior} are linear ones. Its rate is
$Q^{-1}\log_A(\text{number of codewords})$, the information one
codeword carries per symbol. An \textbf{erasure channel} transmits a
codeword one symbol at a time, each symbol independently either
delivered unchanged, with probability $v$, or replaced by a marker
recording only that a symbol was lost. An EXIT function then reports
how much uncertainty is left in one symbol once every other symbol has
passed through such a channel, as a function of the erasure
probability $1-v$; its integral over that probability is the rate,
which is the area theorem of \cite{ashikhmin04} invoked below.
Choose a fresh question $q$ uniformly and reveal each other question
independently with probability $v\in[0,1]$. Think of $v$ as the probability that
the question naturally came up, i.e. $v \approx t$, the macroscopic time. Conditional
on actually revealing $k$ questions, their set is uniform among the $k$-sets
not containing $q$. The mean conditional entropy of the fresh answer
is therefore
\begin{equation}\label{eq:exit-eta}
  \eta_n(v):=\sum_{k=0}^{Q-1}\binom{Q-1}{k}
       v^k(1-v)^{Q-1-k}\,\gamma_{n,k}.
\end{equation}
For binary answers this is the average EXIT function expressed in
reveal probability; the usual erasure probability is $1-v$.
Thus $\gamma_{n,k}$ is its fixed-cardinality counterpart. Integrating
the binomial weights gives its area law,
\[
  \int_0^1\eta_n(v)\,dv
  =\frac1Q\sum_{k=0}^{Q-1}\gamma_{n,k}
  =\frac{H(p^{(n)})}{Q},
\]
which is the code rate for a uniform prior on a binary linear code.
This is the EXIT and area-theorem connection underlying the
Reed--Muller result used in Chapter~\ref{sec:shapeprior}.

We write $\mathbb{E}$ in \eqref{eq:gammank} to emphasize the average over the selected $q$ distinct from $\mathcal{S}$. 
The additional average here: the conditional entropy
averages over answer histories $\phi_j$ under the prior $p_j$ on the random map. 
Conditioning reduces entropy, so coupling a
random $k$-set to a random $(k{+}1)$-set yields a \textbf{staircase}
\begin{equation}
  m \;\ge\; \gamma_{n,0} \;\ge\; \gamma_{n,1} \;\ge\; \cdots
  \;\ge\; \gamma_{n,Q-1} \;\ge\; 0 .
  \label{eq:gammamono}
\end{equation}
This monotonicity is deterministic, the truth and the subset having
already been averaged over. Recall, this is a property of the average, not of any constituent terms.
Concentration of individual learning trajectories would be a separate
result. The two laws \eqref{eq:law-received} and
\eqref{eq:law-remaining} then read
\begin{equation}
  \Big\langle\!\Big\langle \sum_{k=1}^{\ell} s_k \Big\rangle\!\Big\rangle = G_{n,\ell},
  \qquad
  \big\langle\!\big\langle H\big(M^{(\ell)}\big) \big\rangle\!\big\rangle
  = (Q-\ell)\,\gamma_{n,\ell},
  \label{eq:finitelaws}
\end{equation}
for $0 \le \ell < Q$, with zero remaining entropy at $\ell = Q$, and
equation~\eqref{eq:Lk-G} becomes
\begin{equation}
  \mathcal L_{n,\ell}
  = \frac{Q\,G_{n,1} - (Q-\ell)\,\gamma_{n,\ell}}{G_{n,\ell}}
  \qquad (1 \le \ell < Q),
  \qquad
  \mathcal L_{n,Q} = \frac{Q\,G_{n,1}}{G_{n,Q}} .
  \label{eq:finitelev}
\end{equation}

\subsubsection{A precise macroscopic-limit statement}

The next step is to find continuous limits of the functions: $\gamma(t)$ from $\gamma_{n,\ell}$ and $g(t)$ from $G_{n,\ell}$.

The leverage has only these two functions in the numerator. In fact, $G_{n,1}=\gamma_{n,0}$ so we might be
tempted to express everything in terms of the one function $\gamma(t)$, and find the coarse-grained limit
of that. However, that would miss a crucial \emph{delta-like} contribution to the learning.

It is instead useful to keep the initial marginal entropy separate from the
macroscopic profile on the right, since that allows an
$o(Q)$-question boundary layer at $t = 0$. Let
\begin{equation}
  \eta_0 := \lim_{n\to\infty} G_{n,1}, 
\end{equation}
which in general will be allowed to converge to something distinct from $\gamma(t)$.
Suppose the staircases have an
almost-everywhere limit $\gamma$ on $(0,1)$, with
$\gamma_{n,\lfloor tQ\rfloor} \to \gamma(t)$ at any particular $t$
where the remaining entropy is evaluated. At such a point $\gamma(t)$
is the expected entropy of one more answer after a random
$t$-fraction has been observed. Four hypotheses suffice:
\begin{enumerate}
  \item \textbf{Fixed answer width.} The integer $m$ is fixed, which
    supplies the uniform bound $0 \le \gamma_{n,k} \le m$.
  \item \textbf{Initial-entropy convergence.} The limit $\eta_0$ exists.
  \item \textbf{Profile convergence.} The staircase
    $t \mapsto \gamma_{n,\lfloor tQ\rfloor}$ converges almost
    everywhere on $(0,1)$, with direct convergence at any $t$ where
    the remaining entropy is evaluated. At continuity points of a
    selected monotone limit this causes no ambiguity.
  \item \textbf{Positive entropy density.} The limiting area is
    positive, $\int_0^1\gamma(x)\,dx > 0$.
\end{enumerate}
The fourth condition is the nondegenerate form of extensivity, in other words,
learning continues into the bulk and $M$ doesn't settle within a boundary around $t=0$. 
It implies $G_{n,Q}/Q \to \int_0^1\gamma > 0$. 

Extend the finite increments to a staircase on $[0,1]$ by
$\gamma_n(t) := \gamma_{n,\lfloor tQ\rfloor}$.
Telescoping with equation~\eqref{eq:teleskippie} gives an exact Riemann-sum identity at grid points,
\begin{equation}
  \frac{G_{n,\ell}}{Q}
  = \frac1Q\sum_{k=0}^{\ell-1}\gamma_{n,k}
  = \int_0^{\ell/Q}\gamma_n(x)\,dx,
  \label{eq:riemann}
\end{equation}
ignoring the last undefined endpoint in the limit. Since $0 \le \gamma_n \le m$, dominated convergence gives
\begin{equation}
G_{n,\ell_n}/Q \to g(t) := \int_0^t\gamma; \quad 
(Q-\ell_n)\gamma_{n,\ell_n}/Q \to (1-t)\gamma(t),
\end{equation}
using the coordinated construction $\ell_n = \lfloor tQ\rfloor$. Substituting both
into \eqref{eq:finitelev} proves
\begin{equation}
  L(t) \;=\; \frac{\eta_0 - (1-t)\,\gamma(t)}{g(t)},
  \qquad
  g(t) = \int_0^t \gamma(x)\,dx,
  \qquad 0 < t < 1. 
  \label{eq:master}
\end{equation}
Since $\gamma$ is nonnegative and nonincreasing, a positive integral
makes $g(t) > 0$ for every $t > 0$. Let us go over the pieces, for physical intuition:

\begin{itemize}
  \item $\gamma(t)$: the price of the next answer. It is the expected entropy of one fresh answer after a random $t$-fraction of the table has already been observed: what Werner still doesn't know about a question he hasn't asked, given everything he has. It starts at most $m$ and decreases monotonically, because each answer can only explain away some of the next one's novelty. $\gamma$ is the running rate at which surprisal is still available.

\item $g(t)$: the total bill so far. Summing the price of each answer as it was bought gives the cumulative surprisal received by macroscopic time $t$. It is the area under the profile to the left of $t$, and it sits in the denominator of the leverage because it is exactly what Werner ``paid''.

\item $\eta_0$: the size of the job at the outset. It is the marginal entropy of a single answer before anything is learned, so $Q \eta_0$ is the whole table's starting uncertainty $H(M^{(0)})$: the total to be destroyed, by asking or by deducing. It is kept as its own constant rather than read off as $\gamma(0^+)$ because learning finished in $o(Q)$ questions leaves no trace on the profile, and the gap $\eta_0 - \gamma(0^+)$ is precisely that boundary layer.

\item Together: $L(t)$ reads as (everything there was to know) minus (what the unasked columns still hold), over what was paid.
\end{itemize}

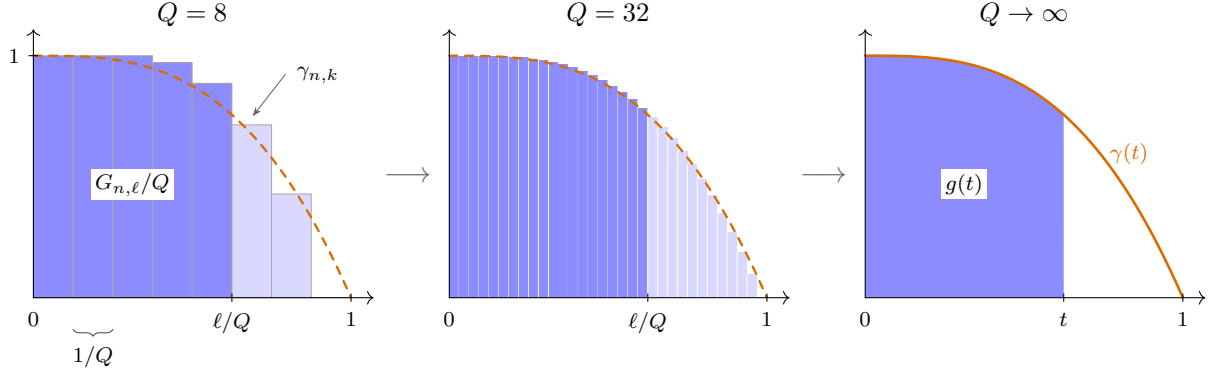
\begin{figure}[H]
\centering
\begin{tikzpicture}[line cap=round]
  \node[font=\small] at (2.1,3.75) {$Q = 8$};
  \foreach \k in {0,...,4}{
    \pgfmathsetmacro\h{3.2*(1 - \k*(\k-1)*(\k-2)/210)}
    \fill[blue!45] ({\k*4.2/8},0) rectangle ({(\k+1)*4.2/8},\h);
    \draw[black!35, line width=0.3pt]
      ({\k*4.2/8},0) rectangle ({(\k+1)*4.2/8},\h);}
  \foreach \k in {5,6,7}{
    \pgfmathsetmacro\h{3.2*(1 - \k*(\k-1)*(\k-2)/210)}
    \fill[blue!15] ({\k*4.2/8},0) rectangle ({(\k+1)*4.2/8},\h);
    \draw[black!35, line width=0.3pt]
      ({\k*4.2/8},0) rectangle ({(\k+1)*4.2/8},\h);}
  \draw[orange!85!black, dashed, line width=0.9pt]
    plot[domain=0:4.2, samples=80] (\x, {3.2*(1-(\x/4.2)^3)});
  \draw[->] (0,0) -- (4.5,0);
  \draw[->] (0,0) -- (0,3.55);
  \draw (-0.06,3.2) -- (0.06,3.2)
    node[left=3pt, font=\scriptsize] {$1$};
  \draw (2.625,-0.05) -- (2.625,0.05);
  \node[font=\scriptsize, below] at (2.625,-0.06) {$\ell/Q$};
  \node[font=\scriptsize, below] at (0,-0.06) {$0$};
  \draw (4.2,-0.05) -- (4.2,0.05);
  \node[font=\scriptsize, below] at (4.2,-0.06) {$1$};
  \node[font=\scriptsize, fill=white, inner sep=1.5pt]
    at (1.31,1.5) {$G_{n,\ell}/Q$};
  \node[font=\scriptsize, anchor=west] at (3.35,2.95) {$\gamma_{n,k}$};
  \draw[black!60, line width=0.4pt, -{Stealth[length=3.5pt]}]
    (3.33,2.95) -- (2.9,2.4);
  \draw[decorate, decoration={brace, mirror, amplitude=2.5pt},
        black!70] (0.525,-0.42) -- (1.05,-0.42);
  \node[font=\scriptsize, below] at (0.79,-0.5) {$1/Q$};

  \begin{scope}[shift={(5.5,0)}]
  \node[font=\small] at (2.1,3.75) {$Q = 32$};
  \foreach \k in {0,...,19}{
    \pgfmathsetmacro\h{3.2*(1 - (\k/31)*((\k-1)/30)*((\k-2)/29))}
    \fill[blue!45] ({\k*4.2/32},0) rectangle ({(\k+1)*4.2/32},\h);
    \draw[white, line width=0.25pt]
      ({(\k+1)*4.2/32},0) -- ({(\k+1)*4.2/32},\h);}
  \foreach \k in {20,...,31}{
    \pgfmathsetmacro\h{3.2*(1 - (\k/31)*((\k-1)/30)*((\k-2)/29))}
    \fill[blue!15] ({\k*4.2/32},0) rectangle ({(\k+1)*4.2/32},\h);
    \draw[white, line width=0.25pt]
      ({\k*4.2/32},0) -- ({\k*4.2/32},\h);}
  \draw[orange!85!black, dashed, line width=0.9pt]
    plot[domain=0:4.2, samples=80] (\x, {3.2*(1-(\x/4.2)^3)});
  \draw[->] (0,0) -- (4.5,0);
  \draw[->] (0,0) -- (0,3.55);
  \draw (2.625,-0.05) -- (2.625,0.05);
  \node[font=\scriptsize, below] at (2.625,-0.06) {$\ell/Q$};
  \node[font=\scriptsize, below] at (0,-0.06) {$0$};
  \draw (4.2,-0.05) -- (4.2,0.05);
  \node[font=\scriptsize, below] at (4.2,-0.06) {$1$};
  \end{scope}
  \node[black!60] at (4.95,1.6) {$\longrightarrow$};

  \begin{scope}[shift={(11,0)}]
  \node[font=\small] at (2.1,3.75) {$Q \to \infty$};
  \fill[blue!45] (0,0) --
    plot[domain=0:2.625, samples=50] (\x, {3.2*(1-(\x/4.2)^3)})
    -- (2.625,0) -- cycle;
  \draw[orange!85!black, line width=1pt]
    plot[domain=0:4.2, samples=80] (\x, {3.2*(1-(\x/4.2)^3)});
  \draw[->] (0,0) -- (4.5,0);
  \draw[->] (0,0) -- (0,3.55);
  \draw (2.625,-0.05) -- (2.625,0.05);
  \node[font=\scriptsize, below] at (2.625,-0.06) {$t$};
  \node[font=\scriptsize, below] at (0,-0.06) {$0$};
  \draw (4.2,-0.05) -- (4.2,0.05);
  \node[font=\scriptsize, below] at (4.2,-0.06) {$1$};
  \node[font=\scriptsize, fill=white, inner sep=1.5pt]
    at (1.31,1.5) {$g(t)$};
  \node[font=\scriptsize, anchor=west, orange!85!black]
    at (3.1,1.9) {$\gamma(t)$};
  \end{scope}
  \node[black!60] at (10.45,1.6) {$\longrightarrow$};
\end{tikzpicture}
\caption{From the finite increments to the profile.
Each bar has width $1/Q$ and height $\gamma_{n,k}$, so the shaded
bars sum, the telescoping identity \eqref{eq:riemann}, to
exactly $G_{n,\ell}/Q$, here at $t=\ell/Q = 5/8$. Refining $Q$ shrinks
the bars onto the staircase's limit, and dominated convergence turns
the shaded sum into the area $g(t) = \int_0^t \gamma$: received
information per question becomes the dark area under the profile.}
\label{fig:riemann}
\end{figure}

Monotonicity and boundedness guarantee convergent subsequences by a
Helly selection argument \cite{rudin}, but they do not guarantee that the full
sequence of priors selects a unique profile: a family can alternate
between two constructions on even and odd $n$. At the prior level one
therefore needs a coherent construction, or must take profile convergence itself as a hypothesis. 
In the next chapter, as well as the appendices, we give examples and explain why they converge neatly.

\section{Polynomial degree as complexity prior}\label{sec:shapeprior}

What could the TDL of an explicit simplicity-based prior look like? Unfortunately, but
not unexpectedly, we also must grapple with no-free-lunch \cite{wolpert96}. It is not possible to find 
general circuit complexities for large $n$ \cite{kabanetscai00}: Appendix~\ref{app:gibbs} builds
a complexity prior that way, weighting each map by the size of the
smallest circuit computing it, and that measure runs out of road. However, Werner has 
another avenue at his disposal: write the boolean map as a polynomial in the question inputs, 
and consider the complexity to be the \emph{degree of that polynomial}. He can construct a prior
of decreasing faith with this degree, which turns out to work elegantly. It is \emph{constructive}: computable from the truth table by a fast
transform, thus tractable at every $n$.

There are two guiding insights in this chapter.
First, we are interested in demonstrating that the leverage in the TDL can be more than
a hyperbola. In order for this to happen, the rate of information $\gamma(t)$ cannot be a
flat constant in the bulk. This flatness is conspicuously difficult to escape. Many naive
constructions are virtually ``finding out what latent set of maps Werner is in''. The simplest
example was ``spike-or-not'' of Section~\ref{sec:spike}. 
Even when this is made richer with more variation, it still costs
 $o(Q)$ bits to learn the set label, and therefore concludes around $t=0^+$. Once the set is
 settled (no pun intended), the rate $\gamma(t)$ inside is often constant. Instead, we
 need a construction that maintains stores of information that don't settle until macroscopic times.
 The polynomial degree prior has that property.

 Second, weighting maps by
complexity through a single temperature, with an exponentially vanishing weight with degree like the Gibbs ensemble does, reduces 
to only effectively consider a thin shell of contributing maps. At low degrees, there are too few maps. At high, there's too little weight.

This may be desirable
in statistical physics to find averages, but as a philosophical choice, Werner won't rule anything out
in his learning journey. The repair
is to \emph{sample} maps from a shared quota of weight per degree, which
keeps every shell, and every single map, at positive weight all
the way into the thermodynamic limit. 

Throughout this chapter the
answer size is $m = 1$, so $u = \tfrac12$ and a fresh uninformed answer
carries at most one bit. As we saw before, larger $A$ mainly changes the rate of
culling, and behaves much like $m$ independent maps being learned simultaneously. 
The single-bit case is fully instructive, and costs little generality.

\subsection{Polynomials over
\texorpdfstring{$\mathbb F_2$}{F2}, and the butterfly}

All linear algebra in this chapter is over the smallest field there
is, the two-element field $\mathbb F_2$, whose two operations are
\begin{equation}
  \mathbb F_2 = \{0, 1\},
  \qquad
  b \oplus b' \;=\; b + b' \bmod 2 \quad (\text{XOR}),
  \qquad
  b \, b' \quad (\text{AND}).
  \label{eq:f2}
\end{equation}
Three observations:
\begin{enumerate}
  \item Every element is its own additive
inverse, $b \oplus b = 0$, so there is no subtraction and there are no
minus signs: a term contributed twice cancels.
\item Every element is
idempotent, $b^2 = b$, so no variable is ever squared, and a product
of input bits is fixed by \emph{which} bits it contains, never by how
often. 
\item All matrices, vectors, ranks and inverses below are read mod
$2$.
\end{enumerate} 
These define the rules of this space and the functions that inhabit it.

As a consequence, we only need to consider \emph{monomials} in the input bits $b_{n-1} \cdots b_0$ of a question.
Each monomial is labeled by the set of bit positions it
contains: $\phi = \prod_i b_i$ is the map answering $1$
exactly when every bit is $1$, the AND of those bits,
and the empty set gives the empty product, the constant map $1$. Write, for example,
$\mathcal B = \{b_1, b_0\}$ for a set of input bits, and the subscript
on a coefficient $\rho\in \{0,1\}$ is tantamount to the monomial itself: $\rho_{b_1 b_0}$ represents $\rho_{b_1 b_0}\cdot b_1\cdot b_0$, the monomial made from
$\mathcal B$, just as $\rho_\emptyset$ belongs to the constant. There
are $Q = 2^n$ bit sets, hence $Q$ monomials. Creating functions in this space is akin
to deciding which monomials are present, which $\rho_{\mathcal B}$ are one (remember property 1: no non-unit coefficients).
There are $2^Q = N$ ways to
switch them on and off: exactly as many as there are maps, and the
correspondence is one to one. 

One more piece of vocabulary. The \textbf{support} of a question is the
set of bit positions where it holds a $1$,
\begin{equation}
  \operatorname{supp}(q) \;=\; \{\, i : b_i(q) = 1 \,\}.
  \label{eq:supp}
\end{equation}

A question and a set of bits are the same data in two manifestations,
and we use the identification freely: $\mathcal B$ names a monomial $\rho_{\mathcal{B}}$
and equally the question whose $1$s sit exactly at $\mathcal B$. The
condition $\operatorname{supp}(q') \subseteq \mathcal B$ then picks out
the \textbf{subcube below} $\mathcal B$: the $2^{|\mathcal B|}$
questions $q'$ that are $0$ everywhere outside $\mathcal B$ and free inside
it.

\begin{example}
  For instance at $n=4$, $q=1010$ has $\operatorname{supp}(q) = \{b_1, b_3\}$. If this is $q\simeq \mathcal{B}$,
  the subcube below is $q'\in \{1010, 1000, 0010, 0000\}$.
\end{example}

Every map now has a unique vector of single-bit coefficients
$\rho_{\mathcal B} \in \{0,1\}$, saying whether monomial $\mathcal B$
is present or absent, and the dictionary between the two descriptions
is the \emph{algebraic normal form},
\begin{equation}
  \phi(q) \;=\; \bigoplus_{\mathcal B} \rho_{\mathcal B}
                \prod_{i \in \mathcal B} b_i(q)
          \;=\; \bigoplus_{\mathcal B \,\subseteq\,
                           \operatorname{supp}(q)} \rho_{\mathcal B}.
  \label{eq:anf}
\end{equation}
It is worth being explicit about which symbol does what. Above,
$q$ is the free variable, the question being asked, and $\mathcal B$ a
dummy running over all $Q$ monomials. This is \emph{evaluation},
coefficients in, an answer out. The middle expression is the same sum
with the dead ($\rho=0$) terms dropped, since $\prod_{i \in \mathcal B} b_i(q)$
equals $1$ precisely when $\mathcal B \subseteq \operatorname{supp}(q)$. How to go the other way?
\begin{equation}
  \rho_{\mathcal B} \;=\;
    \bigoplus_{\operatorname{supp}(q) \,\subseteq\, \mathcal B}
    \phi(q) .
\end{equation}
Conversely, here $\mathcal B$ is the free variable, the coefficient being
extracted, and $q$ the dummy, running over the subcube below
$\mathcal B$: this is \emph{inversion}, truth table in, one coefficient
out. The dictionary is both identities, the second formula being M\"obius inversion. The
\textbf{degree} $\deg\phi$ is the largest $|\mathcal B|$ carrying
$\rho_{\mathcal B} = 1$; it is the complexity notion of this chapter.

Both directions are in fact literally the same matrix. Write $T$ for the
\emph{truth table} of a map read as a vector, $T[q] = \phi(q)$, with
the questions in the descending order of Table~\ref{tab:sixteen}, and
$\rho$ for the coefficient vector with the bit sets in the matching
order. Per bit the kernel is upper triangular, and on $n$ bits its
tensor power is the subset indicator: with rows and columns descending
$11, 10, 01, 00$ (per bit: $1, 0$),
\begin{equation}
  Z = \begin{pmatrix} 1 & 1 \\ 0 & 1 \end{pmatrix},
  \qquad
  \big(Z^{\otimes n}\big)_{\mathcal B, q}
    = \mathbf 1[\, \operatorname{supp}(q) \subseteq \mathcal B \,],
  \qquad
  Z^{\otimes 2} =
  \begin{pmatrix}
    1 & 1 & 1 & 1 \\
    0 & 1 & 0 & 1 \\
    0 & 0 & 1 & 1 \\
    0 & 0 & 0 & 1
  \end{pmatrix},
  \label{eq:zkernel}
\end{equation}
where $\mathbf 1[\cdot]$ evaluates a statement to a bit (1 is True, 0 is False). 
The matrix is called the \emph{XOR-butterfly}, owing to the mod-2 space.
This way, row
$\mathcal B$ is the indicator of the subcube below $\mathcal B$. Under
the identification of bit sets with questions the two halves of
\eqref{eq:anf} are the one matrix multiplying the two vectors,
\begin{equation}
  \rho \;=\; Z^{\otimes n} \, T,
  \qquad
  T \;=\; Z^{\otimes n} \rho ,
  \label{eq:zpair}
\end{equation}
at any $n$: to read off a map's polynomial, transform its truth table;
to tabulate a polynomial, transform its coefficients. The two are the
same operation because over $\mathbb F_2$ the kernel is its own
inverse, as the one-bit case already shows: $Z^2$ has entries
$1, 2, 0, 1$, and $2 = 0$, so $Z^2 = I$ and
$(Z^{\otimes n})^2 = (Z^2)^{\otimes n} = I$ by extension.

In practice, computing $\rho$ is an efficient operation up to high $n$ due to its tensor product nature. See an example
on the $(n,m)=(2,1)$ space.

\begin{example}
Take NAND, map $j = 7$ of Table~\ref{tab:sixteen}, which answers $0$
only when both input bits are $1$, so that
\begin{equation*}
  T = \big(\phi(11), \phi(10), \phi(01), \phi(00)\big) = (0,1,1,1).
\end{equation*}
Its four supports are $\operatorname{supp}(11) = \{b_1, b_0\}$,
$\operatorname{supp}(10) = \{b_1\}$,
$\operatorname{supp}(01) = \{b_0\}$ and
$\operatorname{supp}(00) = \emptyset$: the same four bit sets that
label the four monomials. For each $\mathcal B$, gather the questions
whose support fits inside it. As the set grows, so does its subcube:
\begin{align*}
  \mathcal B &= \emptyset &
    \{q : \operatorname{supp}(q) \subseteq \mathcal B\}
      &= \{00\}, \\
  \mathcal B &= \{b_0\} &
    \{q : \operatorname{supp}(q) \subseteq \mathcal B\}
      &= \{01, 00\}, \\
  \mathcal B &= \{b_1\} &
    \{q : \operatorname{supp}(q) \subseteq \mathcal B\}
      &= \{10, 00\}, \\
  \mathcal B &= \{b_1, b_0\} &
    \{q : \operatorname{supp}(q) \subseteq \mathcal B\}
      &= \{11, 10, 01, 00\} .
\end{align*}
Each coefficient is the XOR of $T$ over its own subcube, so the four
lines below have identical shape and only the subcube grows:
\begin{align*}
  \rho_\emptyset
    &= \phi(00)
    &&= 1
    &&= 1, \\
  \rho_{b_0}
    &= \phi(01) \oplus \phi(00)
    &&= 1 \oplus 1
    &&= 0, \\
  \rho_{b_1}
    &= \phi(10) \oplus \phi(00)
    &&= 1 \oplus 1
    &&= 0, \\
  \rho_{b_1 b_0}
    &= \phi(11) \oplus \phi(10) \oplus \phi(01) \oplus \phi(00)
    &&= 0 \oplus 1 \oplus 1 \oplus 1
    &&= 1 .
\end{align*}
Stacking those four lines is \eqref{eq:zpair} at $n = 2$,
\begin{equation*}
  \begin{pmatrix}
    \rho_{b_1 b_0} \\ \rho_{b_1} \\ \rho_{b_0} \\ \rho_\emptyset
  \end{pmatrix}
  =
  \begin{pmatrix}
    1 & 1 & 1 & 1 \\
    0 & 1 & 0 & 1 \\
    0 & 0 & 1 & 1 \\
    0 & 0 & 0 & 1
  \end{pmatrix}
  \begin{pmatrix}
    \phi(11) \\ \phi(10) \\ \phi(01) \\ \phi(00)
  \end{pmatrix}
  =
  \begin{pmatrix}
    0 \oplus 1 \oplus 1 \oplus 1 \\ 1 \oplus 1 \\ 1 \oplus 1 \\ 1
  \end{pmatrix}
  =
  \begin{pmatrix}
    1 \\ 0 \\ 0 \\ 1
  \end{pmatrix} ,
\end{equation*}
where the two middle rows show the cancellation that makes a monomial
absent.

Read the other way, \eqref{eq:zpair} evaluates. Ask $q = 01$: its
support is $\{b_0\}$, so row $01$ gives
$\phi(01) = \rho_\emptyset \oplus \rho_{b_0} = 1 \oplus 0 = 1$, which
is what $b_1 b_0 \oplus 1$ returns at $b_1 = 0$, $b_0 = 1$. Hence
$\mathrm{NAND} = b_1 b_0 \oplus 1$, of degree $2$: the top monomial is
present, NAND is not affine, and its truth table knows it.
\end{example}

For readers of Appendix~\ref{app:correlators}: $Z$ is the triangular
half of the Walsh--Hadamard kernel \eqref{eq:hadamard}, since in the
same $1,0$ ordering
\begin{equation}
  W = \begin{pmatrix} -1 & 1 \\ 1 & 1 \end{pmatrix}
  = Z \begin{pmatrix} -2 & 0 \\ 0 & 1 \end{pmatrix} Z^{\mathsf T}.
  \label{eq:wfactor}
\end{equation}
The ledger transform is two M\"obius passes with
a $(-2)^{|T|}$ rescaling between them, and reducing mod $2$ leaves
exactly the XOR butterfly.

\begin{example}
The butterfly on every map of the guiding example, truth tables
$T = (\phi(11), \phi(10), \phi(01), \phi(00))$ -- so that $T$ read as
a binary numeral \emph{is} the index $j$ -- and coefficients ordered
$(\rho_{b_1 b_0}, \rho_{b_1}, \rho_{b_0}, \rho_\emptyset)$.

\begin{center}
\begin{tabular}{r l c c l c}
\toprule
$j$ & map & $T$ & $\rho$ & polynomial & $\deg$ \\
\midrule
0 & FALSE & $(0,0,0,0)$ & $(0,0,0,0)$ & $0$ & $0$ \\
1 & NOR & $(0,0,0,1)$ & $(1,1,1,1)$ &
  $b_1b_0 \oplus b_1 \oplus b_0 \oplus 1$ & $2$ \\
2 & $\lnot b_1 \wedge b_0$ & $(0,0,1,0)$ & $(1,0,1,0)$ & $b_1b_0 \oplus b_0$ & $2$ \\
3 & $\lnot b_1$ & $(0,0,1,1)$ & $(0,1,0,1)$ & $b_1 \oplus 1$ & $1$ \\
4 & $b_1 \wedge \lnot b_0$ & $(0,1,0,0)$ & $(1,1,0,0)$ & $b_1b_0 \oplus b_1$ & $2$ \\
5 & $\lnot b_0$ & $(0,1,0,1)$ & $(0,0,1,1)$ & $b_0 \oplus 1$ & $1$ \\
6 & XOR & $(0,1,1,0)$ & $(0,1,1,0)$ & $b_1 \oplus b_0$ & $1$ \\
7 & NAND & $(0,1,1,1)$ & $(1,0,0,1)$ & $b_1b_0 \oplus 1$ & $2$ \\
8 & AND & $(1,0,0,0)$ & $(1,0,0,0)$ & $b_1b_0$ & $2$ \\
9 & XNOR & $(1,0,0,1)$ & $(0,1,1,1)$ & $b_1 \oplus b_0 \oplus 1$ & $1$ \\
10 & $b_0$ & $(1,0,1,0)$ & $(0,0,1,0)$ & $b_0$ & $1$ \\
11 & $b_1 \to b_0$ & $(1,0,1,1)$ & $(1,1,0,1)$ & $b_1b_0 \oplus b_1 \oplus 1$ & $2$ \\
12 & $b_1$ & $(1,1,0,0)$ & $(0,1,0,0)$ & $b_1$ & $1$ \\
13 & $b_0 \to b_1$ & $(1,1,0,1)$ & $(1,0,1,1)$ & $b_1b_0 \oplus b_0 \oplus 1$ & $2$ \\
14 & OR & $(1,1,1,0)$ & $(1,1,1,0)$ & $b_1b_0 \oplus b_1 \oplus b_0$ & $2$ \\
15 & TRUE & $(1,1,1,1)$ & $(0,0,0,1)$ & $1$ & $0$ \\
\bottomrule
\end{tabular}
\end{center}
The guiding example in both bases: every map of
Table~\ref{tab:sixteen} as a truth table $T$ and as a coefficient
vector $\rho$, with its polynomial and its degree.

Read the table from both ends at once: $j$ and $15 - j$ are negations
of each other, and their coefficient vectors differ in $\rho_\emptyset$
alone, negation flipping the constant and nothing else. Four maps are
fixed points of the butterfly, $\rho = T$: FALSE, XOR, AND and OR.
The eight with $\rho_{b_1b_0} = 0$ are the affine ones, class
$\mathrm{RM}(1,2)$, and the other eight are the curved ones that the
top monomial buys.
\end{example}

\subsection{Complexity classes and the sampling rule}

Using the dictionary, every map is in bijection with a polynomial, and each polynomial has a degree.
Grade the hypothesis space by degree. The class of all maps of degree
at most $d$ is a linear space of dimension
\begin{equation}
K_d = \sum_{i \le d} \binom ni.
\end{equation}
$K_d$ counts one coefficient bit $\rho_{\mathcal B}$ per allowed
monomial. It is a linear space because polynomials distribute, and any sum of polynomials, each with degree $\le d$, is again
a polynomial of degree $\le d$, and therefore the space is closed.

The classes run from the two constant polynomials $\rho_\emptyset \in \{0,1\}$
($d = 0$) to the full map space ($d = n$). In coding theory this
class is the Reed-Muller code $\mathrm{RM}(d,n)$ \cite{macwilliams}. We posit two definitions to avoid confusion:
\begin{itemize}
\item the \textbf{class} $d$ is the union of all maps of degree at
most $d$. Classes are nested.
\item A class's outermost layer, the maps of degree
exactly $d$, is the \textbf{shell} $d$. Shells are disjoint. 
\end{itemize} 
The class's \textbf{rate}
is its dimension per question,
\begin{equation}
  R_{n,d} \;=\; \frac{K_d}{Q} \;\in\; (0,1],
  \label{eq:rmrate}
\end{equation}
and it will turn out to be the fraction of a full run at which the
class is ``learned out'', i.e., Werner has settled whether the map $\psi$ is in this class or not. 
Consecutive rates slice $(0,1]$ into
\textbf{rate gaps} $(R_{n,d-1}, R_{n,d}]$ of lengths
\begin{equation}
  R_{n,d} - R_{n,d-1} = \binom nd / 2^n \le \sqrt{2/\pi n}.
  \label{eq:rategap}
\end{equation}
This partition refines as $n$ grows. Rate gaps can be seen as the proportional thickness of the shells.

\begin{example}
  This graphic explains the stochastic process that samples $\psi$. 
  Sampling $\sim F$ determines a class (a maximum degree $d$). Inside that linear space, any polynomial is equally likely.
  Higher classes have exponentially more maps in them.
The graphic illustrates $n = 4$: take one box per monomial, grouped by $d$ into shells. Each box is one coefficient bit, which has
50-50 odds to be included in the true polynomial of $\psi$.
So $K_d$, the number of monomials under consideration
\emph{is} the class $\mathrm{RM}(d,4)$ measured in bits, holding
$2^{K_d}$ maps, and shell $d$ is the $\binom 4d$ boxes the class
adds to the previous. The boxes lie ordered along a sampled line. 
Dividing the position of the edge of a class by
$Q = 16$ puts the class boundaries at the rates $R_{4,d} = K_d/16$,
with the shells inhabiting the rate gaps. 

The shell widths
$1, 4, 6, 4, 1$ below the boxes are the populations' logarithmic footprints, while
the map \emph{counts} $2^{K_d}$ explode left to right.

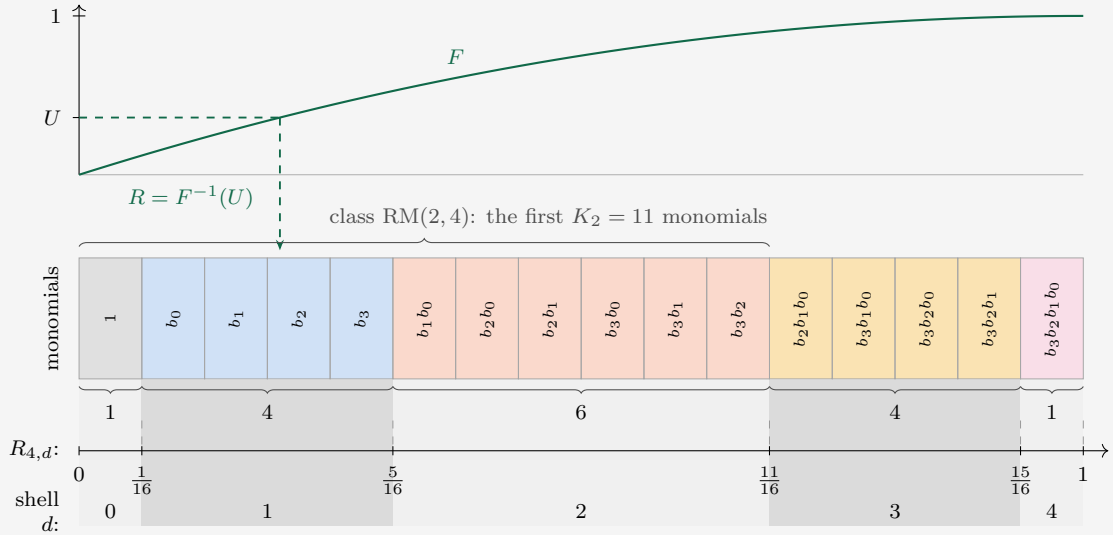
\begin{center}
\begin{tikzpicture}[x=0.83cm, y=1cm]
  \draw[black!30] (0,2.7) -- (16,2.7);
  \draw[->] (0,2.7) -- (0,4.95);
  \draw (-0.1,4.8) -- (0.1,4.8);
  \node[font=\scriptsize, left] at (-0.12,4.8) {$1$};
  \draw[cf1!60!black, thick]
    plot[domain=0:16, samples=60]
    (\x, {2.7 + 2.1*(1-(1-\x/16)^2)});
  \node[font=\scriptsize, cf1!60!black, anchor=south]
    at (6,4.03) {$F$};
  \draw (-0.1,3.456) -- (0.1,3.456);
  \node[font=\scriptsize, left] at (-0.12,3.456) {$U$};
  \draw[cf1!60!black, dashed, line width=0.7pt]
    (0,3.456) -- (3.2,3.456);
  \draw[cf1!60!black, dashed, line width=0.7pt,
    -{Stealth[length=4pt]}] (3.2,3.456) -- (3.2,1.7);
  \node[font=\scriptsize, cf1!60!black, anchor=east]
    at (2.95,2.4) {$R = F^{-1}(U)$};
  \draw[decorate, decoration={brace, amplitude=3pt}, black!70]
    (0,1.75) -- (11,1.75) node[pos=0.68, above=3pt, font=\scriptsize]
    {class $\mathrm{RM}(2,4)$: the first $K_2 = 11$ monomials};
  \node[font=\scriptsize, rotate=90] at (-0.45,0.8) {monomials};
  \fill[black!12]  (0,0)  rectangle (1,1.6);
  \fill[cf2!22]    (1,0)  rectangle (5,1.6);
  \fill[cf3!25]    (5,0)  rectangle (11,1.6);
  \fill[cf4!30]    (11,0) rectangle (15,1.6);
  \fill[cf5!25]    (15,0) rectangle (16,1.6);
  \foreach \i/\lab in {0/{1}, 1/{b_0}, 2/{b_1}, 3/{b_2}, 4/{b_3},
    5/{b_1b_0}, 6/{b_2b_0}, 7/{b_2b_1}, 8/{b_3b_0}, 9/{b_3b_1},
    10/{b_3b_2}, 11/{b_2b_1b_0}, 12/{b_3b_1b_0}, 13/{b_3b_2b_0},
    14/{b_3b_2b_1}, 15/{b_3b_2b_1b_0}}{
    \draw[black!35, line width=0.3pt] (\i,0) rectangle ({\i+1},1.6);
    \node[font=\tiny, rotate=90] at ({\i+0.5},0.8) {$\lab$};}
  \foreach \l/\r/\sh in {0/1/6, 1/5/14, 5/11/6, 11/15/14, 15/16/6}
    \fill[black!\sh] (\l,-0.02) rectangle (\r,-1.95);
  \foreach \l/\r/\lab in {0/1/{$1$}, 1/5/{$4$}, 5/11/{$6$},
    11/15/{$4$}, 15/16/{$1$}}{
    \draw[decorate, decoration={brace, mirror, amplitude=2.5pt},
      black!70] (\l,-0.1) -- (\r,-0.1);
    \node[font=\scriptsize, below] at ({(\l+\r)/2},-0.22) {\lab};}
  \draw[->] (0,-0.95) -- (16.4,-0.95);
  \foreach \x/\lab in {0/{$0$}, 1/{$\tfrac1{16}$}, 5/{$\tfrac5{16}$},
    11/{$\tfrac{11}{16}$}, 15/{$\tfrac{15}{16}$}, 16/{$1$}}{
    \draw (\x,-1.01) -- (\x,-0.89);
    \node[font=\scriptsize, below] at (\x,-1.03) {\lab};}
  \node[font=\scriptsize, anchor=east] at (-0.15,-0.95)
    {$R_{4,d}$:};
  \foreach \x in {1,5,11,15,16}
    \draw[black!40, line width=0.3pt, dashed] (\x,-0.55) -- (\x,-0.89);
  \foreach \l/\r/\lab in {0/1/{$0$}, 1/5/{$1$}, 5/11/{$2$},
    11/15/{$3$}, 15/16/{$4$}}
    \node[font=\scriptsize] at ({(\l+\r)/2},-1.75) {\lab};
  \node[font=\scriptsize, anchor=east, align=right] at (-0.15,-1.75)
    {shell\\$d$:};
\end{tikzpicture}
\end{center}

So far, the structure is fixed and none of it depends on $F$.
The statistic enters only through the dart: draw $U$ uniformly on
the vertical axis, bounce off the graph of $F$, and land at
$R = F^{-1}(U)$, drawn here for $F(x) = 1-(1-x)^2$ (the smaller
of two uniform draws: the $r = 2$ member of the
constant-leverage family \eqref{eq:Lmin}), which favors simple (low-degree) maps.
The rate-gap the dart hits determines $D$, the maximum $d$. For this dart, $D=1$,
and we sample $\psi$ uniformly from polynomials of the form $\rho_\emptyset \oplus \bigoplus_i \rho_{b_i}b_i$.

\end{example}

The exponential explosion in numbers: the share of maps of degree
exactly $d$ is $2^{K_d - Q} - 2^{K_{d-1} - Q}$, and

\begin{table}[H]
\begin{center}
\begin{tabular}{l l cccc}
\toprule
degree & share & $n=2$ & $n=4$ & $n=6$ & $n=8$ \\
\midrule
$n$ & $\tfrac12$
  & $50\%$ & $50\%$ & $50\%$ & $50\%$ \\
$n-1$ & $\tfrac12 - 2^{-(n+1)}$
  & $37.5\%$ & $46.9\%$ & $49.2\%$ & $49.8\%$ \\
$n-2$ & $\approx 2^{-(n+1)}$
  & $12.5\%$ & $3.1\%$ & $0.78\%$ & $0.20\%$ \\
$n-3$ & $\approx 2^{-(n+1+\binom n2)}$
  & - & $4.6\cdot10^{-4}$ & $2.4\cdot10^{-7}$ & $7.3\cdot10^{-12}$ \\
$\le n-4$ & rest
  & - & $3.1\cdot10^{-5}$ & $2.3\cdot10^{-13}$ & $1.0\cdot10^{-28}$ \\
\bottomrule
\end{tabular}
\end{center}
\caption{The share of all $2^Q$ maps carried by each degree. Half sit at full degree $n$ at every size, and the shells below thin out geometrically.}
\label{tab:degreeshare}
\end{table}

Half of all maps sit at full degree, all but a $2^{-(n+1)}$ sliver in
the top two shells, and everything below is doubly exponentially rare:
this is the population contour any per-map weight has to compensate. In our case, we allocate
a macroscopic amount of weight to each shell by construction. Then divide that weight among 
the constituent maps. This scales nicely in the TDL, to never effectively rule out any degree, but also keep nonzero weight on each individual map.
\begin{enumerate}
  \item draw a single random number $R \in [0,1]$ (the
    \textbf{statistic}) from a fixed law with CDF
    $F(x) = \Pr(R \le x)$ and, when it exists, density $F'$;
  \item let $D$ be the class whose rate gap contains $R$;
  \item draw the truth uniformly from class $D$: each \emph{allowed} monomial
    coefficient $\rho$ by a fair coin.
\end{enumerate}
So $R$ and $D$ only decide which monomials are ruled out as too high a degree.
The $R_{n,d}$ form the deterministic grid that $R$ lands in.
Only $d \le D$ are allowed: those with $R_{n,d-1} < R$.

 The class
label $D$ is \emph{latent}, meaning Werner never observes it. It is one value
among $n+1$, worth at most $\log_2(n+1)$ bits. Due to nested classes, a map $\phi_j$ of degree $d$ can result
from any $D \ge d$. To find the probability of that map, we integrate over $D$.

\begin{equation}
  w_d \;=\; F(R_{n,d}) - F(R_{n,d-1}),
  \qquad
  p_j \;=\; \sum_{d \ge \deg\phi_j} w_d\, 2^{-K_d},
  \label{eq:shapepj}
\end{equation}
with $\sum_d w_d = 1$ telescoping. If $F$ increases across every gap
(any positive density), then every $w_d > 0$, and two properties
follow trivially: 
\begin{itemize}
  \item $p_j$ is \emph{strictly decreasing in the degree},
each step up losing a factor of order $2^{-\binom nd}$
\item $p_j \ge w_n 2^{-Q} > 0$ for every single map:
full support, at every $n$ and in the limit.
\end{itemize}
Every shell, and every map, gets a chance, and still it's a genuine Occam prior. 
Semantically, $F$ budgets
the \emph{shells} while nestedness orders the \emph{maps}: $w_d$ is,
up to an exponentially suppressed sliver leaking in from higher classes, the
total weight of the degree-$d$ shell, split evenly within it. 

A judicious choice of $F$ is what puts the correct portion of
aggregate belief on the complexity that Werner expects from nature.
The per-map Occam ordering holds for every $F$, and $F$ decides
only how much aggregate weight each shell carries. The Occam ordering comes 
from the natural explosion of maps at higher degrees, sharing the shell's marginal weight.
The heaviest shell itself is rarely the simplest one: it is whichever
rate gap carries the most $F$-mass. The $d$ maximizing
$w_d$ from equation~\eqref{eq:shapepj} is a contest between the gap width
$\binom nd/2^n$, widest at $d \approx n/2$, and the density of $F$
there. See Example~\ref{ex:shellsplit} at the end of the chapter
for an illustration.

\begin{example}
A geometric interpretation of the prior \eqref{eq:shapepj} at $(n,m) = (2,1)$ with the guiding
statistic $F(x) = 1-(1-x)^2$, which is concave (the $r = 2$ member of \eqref{eq:Fmin}).
The three classes have
\begin{align*}
  (K_0, K_1, K_2) &= (1, 3, 4), \\
  (w_0, w_1, w_2) &= (\tfrac7{16}, \tfrac12, \tfrac1{16}),
\end{align*}
so the rates $K_d/Q$ are $\tfrac14$, $\tfrac34$ and $1$. Class $d$
offers each map it contains a rectangle: width $w_d$, the chance the lottery calls that
class, times height $2^{-K_d}$, the chance the class then calls that map.
A map of degree $d$ lies in every class $D \ge d$, so it collects
every block from $D$ rightward: all the blocks a harder map
collects, and one more besides.

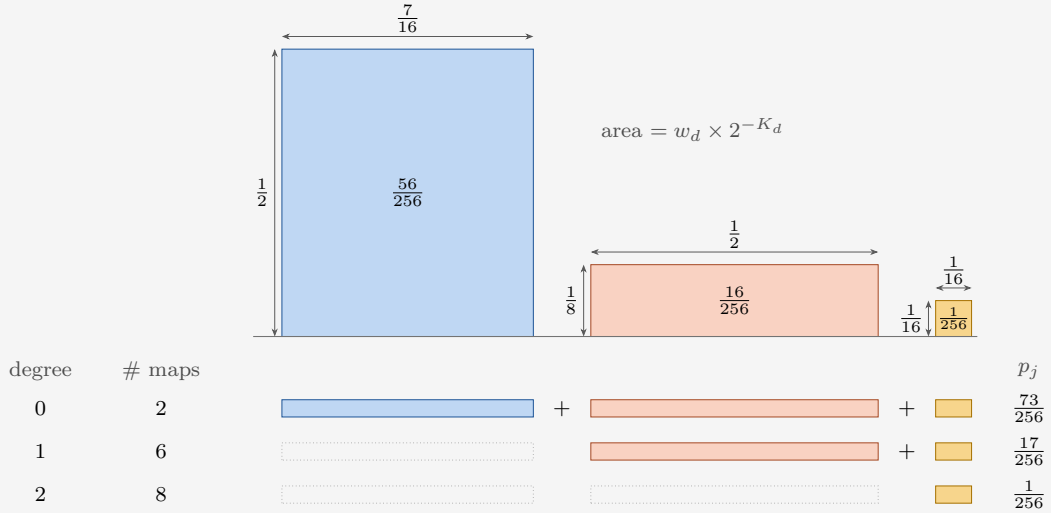
\begin{center}
\begin{tikzpicture}[x=7.6cm, y=7.6cm]
  \fill[cf2!30, draw=cf2!70!black, line width=0.3pt]
    (0,0) rectangle (0.4375,0.5);
  \fill[cf3!30, draw=cf3!70!black, line width=0.3pt]
    (0.5375,0) rectangle (1.0375,0.125);
  \fill[cf4!45, draw=cf4!70!black, line width=0.3pt]
    (1.1375,0) rectangle (1.2,0.0625);
  \node[font=\scriptsize] at (0.21875,0.25) {$\tfrac{56}{256}$};
  \node[font=\scriptsize] at (0.7875,0.0625) {$\tfrac{16}{256}$};
  \node[font=\tiny, inner sep=0pt] at (1.16875,0.03125)
    {$\tfrac{1}{256}$};
  \node[font=\scriptsize, black!70, anchor=west] at (0.5375,0.36)
    {area $= w_d \times 2^{-K_d}$};
  \draw[black!55, line width=0.4pt] (-0.05,0) -- (1.21,0);
  \foreach \xa/\h/\lab in {0/0.5/{\tfrac12}, 0.5375/0.125/{\tfrac18},
      1.1375/0.0625/{\tfrac1{16}}}{
    \draw[{Stealth[length=3pt]}-{Stealth[length=3pt]}, black!65,
      line width=0.35pt] ({\xa-0.0125},0) -- ({\xa-0.0125},\h);
    \node[font=\scriptsize, left, inner sep=1.5pt]
      at ({\xa-0.0125},{\h/2}) {$\lab$};}
  \foreach \xa/\xb/\h/\lab in {0/0.4375/0.5/{\tfrac7{16}},
      0.5375/1.0375/0.125/{\tfrac12}, 1.1375/1.2/0.0625/{\tfrac1{16}}}{
    \draw[{Stealth[length=3pt]}-{Stealth[length=3pt]}, black!65,
      line width=0.35pt] (\xa,{\h+0.022}) -- (\xb,{\h+0.022});
    \node[font=\scriptsize, above, inner sep=1pt]
      at ({(\xa+\xb)/2},{\h+0.024}) {$\lab$};}
  \node[font=\scriptsize, black!70] at (-0.42,-0.06) {degree};
  \node[font=\scriptsize, black!70] at (-0.21,-0.06) {\# maps};
  \node[font=\scriptsize, black!70] at (1.30,-0.06) {$p_j$};
  \fill[cf2!30, draw=cf2!70!black, line width=0.3pt]
    (0,-0.140) rectangle (0.4375,-0.110);
  \fill[cf3!30, draw=cf3!70!black, line width=0.3pt]
    (0.5375,-0.140) rectangle (1.0375,-0.110);
  \fill[cf4!45, draw=cf4!70!black, line width=0.3pt]
    (1.1375,-0.140) rectangle (1.2,-0.110);
  \node[font=\scriptsize] at (0.4875,-0.125) {$+$};
  \node[font=\scriptsize] at (1.0875,-0.125) {$+$};
  \node[font=\scriptsize] at (-0.42,-0.125) {$0$};
  \node[font=\scriptsize] at (-0.21,-0.125) {$2$};
  \node[font=\scriptsize] at (1.30,-0.125) {$\tfrac{73}{256}$};
  \draw[black!35, line width=0.3pt, densely dotted]
    (0,-0.215) rectangle (0.4375,-0.185);
  \fill[cf3!30, draw=cf3!70!black, line width=0.3pt]
    (0.5375,-0.215) rectangle (1.0375,-0.185);
  \fill[cf4!45, draw=cf4!70!black, line width=0.3pt]
    (1.1375,-0.215) rectangle (1.2,-0.185);
  \node[font=\scriptsize] at (1.0875,-0.200) {$+$};
  \node[font=\scriptsize] at (-0.42,-0.200) {$1$};
  \node[font=\scriptsize] at (-0.21,-0.200) {$6$};
  \node[font=\scriptsize] at (1.30,-0.200) {$\tfrac{17}{256}$};
  \draw[black!35, line width=0.3pt, densely dotted]
    (0,-0.290) rectangle (0.4375,-0.260);
  \draw[black!35, line width=0.3pt, densely dotted]
    (0.5375,-0.290) rectangle (1.0375,-0.260);
  \fill[cf4!45, draw=cf4!70!black, line width=0.3pt]
    (1.1375,-0.290) rectangle (1.2,-0.260);
  \node[font=\scriptsize] at (-0.42,-0.275) {$2$};
  \node[font=\scriptsize] at (-0.21,-0.275) {$8$};
  \node[font=\scriptsize] at (1.30,-0.275) {$\tfrac{1}{256}$};
\end{tikzpicture}
\end{center}

Each step up in degree curtails the sum from the left, so $p_j$ falls
with the degree by construction, regardless of $F$: the two constants
of Table~\ref{tab:sixteen} take all three blocks, $\tfrac{73}{256}$,
the six affine maps lose the biggest, $\tfrac{17}{256}$, and the eight
curved maps keep only the last, $\tfrac1{256}$. The blocks are drawn
to scale, illustrating the Occam prior. A concave
$F$ boosts early widths. The small occupancy of low shells boosts early heights. 
The check $2\cdot73 + 6\cdot17 + 8\cdot1 = 256$ closes the distribution.

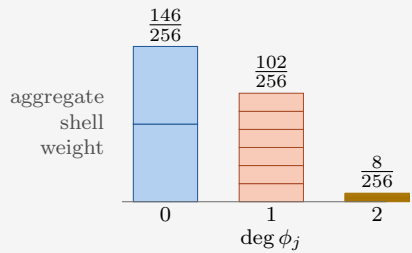
\begin{center}
\begin{tikzpicture}[x=1cm, y=3.6cm]
  \foreach \dg/\nm/\hh/\cc in {0/2/0.28515625/cf2,
      1/6/0.06640625/cf3, 2/8/0.00390625/cf4}{
    \foreach \k in {1,...,\nm}{
      \fill[\cc!35, draw=\cc!70!black, line width=0.25pt]
        ({\dg*1.4},{(\k-1)*\hh}) rectangle ({\dg*1.4+0.85},{\k*\hh});}
    \node[font=\scriptsize, below, inner sep=2pt]
      at ({\dg*1.4+0.425},0) {$\dg$};}
  \draw[black!55, line width=0.4pt] (-0.15,0) -- (3.35,0);
  \node[font=\scriptsize, black!70, align=right, anchor=east]
    at (-0.25,0.285) {aggregate\\shell\\weight};
  \node[font=\small, above, inner sep=2pt]
    at (0.425,0.5703125) {$\tfrac{146}{256}$};
  \node[font=\small, above, inner sep=2pt]
    at (1.825,0.3984375) {$\tfrac{102}{256}$};
  \node[font=\small, above, inner sep=2pt]
    at (3.225,0.03125) {$\tfrac{8}{256}$};
  \node[font=\scriptsize, below, inner sep=2pt]
    at (1.825,-0.08) {$\deg \phi_j$};
\end{tikzpicture}
\end{center}
\end{example}

\subsection{Three terms for the leverage average}

The exact finite law \eqref{eq:Lk-G}, in the TDL chapter's notation
\eqref{eq:finitelev}, needs three ingredients: 

\begin{enumerate}
  \item the prior table entropy $Q\,G_{n,1}$,
  \item the received entropy $G_{n,\ell}$,
  \item the increment $\gamma_{n,\ell}$.
\end{enumerate}
Each reduces to a statement about $F$.

\subsubsection{The table entropy}

A class is a linear code. We'll make the
structure explicit. The \textbf{generator matrix} $\Omega^{(d)}$ of class
$d$ is a $K_d \times Q$ matrix over $\mathbb F_2$. 
Monomials are trivially themselves polynomials, and therefore boolean maps.
$\Omega^{(d)}$'s rows are the monomials $\mathcal B$ that the class contains, evaluated at every question $q$.
\begin{equation}
  \Omega^{(d)}_{\mathcal B,q}
    \;=\; \prod_{i \in \mathcal B} b_i(q)
    \;=\; \mathbf 1[\, \mathcal B \subseteq \operatorname{supp}(q) \,],
  \qquad |\mathcal B| \le d ,
  \label{eq:genmat}
\end{equation}
so that a uniformly random class member is $\psi = \xi^{\mathsf T} \Omega^{(d)}$
with $\xi$ uniformly random on $\{0,1\}^{K_d}$. Every column of $\Omega^{(d)}$ is
nonzero (its constant-monomial entry is $1$), so each single answer is
the image of a uniform $\xi$ under a surjective linear map onto
$\mathbb F_2$: a fair bit, in every class. Any bit that is flipped with 50\% chance is fair, which is what happens when the constant bit is added.
In other words, the initial $M$ is fully uniform.
At every finite $n$, with no limit taken and no reference to $F(x)$,
\begin{equation}
  G_{n,1} = 1,
  \qquad
  H\big(M^{(0)}\big) = Q .
  \label{eq:shapeT1}
\end{equation}
The prior predicts nothing about any single answer; its entire
content is correlational.

\begin{example}
The generator matrix $\Omega$ at $n = 2$: the questions across the top in
descending order, the monomials down the side in degree order, and the
entry where they cross: the value \eqref{eq:genmat} gives. The three boxes are the three classes, each
keeping the first $K_d$ rows, which is what makes them nest:
\begin{center}
\begin{tikzpicture}
  \matrix (m) [matrix of math nodes, nodes={inner xsep=4pt,
    inner ysep=3pt}, row sep=1pt, column sep=16pt] {
    |[minimum height=0.66cm, minimum width=1.15cm]| {}
      & \mathbf{11} & \mathbf{10} & \mathbf{01} & \mathbf{00} \\
    |[minimum height=5pt]| {} & {} & {} & {} & {} \\
    1      & 1 & 1 & 1 & 1 \\
    b_0    & 1 & 0 & 1 & 0 \\
    b_1    & 1 & 1 & 0 & 0 \\
    b_1b_0 & 1 & 0 & 0 & 0 \\
  };
  \coordinate (vx) at ([xshift=5pt]m-1-1.east);
  \coordinate (hy) at ([yshift=-2pt]m-1-1.south);
  \draw[black!55, line width=0.4pt]
    (vx |- m-1-1.north) -- (vx |- m-6-1.south);
  \draw[black!55, line width=0.4pt]
    (m-1-1.west |- hy) -- (m-1-5.east |- hy);
  \coordinate (dtl) at (m-1-1.west |- m-1-1.north);
  \coordinate (dbr) at (vx |- hy);
  \coordinate (dbl) at (m-1-1.west |- hy);
  \coordinate (dtr) at (vx |- m-1-1.north);
  \draw[black!55, line width=0.4pt] (dtl) -- (dbr);
  \node[anchor=south west, inner sep=1.5pt]
    at ($(dbl)!0.2!($(dtl)!(dbl)!(dbr)$)$) {$\mathcal B$};
  \node[anchor=north east, inner sep=1.5pt]
    at ($(dtr)!0.2!($(dtl)!(dtr)!(dbr)$)$) {$\mathbf q$};
  \foreach \rw/\pad/\top/\col in {6/4.5pt/5.5pt/{cf4!65!black},
      5/3pt/4.5pt/{cf3!80!black}, 3/1.5pt/3.5pt/{cf2!80!black}}
    \draw[\col, line width=0.5pt, rounded corners=1.2pt]
      ([shift={(-\pad,-\top)}]m-1-2.west |- m-1-1.south)
      rectangle ([shift={(\pad,-\pad)}]m-1-5.east |- m-\rw-5.south);
  \node[anchor=east] at ([xshift=-6pt]m.west) {$\Omega_{\mathcal B,q} \;=$};
  \node[font=\scriptsize, cf2!80!black, anchor=west]
    at ([xshift=12pt]m-1-5.east |- m-3-5) {$d = 0$};
  \node[font=\scriptsize, cf3!80!black, anchor=west]
    at ([xshift=12pt]m-1-5.east |- m-5-5) {$d = 1$};
  \node[font=\scriptsize, cf4!65!black, anchor=west]
    at ([xshift=12pt]m-1-5.east |- m-6-5) {$d = 2$};
\end{tikzpicture}
\end{center}
A class member is $\psi = \xi^{\mathsf T}\Omega^{(d)}$, the element-wise XOR of the
rows that the coin-sequence $\xi$ switches on. The $K_d$ rows are
independent, so $\xi \mapsto \psi$ is a bijection onto the class:
flipping $K_d$ fair coins, one per allowed monomial, lands on each of
the $2^{K_d}$ maps exactly once, which is step 3 of the sampling rule
and why no map inside a class is favored over another. 
Three examples of coin-sequences yield three maps, at three different class degrees $d$:
\begin{align*}
  \xi &= (1)
    &&\Rightarrow\quad \psi = (1,1,1,1)
    &&= \mathrm{TRUE}, \\
  \xi &= (1,0,1)
    &&\Rightarrow\quad \psi = (1,1,1,1) \oplus (1,1,0,0) = (0,0,1,1)
    &&= \lnot b_1, \\
  \xi &= (1,0,1,1)
    &&\Rightarrow\quad \psi = (0,0,1,1) \oplus (1,0,0,0) = (1,0,1,1)
    &&= b_1 \to b_0 .
\end{align*}
Every column carries the constant row's $1$, so flipping
$\xi_\emptyset$ flips the answer to every question at once: the
$2^{K_d}$ maps pair off into complementary halves and
$M_{a,q} = \tfrac12$ at every question, in every class, for every $F$.
The table starts knowing nothing, $H(M^{(0)}) = Q$. Compare to
\eqref{eq:zkernel}, $\Omega$ is $Z^{\otimes n}$ transposed, its rows
sorted by degree so that every class is a prefix.
\end{example}

\subsubsection{The received entropy}

Let us approach this answer in steps.  
For a straightforward intermediate, suppose \emph{Werner
were told the class}. His prior would instantly collapse to only those maps. 
Inside class $d$ the truth is $\psi = \xi^{\mathsf T}\Omega^{(d)}$ with $\xi$ a string of $K_d$ fair coins,
so his answers are uniform on the image of the columns of $\Omega^{(d)}$ he has asked
and their entropy is that submatrix's rank. Every fresh column (question \& answer)
independent of its predecessors buys one whole bit; once the rank of the 
submatrix seen this way reaches $K_d$, the coins are pinned. 
$\xi$ is known, and every answer
after that is completely determined, and there is no more entropy. Average that behavior over
the classes with weight $w_d = \Pr(D = d)$ and the received entropy
would be settled.

Two things spoil this reasoning at finite $n$, and both dissolve in the TDL.
Asking $K_d$ questions is not the same as reconstructing a full-rank submatrix, because
the columns are not in general position: the smallest dependent set has
$2^{d+1}$ members, the minimum distance of the dual code
$\mathrm{RM}(n{-}d{-}1,n)$ \cite{macwilliams}, and for the middle classes that falls well
short of $K_d$. For instance, four questions against $K_1 = 5$ at $n = 4$, eight
against $K_2 = 37$ at $n = 8$. Four questions whose bit patterns XOR to
zero already determine each other for an affine ($d=1$) map, whatever else is
unsettled. So a question can be redundant early and the rank often is
not full until late: the switch from one bit per question to none is a smeared step,
not a sharp one, though the total entropy encloses area $K_d$ regardless. Only the two
ends escape, the constants ($K_0=1$) and the full space ($K_n = Q$) being the codes for which
every $K_d$ columns really are independent. And Werner is not told the class; he has
to read it off the answers, and that costs information. 

Again, let $\mathcal S$ be the set of questions asked at some point during learning, and $D$ be the class label.
Computing
$H(\psi(\mathcal S), D)$ by the chain rule in both orders prices the
second of these,
\begin{equation}
  H\big(\psi(\mathcal S)\big)
  = \sum_d w_d\, H^{(d)}\big(\psi(\mathcal S)\big)
  + I\big(D; \psi(\mathcal S)\big),
  \qquad
  0 \le I \le \log_2(n+1):
  \label{eq:ungroup}
\end{equation}
mixture entropy is mean class entropy plus what the answers reveal
about the label, and the label is cheap: $o(1)$ bits per question.
Within class $d$ the answers on $\mathcal S$ are
$\xi^{\mathsf T}\Omega^{(d)}_{\cdot,\mathcal S}$, uniform on the image of
the column submatrix, so
$H^{(d)}(\psi(\mathcal S)) = \operatorname{rank}(\Omega^{(d)}_{\cdot,\mathcal S})$
exactly: the number of $\mathbb F_2$-linearly
independent columns, the number of bits needed to describe those
answers.

Averaging \eqref{eq:ungroup} over the $\binom Q\ell$ sets of size $\ell$
turns the left side into $G_{n,\ell}$ of \eqref{eq:Gnk} and the right
side into a mean rank, one per class:
\begin{equation}
  \underbrace{G_{n,\ell}}_{\text{Average}}
  \;=\; \underbrace{\sum_d w_d\, G^{(d)}_{n,\ell}}_{\text{told the class}}
  \;+\; \underbrace{I_{n,\ell}}_{\le\,\log_2(n+1)},
  \qquad
  G^{(d)}_{n,\ell}
  \;:=\; \binom Q\ell^{-1} \sum_{|\mathcal S| = \ell}
    H^{(d)}\big(\psi(\mathcal S)\big),
  \label{eq:Gmix}
\end{equation}
where $I_{n,\ell}$ is the label term of \eqref{eq:ungroup} averaged
over the same sets, inheriting its bound. Only $G_{n,\ell}/Q$ ever
enters the leverage, and $\log_2(n+1)/Q \to 0$, so the label costs a
vanishing fraction of the run: $G_{n,\lfloor tQ\rfloor}/Q$ and
$\sum_d w_d\, G^{(d)}_{n,\lfloor tQ\rfloor}/Q$ have the same limit at
every $t$. It is enough to find the class averages.

\begin{example}
Say Werner knows $d=1$. This means:
\begin{equation*}
  \psi(q) = \rho_\emptyset \oplus \rho_{b_0}b_0(q) \oplus \rho_{b_1}b_1(q)
\end{equation*}
In the matrix below, see how the sum of all questions vanishes for such affine maps (a linear dependence).
\begin{equation*}
  \psi(11) \oplus \psi(10) \oplus \psi(01) \oplus \psi(00) = 4\rho_\emptyset \oplus 2\rho_{b_0} \oplus 2\rho_{b_1} = 0
\end{equation*}
Ask any three questions $\mathcal S$, the covered submatrix
$\Omega^{(1)}_{\cdot,\mathcal S}$ is square and of full rank $3$.
\begin{center}
\begin{tikzpicture}
  \matrix (r) [matrix of math nodes, nodes={inner xsep=4pt,
    inner ysep=3pt}, row sep=1pt, column sep=16pt] {
    |[minimum height=0.66cm, minimum width=1.15cm]| {}
      & \mathbf{11} & \mathbf{10} & \mathbf{01} & \mathbf{00} \\
    |[minimum height=5pt]| {} & {} & {} & {} & {} \\
    1   & 1 & 1 & 1 & 1 \\
    b_0 & 1 & 0 & 1 & 0 \\
    b_1 & 1 & 1 & 0 & 0 \\
  };
  \coordinate (rvx) at ([xshift=5pt]r-1-1.east);
  \coordinate (rhy) at ([yshift=-2pt]r-1-1.south);
  \draw[black!55, line width=0.4pt]
    (rvx |- r-1-1.north) -- (rvx |- r-5-1.south);
  \draw[black!55, line width=0.4pt]
    (r-1-1.west |- rhy) -- (r-1-5.east |- rhy);
  \coordinate (dtl) at (r-1-1.west |- r-1-1.north);
  \coordinate (dbr) at (rvx |- rhy);
  \coordinate (dbl) at (r-1-1.west |- rhy);
  \coordinate (dtr) at (rvx |- r-1-1.north);
  \draw[black!55, line width=0.4pt] (dtl) -- (dbr);
  \node[anchor=south west, inner sep=1.5pt]
    at ($(dbl)!0.2!($(dtl)!(dbl)!(dbr)$)$) {$\mathcal B$};
  \node[anchor=north east, inner sep=1.5pt]
    at ($(dtr)!0.2!($(dtl)!(dtr)!(dbr)$)$) {$\mathbf q$};
  \node[anchor=east] at ([xshift=-6pt]r.west) {$\Omega^{(1)} \;=$};
  \draw[cf3!75!black, line width=0.6pt, rounded corners=1.2pt]
    ([shift={(-3pt,-3.5pt)}]r-1-2.west |- r-1-1.south)
    rectangle ([shift={(3pt,-3pt)}]r-1-4.east |- r-5-4.south);
\end{tikzpicture}
\end{center}
That means the first three questions all have $\gamma_{2,\ell}=1$, they carry a bit
of surprisal. After that, at $t=R_{2,1} = \tfrac34$, through
a system of equations we can infer which monomials are present in the polynomial of $\psi$. The final $q$ cannot have 
any uncertainty left, and $\gamma_{2,3}=0$. In total then, the received entropy is capped at $K_1=3$.
\begin{equation*}
  G_{2,\ell} = \min(\ell, 3)
\end{equation*}

\end{example}

Three facts pin the limit of the class average
$G^{(d)}_{n,\ell} = \mathbb E\,\mathrm{rank} \Omega^{(d)}$. The mean increment
$\gamma^{(d)}_{n,\ell} = \Pr[\text{a fresh column is independent of }
\ell \text{ random ones}]$ is nonincreasing in $\ell$, by rank
submodularity (what is dependent on few columns is dependent on
more). The increments telescope to the dimension,
\begin{equation}
  \sum_{\ell < Q} \gamma^{(d)}_{n,\ell} = K_d,
\end{equation}
the finite-$n$ area theorem. Inside a class $d$, there can be at most $K_d$ linearly independent questions, and at most $K_d$ bits to learn. 
And we must invoke one external result:
Reed-Muller codes achieve capacity on the binary erasure channel.
Kudekar, Kumar, Mondelli, Pfister, \c Sa\c so\u glu and Urbanke
\cite{kkmpsu} showed that revealing any fraction beyond the rate
determines a fresh answer with probability tending to one:
\begin{equation}
  R_{n,d_n} \to R \in (0,1)
  \quad\Longrightarrow\quad
  \lim_{n\to\infty} \gamma^{(d_n)}_{n,\lfloor tQ\rfloor} = 0
  \quad\text{for every fixed } t > R .
  \label{eq:kkmpsu}
\end{equation}
So if the class rate tends to the statistic $R$, the support of the entropy rate tends to below $R$.
For the approach to this limit, see Example~\ref{ex:rmstep}.
Here $m = 1$, which bounds the increment: $\gamma \in [0,1]$.
Such a nonincreasing sequence with area $R_{n,d_n}$ and no mass to the
right of it has no freedom. These characteristics force the shape, so
along that same sequence of classes the increments converge pointwise
away from $t=R$ to a pure step, written with the Heaviside function $\theta$, which
is $1$ when its argument is positive and $0$ otherwise:
\begin{equation}
  \lim_{n\to\infty} \gamma^{(d_n)}_{n,\lfloor tQ\rfloor}
  \;=\; \theta(R - t),
  \qquad
  \frac{G^{(d_n)}_{n,\lfloor tQ\rfloor}}{Q}
  \;\longrightarrow\;
  \int_0^t \theta(R - x)\, dx = \min(t, R):
  \label{eq:heavistep}
\end{equation}
each class pays one full bit per question until its tank runs out at
its own rate, and nothing after. No pointwise limit is asserted at
the step discontinuity $t=R$; its assigned value does not affect the
integral.

So, for a given sampled $R$, class $D$ is pinned, and it is exactly a
sequence of the kind \eqref{eq:heavistep} asks for: $R_{n,D}$ is $R$
rounded up to the top of its gap, and by \eqref{eq:rategap} that gap is
at most $\sqrt{2/\pi n}$ wide, so $R_{n,D} \to R$ and
$G^{(D)}_{n,\lfloor tQ\rfloor}/Q \to \min(t,R)$. All that is left is to
integrate over $R$ to get the true prior average.

Since $\min(t,\cdot)$ is $1$-Lipschitz the same gap bound controls the
expectation, and the class
average converges to an average over the statistic,
\begin{equation}
  \frac{G_{n,\lfloor tQ\rfloor}}{Q}
  \;\longrightarrow\;
  \mathbb E_F\big[\min(t, R)\big]
  \;=\; \int_0^t \big(1 - F(x)\big)\, dx \;=:\; g(t).
  \label{eq:gF}
\end{equation}
The last step is achieved through integration by parts. Pointwise
\begin{equation}
  \min(t,R) = \int_0^t \theta(R - x)\,dx;\quad \mathbb E\,\theta(R - x) = 1 - F(x),
\end{equation}
by the defining sampling process $R \sim F'$.
Received information per question is a mean of the rates per class, weighted by the expected contribution of those classes. 
The correction from the mutual information term from equation \eqref{eq:ungroup} vanishes in the TDL.

\begin{example}[label=ex:rmstep]
Where the step in $\gamma$ comes from. Conditional on the
latent class $D$, the answers on $\mathcal S$ are
$\xi^{\mathsf T}\Omega^{(d)}_{\cdot,\mathcal S}$, so their entropy is the rank
of that column submatrix, and $\gamma^{(d)}_{n,\ell}$ is the chance a
fresh column is still independent of the $\ell$ already asked. 

To see the limit approach, hold the draw fixed at $R = 0.6$ and let $n$ grow. 
The sampling rule hands back whichever class's rate gap contains it, so the class index
climbs with $n$, taking $d = 3, 4, 5, 6$ below, while $R$ itself stays
put; by \eqref{eq:kkmpsu} the increment tends to the black step at
$R$. Outside the window $\gamma$ is flat: one to the left, nothing to
the right.

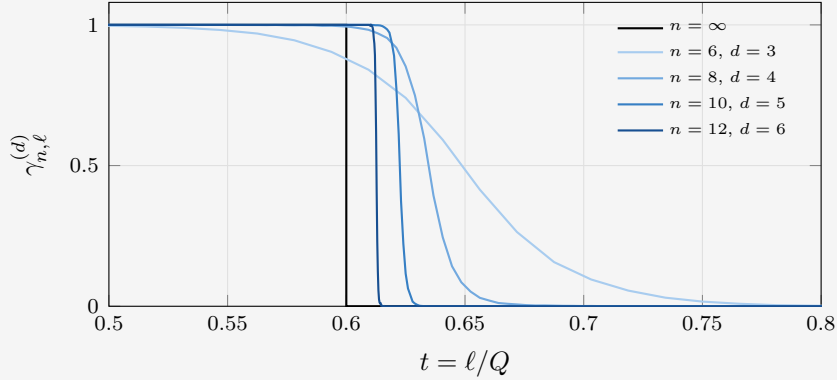
\begin{center}
\begin{tikzpicture}
\begin{axis}[width=11cm, height=5.6cm,
  xlabel={$t = \ell/Q$},
  ylabel={$\gamma^{(d)}_{n,\ell}$},
  xmin=0.5, xmax=0.8, ymin=0, ymax=1.08,
  tick label style={font=\scriptsize}, label style={font=\small},
  grid=major, grid style={black!12},
  legend style={font=\tiny, at={(0.97,0.97)}, anchor=north east,
    draw=none, fill=none},
  legend cell align=left]
\addplot[black, thick] coordinates {(0.5,1) (0.6,1) (0.6,0) (0.8,0)};
\dataplot{cn1, thick}{t}{gamma}{data/rmgamma_n6.dat}
\dataplot{cn2, thick}{t}{gamma}{data/rmgamma_n8.dat}
\dataplot{cn3, thick}{t}{gamma}{data/rmgamma_n10.dat}
\dataplot{cn4, thick}{t}{gamma}{data/rmgamma_n12.dat}
\datalegend{{$n=\infty$}, {$n=6$, $d=3$}, {$n=8$, $d=4$},
  {$n=10$, $d=5$}, {$n=12$, $d=6$}}
\end{axis}
\end{tikzpicture}
\end{center}

Two things converge, at very different speeds. The \emph{shape} goes
fast: the drop from $\gamma = 0.9$ down to $0.1$ measures six
questions at every $n$ here, so as a fraction of the run it closes like
$1/Q$, from $0.094$ at $n = 6$ to $0.0015$ at $n = 12$. The
\emph{position} of the transition goes slowly with $n$: each curve steps at its own rate, which is
$R$ rounded up to the grid and so overshoots by at most the gap width
$\binom nd/2^n \le \sqrt{2/\pi n}$. Here $0.056, 0.037, 0.023,
0.013$. Whatever the
overshoot, each curve encloses exactly its own rate: the increments
telescope to the dimension, $\sum_{\ell<Q}\gamma^{(d)}_{n,\ell} = K_d$,
at every $n$ and with no approximation.
\end{example}

\subsubsection{The increment}

This one now follows for free. The sequence $\ell \mapsto G_{n,\ell}$ is
concave (increments nonincreasing), and slopes of concave functions
converge wherever the limit is differentiable, so at every continuity
point of $F$
\begin{equation}
  \gamma_{n,\lfloor tQ\rfloor}
  \;\longrightarrow\;
  g'(t) = 1 - F(t) = \Pr(R > t).
  \label{eq:gammaF}
\end{equation}
This caps the construction: the mean conditional entropy
of a fresh answer is the probability that the true class is not yet
exhausted. The survival function of the statistic is not a postulate;
it is the increment term of the finite law.
The relative contribution of the label to $\gamma_{n,\ell}$ vanishes with $n$ as
$\log_2(n+1)/Q$. As $n$ grows, the sliver of $I_{n,\ell}$ scales away.

Initially, as we increase $n$, steps emerge. This is because at a given time a class has
either been settled, and contributes no uncertainty, or is still running, and
contributes its full bit (scaled by the probability of that class: the step height) at every question, until the next class
boundary is reached. The number of questions in which $K_d$ rank (total information) is reached inside a class, statistically, is $o(Q)$, 
which rounds the step edges. At low $n$, they bleed together, but at high $n$, the steps become distinct.
The steps hug $1 - F$ more tightly, as $n \to \infty$, by construction of the sampling protocol \eqref{eq:shapepj}.

\begin{example}
This example is meant to illustrate how equation~\eqref{eq:ungroup} behaves in the TDL.
Concisely, the left sum leads to a staircase in $\gamma(t)$, and the mutual information $I$ adds
an extra dusting of uncertainty that lies on top.

The shaded region has height $\gamma(t) = 1 - F(t)$: the sampling distribution. The average over $D$ has contributions from all classes $d\le n$ present.
Dashed vertical lines are the class rate boundaries $R_{n,d}$. 
The horizontal lines are the aggregate class weights, largest below, adding 
lower classes as we rise. This puts the shortest-lived ones on top, so as we move
to the right through time, more and more classes ``settle'', and stop contributing uncertainty.
Increasing $n$ cuts the region into more filigree rate gaps horizontally, and more
class weights vertically.
The two jagged functions on top are two exact increments, the entropy of a single answer, under two
states of knowledge. Dashed: what a fresh answer is worth to someone
already \emph{told} the class, $\sum_d w_d\gamma^{(d)}_{n,\ell}$.
Solid: $\gamma_{n,\ell}$ is what it's worth to Werner, who is not. 
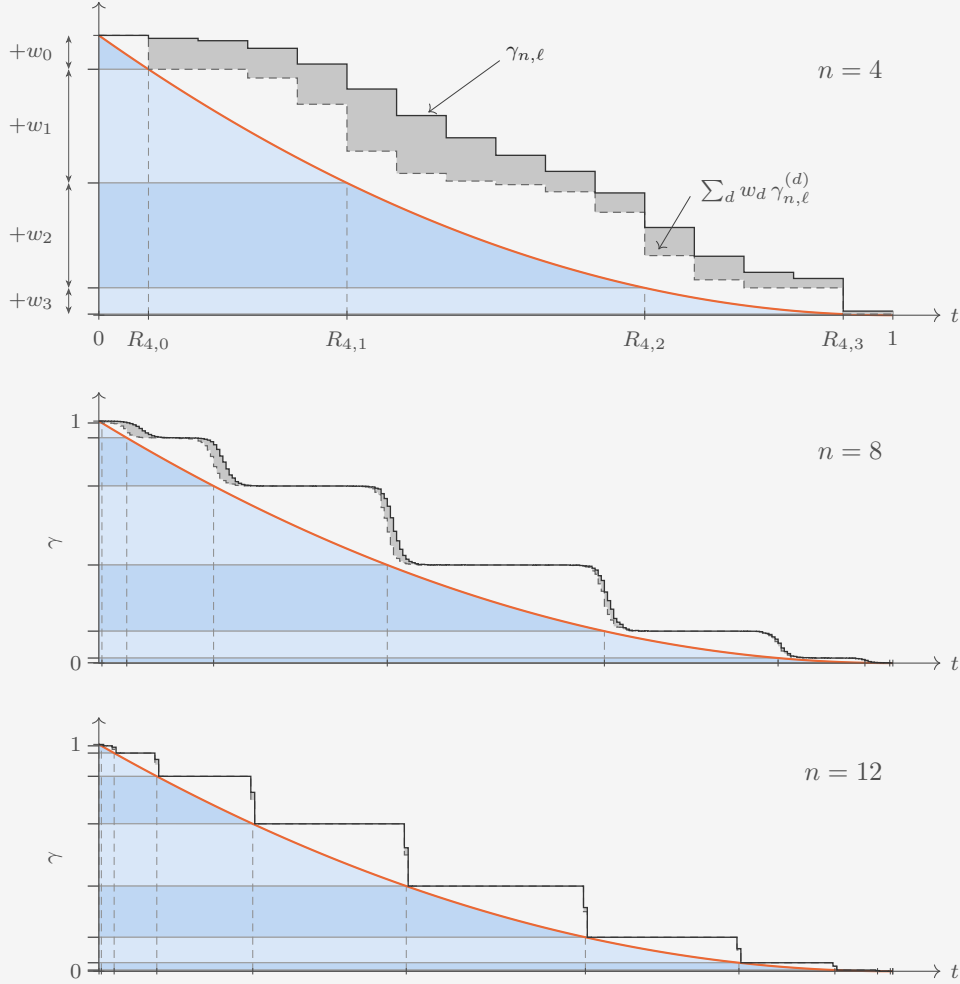
\begin{center}

\end{center}
By \eqref{eq:ungroup}
the mixture is more uncertain than its components by exactly what the
answers say about the label, so the gray shaded region between the two integrates to $I_{n,\ell}$:
the label, and the white between the dashed curve and the region is due to
the coarseness of the rate grid.
Pretending the steps are smooth is \eqref{eq:gF}. Ignoring the gray is \eqref{eq:Gmix}.
\end{example}

\subsection{The macroscopic curve}

Substitute \eqref{eq:shapeT1}, \eqref{eq:gF} and \eqref{eq:gammaF}
into \eqref{eq:Lk-G} and cancel the common factor $Q$; the numerator
tidies from $1 - (1-t)(1-F(t))$ to
\begin{equation}
  L_F(t)
  \;=\; \frac{t + (1-t)\,F(t)}
  {\displaystyle\int_0^t \big(1 - F(x)\big)\, dx} .
  \label{eq:LF}
\end{equation}
This is the master law \eqref{eq:master} realized with 
\begin{equation}
  \eta_0 = 1;\quad \gamma(t) = 1 - F(t); \quad g(t) = \int_0^t(1-F),
\end{equation}
Here $F$ must be a monotone (non-constant) CDF, so $\gamma$ is bounded and
decreasing by construction. The sequence of increments, and hence
the leverage where its denominator is positive, converges at
continuity points of $F$. Step discontinuities are excluded from
this pointwise claim; assigning a value at those does not change the
integral. Degenerate zero-entropy laws and singular endpoint behavior
require separate treatment. The displayed initial endpoint below assumes
$F(0)=0$ and a finite right derivative there. The numerator has its own reading: destroyed uncertainty density is
the asked fraction $t$ plus $F(t)$ per unasked question (column), since an
unasked answer is deducible exactly for maps whose class died before $t$.
Asked plus deduced, over received. Zooming in on the endpoints,
\begin{equation}
  L_F(0) = 1 + F'(0);\quad L_F(1) = \frac 1{\mathbb E_F[R]},
\end{equation}
the curve
starts at baseline plus the density of the statistic at zero, and
completes at the reciprocal mean class rate. 

Conversely, any nonincreasing
profile with values in $[0,1]$ is the survival function $\Pr(R>x)$ of exactly
one law $F(x)$. This means \eqref{eq:LF} is invertible: pick any desired increment profile, and
\eqref{eq:master} at $m = 1$ can achieve it with $F(x) = 1- \gamma(x)$.

A parallel construction, in which we build priors by postulating that blocks of questions have exchangeable answers,
is mathematically complete in the same sense. However, as it is less philosophically aligned with
modeling a learner, it is collected in Appendix~\ref{app:blockpriors}.

\subsection{Toy statistics: min and max draws}

\begin{figure}[H]
\centering
\begin{tikzpicture}
\begin{axis}[width=15.0cm, height=8.0cm,
  xlabel={$t$}, ylabel={$L_F(t)$},
  xmin=0, xmax=1, ymin=0.9, ymax=3.1, domain=0.001:1, samples=200,
  ytick={1,1.5,2,2.5,3},
  tick label style={font=\small}, label style={font=\small},
  grid=major, grid style={black!12},
  legend style={font=\footnotesize, at={(0.5,0.841)}, anchor=center,
    legend columns=3, draw=black!25, fill=white,
    /tikz/every even column/.append style={column sep=8pt}},
  legend cell align=left]
\addlegendimage{cf2, thick}
\addlegendentry{$F_1(x)=x$}
\addlegendimage{cf3, thick}
\addlegendentry{$F^+_2(x)=x^2$}
\addlegendimage{cf1, thick}
\addlegendentry{$F^-_2(x)=1-(1-x)^2$}
\addlegendimage{black!70, only marks, mark=*, mark size=1.0}
\addlegendentry{$n=4$}
\addlegendimage{black!70, only marks, mark=o, mark size=1.1}
\addlegendentry{$n=6$}
\addlegendimage{black!70, only marks, mark=triangle, mark size=1.2}
\addlegendentry{$n=8$}
\addplot[cf2, thick] {2};
\addplot[cf3, thick] {3*(1+x-x^2)/(3-x^2)};
\addplot[cf1, thick] {3};
\addplot[black, dotted, forget plot] {1};
\addplot[cf2, only marks, mark=*, mark size=0.9] table[x=t, y=lev]
  {data/rmfinite_unif_n4.dat};
\addplot[cf2, only marks, mark=o, mark size=1.0] table[x=t, y=lev]
  {data/rmfinite_unif_n6.dat};
\addplot[cf2, only marks, mark=triangle, mark size=1.1] table[x=t, y=lev]
  {data/rmfinite_unif_n8.dat};
\addplot[cf3, only marks, mark=*, mark size=0.9] table[x=t, y=lev]
  {data/rmfinite_max_n4.dat};
\addplot[cf3, only marks, mark=o, mark size=1.0] table[x=t, y=lev]
  {data/rmfinite_max_n6.dat};
\addplot[cf3, only marks, mark=triangle, mark size=1.1] table[x=t, y=lev]
  {data/rmfinite_max_n8.dat};
\addplot[cf1, only marks, mark=*, mark size=0.9] table[x=t, y=lev]
  {data/rmfinite_min2_n4.dat};
\addplot[cf1, only marks, mark=o, mark size=1.0] table[x=t, y=lev]
  {data/rmfinite_min2_n6.dat};
\addplot[cf1, only marks, mark=triangle, mark size=1.1] table[x=t, y=lev]
  {data/rmfinite_min2_n8.dat};
\end{axis}
\end{tikzpicture}
\caption{The two families of \eqref{eq:Lmin} and \eqref{eq:Lmax} at
$r = 2$, agree at $r = 1$; their profiles are in
Figure~\ref{fig:Fprofiles}. The larger of two draws
($F^+_2(x) = x^2$, orange) delays its deductions and climbs from the
baseline to $L^+_2(1) = \tfrac32$; the smaller of two draws
($F^-_2(x) = 1-(1-x)^2$, green) holds the constant $L^-_2 \equiv 3$;
between them lies the single uniform draw ($F_1(x) = x$, blue), in both families, at $L \equiv 2$. Dots are
finite-$n$ leverage: filled $n = 4$ (all $2^{16}$ question sets
enumerated, exact), circles $n = 6$ and triangles $n = 8$ (Monte
Carlo over question orders). Convergence is slow
from below: the label costs $O(\log n)$ of the $Q$ available bits, so
the mean rate approaches
its limit only as $1/\sqrt n$. The left becomes a boundary layer:
$L_F(0) = 1 + F'(0)$ presumes $1 \ll \ell \ll Q$, while at
$\ell = 1$ the only veritable exhaustible class is the constants.
Then $\mathcal L_1 = 1 + (Q-1)\,w_0^2/(2\ln 2) + O(w_0^3)$, the
early deductions being spent on identifying the latent label. The
finite curves scallop, most visibly at $n = 8$: each surge is
one class being learned out near its rate $K_d/Q$, a discreteness the
limit scrunches leftward as the rate grid refines.}
\label{fig:Fexamples}
\end{figure}
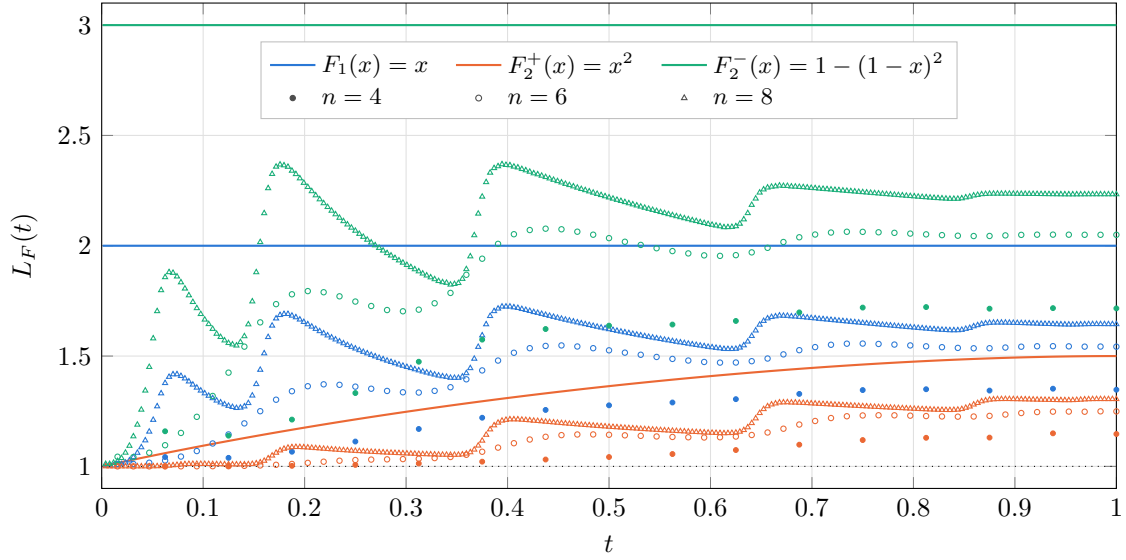

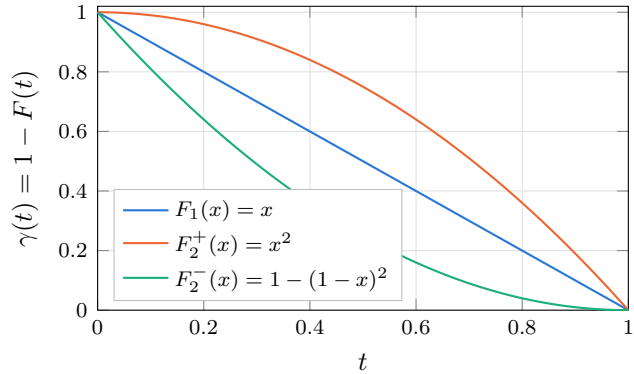
\begin{figure}[H]
\centering
\begin{tikzpicture}
\begin{axis}[width=8.6cm, height=5.6cm,
  xlabel={$t$}, ylabel={$\gamma(t) = 1 - F(t)$},
  xmin=0, xmax=1, ymin=0, ymax=1.02, domain=0:1, samples=200,
  tick label style={font=\scriptsize}, label style={font=\small},
  grid=major, grid style={black!12},
  legend style={font=\scriptsize, at={(0.03,0.03)}, anchor=south west,
    draw=black!25, fill=white},
  legend cell align=left]
\addplot[cf2, thick] {1-x};
\addplot[cf3, thick] {1-x^2};
\addplot[cf1, thick] {(1-x)^2};
\legend{{$F_1(x)=x$}, {$F^+_2(x)=x^2$}, {$F^-_2(x)=1-(1-x)^2$}}
\end{axis}
\end{tikzpicture}
\caption{The increment profiles behind Figure~\ref{fig:Fexamples}: the survival
function $\gamma(t) = 1 - F(t)$ of each statistic. By
\eqref{eq:gammaF} it is the mean entropy (expected surprisal) of one fresh answer at
macroscopic time $t$. Convex, straight and concave $F$ give concave,
straight and convex profiles respectively; the smaller of two draws spends its
predictive power earliest, the larger of two hoards it.}
\label{fig:Fprofiles}
\end{figure}

Two one-parameter families are enough to see the law work. Draw $r$
independent uniformly random values on $[0,1]$ and keep one of them: keeping the
smallest suppresses it toward zero, keeping the largest inflates it
toward one, $r$ controls how hard. Every curve in
Figure~\ref{fig:Fexamples} comes from this extended family.

The smaller of $r$ draws: All $r$ land above $x$ with
probability $(1-x)^r$, so
\begin{equation}
  F^-_r(x) = 1 - (1-x)^r,
  \qquad
  \gamma^-_r(t) = (1-t)^r,
  \qquad
  \mathbb E[R] = \frac1{r+1} .
  \label{eq:Fmin}
\end{equation}
The density $r(1-x)^{r-1}$ is a decreasing function, so budget (weight) is pulled into the low
shells. Substituting into \eqref{eq:LF}, the numerator factorizes and
the denominator ends up carrying the same bracket,
\begin{equation}
  L^-_r(t)
  \;=\; \frac{t + (1-t)\big(1-(1-t)^r\big)}
             {\displaystyle\int_0^t (1-x)^r\, dx}
  \;=\; \frac{1-(1-t)^{r+1}}{\big(1-(1-t)^{r+1}\big)\big/(r+1)}
  \;=\; r+1
  \label{eq:Lmin}
\end{equation}
at every macroscopic time: a whole ladder of constant curves, one rung
per $r$, $r$ deduced bits accompanying each received one. Both endpoint
identities below \eqref{eq:LF} return that same number, $1 + F'(0^+) =
1 + r$ at the start and $1/\mathbb E_F[R] = r+1$ at the finish, and with
a constant curve they have no choice but to agree.

The larger of $r$ draws: All $r$ land below $x$ with
probability $x^r$, so
\begin{equation}
  F^+_r(x) = x^r,
  \qquad
  \gamma^+_r(t) = 1 - t^r,
  \qquad
  \mathbb E[R] = \frac r{r+1} ,
  \label{eq:Fmax}
\end{equation}
an increasing density, budget on the high shells. Nothing telescopes
now and the leverage varies with $t$,
\begin{equation}
  L^+_r(t)
  \;=\; \frac{t + (1-t)\,t^r}{t - t^{r+1}/(r+1)}
  \;=\; \frac{(r+1)\big(t + t^r(1-t)\big)}{(r+1)\,t - t^{r+1}},
  \label{eq:Lmax}
\end{equation}
rising monotonically across the run. At the start the density vanishes
and $L^+_r(0) = 1 + F'(0) = 1$ for every $r > 1$: no leverage at
all, each early answer releases no direct deductions. At the finish
$L^+_r(1) = 1/\mathbb E_F[R] = (r+1)/r$, so the deductions are
late-loaded, but even the last of them buys only $1/r$ extra. The two
finishes are complements, $1/L^-_r(1) + 1/L^+_r(1) = 1$, because the
mean of the smaller and the mean of the larger of the same $r$ draws
sum to one. Pushing $r$ below $1$ gives up the min-max reading and
flips the tilt of $F^+_r$ back to concave, but with an infinite
density at the origin ($F = \sqrt x$ at $r = \tfrac12$), and
$L^+_r(0)$ diverges: front-loaded deduction, the shape of a boundary
layer, no longer a constant.

Where do the two choices meet? At $r = 1$ the min is the max,
and both families collapse onto the single uniform draw $F(x) = x$,
whose numerator $t(2-t)$ cancels its denominator $t(2-t)/2$ to leave
$L \equiv 2$ exactly. That is the hinge of
Figure~\ref{fig:Fexamples}: the blue line is the shared $r = 1$ case,
and the curves either side of it are the $r = 2$ members of the two
families, $L^+_2(t)$ climbing from $1$ to $\tfrac32$ and $L^-_2(t) \equiv 3$.
The uniform draw is also the flattest Occam on offer, each shell
receiving exactly the share of coefficient space its degree inhabits,
$w_d = \binom nd/2^n$. An observed flat leverage does not mean we had this microscopic prescription:
Appendix~\ref{app:blockpriors} produces the same constant $2$ by a completely different mechanism.

The concave direction may look like the
natural one, a simplicity prior ought to favor simple worlds, but
$F$ never decides \emph{whether} Werner is an Occamist. The classes
are nested, so the weight $p_j$ of a single map falls with its degree
for every $F$ whatsoever, and even the convex $F^+_r$, which loads the
high shells, still ranks each complicated (high-degree, highly composite) map as less likely than each simple one.
What $F$ moves is the aggregate: which shell carries the belief, and
with it the leverage, from $r$ deduced bits per received one under
$F^-_r$ down to at most $1/r$ under $F^+_r$.

\begin{example}[label=ex:shellsplit]
How is belief split over the aggregate degree shells? We will illustrate for $n = 4$, $m = 1$, with the
statistic: $F(x) = 1 - (1-x)^2$, the smaller of
two uniform draws. The five classes have dimensions
\begin{align*}
  K_d &\in \left\{1, 5, 11, 15, 16\right\},\\
  w_d & \in \left\{\tfrac{31}{256},\, \tfrac{104}{256},\, \tfrac{96}{256},\,
\tfrac{24}{256},\, \tfrac1{256} \right\},
\end{align*}
using \eqref{eq:shapepj} with rates $R_{4,d} = K_d/16$. We can visualize what prior weight maps
of each polynomial degree receive.

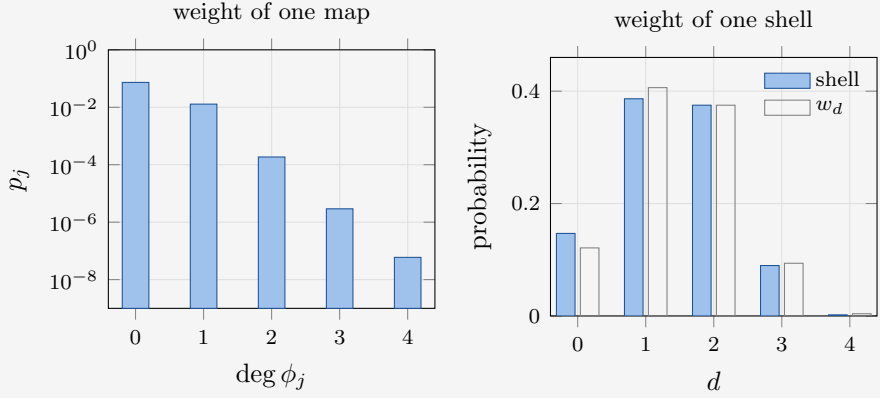
\begin{center}
\begin{tikzpicture}
\begin{axis}[width=5.9cm, height=5.0cm,
  ybar, bar width=10pt, ymode=log, log origin=infty,
  title={weight of one map}, title style={font=\footnotesize},
  xlabel={$\deg \phi_j$}, ylabel={$p_j$},
  xtick={0,1,2,3,4}, ymin=1e-9, ymax=1,
  ytick={1e-8,1e-6,1e-4,1e-2,1},
  grid=major, grid style={black!12},
  tick label style={font=\scriptsize}, label style={font=\small}]
\addplot[fill=cf2!45, draw=cf2!70!black]
  coordinates {(0,7.3428e-2) (1,1.2881e-2) (2,1.8603e-4)
               (3,2.9206e-6) (4,5.9605e-8)};
\end{axis}
\end{tikzpicture}\hspace{3mm}%
\begin{tikzpicture}
\begin{axis}[width=5.9cm, height=5.0cm,
  ybar, bar width=7pt,
  title={weight of one shell}, title style={font=\footnotesize},
  xlabel={$d$}, ylabel={probability},
  xtick={0,1,2,3,4}, ymin=0, ymax=0.46, area legend,
  grid=major, grid style={black!12},
  tick label style={font=\scriptsize}, label style={font=\small},
  legend style={font=\scriptsize, draw=none, fill=none,
    at={(0.99,0.99)}, anchor=north east, cells={anchor=west}}]
\addplot[fill=cf2!45, draw=cf2!70!black]
  coordinates {(0,0.14686) (1,0.38644) (2,0.37503)
               (3,0.08972) (4,0.00195)};
\addplot[fill=none, draw=black!50]
  coordinates {(0,0.12109) (1,0.40625) (2,0.37500)
               (3,0.09375) (4,0.00391)};
\legend{shell, $w_d$}
\end{axis}
\end{tikzpicture}
\end{center}

On the left, per map: the weight drops by a factor $5.7$, then $69$, $64$, $49$,
matching the factor $2^{\binom 4d}$ that the next class costs in dimension,
tilted by the budget ratio $w_{d-1}/w_d$. Six decades across five
degrees, which is why the axis is logarithmic. Occam is priced per
map, and it is highly undemocratic. On the right, one shell: the
$2^{K_d} - 2^{K_{d-1}}$ maps of each degree carry the belief in blue on
a linear axis, peaking at the affine ($d = 1$) maps.

\begin{center}
\begin{tabular}{crrr}
\toprule
$d$ & maps in shell & shell weight & budget $w_d$ \\
\midrule
0 & $2$        & $14.7\%$ & $12.1\%$ \\
1 & $30$       & $38.6\%$ & $40.6\%$ \\
2 & $2{,}016$  & $37.5\%$ & $37.5\%$ \\
3 & $30{,}720$ & $9.0\%$  & $9.4\%$ \\
4 & $32{,}768$ & $0.2\%$  & $0.4\%$ \\
\midrule
total & $65{,}536$ & $100\%$ & $100\%$ \\
\bottomrule
\end{tabular}
\end{center}

In general, population can beat simplicity in the aggregate,
while the per-map ordering never wavers. The white bars are the
budgets $w_d$, and the slight difference between the two is the
$2^{-\binom 4d}$ leakage into a degree from higher classes.
\end{example}

\section{Discussion and Next Steps}\label{sec:next}

\subsection{Stochastic truth maps}

To all questions in this essay, nature answers deterministically: $\psi$ is a map, and an answer
that disagrees with a candidate kills it outright. Replace the truth by
a conditional law $\Psi(a \mid q)$ and the update rule
\eqref{eq:update} becomes genuine Bayes conditioning, reweighting each candidate by
its likelihood instead of eliminating it. The hypothesis space for the single truth becomes a
simplex rather than the finite set of $N$ maps. The chain rule still
gives \eqref{eq:telescope} and the received-entropy law for the
observed answer string $y$. The remaining-entropy law
\eqref{eq:law-remaining}, however, needs a revised prediction target:
an already asked question need not have zero entropy for a fresh
response $a'$. If $M$ predicts fresh responses at all questions,
the asked columns must still be included; if the target consists
only of unasked questions, that restriction must be stated.
Another identification that breaks is $G_Q = H(p)$. Answers no longer pin down
the truth fully, so a complete run does not end with the belief settled. 
The profile inherits that: no class is ever finished, so $\gamma$ stays positive
forever, as faith in individual answers grows asymptotically. The question
is what becomes of the master law \eqref{eq:master} when the remaining
predictive entropy need not reach zero.

The denominator needs reconsidering as well. Under elimination the
surprisal of an answer is also the relative entropy between prior and
posterior, so one number both prices the answer and measures how much it shifted 
the Bayesian belief. A soft likelihood (not $0$ or $1$) separates them, and this is exactly the
distinction Baldi and Itti draw between actual informational content and
surprise~\cite{baldiitti10}: a datapoint can carry a large $-\log_2 M_{a,q}$
and barely disturb the belief, or be unremarkable and shift the posterior by a landslide.
Leverage then has a choice of denominator, and the two choices are
different theories. Pricing by surprisal keeps the accounting of
Chapter~\ref{sec:general} and the telescoping identity
\eqref{eq:telescope} with it; pricing by belief movement measures what the
learner absorbed rather than what it paid, and gives up that identity.
It would be interesting to research how to reconcile these methods.

\subsection{A prior that is wrong}

Every average in this work assumes Werner's prior is that from which nature samples.
This assumption is woven into all results. It is what makes the expected surprisal
of a question equal to the entropy of its own column. Let nature draw
from $\{p'_j\}$ while Werner believes $\{p_j\}$, and his expected surprisal at
$q$ becomes a cross-entropy, $H(\pi_q) + D(\pi_q \Vert M_{\cdot,q})$
\cite{covthomas}:
costing him the true uncertainty plus the divergence of his belief from it.
The numerator of \eqref{eq:leverage} is still measured in his own
units, so the ratio falls twice over, once for the inflated bill and
once for deductions drawn on correlations that are not there. Some quantities analogous 
to the KL-divergence could be derived.
The interesting
assumption is a prior with full support, so under elimination Werner can't rule out the truth outright, 
which would leave $M_{a,q} = 0$ at some later
question: an infinite surprisal and an undefined update. Two questions in particular: does aggregate leverage
$\mathcal L_\ell$ ever fall below one? If so, is
the deficit a divergence, or a sum of per-question KL terms? For the
denominator alone the answer is already known: the cumulative excess
surprisal of a wrong model is exactly the relative entropy between the
two joint laws, which is the redundancy of universal
prediction~\cite{merhav98}. What that literature does not price is the
numerator.

\subsection{Designing a curve, shell by shell}

The truncation remark of Chapter~\ref{sec:expectations} cuts both ways.
Read forward it says the first $\ell$ rounds cannot see correlators
spanning more than $\ell+1$ questions; read backward it says the shell
of order $\ell+1$ is the first place that has an influence on learning at
step $\ell$. Suppose then that we want a particular sequence of
leverages. Equation \eqref{eq:Lk-G} inverts,
\begin{equation}
  G_{\ell+1} \;=\; G_\ell
  \;+\; \frac{Q\,G_1 - \mathcal L_\ell\, G_\ell}{Q - \ell},
  \label{eq:designrec}
\end{equation}
so the demanded curve fixes the whole block-entropy sequence
recursively from $G_1$. Each $G_{\ell+1}$ in that recursion, with the previously
fixed correlators, forms a set of conditions on shell $\ell+1$. 
One could then
try to build the prior a shell at a time: shell $1$ to fix $G_1$, shell
$2$ to fix $G_2$, and onward, leaving the earlier rounds untouched. 
One constraint is submodularity on $G$, which forces it to be concave with
increments in $[0,m]$. What makes this nontrivial is that not every set of correlators $\chi_{\mathcal S}$ is
admissible. The shells cannot be chosen freely: they must be the
transform of an actual distribution. When all correlators are
specified, the inverse Walsh transform recovers the candidate prior,
$p=N^{-1}W^{\otimes mQ}\chi$; with $\chi_{\emptyset}=1$,
admissibility is checked by testing every recovered $p_j\geq0$.
The challenge in a shell-by-shell construction is completing the
as-yet unspecified higher correlators while preserving these global
inequalities. Constraints on the already chosen shells alone need
not establish that such a completion exists. The geometry of partial
correlation constraints is discussed in~\cite{pitowsky91}.
The same language is worth carrying into the limit. To
second order the block entropy reads
\begin{equation}
  G_k \;=\; km \;-\; \frac{1}{2\ln 2}
  \sum_{\mathcal V \neq \emptyset}
  \Pr\big[\mathcal V \subseteq \mathcal S\big]\, \chi_{\mathcal V}^2
  \;+\; O(\chi^3),
  \label{eq:Gshell}
\end{equation}
where the remainder begins with cubic products of correlators in
general. The inclusion probability of a correlator touching $r$ questions
tends to $t^{\,r}$ at macroscopic time. Each shell enters the
profile suppressed by its own power of $t$. This codifies the idea of
storing knowledge in deep correlations that do not surface until late.
For this approach to work, along the way, we would develop TDLs of the 
correlators themselves, which would be an interesting exercise in its own right.

\subsection{Measuring leverage in a real learner}

The protocol of Chapter~\ref{sec:general} is an experiment, not only a
calculation, and a language model is a convenient Werner:
its answer matrix is directly observable, the token probabilities in response to a prompt being the column $M_{\cdot,q}$ itself, and its in-context learning has
itself been read as implicit Bayesian inference \cite{xie22}. 
Fix a finite
universe of questions with a latent rule, ask them one at a time,
record the surprisal of each answer and the drop in total predictive
entropy over the questions not yet asked, and both halves of
\eqref{eq:leverage} are measured rather than modeled. The same
protocol runs on a network during training, reading the predictive
entropy off a held-out question set between updates. Tracking a
learner's predictive entropy is established ground, and its
sub-extensive part is already the standard measure of how much there
was to learn~\cite{bnt01}. That work models learning as mutual information between
past and future, in a statistical physics paradigm. It reproduces the de Finetti result
for $o(Q)$ settling of the label. In our language, that would emerge as hyperbolic leverage. 
Let us now apply these ideas to LLMs. Perhaps they store their correlations deeper.
The obstacle is
scale: $t = \ell/Q$ needs a question universe small enough to exhaust,
or the whole experiment lives in the boundary layer at $t = 0^+$ and
reports nothing about macroscopic learning. Nonetheless, we think the 
TDL is the correct regime to understand neural networks, as the scale should be large
enough to coarse-grain away the microscopic details of boolean maps.

\appendix

\clearpage
\phantomsection
\section*{Appendices}
\addcontentsline{toc}{section}{Appendices}

They follow the order in which the argument first calls on them, except
where one appendix serves another and is placed beside it.
Appendix~\ref{app:correlators} changes basis so that the correlation
store $C$ becomes one coordinate per set of questions, and the update
rule a two-term convolution. Appendix~\ref{app:playing} plays the
guiding example through all four questions, one round at a time.
Appendix~\ref{app:tensor-marginals} arranges the elimination rules as a
tensor and takes every block entropy $G_k$ off a single sweep of it.
Appendix~\ref{app:blockpriors} collects the block-prior construction and
its Bernstein profile language, a block-based route to the same family of
macroscopic curves. Appendix~\ref{app:spike} continues the spike-prior
example of Section~\ref{sec:spike}. The next three go together:
Appendix~\ref{app:strategies} asks what Werner can do when he chooses the
questions rather than taking them as they come,
Appendix~\ref{app:gibbs} builds a prior from circuit complexity as a Gibbs
ensemble and puts those strategies to work on it, and
Appendix~\ref{app:aig} records how the circuit complexities themselves were
obtained, the symmetry reduction and the satisfiability encoding that make
a minimum over all circuits computable at all. The last two are for reference:
Appendix~\ref{app:symbols} collects the symbols on one page,
and Appendix~\ref{app:scripts} lists the scripts that produce the numbers
and the curves.

\section{Correlators}\label{app:correlators}

The body refers to this appendix in a few asides, but does not
depend on it.

What kind of operator is the update rule~\eqref{eq:update}? In map
space it is as simple as an operator gets: multiplication by a
diagonal $0/1$ mask, then renormalization. But there is a change of
basis in which the same operation becomes a \emph{convolution} with a
two-term kernel, and in which the correlation store $C$ stops being a
single number and becomes visible coordinate by coordinate. The main advantage is 
that the effect of each update is well tabulated beforehand into a small number of targeted parameters.

This
appendix builds that basis for $m = 1$; its last section deals
with general $m$.

First, re-encode each answer as a spin,
\begin{equation}\label{eq:spin}
  \sigma_q(\phi_j) \;:=\; (-1)^{\phi_j(q)} \;\in\; \{+1, -1\},
\end{equation}
so answer $0 \mapsto +1$ and $1 \mapsto -1$. A pure relabeling, but
it buys the one algebraic fact everything below turns on:
$\sigma_q^2 = 1$.

The \textbf{Walsh--Hadamard transform}, the Fourier transform of
Boolean-function analysis \cite{odonnell14}, is built from a single
matrix. For one bit,
\begin{equation}\label{eq:hadamard}
  W \;=\; \begin{pmatrix} 1 & 1 \\ 1 & -1 \end{pmatrix},
  \qquad W_{c,b} \;=\; (-1)^{c\,b}, \quad c, b \in \{0,1\}.
\end{equation}
\begin{example}
Acting on a one-bit distribution,
\[
  W \binom{p_0}{p_1}
  \;=\; \binom{p_0 + p_1}{p_0 - p_1}
  \;=\; \binom{1}{\langle \sigma \rangle}:
\]
row $c = 0$ computes the normalization, row $c = 1$ the spin bias.
Distribution in, correlators out.
\end{example}

The normalization riding along is a redundancy. The original distribution also carries one variable more than its degrees of freedom.

\begin{example}
Tensoring two copies (the $n=1$ case) does the same for two bits at once:
\begin{equation}\label{eq:HH}
  W \otimes W \;=\;
  \begin{pmatrix}
    W & W \\
    W & -W
  \end{pmatrix}
  \;=\;
  \begin{pmatrix}
    1 & 1 & 1 & 1 \\
    1 & -1 & 1 & -1 \\
    1 & 1 & -1 & -1 \\
    1 & -1 & -1 & 1
  \end{pmatrix}
  \quad
  \begin{array}{l}
    \leftarrow c = 00\colon\ \text{sums to } 1 \\
    \leftarrow c = 01\colon\ \langle \sigma_0 \rangle \\
    \leftarrow c = 10\colon\ \langle \sigma_1 \rangle \\
    \leftarrow c = 11\colon\ \langle \sigma_1 \sigma_0 \rangle
  \end{array}
\end{equation}
\end{example}

The Kronecker product multiplies signs, so for any number $l$ of
tensor factors the entry at row $c$, column $j$ is
\begin{equation}\label{eq:kron}
  \big(W^{\otimes l}\big)_{c,j}
  \;=\; \prod_{i=0}^{l-1} (-1)^{c_i\, j_i}
  \;=\; (-1)^{\sum_i c_i\, j_i},
\end{equation}
with $c_i$, $j_i$ the $i$-th bits of the row and column indices;
every further tensor factor appends one more bit.

For $m = 1$ a map \emph{is} its truth table of $Q$ bits (the
numeral $j$) so Werner's belief is a vector of length
$N = 2^{Q}$, and one transform computes every correlator at once:
\begin{equation}\label{eq:wht}
  \chi \;=\; W^{\otimes Q} p,
  \qquad
  \chi_{\mathcal S} \;=\; \Big\langle \prod_{q \in \mathcal S} \sigma_q \Big\rangle
  \;=\; \sum_{j=0}^{N-1} (-1)^{\sum_{q \in \mathcal S} \phi_j(q)}\, p_j.
\end{equation}
Each row computes the \textbf{correlator} of one subset
$\mathcal S \subseteq \mathcal Q$ of questions. In the subscript we may encode the set as
its \textbf{mask}: written in the same descending question order as everywhere else, bit $q$ of the subscript
records whether $q \in \mathcal S$. The row index of the matrix, read as a
numeral, is the mask. 

\begin{example}
For $n = 2$,
the question order is $(11, 10, 01, 00)$ so $\chi_{0110}$ is the correlator of
$\mathcal S = \{10, 01\}$.

the sixteen rows of $W^{\otimes 4}$ probe the sixteen
subsets of $Q = \{11, 10, 01, 00\}$. A few entries of the
dictionary:
\begin{center}
\begin{tabular}{c c l}
\toprule
mask & set $\mathcal S$ & correlator \\
\midrule
$0000$ & $\emptyset$ & $\chi_{0000} = 1$ (normalization) \\
$0001$ & $\{00\}$ & $\chi_{0001} = \langle \sigma_{00} \rangle$ \\
$1000$ & $\{11\}$ & $\chi_{1000} = \langle \sigma_{11} \rangle$ \\
$0110$ & $\{10, 01\}$ & $\chi_{0110} = \langle \sigma_{10}\, \sigma_{01} \rangle$ \\
$1111$ & $\{11, 10, 01, 00\}$ & $\chi_{1111} = \langle \sigma_{11}\, \sigma_{10}\, \sigma_{01}\, \sigma_{00} \rangle$, the full parity \\
\bottomrule
\end{tabular}
\end{center}
\end{example}

Sort the $N$ correlators by $|\mathcal S|$, the number
of questions probed, into \textbf{shells}. Sorting a Boolean function's
Fourier weight by degree in the same way is the basis of Fourier-analytic
learning \cite{linial93}. Shell $0$ is the normalization, $\chi_\emptyset = 1$, riding along
as a correlator. Shell $1$ is $M$ in different clothes:
\begin{equation}\label{eq:shell1}
  \chi_{\{q\}} \;=\; M_{0,q} - M_{1,q},
  \qquad
  M_{a,q} \;=\; \tfrac12\big(1 + (-1)^a\, \chi_{\{q\}}\big),
\end{equation}
the bias of column $q$. For $m=1$, we see that, elegantly, the first shell correlators are the distance from fair coins. 
A prior invariant under negating every answer simultaneously sends
$\chi_{\mathcal S} \mapsto (-1)^{|\mathcal S|}\chi_{\mathcal S}$, so all
its odd shells vanish identically, shell $1$ among them: such a
prior has an agnostic matrix $M$.

Shell $2$ holds the agreement biases of
pairs of questions, knowledge $M$ cannot see: two priors can
share every column of $M$ and differ here. Higher shells store
subtler and subtler parity knowledge. Nothing is lost in the
sorting: $W^2 = 2I$, so $(W^{\otimes Q})^2 = N I$. The
transform is its own inverse up to a factor $N$, and knowing every
correlator is knowing the belief. The ledger also gives $C$ a
signature: a product prior ($C = 0$) factorizes every correlator,
$\chi_{\mathcal S} = \prod_{q \in \mathcal S} \chi_{\{q\}}$, so its upper shells
repeat what shell $1$ already said; correlated knowledge is
precisely \emph{non-factorizing} correlators.

\begin{example}
The prior of Table~\ref{tab:sixteen} transforms to the ledger
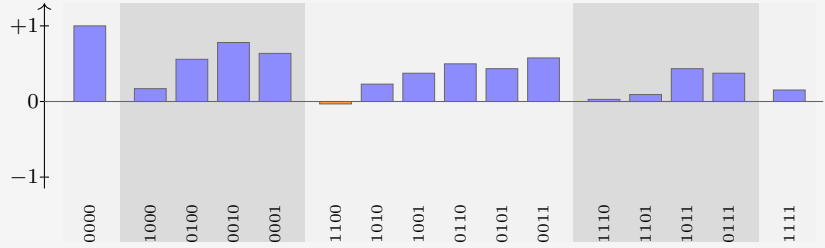
\begin{center}
\begin{tabular}{c l}
\toprule
shell & correlators \\
\midrule
0 & $\chi_{0000} = 1$ \\[2pt]
1 & $\chi_{1000} = +0.170$,\ \ $\chi_{0100} = +0.558$,\ \
    $\chi_{0010} = +0.780$,\ \ $\chi_{0001} = +0.636$ \\[2pt]
2 & $\chi_{1100} = -0.032$,\ \ $\chi_{1010} = +0.230$,\ \
    $\chi_{1001} = +0.374$, \\
  & $\chi_{0110} = +0.498$,\ \ $\chi_{0101} = +0.434$,\ \
    $\chi_{0011} = +0.576$ \\[2pt]
3 & $\chi_{1110} = +0.028$,\ \ $\chi_{1101} = +0.092$,\ \
    $\chi_{1011} = +0.434$,\ \ $\chi_{0111} = +0.374$ \\[2pt]
4 & $\chi_{1111} = +0.152$ \\
\bottomrule
\end{tabular}
\end{center}
\ledgerbars{0/1/0000/blue!45,
  0.80/0.170/1000/blue!45, 1.35/0.558/0100/blue!45,
  1.90/0.780/0010/blue!45, 2.45/0.636/0001/blue!45,
  3.25/-0.032/1100/orange!80, 3.80/0.230/1010/blue!45,
  4.35/0.374/1001/blue!45, 4.90/0.498/0110/blue!45,
  5.45/0.434/0101/blue!45, 6.00/0.576/0011/blue!45,
  6.80/0.028/1110/blue!45, 7.35/0.092/1101/blue!45,
  7.90/0.434/1011/blue!45, 8.45/0.374/0111/blue!45,
  9.25/0.152/1111/blue!45}
Shell $1$ reproduces the columns of equation~\eqref{eq:M16}:
$(1 + 0.170)/2 = 0.585 = M_{0,11}$, and likewise $0.779$, $0.890$,
$0.818$. The negative entry $\chi_{1100}$ says questions $11$
and $10$ disagree slightly more often than they agree; every other
pair leans toward agreement. This is a reflection of the strong coherent $0$-preference encoded in the prior. A prior leaning toward $1$ would give a negative shell $1$.
\end{example}

Now look at the \emph{rows} of $W^{\otimes Q}$ as functions of
$j$. They are \textbf{block waves}. The observation mask is built of the same material:
\begin{equation}\label{eq:mask}
  \delta\big(\phi_j(q), a\big)
  \;=\; \frac12\left(1 + (-1)^{a+\phi_j(q)}\right),
\end{equation}
a $0/1$ block wave: one constant plus one Hadamard row. And
multiplication and convolution swap under this transform exactly as
under Fourier: multiplying two functions of $j$ pointwise convolves
their ledgers. And the observation mask's ledger has only two
nonzero entries - at $\emptyset$ and at $\{q\}$ - so this
convolution is no great sum: it combines each correlator with
exactly one other.

The update rule, carried out entirely in the ledger, is therefore
that two-term combination, followed by renormalization by the new
shell $0$. In index notation, learning $a = \psi(q)$ sends every
set $\mathcal S$ to
\begin{equation}\label{eq:corr-update}
  \chi'_{\mathcal S} \;=\;
  \frac{\chi_{\mathcal S} + (-1)^a\, \chi_{\mathcal S \triangle \{q\}}}
       {1 + (-1)^a\, \chi_{\{q\}}},
\end{equation}
where $\mathcal S \triangle \{q\}$ is $\mathcal S$ with the membership of $q$
toggled: $q$ is added if absent, removed if present (on masks, bit
$q$ is flipped). The numerator is the monomial
algebra of $\sigma_q^2 = 1$: multiplying
$\prod_{r \in \mathcal S} \sigma_r$ by
$\sigma_q$ toggles $q$'s membership of the product, so every correlator
is averaged with its \textbf{partner}: the correlator differing
by $q$, signed by the answer. The denominator is the
$\mathcal S = \emptyset$ instance of the numerator, and by
equation~\eqref{eq:shell1} it equals $2 M_{a,q}$: the same number
that priced the answer in equation~\eqref{eq:update} renormalizes
the ledger. 
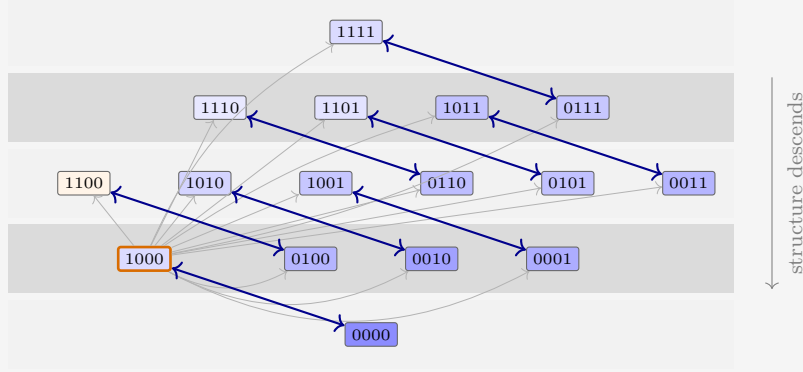
\begin{example}
\begin{center}
\begin{tikzpicture}[x=1cm, y=1cm,
  lnode/.style={draw=black!60, rounded corners=1pt,
                inner sep=2.5pt, font=\tiny, align=center},
  pair/.style={<->, blue!55!black, line width=0.8pt},
  denom/.style={->, gray!60, line width=0.35pt}]
  \fill[gray!28] (-0.3,0.55) rectangle (9.3,1.45);
  \fill[gray!10] (-0.3,1.55) rectangle (9.3,2.45);
  \fill[gray!28] (-0.3,2.55) rectangle (9.3,3.45);
  \fill[gray!10] (-0.3,3.55) rectangle (9.3,4.45);
  \fill[gray!10] (-0.3,-0.45) rectangle (9.3,0.45);
  \node[lnode, fill=blue!14] (n1111) at (4.3,4) {1111};
  \node[lnode, fill=blue!9]  (n1110) at (2.5,3) {1110};
  \node[lnode, fill=blue!11] (n1101) at (4.1,3) {1101};
  \node[lnode, fill=blue!24] (n1011) at (5.7,3) {1011};
  \node[lnode, fill=blue!22] (n0111) at (7.3,3) {0111};
  \node[lnode, fill=orange!9] (n1100) at (0.7,2) {1100};
  \node[lnode, fill=blue!17]  (n1010) at (2.3,2) {1010};
  \node[lnode, fill=blue!22]  (n1001) at (3.9,2) {1001};
  \node[lnode, fill=blue!26]  (n0110) at (5.5,2) {0110};
  \node[lnode, fill=blue!24]  (n0101) at (7.1,2) {0101};
  \node[lnode, fill=blue!29]  (n0011) at (8.7,2) {0011};
  \node[lnode, fill=blue!14, draw=orange!85!black, line width=1pt]
    (n1000) at (1.5,1) {1000};
  \node[lnode, fill=blue!29] (n0100) at (3.7,1) {0100};
  \node[lnode, fill=blue!37] (n0010) at (5.3,1) {0010};
  \node[lnode, fill=blue!32] (n0001) at (6.9,1) {0001};
  \node[lnode, fill=blue!45] (n0000) at (4.5,0) {0000};
  \draw[denom] (n1000) to (n1100);
  \draw[denom] (n1000) to (n1010);
  \draw[denom] (n1000) to (n1001);
  \draw[denom] (n1000) to (n0110);
  \draw[denom] (n1000) to (n0101);
  \draw[denom] (n1000) to (n0011);
  \draw[denom] (n1000) to (n1110);
  \draw[denom] (n1000) to (n1101);
  \draw[denom, bend left=8] (n1000) to (n1011);
  \draw[denom, bend right=8] (n1000) to (n0111);
  \draw[denom, bend left=18] (n1000) to (n1111);
  \draw[denom, bend right=28] (n1000) to (n0100);
  \draw[denom, bend right=28] (n1000) to (n0010);
  \draw[denom, bend right=28] (n1000) to (n0001);
  \draw[pair] (n1000) -- (n0000);
  \draw[pair] (n1100) -- (n0100);
  \draw[pair] (n1010) -- (n0010);
  \draw[pair] (n1001) -- (n0001);
  \draw[pair] (n1110) -- (n0110);
  \draw[pair] (n1101) -- (n0101);
  \draw[pair] (n1011) -- (n0011);
  \draw[pair] (n1111) -- (n0111);
  \draw[->, gray] (9.8,3.4) -- (9.8,0.6);
  \node[rotate=90, gray, font=\scriptsize] at (10.1,2)
    {structure descends};
\end{tikzpicture}
\end{center}
The update rule in the ledger, drawn for the
guiding example's first question ($q = 11$). Every set is paired
with the partner whose membership of $11$ is toggled (arrows);
each pair is averaged with the answer's sign and renormalized,
per equation~\eqref{eq:corr-update}. The pair
$\emptyset \leftrightarrow \{11\}$ at the orange-bordered asked
set fixes the normalization and
saturates the asked bias; each blue double arrow averages a pair
into both of its members, and the net flow of structure is one
shell down, toward $M$. The thin gray arrows from the
asked $q$ to every other pill are a reminder that this sets the
renormalization. The fill encodes each prior value to
scale: darker is larger, orange is negative.
\end{example}

Read formula~\eqref{eq:corr-update} by cases:
\begin{itemize}
  \item $\mathcal S = \emptyset$: partner $\{q\}$, giving
    $(1 + (-1)^a \chi_{\{q\}})/(1 + (-1)^a \chi_{\{q\}}) = 1$.
    Normalization is preserved.
  \item $\mathcal S = \{q\}$, the asked question: partner $\emptyset$, giving
    $(\chi_{\{q\}} + (-1)^a)/(1 + (-1)^a \chi_{\{q\}}) = (-1)^a$.
    The bias saturates to certainty - the one-hot collapse of
    column $q$, derived rather than decreed.
  \item $\mathcal S = \{q'\}$, an unasked question: the partner is the pair
    $\{q', q\}$, and one line of algebra puts the
    shift in terms of the \emph{connected} correlator:
    \begin{equation}\label{eq:cov-shift}
      \chi'_{\{q'\}} - \chi_{\{q'\}}
      \;=\; \frac{(-1)^a \big(\chi_{\{q',q\}}
        - \chi_{\{q'\}}\, \chi_{\{q\}}\big)}{2 M_{a,q}}
      \;=\; \frac{(-1)^a\, \mathrm{Cov}\big(\sigma_{q'}, \sigma_q\big)}
        {2 M_{a,q}}.
    \end{equation}
    Unasked columns move in proportion to their covariance with the
    asked question: knowledge stored one shell up descends into the
    predictions. A product prior has zero covariance everywhere, so
    its unasked columns are frozen - the ``$C = 0$ means no
    transfer'' fact, now a one-line consequence.
  \item Higher shells: the same cascade. Each update marries
    shell-$k$ sets containing $q$ to shell-$(k{-}1)$ sets without
    it; whatever structure was stored jointly with $q$ moves one
    shell down, toward $M$.
\end{itemize}

\begin{example}
Question 1 of the walkthrough asked $q = 11$ and received $a = 1$:
sign $(-1)^a = -1$, toggled set $\{11\}$, mask $1000$.
Equation~\eqref{eq:corr-update}
maps the ledger of the previous box to
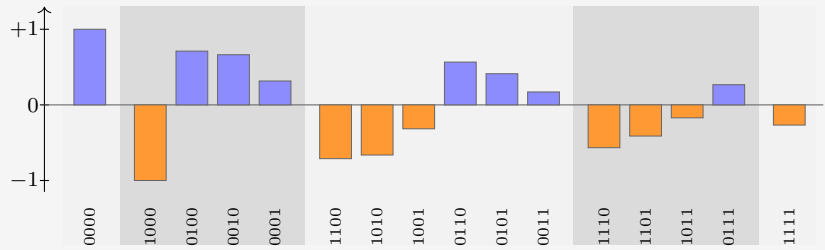
\begin{center}
\begin{tabular}{c l}
\toprule
shell & updated correlators \\
\midrule
0 & $\chi'_{0000} = 1$ \\[2pt]
1 & $\chi'_{1000} = \mathbf{-1}$,\ \ $\chi'_{0100} = +0.711$,\ \
    $\chi'_{0010} = +0.663$,\ \ $\chi'_{0001} = +0.316$ \\[2pt]
2 & $\chi'_{1100} = -0.711$,\ \ $\chi'_{1010} = -0.663$,\ \
    $\chi'_{1001} = -0.316$, \\
  & $\chi'_{0110} = +0.566$,\ \ $\chi'_{0101} = +0.412$,\ \
    $\chi'_{0011} = +0.171$ \\[2pt]
3 & $\chi'_{1110} = -0.566$,\ \ $\chi'_{1101} = -0.412$,\ \
    $\chi'_{1011} = -0.171$,\ \ $\chi'_{0111} = +0.267$ \\[2pt]
4 & $\chi'_{1111} = -0.267$ \\
\bottomrule
\end{tabular}
\end{center}
\ledgerbars{0/1/0000/blue!45,
  0.80/-1/1000/orange!80, 1.35/0.711/0100/blue!45,
  1.90/0.663/0010/blue!45, 2.45/0.316/0001/blue!45,
  3.25/-0.711/1100/orange!80, 3.80/-0.663/1010/orange!80,
  4.35/-0.316/1001/orange!80, 4.90/0.566/0110/blue!45,
  5.45/0.412/0101/blue!45, 6.00/0.171/0011/blue!45,
  6.80/-0.566/1110/orange!80, 7.35/-0.412/1101/orange!80,
  7.90/-0.171/1011/orange!80, 8.45/0.267/0111/blue!45,
  9.25/-0.267/1111/orange!80}
The asked bias saturates:
$\chi'_{1000} = -1$, directly storing the answer.
Other singletons move by equation~\eqref{eq:cov-shift}: for $q' = 10$ the
covariance is $-0.032 - (0.558)(0.170) = -0.127$, so the shift is
$(-1)(-0.127)/(2 \times 0.415) = +0.153$, landing on $+0.711$,
and indeed $(1 + 0.711)/2 = 0.855$, the column $M_{0,10}$ of the
Question 2 box. With $\sigma_{11}$ pinned, every mask containing the asked bit
is now just $(-1)^a$ times its partner
($\chi'_{1010} = -\chi'_{0010}$
etc.).

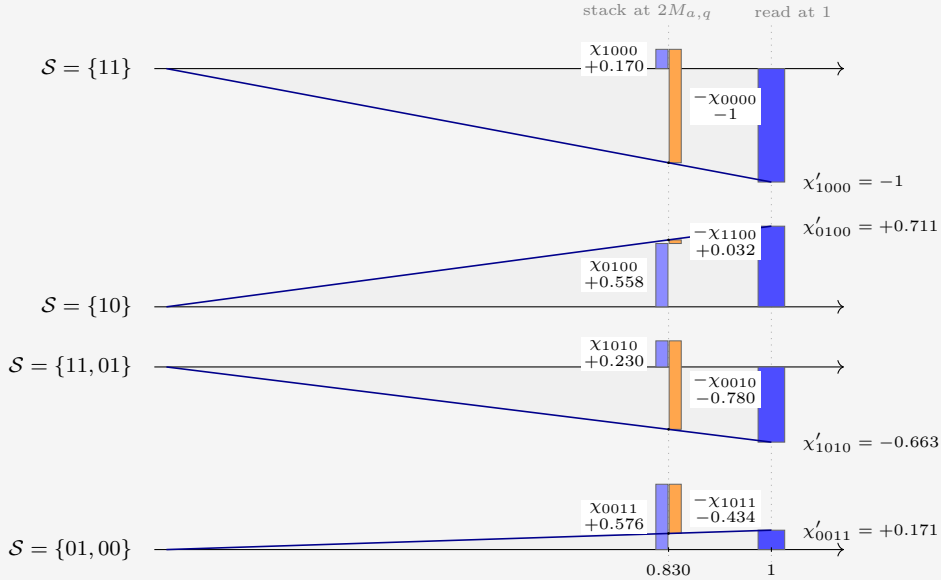
\begin{center}
\begin{tikzpicture}[x=8cm, y=1.5cm]
  \draw[dotted, gray!60] (0.830,0.38) -- (0.830,-4.35);
  \draw[dotted, gray!60] (1,0.38) -- (1,-4.35);
  \node[font=\tiny, gray, anchor=east] at (0.92,0.50)
    {stack at $2M_{a,q}$};
  \node[font=\tiny, gray, anchor=west] at (0.955,0.50)
    {read at $1$};
  \node[font=\scriptsize, anchor=east] at (-0.04,0)
    {$\mathcal S = \{11\}$};
  \node[font=\scriptsize, anchor=east] at (-0.04,-2.1)
    {$\mathcal S = \{10\}$};
  \node[font=\scriptsize, anchor=east] at (-0.04,-2.63)
    {$\mathcal S = \{11, 01\}$};
  \node[font=\scriptsize, anchor=east] at (-0.04,-4.24)
    {$\mathcal S = \{01, 00\}$};
  \fill[gray!12] (0,0) -- (1,0) -- (1,-1) -- cycle;
  \draw[->] (-0.02,0) -- (1.12,0);
  \draw[fill=blue!45, draw=black!60, line width=0.3pt]
    (0.809,0) rectangle (0.829,0.170);
  \draw[fill=orange!70, draw=black!60, line width=0.3pt]
    (0.831,0.170) rectangle (0.851,-0.830);
  \draw[fill=blue!70, draw=black!60, line width=0.3pt]
    (0.978,0) rectangle (1.022,-1);
  \draw[blue!55!black, line width=0.6pt] (0,0) -- (1,-1);
  \fill[black] (0.830,-0.830) circle (0.5pt);
  \node[font=\tiny, fill=white, inner sep=1pt, anchor=east,
    align=center] at (0.795,0.085) {$\chi_{1000}$\\$+0.170$};
  \node[font=\tiny, fill=white, inner sep=1pt, anchor=west,
    align=center] at (0.865,-0.33) {$-\chi_{0000}$\\$-1$};
  \node[font=\tiny, anchor=west, align=center] at (1.035,-1)
    {$\chi'_{1000} = -1$};
  \fill[gray!12] (0,-2.1) -- (1,-2.1) -- (1,-1.389) -- cycle;
  \draw[->] (-0.02,-2.1) -- (1.12,-2.1);
  \draw[fill=blue!45, draw=black!60, line width=0.3pt]
    (0.809,-2.1) rectangle (0.829,-1.542);
  \draw[fill=orange!70, draw=black!60, line width=0.3pt]
    (0.831,-1.542) rectangle (0.851,-1.510);
  \draw[fill=blue!70, draw=black!60, line width=0.3pt]
    (0.978,-2.1) rectangle (1.022,-1.389);
  \draw[blue!55!black, line width=0.6pt] (0,-2.1) -- (1,-1.389);
  \fill[black] (0.830,-1.510) circle (0.5pt);
  \node[font=\tiny, fill=white, inner sep=1pt, anchor=east,
    align=center] at (0.795,-1.821) {$\chi_{0100}$\\$+0.558$};
  \node[font=\tiny, fill=white, inner sep=1pt, anchor=west,
    align=center] at (0.865,-1.526) {$-\chi_{1100}$\\$+0.032$};
  \node[font=\tiny, anchor=west, align=center] at (1.035,-1.389)
    {$\chi'_{0100} = +0.711$};
  \fill[gray!12] (0,-2.63) -- (1,-2.63) -- (1,-3.293) -- cycle;
  \draw[->] (-0.02,-2.63) -- (1.12,-2.63);
  \draw[fill=blue!45, draw=black!60, line width=0.3pt]
    (0.809,-2.63) rectangle (0.829,-2.40);
  \draw[fill=orange!70, draw=black!60, line width=0.3pt]
    (0.831,-2.40) rectangle (0.851,-3.18);
  \draw[fill=blue!70, draw=black!60, line width=0.3pt]
    (0.978,-2.63) rectangle (1.022,-3.293);
  \draw[blue!55!black, line width=0.6pt] (0,-2.63) -- (1,-3.293);
  \fill[black] (0.830,-3.18) circle (0.5pt);
  \node[font=\tiny, fill=white, inner sep=1pt, anchor=east,
    align=center] at (0.795,-2.515) {$\chi_{1010}$\\$+0.230$};
  \node[font=\tiny, fill=white, inner sep=1pt, anchor=west,
    align=center] at (0.865,-2.83) {$-\chi_{0010}$\\$-0.780$};
  \node[font=\tiny, anchor=west, align=center] at (1.035,-3.293)
    {$\chi'_{1010} = -0.663$};
  \fill[gray!12] (0,-4.24) -- (1,-4.24) -- (1,-4.069) -- cycle;
  \draw[->] (-0.02,-4.24) -- (1.12,-4.24);
  \draw[fill=blue!45, draw=black!60, line width=0.3pt]
    (0.809,-4.24) rectangle (0.829,-3.664);
  \draw[fill=orange!70, draw=black!60, line width=0.3pt]
    (0.831,-3.664) rectangle (0.851,-4.098);
  \draw[fill=blue!70, draw=black!60, line width=0.3pt]
    (0.978,-4.24) rectangle (1.022,-4.069);
  \draw[blue!55!black, line width=0.6pt] (0,-4.24) -- (1,-4.069);
  \fill[black] (0.830,-4.098) circle (0.5pt);
  \node[font=\tiny, fill=white, inner sep=1pt, anchor=east,
    align=center] at (0.795,-3.952) {$\chi_{0011}$\\$+0.576$};
  \node[font=\tiny, fill=white, inner sep=1pt, anchor=west,
    align=center] at (0.865,-3.881) {$-\chi_{1011}$\\$-0.434$};
  \node[font=\tiny, anchor=west, align=center] at (1.035,-4.069)
    {$\chi'_{0011} = +0.171$};
  \draw (0.830,-4.26) -- (0.830,-4.22);
  \draw (1,-4.26) -- (1,-4.22);
  \node[font=\tiny, below] at (0.830,-4.28) {$0.830$};
  \node[font=\tiny, below] at (1,-4.28) {$1$};
\end{tikzpicture}
\end{center}
The update rule of
equation~\eqref{eq:corr-update}, drawn for the
first update. One row per set $\mathcal S$. $x = 2M_{a,q} = 1 + (-1)^a
\chi_{1000} = 0.830$ is the denominator. Each numerator is a waterfall: the bar
of $\chi_{\mathcal S}$ (blue), plus its partner's (with first bit flipped) signed
contribution $(-1)^a \chi_{\mathcal S \triangle \{q\}}$ (orange). The ray from the origin through the
waterfall's endpoint rescales it to a similar triangle with base
$x = 1$, so its height is the updated correlator $\chi'_{\mathcal S}$.
\end{example}

As Werner moves through the questions, more correlators settle to $\pm 1$. The 1-shell models the belief of the bias of each question. 
The 2-shell answers questions such as ``if $\psi(10) = 0$, then $\psi(01) = 1$''. And the higher orders are even more conditional, measuring belief about such statements, in the case of certain answers, etc. As we learn more, the conditions become satisfied or not, and the information becomes less nested. 

\begin{example}
The remaining
questions of the run ($00$, $01$, $10$, all answered $0$, so with
sign $+1$) each saturate their own bit the same way, and since
XOR toggles commute, the order does not matter - the
commutativity of updating, seen again. After all four, the ledger
is pure signs, $\chi_{\mathcal S} = (-1)^{|\mathcal S \cap \{11\}|}$, which is $-1$ on every
set containing $11$, $+1$ on the rest: the transform
of the point mass on $\psi = \phi_8$:
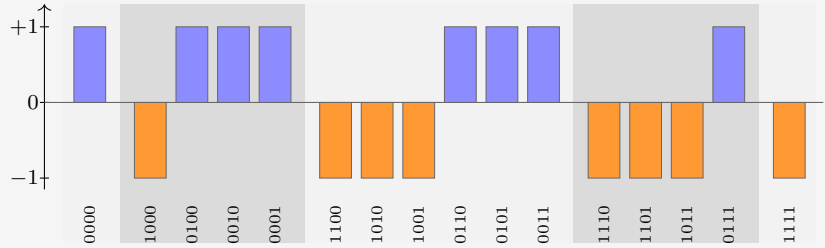
\begin{center}
\begin{tabular}{c l}
\toprule
shell & final correlators \\
\midrule
0 & $\chi_{0000} = 1$ \\[2pt]
1 & $\chi_{1000} = -1$,\ \ $\chi_{0100} = +1$,\ \
    $\chi_{0010} = +1$,\ \ $\chi_{0001} = +1$ \\[2pt]
2 & $\chi_{1100} = -1$,\ \ $\chi_{1010} = -1$,\ \
    $\chi_{1001} = -1$, \\
  & $\chi_{0110} = +1$,\ \ $\chi_{0101} = +1$,\ \
    $\chi_{0011} = +1$ \\[2pt]
3 & $\chi_{1110} = -1$,\ \ $\chi_{1101} = -1$,\ \
    $\chi_{1011} = -1$,\ \ $\chi_{0111} = +1$ \\[2pt]
4 & $\chi_{1111} = -1$ \\
\bottomrule
\end{tabular}
\end{center}
\ledgerbars{0/1/0000/blue!45,
  0.80/-1/1000/orange!80, 1.35/1/0100/blue!45,
  1.90/1/0010/blue!45, 2.45/1/0001/blue!45,
  3.25/-1/1100/orange!80, 3.80/-1/1010/orange!80,
  4.35/-1/1001/orange!80, 4.90/1/0110/blue!45,
  5.45/1/0101/blue!45, 6.00/1/0011/blue!45,
  6.80/-1/1110/orange!80, 7.35/-1/1101/orange!80,
  7.90/-1/1011/orange!80, 8.45/1/0111/blue!45,
  9.25/-1/1111/orange!80}
\end{example}

\subsection{General $m$}\label{sec:generalm}

It turns out that taking $m > 1$ is more
general but not much more instructive. With multiple output bits it
is as if we have $m$ a priori independent maps per question; they
just happen always to be asked and answered simultaneously, in
unison. Therefore we do not have to think about what one of them
teaches us about another. It would mean we have more than $2$
options every question, and therefore the factor by which every
update thins the hypotheses is not $2$ but $A$. The guiding
example helps because the maps are represented as binary numbers,
which are more familiar.

For completeness, the main formulas. The truth table is now
$mQ$ bits: one \textbf{site} $(q, i)$ per bit $i$ of each
answer, each carrying a spin
$\sigma_{q,i}(\phi_j) := (-1)^{\phi_j(q)_i}$, where $\phi_j(q)_i$
is the $i$-th bit of the answer. Correlators are indexed by sets
$\mathcal S$ of sites: masks of $mQ$ bits under the same numeral
convention, $m$ of them per question:
\begin{equation}\label{eq:wht-m}
  \chi_{\mathcal S} \;=\; \Big\langle \prod_{(q,i) \in \mathcal S} \sigma_{q,i}
  \Big\rangle
  \;=\; \sum_{j=0}^{N-1} (-1)^{\sum_{(q,i) \in \mathcal S} \phi_j(q)_i}\,
  p_j.
\end{equation}

\begin{example}
For $n = 3$, $m = 2$ the sites form a $2 \times 8$ grid: one column
per question, one row per bit of the answer. A set $\mathcal S$ is a pattern
of hot cells in this grid, and $\chi_{\mathcal S}$ multiplies the
corresponding spins:
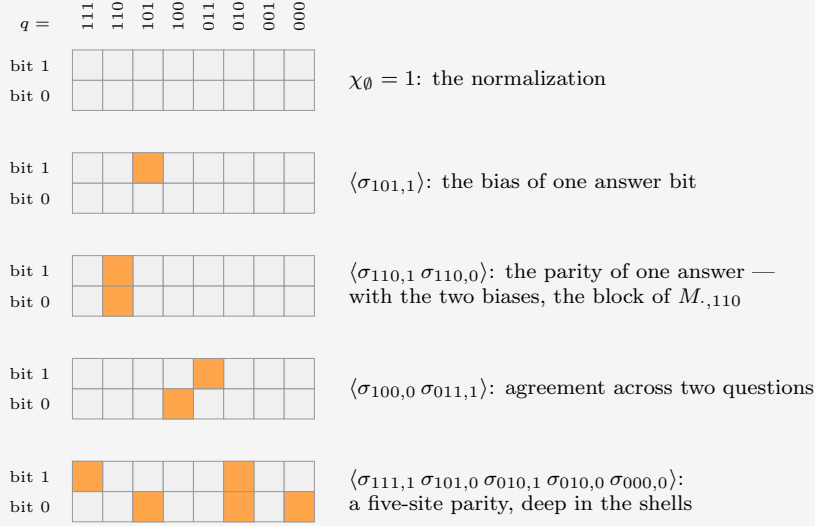
\begin{center}
\begin{tikzpicture}[x=0.40cm, y=0.40cm]
  \node[font=\tiny, anchor=east] at (-0.4,0.8) {$q =$};
  \foreach \c/\q in {0/111, 1/110, 2/101, 3/100,
                     4/011, 5/010, 6/001, 7/000}
    \node[font=\tiny, rotate=90, anchor=west] at ({\c+0.5},0.3)
      {$\q$};
  \foreach \g in {0, 1, 2, 3, 4} {
    \node[font=\tiny, anchor=east] at (-0.4,{-3.4*\g - 0.5})
      {bit $1$};
    \node[font=\tiny, anchor=east] at (-0.4,{-3.4*\g - 1.5})
      {bit $0$};
  }
  \foreach \g in {0, 1, 2, 3, 4}
    \foreach \c in {0,...,7}
      \foreach \r in {0, 1}
        \draw[black!40, line width=0.3pt, fill=gray!12]
          (\c,{-3.4*\g - 2 + \r}) rectangle
          ({\c+1},{-3.4*\g - 1 + \r});
  \foreach \g/\c/\r in {1/2/1, 2/1/0, 2/1/1, 3/3/0, 3/4/1,
                        4/0/1, 4/2/0, 4/5/0, 4/5/1, 4/7/0}
    \draw[black!40, line width=0.3pt, fill=orange!70]
      (\c,{-3.4*\g - 2 + \r}) rectangle
      ({\c+1},{-3.4*\g - 1 + \r});
  \node[font=\scriptsize, anchor=west] at (8.8,-1)
    {$\chi_\emptyset = 1$: the normalization};
  \node[font=\scriptsize, anchor=west] at (8.8,-4.4)
    {$\langle \sigma_{101,1} \rangle$: the bias of one answer bit};
  \node[font=\scriptsize, anchor=west, align=left] at (8.8,-7.8)
    {$\langle \sigma_{110,1}\, \sigma_{110,0} \rangle$: the parity
     of one answer ---\\with the two biases, the block of
     $M_{\cdot,110}$};
  \node[font=\scriptsize, anchor=west] at (8.8,-11.2)
    {$\langle \sigma_{100,0}\, \sigma_{011,1} \rangle$: agreement
     across two questions};
  \node[font=\scriptsize, anchor=west, align=left] at (8.8,-14.6)
    {$\langle \sigma_{111,1}\, \sigma_{101,0}\, \sigma_{010,1}\,
      \sigma_{010,0}\, \sigma_{000,0} \rangle$:\\a five-site
      parity, deep in the shells};
\end{tikzpicture}
\end{center}
Any of the $2^{16} = 65536$ hot/cold patterns is a correlator, and
sorting by the number of hot cells gives the shells. The mask of
$\mathcal S$ is the grid read as a $16$-bit numeral.
\end{example}

Write $\mathcal B_q$ for the block of $q$'s $m$ sites, and
$a \cdot \mathcal T := \sum_{i \in \mathcal T} a_i$ for the parity of the answer $a$
on a subset $\mathcal T \subseteq \mathcal B_q$. The answer matrix is read off the
correlators supported on one block. There are $A$ of them per column:
\begin{equation}\label{eq:M-m}
  M_{a,q} \;=\; \frac{1}{A} \sum_{\mathcal T \subseteq \mathcal B_q}
  (-1)^{a \cdot \mathcal T}\, \chi_{\mathcal T}.
\end{equation}
The observation mask at $(q, a)$ is a function of the $m$ bits of
one block, so its ledger has $A$ nonzero entries, and hearing
$a = \psi(q)$ combines every correlator with its $A$ partners ---
one per toggle $\mathcal T$ within the asked block:
\begin{equation}\label{eq:corr-update-m}
  \chi'_{\mathcal S} \;=\; \frac{\sum_{\mathcal T \subseteq \mathcal B_q} (-1)^{a \cdot \mathcal T}\,
    \chi_{\mathcal S \triangle \mathcal T}}{A\, M_{a,q}},
\end{equation}
the denominator once more the $\mathcal S = \emptyset$ instance of the
numerator. For $m = 1$ the three
formulas collapse to
equations~\eqref{eq:wht}, \eqref{eq:shell1}
and~\eqref{eq:corr-update}.

\begin{example}
To draw the update, take a sparse prior at $n = 3$, $m = 2$: four
maps, $\phi(q) = q \bmod 4$ (weight $0.45$), constant $00$
($0.35$), ``$11$ iff $q \geq 4$'' ($0.12$), constant $11$
($0.08$). Werner asks $q = 110$ and hears $a = 11$: toggle signs
$(+, -, -, +)$ on $(\emptyset, 110_0, 110_1, \text{both})$, where
$q_i$ denotes site $(q, i)$, and denominator
$A M_{11,110} = 0.8$. The pairs of the $m = 1$ figure become
\emph{quartets}:
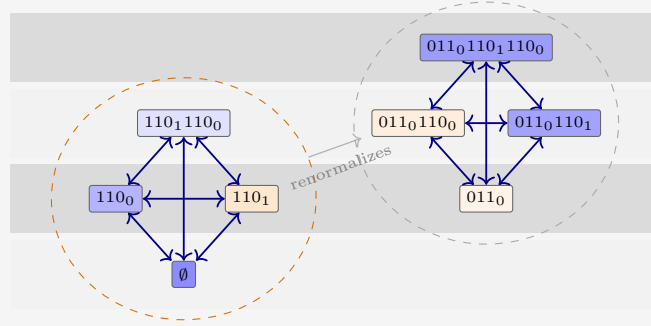
\begin{center}
\begin{tikzpicture}[x=1cm, y=1cm,
  lnode/.style={draw=black!60, rounded corners=1pt,
                inner sep=2.5pt, font=\tiny, align=center},
  pair/.style={<->, blue!55!black, line width=0.7pt}]
  \fill[gray!10] (-0.3,-0.45) rectangle (8.3,0.45);
  \fill[gray!28] (-0.3,0.55) rectangle (8.3,1.45);
  \fill[gray!10] (-0.3,1.55) rectangle (8.3,2.45);
  \fill[gray!28] (-0.3,2.55) rectangle (8.3,3.45);
  \node[lnode, fill=blue!45]  (qa0) at (2,0)   {$\emptyset$};
  \node[lnode, fill=blue!30]  (qa1) at (1.1,1) {$110_0$};
  \node[lnode, fill=orange!19] (qa2) at (2.9,1) {$110_1$};
  \node[lnode, fill=blue!12]  (qa3) at (2,2)   {$110_1 110_0$};
  \draw[pair] (qa0) -- (qa1); \draw[pair] (qa0) -- (qa2);
  \draw[pair] (qa0) -- (qa3); \draw[pair] (qa1) -- (qa2);
  \draw[pair] (qa1) -- (qa3); \draw[pair] (qa2) -- (qa3);
  \node[lnode, fill=orange!10] (qb0) at (6,1)   {$011_0$};
  \node[lnode, fill=orange!13] (qb1) at (5.1,2) {$011_0 110_0$};
  \node[lnode, fill=blue!36]   (qb2) at (6.9,2) {$011_0 110_1$};
  \node[lnode, fill=blue!39]   (qb3) at (6,3)   {$011_0 110_1 110_0$};
  \draw[pair] (qb0) -- (qb1); \draw[pair] (qb0) -- (qb2);
  \draw[pair] (qb0) -- (qb3); \draw[pair] (qb1) -- (qb2);
  \draw[pair] (qb1) -- (qb3); \draw[pair] (qb2) -- (qb3);
  \draw[dashed, orange!85!black] (2,1) ellipse (1.75 and 1.6);
  \draw[dashed, gray!70] (6,2) ellipse (1.75 and 1.6);
  \draw[->, gray!60, line width=0.5pt] (3.65,1.55) -- (4.35,1.8);
  \node[font=\tiny, gray, below, rotate=20] at (4.0,1.62)
    {renormalizes};
\end{tikzpicture}
\end{center}
For $m = 2$, toggling any subset of the asked
block's two sites ties four correlators into a clique that
averages jointly, per equation~\eqref{eq:corr-update-m}. The
on-block quartet (orange hull) contains the normalization and the
entire block of $M_{\cdot,110}$; its signed sum is the denominator
that renormalizes every other quartet. Fills to scale for the
sparse prior, as before.

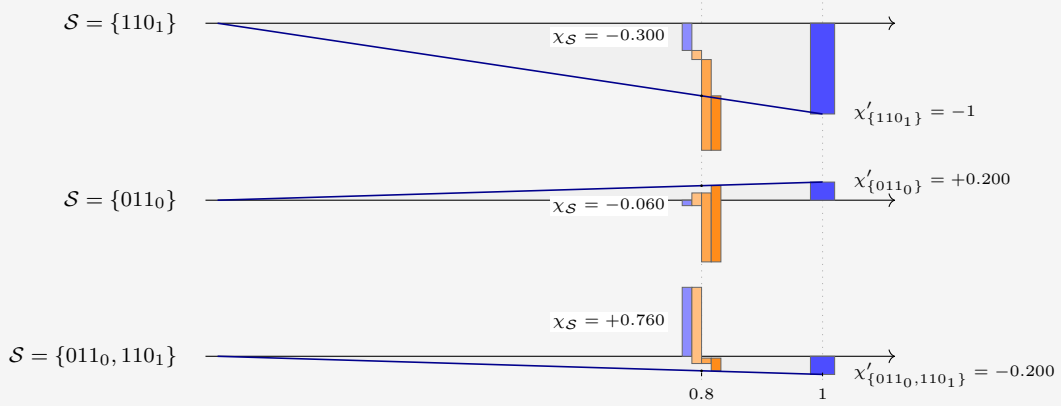
\begin{center}

\begin{tikzpicture}[x=8cm, y=1.2cm]
  \draw[dotted, gray!60] (0.8,0.3) -- (0.8,-3.95);
  \draw[dotted, gray!60] (1,0.3) -- (1,-3.95);
  \fill[gray!12] (0,0) -- (1,0) -- (1,-1) -- cycle;
  \draw[->] (-0.02,0) -- (1.12,0);
  \draw[fill=blue!45, draw=black!60, line width=0.3pt]
    (0.768,0) rectangle (0.784,-0.300);
  \draw[fill=orange!50, draw=black!60, line width=0.3pt]
    (0.784,-0.300) rectangle (0.800,-0.400);
  \draw[fill=orange!70, draw=black!60, line width=0.3pt]
    (0.800,-0.400) rectangle (0.816,-1.400);
  \draw[fill=orange!90, draw=black!60, line width=0.3pt]
    (0.816,-1.400) rectangle (0.832,-0.800);
  \draw[fill=blue!70, draw=black!60, line width=0.3pt]
    (0.98,0) rectangle (1.02,-1);
  \draw[blue!55!black, line width=0.6pt] (0,0) -- (1,-1);
  \fill[black] (0.8,-0.8) circle (0.5pt);
  \node[font=\tiny, fill=white, inner sep=1pt, anchor=east]
    at (0.745,-0.15) {$\chi_{\mathcal S} = -0.300$};
  \node[font=\tiny, anchor=west] at (1.035,-1)
    {$\chi'_{\{110_1\}} = -1$};
  \node[font=\scriptsize, anchor=east] at (-0.05,0)
    {$\mathcal S = \{110_1\}$};
  \fill[gray!12] (0,-1.95) -- (1,-1.95) -- (1,-1.75) -- cycle;
  \draw[->] (-0.02,-1.95) -- (1.12,-1.95);
  \draw[fill=blue!45, draw=black!60, line width=0.3pt]
    (0.768,-1.95) rectangle (0.784,-2.010);
  \draw[fill=orange!50, draw=black!60, line width=0.3pt]
    (0.784,-2.010) rectangle (0.800,-1.870);
  \draw[fill=orange!70, draw=black!60, line width=0.3pt]
    (0.800,-1.870) rectangle (0.816,-2.630);
  \draw[fill=orange!90, draw=black!60, line width=0.3pt]
    (0.816,-2.630) rectangle (0.832,-1.790);
  \draw[fill=blue!70, draw=black!60, line width=0.3pt]
    (0.98,-1.95) rectangle (1.02,-1.75);
  \draw[blue!55!black, line width=0.6pt] (0,-1.95) -- (1,-1.75);
  \fill[black] (0.8,-1.79) circle (0.5pt);
  \node[font=\tiny, fill=white, inner sep=1pt, anchor=east]
    at (0.745,-2.01) {$\chi_{\mathcal S} = -0.060$};
  \node[font=\tiny, anchor=west] at (1.035,-1.75)
    {$\chi'_{\{011_0\}} = +0.200$};
  \node[font=\scriptsize, anchor=east] at (-0.05,-1.95)
    {$\mathcal S = \{011_0\}$};
  \fill[gray!12] (0,-3.67) -- (1,-3.67) -- (1,-3.87) -- cycle;
  \draw[->] (-0.02,-3.67) -- (1.12,-3.67);
  \draw[fill=blue!45, draw=black!60, line width=0.3pt]
    (0.768,-3.67) rectangle (0.784,-2.910);
  \draw[fill=orange!50, draw=black!60, line width=0.3pt]
    (0.784,-2.910) rectangle (0.800,-3.750);
  \draw[fill=orange!70, draw=black!60, line width=0.3pt]
    (0.800,-3.750) rectangle (0.816,-3.690);
  \draw[fill=orange!90, draw=black!60, line width=0.3pt]
    (0.816,-3.690) rectangle (0.832,-3.830);
  \draw[fill=blue!70, draw=black!60, line width=0.3pt]
    (0.98,-3.67) rectangle (1.02,-3.87);
  \draw[blue!55!black, line width=0.6pt] (0,-3.67) -- (1,-3.87);
  \fill[black] (0.8,-3.83) circle (0.5pt);
  \node[font=\tiny, fill=white, inner sep=1pt, anchor=east]
    at (0.745,-3.29) {$\chi_{\mathcal S} = +0.760$};
  \node[font=\tiny, anchor=west] at (1.035,-3.87)
    {$\chi'_{\{011_0, 110_1\}} = -0.200$};
  \node[font=\scriptsize, anchor=east] at (-0.05,-3.67)
    {$\mathcal S = \{011_0, 110_1\}$};
  \draw (0.8,-3.89) -- (0.8,-3.85);
  \draw (1,-3.89) -- (1,-3.85);
  \node[font=\tiny, below] at (0.8,-3.91) {$0.8$};
  \node[font=\tiny, below] at (1,-3.91) {$1$};
\end{tikzpicture}
\end{center}
The $m = 2$ update, three sets from the sparse
prior. The waterfall now has four segments: $\chi_{\mathcal S}$ (blue) and
its three signed quartet partners, in toggle order $110_0$,
$110_1$, both (light to dark orange). The ray from the origin
rescales the endpoint from $x = A M_{a,q} = 0.8$ to $x = 1$.
The asked site saturates to $-1$; the foreign site $011_0$ flips
sign, $-0.060 \to +0.200$; and $\{011_0, 110_1\}$ lands on
$-0.200$, the renormalized value $-\chi'_{\{011_0\}}$.
\end{example}

\section{Playing the guiding example}\label{app:playing}

This appendix plays the guiding example of Table~\ref{tab:sixteen}
through all four questions, one round per example, so that the ledgers
of Subsection~\ref{sec:bookkeeping} can be watched in motion. Before
each round Werner computes, for every question still unasked, the two
forecasts of Subsection~\ref{sec:forecasts}: the expected surprisal
$\langle s \rangle(q)$, which is the entropy of that question's
column, and the expected drop $\langle \Delta H(M) \rangle(q)$ of the
whole table. We append both forecasts as extra rows under $M$, so every
table below carries beneath it what Werner expected before he asked.

\begin{example}

Werner's belief is the prior of Table~\ref{tab:sixteen} as before. He plays greedily: always asking the question with the largest
forecast reduction in matrix entropy. The truth is AND ($\psi = \phi_8$).

\textbf{Question 1.} The starting state:
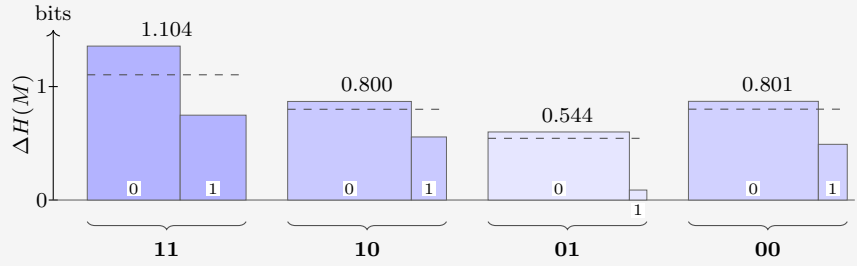
\[
  \begin{array}{c|cccc}
    \dgc[\langle \Delta H(M) \rangle]{\mathbf a}{\mathbf q}
      & \mathbf{11} & \mathbf{10} & \mathbf{01} & \mathbf{00} \\
    \hline
    \mathbf{0} & \hm{20} 0.585 & \hm{27} 0.779 & \hm{31} 0.890 & \hm{29} 0.818 \\[2pt]
    \mathbf{1} & \hm{15} 0.415 & \hm{8} 0.221 & \hm{4} 0.110 & \hm{6} 0.182 \\
    \hline
    \langle s \rangle & \hm{30} 0.979 & \hm{21} 0.762 & \hm{10} 0.500 & \hm{18} 0.684 \\[2pt]
    \langle \Delta H(M) \rangle & \hm{35} 1.104 & \hm{22} 0.800 & \hm{12} 0.544 & \hm{22} 0.801
  \end{array}
\]
\begin{center}
\begin{tikzpicture}[x=1cm, y=1.5cm]
  \faxis
  \fslot{0}{30}{0.585}{1.357}{0.748}{1.104}
  \fslot{2.65}{21}{0.779}{0.869}{0.556}{0.800}
  \fslot{5.30}{10}{0.890}{0.600}{0.088}{0.544}
  \fslot{7.95}{18}{0.818}{0.870}{0.491}{0.801}
  \fbrace{0}{11}\fbrace{2.65}{10}\fbrace{5.30}{01}\fbrace{7.95}{00}
\end{tikzpicture}
\end{center}
To see where the forecast entries come from, take $q = 11$: the
answer $a = 0$ arrives with probability $M_{0,11} = 0.585$ and would
drop $H(M)$ from $2.925$ to $1.569$; the answer $a = 1$ arrives with
probability $0.415$ and drops it only to $2.178$, so
\begin{align*}
  \langle s \rangle(11)
  &= 0.585\, (0.773) + 0.415\, (1.269) = 0.979, \\
  \langle \Delta H(M) \rangle(11)
  &= 0.585\, (1.357) + 0.415\, (0.748) = 1.104.
\end{align*}
\end{example}

\begin{example}
Both forecast rows point to $q = 11$: it is the closest thing to a
coin on offer ($\langle s\rangle = 0.979$) and moves the whole
table most, by a wide margin. Part of $\Delta H(M)$ is just $\langle s \rangle$ itself.
Werner asks it, and nature answers $a = 1$ (probability $0.415$,
surprisal $s_1 = 1.269$). Eight maps die:
\begin{center}
\begin{tabular}{c c l c c c}
\toprule
$j$ & $\phi_j(11)$ & name & $p_j$ & $p'_j$ & \\
\midrule
\tikzmark{w1a}0 & 0 & FALSE (constant $0$) & 0.394 & - & \tikzmark{w1ae} \\
\tikzmark{w1b}1 & 0 & $\lnot(b_1 \vee b_0)$ (NOR) & 0.010 & - & \tikzmark{w1be} \\
\tikzmark{w1c}2 & 0 & $\lnot b_1 \wedge b_0$ & 0.010 & - & \tikzmark{w1ce} \\
\tikzmark{w1d}3 & 0 & $\lnot b_1$ & 0.010 & - & \tikzmark{w1de} \\
\tikzmark{w1e}4 & 0 & $b_1 \wedge \lnot b_0$ & 0.131 & - & \tikzmark{w1ee} \\
\tikzmark{w1f}5 & 0 & $\lnot b_0$ & 0.010 & - & \tikzmark{w1fe} \\
\tikzmark{w1g}6 & 0 & $b_1 \oplus b_0$ (XOR) & 0.010 & - & \tikzmark{w1ge} \\
\tikzmark{w1h}7 & 0 & $\lnot(b_1 \wedge b_0)$ (NAND) & 0.010 & - & \tikzmark{w1he} \\
8 & 1 & $b_1 \wedge b_0$ (AND) & 0.213 & 0.513 & \pbar{0.513} \\
9 & 1 & $b_1 = b_0$ (XNOR) & 0.092 & 0.222 & \pbar{0.222} \\
10 & 1 & $b_0$ (projection) & 0.040 & 0.096 & \pbar{0.096} \\
11 & 1 & $b_1 \to b_0$ (IF THEN) & 0.010 & 0.024 & \pbar{0.024} \\
12 & 1 & $b_1$ (projection) & 0.010 & 0.024 & \pbar{0.024} \\
13 & 1 & $b_0 \to b_1$ (IF THEN) & 0.030 & 0.072 & \pbar{0.072} \\
14 & 1 & $b_1 \vee b_0$ (OR) & 0.010 & 0.024 & \pbar{0.024} \\
15 & 1 & TRUE (constant $1$) & 0.010 & 0.024 & \pbar{0.024} \\
\bottomrule
\end{tabular}
\begin{tikzpicture}[overlay, remember picture]
  \foreach \s/\e in {w1a/w1ae, w1b/w1be, w1c/w1ce, w1d/w1de,
                     w1e/w1ee, w1f/w1fe, w1g/w1ge, w1h/w1he}
    \draw[line width=0.5pt]
      ([shift={(-2pt,0.55ex)}]pic cs:\s) --
      ([shift={(2pt,0.55ex)}]pic cs:\e);
\end{tikzpicture}
\end{center}

The ledger:
\[
  H(M)\colon 2.925 \to 2.178, \qquad
  H(p)\colon 2.707 \to 2.093, \qquad
  C\colon 0.218 \to 0.085.
\]
Werner received $1.269$ bits but destroyed only $0.748$ bits of table.
Correlations point the right way only on average, not on every draw. In this case, Werner got the much less informative branch, and paid the premium price for it.
The destroyed-per-received ratio of this round is $L_1 = 0.589$: well
below $1$, anti-leverage.
\end{example}

Question 1 is a reminder that the forecast rows are
\emph{expectations}: they hold on average over many repetitions of
the experiment, and only if the world really does draw its truth $\psi$
from Werner's prior. On any single run the realized numbers are free
to deviate.

\begin{example}
\textbf{Question 2.} The state after the collapse, with fresh
forecasts:
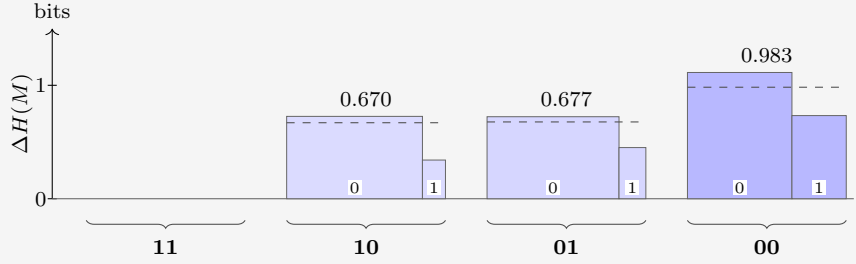
\[
  \begin{array}{c|cccc}
    \dgc[\langle \Delta H(M) \rangle]{\mathbf a}{\mathbf q}
      & \mathbf{11} & \mathbf{10} & \mathbf{01} & \mathbf{00} \\
    \hline
    \mathbf{0} & \hm{0} 0 & \hm{30} 0.855 & \hm{29} 0.831 & \hm{23} 0.658 \\[2pt]
    \mathbf{1} & \hm{35} 1 & \hm{5} 0.145 & \hm{6} 0.169 & \hm{12} 0.342 \\
    \hline
    \langle s \rangle & \text{---} & \hm{14} 0.596 & \hm{16} 0.655 & \hm{28} 0.927 \\[2pt]
    \langle \Delta H(M) \rangle & \text{---} & \hm{17} 0.670 & \hm{17} 0.677 & \hm{30} 0.983
  \end{array}
\]
\begin{center}
\begin{tikzpicture}[x=1cm, y=1.5cm]
  \faxis
  \fslot{2.65}{14}{0.855}{0.726}{0.341}{0.670}
  \fslot{5.30}{16}{0.831}{0.723}{0.451}{0.677}
  \fslot{7.95}{28}{0.658}{1.113}{0.733}{0.983}
  \fbrace{0}{11}\fbrace{2.65}{10}\fbrace{5.30}{01}\fbrace{7.95}{00}
\end{tikzpicture}
\end{center}
Greedy Werner picks $q = 00$. Nature answers
$a = 0$, which is the answer Werner favored (probability $0.658$), so the
surprisal is only $s_2 = 0.604$ bits. Four of the eight die:
\begin{center}
\begin{tabular}{c c l c c c}
\toprule
$j$ & $\phi_j(00)$ & name & $p_j$ & $p'_j$ & \\
\midrule
8 & 0 & $b_1 \wedge b_0$ (AND) & 0.513 & 0.780 & \pbar{0.780} \\
\tikzmark{w2a}9 & 1 & $b_1 = b_0$ (XNOR) & 0.222 & - & \tikzmark{w2ae} \\
10 & 0 & $b_0$ (projection) & 0.096 & 0.147 & \pbar{0.147} \\
\tikzmark{w2b}11 & 1 & $b_1 \to b_0$ (IF THEN) & 0.024 & - & \tikzmark{w2be} \\
12 & 0 & $b_1$ (projection) & 0.024 & 0.037 & \pbar{0.037} \\
\tikzmark{w2c}13 & 1 & $b_0 \to b_1$ (IF THEN) & 0.072 & - & \tikzmark{w2ce} \\
14 & 0 & $b_1 \vee b_0$ (OR) & 0.024 & 0.037 & \pbar{0.037} \\
\tikzmark{w2d}15 & 1 & TRUE (constant $1$) & 0.024 & - & \tikzmark{w2de} \\
\bottomrule
\end{tabular}
\begin{tikzpicture}[overlay, remember picture]
  \foreach \s/\e in {w2a/w2ae, w2b/w2be, w2c/w2ce, w2d/w2de}
    \draw[line width=0.5pt]
      ([shift={(-2pt,0.55ex)}]pic cs:\s) --
      ([shift={(2pt,0.55ex)}]pic cs:\e);
\end{tikzpicture}
\end{center}

The ledger:
\[
  H(M)\colon 2.178 \to 1.065, \qquad
  H(p)\colon 2.093 \to 1.035, \qquad
  C\colon 0.085 \to 0.030.
\]
This time $1.113$ bits are destroyed for $0.604$ received. Where did
the extra $0.509$ come from? Partly from the favored answer arriving
- the belief dropped $1.058$ against only $0.604$ paid - and
partly from the store: $C$ fell by $0.055$.
Step ratio $1.842$. Cumulatively, however, $1.860$ destroyed per $1.873$
received gives $L_2 = 0.993$: still a hair
below $1$. Even a strong second round has not quite repaid the
expensive shock of question 1.
\end{example}

Why track $C$, when its step-by-step changes visibly do not match
the amounts deduced? Because on a single realized run the deduced
part has \emph{two} sources, and $C$ is the only quantity that
separates them. Writing $\Delta$ for the realized drops in one step,
\begin{equation}\label{eq:windfall-tower}
  \underbrace{\Delta H(M) - s}_{\text{deduced}}
  \;=\;
  \underbrace{\big[\Delta H(p) - s\big]}_{\text{windfall}}
  \;+\;
  \underbrace{\big[{-}\Delta C\big]}_{\text{tower}},
\end{equation}
an identity, since $H(M) = H(p) + C$ at every stage. The windfall is
luck: the realized surprisal undershooting (or overshooting) the
realized drop in belief entropy. Windfall is zero in expectation. The tower
term is structure: a genuine withdrawal from the correlation store,
and the channel with a global guarantee. Over a complete run its
withdrawals sum to exactly the initial $C$, while the windfalls
average away. 

\begin{example}
\textbf{Question 3.} Two questions remain:
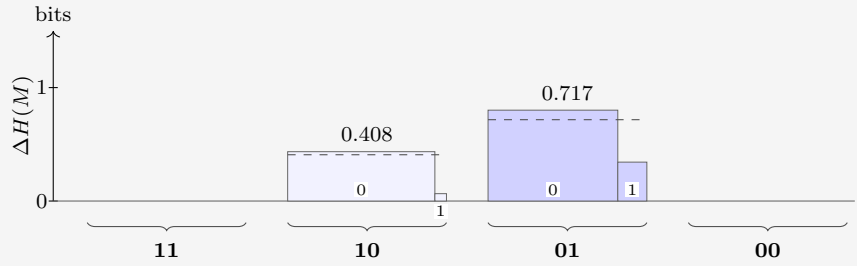
\[
  \begin{array}{c|cccc}
    \dgc[\langle \Delta H(M) \rangle]{\mathbf a}{\mathbf q}
      & \mathbf{11} & \mathbf{10} & \mathbf{01} & \mathbf{00} \\
    \hline
    \mathbf{0} & \hm{0} 0 & \hm{32} 0.927 & \hm{29} 0.817 & \hm{35} 1 \\[2pt]
    \mathbf{1} & \hm{35} 1 & \hm{3} 0.073 & \hm{6} 0.183 & \hm{0} 0 \\
    \hline
    \langle s \rangle & \text{---} & \hm{5} 0.378 & \hm{18} 0.687 & \text{---} \\[2pt]
    \langle \Delta H(M) \rangle & \text{---} & \hm{6} 0.408 & \hm{19} 0.717 & \text{---}
  \end{array}
\]
\begin{center}
\begin{tikzpicture}[x=1cm, y=1.5cm]
  \faxis
  \fslot{2.65}{5}{0.927}{0.435}{0.065}{0.408}
  \fslot{5.30}{18}{0.817}{0.801}{0.343}{0.717}
  \fbrace{0}{11}\fbrace{2.65}{10}\fbrace{5.30}{01}\fbrace{7.95}{00}
\end{tikzpicture}
\end{center}
Greedy Werner picks $q = 01$; nature answers $a = 0$ (probability $0.817$,
surprisal $s_3 = 0.292$). Two die:
\begin{center}
\begin{tabular}{c c l c c c}
\toprule
$j$ & $\phi_j(01)$ & name & $p_j$ & $p'_j$ & \\
\midrule
8 & 0 & $b_1 \wedge b_0$ (AND) & 0.780 & 0.955 & \pbar{0.955} \\
\tikzmark{w3a}10 & 1 & $b_0$ (projection) & 0.147 & - & \tikzmark{w3ae} \\
12 & 0 & $b_1$ (projection) & 0.037 & 0.045 & \pbar{0.045} \\
\tikzmark{w3b}14 & 1 & $b_1 \vee b_0$ (OR) & 0.037 & - & \tikzmark{w3be} \\
\bottomrule
\end{tabular}
\begin{tikzpicture}[overlay, remember picture]
  \foreach \s/\e in {w3a/w3ae, w3b/w3be}
    \draw[line width=0.5pt]
      ([shift={(-2pt,0.55ex)}]pic cs:\s) --
      ([shift={(2pt,0.55ex)}]pic cs:\e);
\end{tikzpicture}
\end{center}

The ledger:
\[
  H(M)\colon 1.065 \to 0.264, \qquad
  H(p)\colon 1.035 \to 0.264, \qquad
  C\colon 0.030 \to 0.
\]
Again a bargain: $0.801$ destroyed for a mere $0.292$ received, and
$L_3 = 1.229$ finally climbs above unity. And note that $C$
has dropped to zero.
\end{example}

$C$ measures correlations that are not visible from $M$. If there's only one unknown column, there's nothing to infer. 
Put another way, usually, the binary maps are an overcomplete basis for $M$, because their number $N \gg \dim(M) = Q(A-1)$. Expressing $M$ as a linear combination of maps drops knowledge.
That's not the case once learning has reduced the number of free parameters to one each.

\begin{example}[label=ex:run21]
\textbf{Question 4.} One question, one heavily loaded coin:
\[
  \begin{array}{c|cccc}
    \dgc[\langle \Delta H(M) \rangle]{\mathbf a}{\mathbf q}
      & \mathbf{11} & \mathbf{10} & \mathbf{01} & \mathbf{00} \\
    \hline
    \mathbf{0} & \hm{0} 0 & \hm{33} 0.955 & \hm{35} 1 & \hm{35} 1 \\[2pt]
    \mathbf{1} & \hm{35} 1 & \hm{2} 0.045 & \hm{0} 0 & \hm{0} 0 \\
    \hline
    \langle s \rangle & \text{---} & \hm{0} 0.264 & \text{---} & \text{---} \\[2pt]
    \langle \Delta H(M) \rangle & \text{---} & \hm{0} 0.264 & \text{---} & \text{---}
  \end{array}
\]
The two forecasts now coincide: with $C = 0$ there is no transfer
left to hope for, and the last question can only be observed, not
leveraged. Nature answers $a = 0$ (probability $0.955$, surprisal
$s_4 = 0.066$):
\begin{center}
\begin{tabular}{c c l c c c}
\toprule
$j$ & $\phi_j(10)$ & name & $p_j$ & $p'_j$ & \\
\midrule
8 & 0 & $b_1 \wedge b_0$ (AND) & 0.955 & 1.000 & \pbar{1.000} \\
\tikzmark{w4a}12 & 1 & $b_1$ (projection) & 0.045 & - & \tikzmark{w4ae} \\
\bottomrule
\end{tabular}
\begin{tikzpicture}[overlay, remember picture]
  \draw[line width=0.5pt]
    ([shift={(-2pt,0.55ex)}]pic cs:w4a) --
    ([shift={(2pt,0.55ex)}]pic cs:w4ae);
\end{tikzpicture}
\end{center}

The ledger closes:
\[
  H(M)\colon 0.264 \to 0, \qquad
  H(p)\colon 0.264 \to 0, \qquad
  C\colon 0 \to 0.
\]
The truth is AND. The step destroyed $0.264$ for $0.066$ received
(ratio $3.990$) - not deduction this time, but luck: the destroyed
amount was fixed in advance (the column's full entropy), while the
realized surprisal undershot it because the likelier answer arrived.
In the split of equation~\eqref{eq:windfall-tower}: pure windfall,
tower $0$.

\textbf{Summary.} Cumulatively: $2.925$ bits of table
destroyed, all of it, for $2.231$ bits of
surprisal received: $L_4 = 1.311$. Most of that was luck, not
intelligence. The figure below
stacks the bits after each question: what uncertainty remains, plus what was
received. Once it dips below the starting entropy, our prior has done net work. 

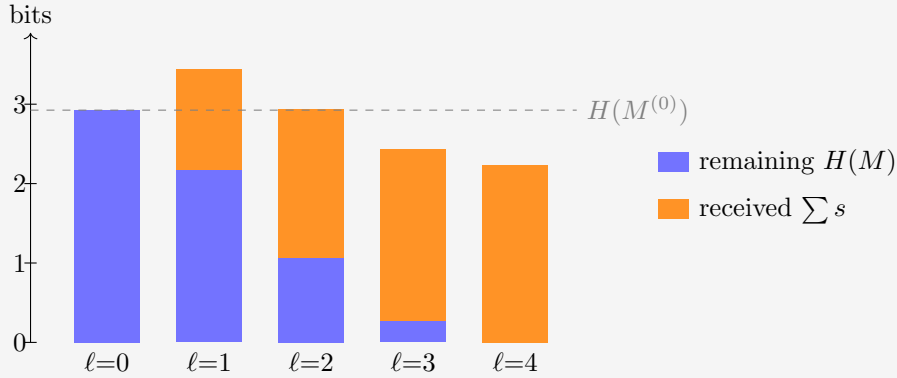
\begin{center}
\begin{tikzpicture}[x=1.35cm, y=1.05cm]
  \draw[->] (-0.75,0) -- (-0.75,3.9);
  \node[above, font=\small] at (-0.75,3.9) {bits};
  \foreach \y in {0,1,2,3}
    \draw (-0.81,\y) -- (-0.69,\y) node[left, font=\small] {$\y$};
  \fill[blue!55] (-0.32,0) rectangle (0.32,2.925);
  \fill[blue!55] (0.68,0) rectangle (1.32,2.178);
  \fill[orange!85] (0.68,2.178) rectangle (1.32,3.446);
  \fill[blue!55] (1.68,0) rectangle (2.32,1.065);
  \fill[orange!85] (1.68,1.065) rectangle (2.32,2.938);
  \fill[blue!55] (2.68,0) rectangle (3.32,0.264);
  \fill[orange!85] (2.68,0.264) rectangle (3.32,2.429);
  \fill[orange!85] (3.68,0) rectangle (4.32,2.231);
  \draw[dashed, gray] (-0.75,2.925) -- (4.6,2.925);
  \node[right, font=\small, gray] at (4.6,2.925) {$H(M^{(0)})$};
  \foreach \l in {0,...,4}
    \node[below, font=\small] at (\l,0) {$\ell{=}\l$};
  \fill[blue!55] (5.4,2.15) rectangle (5.7,2.4);
  \node[right, font=\small] at (5.7,2.27) {remaining $H(M)$};
  \fill[orange!85] (5.4,1.55) rectangle (5.7,1.8);
  \node[right, font=\small] at (5.7,1.67) {received $\sum s$};
\end{tikzpicture}
\end{center}
The realized run through the guiding example:
remaining table entropy plus cumulative surprisal received, after
each question, against the initial total (dashed). The gap between
each stack and the line is the cumulative deduced part. At
$\ell = 1$ the stack tops out $0.521$ bits \emph{above} the
line - the negative deduction of question 1, plainly visible. At
$\ell = 2$ the stack still grazes the line from above ($0.013$ bits);
only from $\ell = 3$ on do the stacks fall short of it, as genuine
deduction overtakes the early loss.
\end{example}

\section{Block entropies by elimination}\label{app:tensor-marginals}

A question eliminates maps whose answers disagree with the answer received.
For any fixed set of questions and answer string, the probability of that
string is the total prior weight of the maps that survive. Calculating
$G_k$ of \eqref{eq:Gk} therefore involves three sums: over question sets, over their possible
answer strings, and over the maps that elimination has not excluded.
The last sum can reuse many of the same additions. We arrange these
elimination rules in a tensor, so that all the surviving weights are
calculated together. The construction requires no symmetry or independence
of the prior.

\subsection{Map survival under binary answers}

In order to discover the mechanism, it suffices to take $m=1$. For one question, there are three rules we want to represent:
keep maps answering $0$, keep maps answering $1$, or keep both. We denote the third rule by $\wild$, a wildcard standing for either
answer. It means that \emph{this question has not excluded the map}: because it hasn't been asked yet.
The symbol labels an elimination rule, not an additional possible answer.
In particular, it makes no assertion about whether the answer could be
deduced from other questions.

The local matrix is
\begin{equation}\label{eq:Xi-binary}
  \Xi\;:=\;
  \begin{array}{c|cc}
    &a=0&a=1\\ \hline
    0&1&0\\
    1&0&1\\
    \wild&1&1
  \end{array}.
\end{equation}
The column is the map's answer to this question. The row is the rule being
applied. An entry $1$ says that elimination has not excluded this map;
an entry $0$ says that it has. Thus the first row keeps only maps answering
$0$, the second keeps only maps answering $1$, and the third keeps both.
If the two columns carry total prior weights $p_0,p_1$, multiplication gives
\begin{equation}
  \Xi\begin{pmatrix}p_0\\p_1\end{pmatrix}
  =\begin{pmatrix}p_0\\p_1\\p_0+p_1\end{pmatrix}.
\end{equation}
The third component adds the weights of both surviving groups.

A map has $Q$ truth-table digits. In descending question order, we reshape its
prior weight as
\begin{equation}
  p_{a_{Q-1}\cdots a_0}:=p_j,
  \qquad j=\sum_{q=0}^{Q-1}a_q2^q,
  \qquad \mathbb R^{2^Q}\cong(\mathbb R^2)^{\otimes Q}.
\end{equation}
There is one tensor factor per question. This reshaping does not assume
that the answers are independent, it is merely a representation of index $j$. Applying the local rule at every
question gives the full matrix and its output:
\begin{equation}\label{eq:Xi-transform}
  \mu=\Xi^{\otimes Q}p,
  \qquad z\in\{0,1,\wild\}^Q.
\end{equation}
A row $z$ chooses one rule at each question. Its entry in column $j$ is
\begin{equation}\label{eq:Xi-survival}
  (\Xi^{\otimes Q})_{z,j}
  =\prod_{q=0}^{Q-1}\Xi_{z_q,\phi_j(q)}.
\end{equation}
The product is one precisely when the map survives every rule. A single
incompatible answer makes it zero. The tensor product includes every
combination of the local rules.

Let $\mathcal S(z)=\{q:z_q\ne\wild\}$ be the questions on which a
particular answer is required, and let $y$ be those answers in the inherited
question order. The resulting component is
\begin{equation}\label{eq:Xi-components}
  \mu_z=\sum_{j:\,\phi_j(\mathcal S(z))=y}p_j
      =P_{\mathcal S(z),y}.
\end{equation}
It is the total prior weight that elimination leaves. It is not yet a
posterior distribution over the surviving maps; that posterior would
normalize their individual weights by $\mu_z$, when $\mu_z>0$.

In particular, the component $\mu_{\wild\wild\ldots\wild} = 1$ is the normalization
condition, with no maps excluded.

\begin{example}
On the guiding space $n=2, m=1$, with question order $(11,10,01,00)$, the same
rule gives
\begin{equation}
  \mu_{1\wild0\wild}
  =\mu_{100\wild}+\mu_{110\wild}
  =(p_8+p_9)+(p_{12}+p_{13}).
\end{equation}
Only questions $11$ and $01$ restrict the maps. The other two questions
exclude none. The intermediate sums can also be used by other rows.
\end{example}

Collecting the averages:
we write $|z|:=|\mathcal S(z)|$ for the number of single-answer restrictions,
and $f(u):=-u\log_2u$, with $f(0):=0$. Each pair $(\mathcal S,y)$ occurs
exactly once among the rows: we put $y$ on $\mathcal S$ and $\wild$
everywhere else. Therefore
\begin{equation}\label{eq:Xi-Gk}
  G_k=\binom Qk^{-1}
       \sum_{\substack{z\in\{0,1,\wild\}^Q\\|z|=k}}f(\mu_z).
\end{equation}
There are $\binom Qk$ choices of the restricted questions, each with $2^k$
answer strings. We sum the entropy contributions over those strings, then
average the resulting block entropies over question sets. The denominator
is $\binom Qk$, not $\binom Qk2^k$: the probabilities already supply the
weights for averaging over answers. For $Q=4,k=2$, this collects $24$
contributions, grouped into six normalized four-answer distributions,
and divides by six.

The whole array $\mu$ is not one normalized distribution. Each fixed choice
of restricted questions gives its own distribution over answer strings:
\begin{equation}
  \sum_{|z|=k}\mu_z=\binom Qk,
  \qquad \sum_z \mu_z=2^Q.
\end{equation}
The row colors in Example~\ref{ex:xi-q2} identify which entries
contribute to each $G_k$.

\begin{example}[label=ex:xi-q2]
For $n=1,m=1$, there are two questions, printed in order $(1,0)$, and four
maps with truth tables $00,01,10,11$. The full matrix is $9\times4$.
Its row $z=1\wild$ keeps $p_2$ and $p_3$: both maps answer $1$ to
question $1$, and question $0$ eliminates neither. Thus
\begin{equation*}
  \mu_{1\wild}=p_2+p_3.
\end{equation*}
The row $10$ additionally requires answer $0$ at question $0$, leaving
only $p_2$. The row $\wild\wild$ eliminates no maps and gives
$\sum_jp_j=1$. 

Keep $0$, keep $1$, or keep either, independently for each question. All $3^2$ combinations appear. 

\begin{center}
\begingroup
\colorlet{ink}{black}
\colorlet{zero}{cok}
\colorlet{one}{cpre}
\colorlet{two}{clim}
\resizebox{0.9\linewidth}{!}{%
\begin{tikzpicture}[text=ink,>=Stealth]
\node[font=\small] at (-.35,2.65) {$\underbrace{\begin{array}{r|cc} &a=0&a=1\\\hline a=0&1&0\\a=1&0&1\\a=\wild{}&1&1\end{array}}_{\Xi\text{ for question }1}$};
\node at (1.75,2.65) {$\otimes$};
\node[font=\small] at (3.85,2.65) {$\underbrace{\begin{array}{r|cc} &a=0&a=1\\\hline a=0&1&0\\a=1&0&1\\a=\wild{}&1&1\end{array}}_{\Xi\text{ for question }0}$};
\node[font=\small] at (-.95,.65) {Questions $q$};
\node[font=\small] at (2.7,.9) {Columns: map $j$ / truth table};
\node at (-1.35,0) {$1$};\node at (-.45,0) {$0$};
\node[font=\small] at (5.0,0) {$k$};
\node[font=\scriptsize,text=black!55] at (0.9,.42) {0};
\node at (0.9,0) {$00$};
\node[font=\scriptsize,text=black!55] at (1.9500000000000002,.42) {1};
\node at (1.9500000000000002,0) {$01$};
\node[font=\scriptsize,text=black!55] at (3.0,.42) {2};
\node at (3.0,0) {$10$};
\node[font=\scriptsize,text=black!55] at (4.050000000000001,.42) {3};
\node at (4.050000000000001,0) {$11$};
\fill[two!15] (-1.8,-0.845) rectangle (5.27,-0.355);
\fill[two!75] (5.19,-0.845) rectangle (5.27,-0.355);
\node at (-1.35,-0.6) {$0$};
\node at (-0.45000000000000007,-0.6) {$0$};
\node[text=ink] at (0.9,-0.6) {1};
\node[text=black!35] at (1.9500000000000002,-0.6) {0};
\node[text=black!35] at (3.0,-0.6) {0};
\node[text=black!35] at (4.050000000000001,-0.6) {0};
\node[text=two!80!black] at (5.0,-0.6) {2};
\fill[two!15] (-1.8,-1.395) rectangle (5.27,-0.9049999999999999);
\fill[two!75] (5.19,-1.395) rectangle (5.27,-0.9049999999999999);
\node at (-1.35,-1.15) {$0$};
\node at (-0.45000000000000007,-1.15) {$1$};
\node[text=black!35] at (0.9,-1.15) {0};
\node[text=ink] at (1.9500000000000002,-1.15) {1};
\node[text=black!35] at (3.0,-1.15) {0};
\node[text=black!35] at (4.050000000000001,-1.15) {0};
\node[text=two!80!black] at (5.0,-1.15) {2};
\fill[one!15] (-1.8,-1.9450000000000003) rectangle (5.27,-1.455);
\fill[one!75] (5.19,-1.9450000000000003) rectangle (5.27,-1.455);
\node at (-1.35,-1.7000000000000002) {$0$};
\node at (-0.45000000000000007,-1.7000000000000002) {$\wild{}$};
\node[text=ink] at (0.9,-1.7000000000000002) {1};
\node[text=ink] at (1.9500000000000002,-1.7000000000000002) {1};
\node[text=black!35] at (3.0,-1.7000000000000002) {0};
\node[text=black!35] at (4.050000000000001,-1.7000000000000002) {0};
\node[text=one!80!black] at (5.0,-1.7000000000000002) {1};
\fill[two!15] (-1.8,-2.495) rectangle (5.27,-2.005);
\fill[two!75] (5.19,-2.495) rectangle (5.27,-2.005);
\node at (-1.35,-2.25) {$1$};
\node at (-0.45000000000000007,-2.25) {$0$};
\node[text=black!35] at (0.9,-2.25) {0};
\node[text=black!35] at (1.9500000000000002,-2.25) {0};
\node[text=ink] at (3.0,-2.25) {1};
\node[text=black!35] at (4.050000000000001,-2.25) {0};
\node[text=two!80!black] at (5.0,-2.25) {2};
\fill[two!15] (-1.8,-3.0450000000000004) rectangle (5.27,-2.555);
\fill[two!75] (5.19,-3.0450000000000004) rectangle (5.27,-2.555);
\node at (-1.35,-2.8000000000000003) {$1$};
\node at (-0.45000000000000007,-2.8000000000000003) {$1$};
\node[text=black!35] at (0.9,-2.8000000000000003) {0};
\node[text=black!35] at (1.9500000000000002,-2.8000000000000003) {0};
\node[text=black!35] at (3.0,-2.8000000000000003) {0};
\node[text=ink] at (4.050000000000001,-2.8000000000000003) {1};
\node[text=two!80!black] at (5.0,-2.8000000000000003) {2};
\fill[one!15] (-1.8,-3.595) rectangle (5.27,-3.105);
\fill[one!75] (5.19,-3.595) rectangle (5.27,-3.105);
\node at (-1.35,-3.35) {$1$};
\node at (-0.45000000000000007,-3.35) {$\wild{}$};
\node[text=black!35] at (0.9,-3.35) {0};
\node[text=black!35] at (1.9500000000000002,-3.35) {0};
\node[text=ink] at (3.0,-3.35) {1};
\node[text=ink] at (4.050000000000001,-3.35) {1};
\node[text=one!80!black] at (5.0,-3.35) {1};
\fill[one!15] (-1.8,-4.1450000000000005) rectangle (5.27,-3.6550000000000002);
\fill[one!75] (5.19,-4.1450000000000005) rectangle (5.27,-3.6550000000000002);
\node at (-1.35,-3.9000000000000004) {$\wild{}$};
\node at (-0.45000000000000007,-3.9000000000000004) {$0$};
\node[text=ink] at (0.9,-3.9000000000000004) {1};
\node[text=black!35] at (1.9500000000000002,-3.9000000000000004) {0};
\node[text=ink] at (3.0,-3.9000000000000004) {1};
\node[text=black!35] at (4.050000000000001,-3.9000000000000004) {0};
\node[text=one!80!black] at (5.0,-3.9000000000000004) {1};
\fill[one!15] (-1.8,-4.695) rectangle (5.27,-4.205);
\fill[one!75] (5.19,-4.695) rectangle (5.27,-4.205);
\node at (-1.35,-4.45) {$\wild{}$};
\node at (-0.45000000000000007,-4.45) {$1$};
\node[text=black!35] at (0.9,-4.45) {0};
\node[text=ink] at (1.9500000000000002,-4.45) {1};
\node[text=black!35] at (3.0,-4.45) {0};
\node[text=ink] at (4.050000000000001,-4.45) {1};
\node[text=one!80!black] at (5.0,-4.45) {1};
\fill[zero!15] (-1.8,-5.245) rectangle (5.27,-4.755);
\fill[zero!75] (5.19,-5.245) rectangle (5.27,-4.755);
\node at (-1.35,-5.0) {$\wild{}$};
\node at (-0.45000000000000007,-5.0) {$\wild{}$};
\node[text=ink] at (0.9,-5.0) {1};
\node[text=ink] at (1.9500000000000002,-5.0) {1};
\node[text=ink] at (3.0,-5.0) {1};
\node[text=ink] at (4.050000000000001,-5.0) {1};
\node[text=zero!80!black] at (5.0,-5.0) {0};

\draw[black!60] (.47,-.28)--(.32,-.28)--(.32,-5.28)--(.47,-5.28);
\draw[black!60] (4.47,-.28)--(4.62,-.28)--(4.62,-5.28)--(4.47,-5.28);
\node[anchor=west,font=\small] at (5.7,-.55) {Row colors};
\fill[zero!25] (5.75,-1.28) rectangle (6.05,-0.9600000000000001);
\node[anchor=west,font=\small] at (6.2,-1.12) {$k=0$: $1$ row $\to G_0$};
\fill[one!25] (5.75,-1.8699999999999999) rectangle (6.05,-1.55);
\node[anchor=west,font=\small] at (6.2,-1.71) {$k=1$: $4$ rows $\to G_1$};
\fill[two!25] (5.75,-2.46) rectangle (6.05,-2.1399999999999997);
\node[anchor=west,font=\small] at (6.2,-2.3) {$k=2$: $4$ rows $\to G_2$};

\draw[ink,thick] (-1.8,-3.595) rectangle (5.27,-3.105);
\draw[->] (5.30,-3.35)--(5.72,-3.35);
\node[anchor=north west,font=\small,align=left] at (5.75,-3.0) {Row $z=1\wild{}$\\[5pt]Keep maps with $\phi_j(1)=1$;\\question $0$ excludes no maps.\\[5pt]$\mu_{1\wild{}}=p_2+p_3$\\This contributes to $G_1$.};

\end{tikzpicture}%
}
\endgroup
\end{center}

$\wild$ allows either answer: this question excludes no maps. Multiply by $p$, apply $f(u)=-u\log_2u$ elementwise, then collect each color:
\begin{align*}
  G_0&=f(\mu_{\wild{}\wild{}})=0,\\
  G_1&=\tfrac12\big[f(\mu_{0\wild{}})+f(\mu_{1\wild{}})+f(\mu_{\wild{}0})+f(\mu_{\wild{}1})\big],\\
  G_2&=f(p_0)+f(p_1)+f(p_2)+f(p_3)=H(p).
\end{align*}
A $1$ keeps the map and a $0$ excludes it. The row colors group the
surviving weights by the number $k$ of single-answer restrictions.
\end{example}

Equivalently, the operation of calculating $G_k$ is linear, then elementwise nonlinear, then
linear. We define the constant weights
\begin{equation}
  (\mathbf c_k)_z:=\frac{\mathbf1_{\{|z|=k\}}}{\binom Qk},
\end{equation}
which pick out exactly those rows of the tensor product with $Q-k$ times the symbolic $\wild$ in the margin. Then
\begin{equation}\label{eq:Xi-linear-nonlinear-linear}
  G_k=\mathbf c_k^{\mathsf T}f(\Xi^{\otimes Q}p).
\end{equation}
Stacking these row vectors, for increasing $k$, into an averaging matrix $\Gamma$ gives
\begin{equation}\label{eq:Xi-all-G}
  \begin{pmatrix}G_0\\G_1\\\vdots\\G_Q\end{pmatrix}
  =\Gamma\,f(\Xi^{\otimes Q}p).
\end{equation}

\begin{example}[label=ex:gamma-q2]
Continuing Example~\ref{ex:xi-q2}: at $Q=2$ the vector $f(\Xi^{\otimes2}p)$
has nine entries, one per elimination rule, so $\Gamma$ is $3\times9$, one
row per block size $k$. Taking the columns in the row order of that
example, and with $\binom20=\binom22=1$ and $\binom21=2$,
\begin{equation*}
  \Gamma =
  \begin{array}{c|ccccccccc}
    \dgc{k}{z}
      & 00 & 01 & 0\wild & 10 & 11 & 1\wild & \wild0 & \wild1 & \wild\wild \\
    \hline
    0 & 0 & 0 & 0 & 0 & 0 & 0 & 0 & 0 & 1 \\[2pt]
    1 & 0 & 0 & \tfrac12 & 0 & 0 & \tfrac12 & \tfrac12 & \tfrac12 & 0 \\[2pt]
    2 & 1 & 1 & 0 & 1 & 1 & 0 & 0 & 0 & 0
  \end{array}.
\end{equation*}
Each rule $z$ belongs to exactly one block size $|z|$, so every column
carries a single nonzero entry, and the three row colors of
Example~\ref{ex:xi-q2} are the three rows of $\Gamma$. Row $k$ sums to
$2^k$ rather than to one: the $2^k$ answer strings of a block already
carry their own probabilities, which is why the denominator in
\eqref{eq:Xi-Gk} is $\binom Qk$ and not $\binom Qk2^k$.
\end{example}

Both linear maps are independent of $p$; $f$ acts elementwise. In an
implementation, the first map is a tensor sweep and the last is an
accumulation by $|z|$. Neither requires constructing a dense matrix.

Sharing the elimination sums:
at each tensor axis we retain the two answer slices and append their sum.
After $r$ axes the array has $3^r2^{Q-r}$ entries. The full sweep uses
\begin{equation}\label{eq:Xi-binary-cost}
  \sum_{r=1}^Q3^{r-1}2^{Q-r}=3^Q-2^Q
\end{equation}
scalar additions. Including the entropy accumulation, the work is order
$3^Q$. A separate scan of all $2^Q$ prior entries for each of the $2^Q$
question sets takes order $4^Q$ work. The saving comes from reusing sums
of surviving weights. The calculation is still exponential in $Q$, but at
the sizes this work tabulates the reduction is worth having: about a
hundredfold at $(4,1)$ and fortyfold at $(3,2)$.

\subsection{General answer alphabet}

For $m$ answer bits, let $A=2^m$. Each question now has $A$ possible answers,
so its tensor factor has $A$ columns. There is one row that keeps each
particular answer, and a final row that keeps every answer:
\begin{equation}\label{eq:Xi-alphabet}
  \Xi=\begin{pmatrix}I_A\\\mathbf1_A^{\mathsf T}\end{pmatrix},
  \qquad
  \Xi_{z,a}=\begin{cases}
    1,&z=a\text{ or }z=\wild,\\
    0,&\text{otherwise}.
  \end{cases}
\end{equation}
We retain $\wild$ for the rule that this question excludes no maps.
For a larger alphabet it allows \emph{all} $A$ answers. Its meaning is
therefore the same elimination rule, even though there are more than two
answers to keep.

\begin{example}
For $m=2$, each answer is one of $00,01,10,11$, and the local matrix is
\begin{equation}\label{eq:Xi-m2}
  \Xi=
  \begin{array}{c|cccc}
    &a=00&a=01&a=10&a=11\\ \hline
    00&1&0&0&0\\
    01&0&1&0&0\\
    10&0&0&1&0\\
    11&0&0&0&1\\
    \wild&1&1&1&1
  \end{array}.
\end{equation}
The first row keeps maps answering $00$; the next three keep maps answering
$01$, $10$, or $11$, respectively. The last row keeps all four groups.
Each answer is treated as a whole symbol: a question either requires its
complete answer or excludes no maps. We do not add rows for specifying
only one bit of an answer.

For $n=1$, the full matrix $\Xi^{\otimes2}$ is consequently $25\times16$.
There are $1,8,16$ rows contributing to $G_0,G_1,G_2$. Using decimal answer
symbols $0,1,2,3$ and question order $(1,0)$, we have $j=4a_1+a_0$ and
\begin{equation*}
  \mu_{2\wild}=p_8+p_9+p_{10}+p_{11},
  \qquad \mu_{\wild1}=p_1+p_5+p_9+p_{13}.
\end{equation*}
The first rule keeps answer $2$ at question $1$ and every answer at
question $0$. The second keeps answer $1$ at question $0$ and every answer
at question $1$. The two four-answer distributions give
\begin{equation*}
  G_1=\tfrac12\left[\sum_{a=0}^3f(\mu_{a\wild})
                         +\sum_{a=0}^3f(\mu_{\wild a})\right].
\end{equation*}
\end{example}

In general the full matrix is $(A+1)^Q\times A^Q$, and
\begin{equation}\label{eq:Xi-Gk-alphabet}
  \mu=\Xi^{\otimes Q}p,
  \qquad
  G_k=\binom Qk^{-1}
       \sum_{\substack{z\in(\mathcal A\cup\{\wild\})^Q\\|z|=k}}f(\mu_z).
\end{equation}
There are $\binom Qk A^k$ contributions at size $k$. The averaging matrix
in \eqref{eq:Xi-all-G} has entries
$\mathsf C_{k,z}=\mathbf1_{\{|z|=k\}}/\binom Qk$ as before.
Appending each all-answers component takes $A-1$ additions, giving
\begin{equation}\label{eq:Xi-alphabet-cost}
  (A-1)\sum_{r=1}^Q(A+1)^{r-1}A^{Q-r}
  =(A-1)\big[(A+1)^Q-A^Q\big].
\end{equation}
For fixed $A$, the complete construction and entropy accumulation take
order $(A+1)^Q$ work, compared with order $(2A)^Q$ for scanning the prior
separately for every question set.

\subsection{Stopping at $\ell$ and the cost of another $G_k$}
If only $G_0,\ldots,G_\ell$ are wanted, we retain only branches with
at most $\ell$ single-answer restrictions among the axes already processed.
A single-answer restriction at a processed axis never becomes an either-answer row later
in the sweep, so discarded branches cannot contribute to the
requested output. That output has
\begin{equation}
  \sum_{k=0}^{\ell}\binom Qk A^k
\end{equation}
entries. For this pruned sweep the exact addition count is
\begin{equation}\label{eq:Xi-truncated-cost}
  (A-1)\sum_{r=1}^Q A^{Q-r}
    \sum_{s=0}^{\min(\ell,r-1)}\binom{r-1}{s}A^s.
\end{equation}
Here $s$ counts single-answer restrictions among the first $r-1$ axes.
This shares the work for all requested sizes. The output count
alone is not the construction cost: an arbitrary explicitly
supplied prior still has $A^Q$ entries to read.

Once the marginals are available, extracting a further $G_k$
requires $\binom Qk A^k$ evaluations of $f$ and their accumulation,
with no further marginalization. Better still, we can accumulate one
entropy sum per $k$ as the final components are produced; then
all requested $G_k$ are ready at the end. If the sweep was pruned
at $\ell$, this does not make $G_{\ell+1}$ free: its discarded
branches must also be computed. Nor do the scalar values
$G_0,\ldots,G_\ell$ alone generally determine the next one.

\section{Block priors and the Bernstein profile language}\label{app:blockpriors}

Subsection~\ref{sec:spikedefinetti} read the spike prior as a mixture over
two answer laws. Nothing there forced the number to be two, and the
whole exchangeable family is worth seeing before it is left behind.
Let $\Lambda$ be any distribution over answer laws $\nu$ on
$\mathcal A$. Reading \eqref{eq:definetti} at $k = Q$, the prior it
induces on maps is
\begin{equation}
  p_j \;=\; \int \prod_{q \in \mathcal Q} \nu\big(\phi_j(q)\big)\;
  d\Lambda(\nu),
  \label{eq:definettiprior}
\end{equation}
a map drawn by fixing one answer law and then answering every question
independently from it, and by de Finetti's theorem
\cite{definetti37,hewittsavage} every projectively consistent
exchangeable prior is of this form and no other.

The construction is generous. Support is full wherever $\Lambda$'s is,
with no sliver to argue about. Consistency across sizes holds by
definition, since $\Lambda$ carries no reference to $n$, so the
sequence of priors converges without the separate hypothesis that
Section~\ref{sec:tdlcalc} otherwise has to assume. And every block law
depends on $\mathcal S$ through its size alone, so $G_k$ is the entropy
of $A^k$ block probabilities, each a moment of $\Lambda$, and the limit
closes in form:
\begin{equation}
  \gamma(t) \;\equiv\; \mathbb E_\Lambda\big[H(\nu)\big],
  \qquad
  \eta_0 \;=\; H\big(\mathbb E_\Lambda\, \nu\big),
  \qquad
  L(t) \;=\; 1 + \frac{I(\nu\,;\,\xi_1)}
                      {\mathbb E_\Lambda\big[H(\nu)\big]\; t},
  \label{eq:definettilimit}
\end{equation}
the numerator once more a mutual information, as at
\eqref{eq:spikejensen}. Take $m = 1$ and $\Lambda$ uniform on the coin
biases, for instance. The prior is then
$p_j = \big((Q+1)\binom{Q}{|j|}\big)^{-1}$, with $|j|$ the number of
ones in the truth table, positive on every one of the $N$ maps, and
$\eta_0 = 1$ against
$\mathbb E_\Lambda[H(\nu)] = \int_0^1 h_2 = 1/(2\ln 2)$ gives
\begin{equation}
  L(t) \;=\; 1 + \frac{2\ln 2 - 1}{t} \;=\; 1 + \frac{0.3863}{t}
  \label{eq:definettiuniform}
\end{equation}
exactly: a full-support prior whose thermodynamic limit is closed form,
with no external theorem needed to take it.

And there the family stops. Every member has a flat interior profile
and therefore a hyperbola, so a curve that changes throughout
$0 < t < 1$ needs a prior that retains structure tied to particular
groups or addresses of questions, and a fixed partition into local
blocks is the simplest solvable construction. The escape is narrower
than it looks. Failing to be a mixture of independent draws is not
enough: a prior supported on the truth tables with an even number of
ones is exchangeable and is no such mixture at any $Q \ge 4$, since
giving zero weight to a single one forces the mixing law onto the two
constant answers and thereby kills every other count as well, yet its
frequencies still converge and its profile still goes flat. What does
escape is a family whose count law never settles, alternating between
constructions on even and odd $n$; that sequence has no limiting
frequency law, and so no thermodynamic limit to speak of.

\subsection{Independent blocks of questions}

Partition $\mathcal Q$ into independent blocks drawn from a fixed
finite set of types, writing $\mathcal P_n$ for the partition and
indexing types by $\kappa$. At finite $n$ let $K_{\kappa,n}$ count
the type-$\kappa$ blocks and set
$w_{\kappa,n} := r_\kappa K_{\kappa,n}/Q$, the fraction of all
questions belonging to them, assuming $w_{\kappa,n}\to w_\kappa$ with
$\sum_\kappa w_\kappa = 1$. A type $\kappa$ carries a size
$r_\kappa$, a fixed joint distribution on its $r_\kappa$ answers
shared by every block of that type up to relabeling, and a mean
subset-entropy profile defined as follows. Let
$\mathcal V \subseteq \mathcal Q$ be one particular block of type
$\kappa$, fix an ordering of it once and for all, and give every
subset $\mathcal J \subseteq \mathcal V$ the inherited order, so that
$\psi(\mathcal J) \in \mathcal A^{|\mathcal J|}$ is its random answer
vector. For $j \in \{0,\ldots,r_\kappa\}$ define
\begin{equation}
  H_\kappa(j)
  := \binom{r_\kappa}{j}^{-1}
  \sum_{\substack{\mathcal J\subseteq\mathcal V\\ |\mathcal J| = j}}
  H\big(\psi(\mathcal J)\big).
  \label{eq:blockprofile}
\end{equation}
Here $j$ is the \emph{number of questions selected from the block},
not a question label, and $H(\psi(\mathcal J))$ measures the
learner's uncertainty in their \emph{joint answers}, not uncertainty
about which questions were selected: the set $\mathcal J$ is known.
No symmetry is required for this averaged definition, although every
zoo member below has the stronger property of \textbf{entropy
homogeneity}, meaning that $H(\psi(\mathcal J))$ itself depends only
on $|\mathcal J|$. Full permutation invariance is sufficient for that
but not necessary: MDS code blocks \cite{macwilliams} are entropy homogeneous even
though their coordinate distribution is not invariant under every
permutation.

The profile records how much answer uncertainty remains visible after
looking at $j$ locations inside a block, and its purpose is that
successive differences have an operational meaning. Averaging the
chain rule over a uniform $i$-subset $\mathcal J$ and then a fresh
question drawn uniformly from $\mathcal V\setminus\mathcal J$ gives
the within-block increments
\begin{equation}
  \zeta_\kappa(i) := H_\kappa(i+1)-H_\kappa(i)
  = \mathbb E_{\mathcal J, q}
  \big[H\big(\psi(q)\mid\psi(\mathcal J)\big)\big],
  \qquad 0 \le i \le r_\kappa - 1,
  \label{eq:zetakappa}
\end{equation}
the mean uncertainty in one fresh answer after $i$ of its block
partners have been answered. It's $\gamma$ restricted to a block type. 
This is why the block-entropy profile is
useful: it converts a joint law on an entire block into exactly the
conditional entropies the learning curve needs. The endpoints make
the meaning concrete, $H_\kappa(0) = 0$ while
$H_\kappa(r_\kappa) = H(\psi(\mathcal V))$ is the entropy of the
whole block taken jointly. Two independent fair binary answers have
profile $(0,1,2)$ and increments $(1,1)$; the same fair bit copied
into both questions has profile $(0,1,1)$ and increments $(1,0)$,
since after either answer is known the other carries no remaining
uncertainty.

For a fresh question in a block of size $r_\kappa$, the number of its
$r_\kappa - 1$ partners already present in a uniformly random
$\ell$-set is hypergeometric, so exactly at finite $Q$, for
$0 \le \ell < Q$,
\begin{equation}
  \gamma_{n,\ell}
  = \sum_\kappa w_{\kappa,n}\sum_{i=0}^{r_\kappa-1}
  \frac{\binom{r_\kappa-1}{i}\binom{Q-r_\kappa}{\ell-i}}
       {\binom{Q-1}{\ell}}\,\zeta_\kappa(i).
  \label{eq:blockfinitegamma}
\end{equation}
Holding every $r_\kappa$ fixed, taking $\ell = \lfloor tQ\rfloor$ and
assuming $w_{\kappa,n}\to w_\kappa$, sampling without replacement
approaches independent sampling,
\begin{equation}
  \frac{\binom{r-1}{i}\binom{Q-r}{\ell-i}}{\binom{Q-1}{\ell}}
  \;\longrightarrow\;
  \underbrace{\binom{r-1}{i}t^i(1-t)^{r-1-i}}_{\textstyle
    \beta_{r-1,i}(t)},
  \label{eq:hyptobin}
\end{equation}
which simultaneously defines the Bernstein basis polynomial
$\beta_{r-1,i}$ and, probabilistically, is the limiting probability
that exactly $i$ of the fresh question's $r-1$ partners have been
asked by macroscopic time $t$. The limiting profile is therefore
\begin{equation}
  \gamma(t)
  = \sum_\kappa w_\kappa\sum_{i=0}^{r_\kappa-1}
  \beta_{r_\kappa-1,i}(t)\,\zeta_\kappa(i). 
  \label{eq:bernsteinprofile}
\end{equation}
At the start of the run no partner has been asked, so only the
$i = 0$ term survives and
$\eta_0 = \gamma(0) = \sum_\kappa w_\kappa\zeta_\kappa(0)
= \sum_\kappa w_\kappa H_\kappa(1)$: the same block-entropy profile
supplies both the initial marginal entropy and the entire macroscopic
conditional-entropy curve.

To construct a sequence with prescribed fractions, choose integer
block counts with $r_\kappa K_{\kappa,n}/Q \to w_\kappa$ and
$\sum_\kappa r_\kappa K_{\kappa,n}\le Q$, assigning only $o(Q)$
leftover sites, whose treatment does not affect the limit; at finite
$n$ any leftovers can be recorded as an additional block type, so
\eqref{eq:blockfinitegamma} still includes every question and remains
exact. If block sizes grow with $n$ the proof changes, a useful
sufficient condition for the hypergeometric-to-binomial approximation
being
$\big(\max_{\mathcal V\in\mathcal P_n}|\mathcal V|\big)^2/Q \to 0$.
For countably many types, pointwise frequency convergence must be
replaced by an $\ell^1$ or tightness condition.

\begin{example}
Two types tile the growing table: pairs ($r_1 = 2$, blue) and quads
($r_2 = 4$, orange), aiming at $w_1 = 3/8$ and $w_2 = 5/8$. Integer
counts force detours. At $n = 1$ only a pair fits and at $n = 2$
only a quad, while at $Q = 8$ and $Q = 16$ the best leftover-free
tilings still miss the target shares by $1/8$.

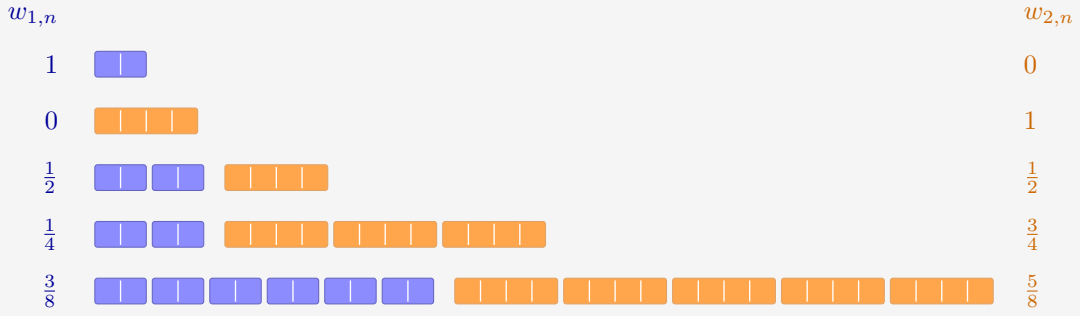
\begin{center}
\begin{tikzpicture}[x=1cm, y=1cm,
  pairblk/.style={fill=blue!45, draw=blue!60!black!60,
    line width=0.4pt, rounded corners=1pt},
  quadblk/.style={fill=orange!70, draw=orange!80!black!60,
    line width=0.4pt, rounded corners=1pt},
  cellsep/.style={white, line width=0.5pt},
  wl/.style={font=\small, blue!60!black, anchor=east},
  wr/.style={font=\small, orange!80!black, anchor=west}]

  \node[wl] at (-0.35, 0.79) {$w_{1,n}$};
  \node[wr] at (12.15, 0.79) {$w_{2,n}$};

  \draw[pairblk] (0,0) rectangle (0.68,0.34);
  \draw[cellsep] (0.34,0.03) -- (0.34,0.31);
  \node[wl] at (-0.35,0.17) {$1$};
  \node[wr] at (12.15,0.17) {$0$};

  \draw[quadblk] (0,-0.75) rectangle (1.36,-0.41);
  \foreach \d in {0.34,0.68,1.02}
    \draw[cellsep] (\d,-0.72) -- (\d,-0.44);
  \node[wl] at (-0.35,-0.58) {$0$};
  \node[wr] at (12.15,-0.58) {$1$};

  \foreach \i in {0,1}{
    \draw[pairblk] ({\i*0.76},-1.5) rectangle ({\i*0.76+0.68},-1.16);
    \draw[cellsep] ({\i*0.76+0.34},-1.47) -- ({\i*0.76+0.34},-1.19);}
  \draw[quadblk] (1.72,-1.5) rectangle (3.08,-1.16);
  \foreach \d in {0.34,0.68,1.02}
    \draw[cellsep] ({1.72+\d},-1.47) -- ({1.72+\d},-1.19);
  \node[wl] at (-0.35,-1.33) {$\tfrac12$};
  \node[wr] at (12.15,-1.33) {$\tfrac12$};

  \foreach \i in {0,1}{
    \draw[pairblk] ({\i*0.76},-2.25) rectangle ({\i*0.76+0.68},-1.91);
    \draw[cellsep] ({\i*0.76+0.34},-2.22) -- ({\i*0.76+0.34},-1.94);}
  \foreach \i in {0,1,2}{
    \draw[quadblk] ({1.72+\i*1.44},-2.25)
      rectangle ({1.72+\i*1.44+1.36},-1.91);
    \foreach \d in {0.34,0.68,1.02}
      \draw[cellsep] ({1.72+\i*1.44+\d},-2.22)
        -- ({1.72+\i*1.44+\d},-1.94);}
  \node[wl] at (-0.35,-2.08) {$\tfrac14$};
  \node[wr] at (12.15,-2.08) {$\tfrac34$};

  \foreach \i in {0,...,5}{
    \draw[pairblk] ({\i*0.76},-3.0) rectangle ({\i*0.76+0.68},-2.66);
    \draw[cellsep] ({\i*0.76+0.34},-2.97) -- ({\i*0.76+0.34},-2.69);}
  \foreach \i in {0,...,4}{
    \draw[quadblk] ({4.76+\i*1.44},-3.0)
      rectangle ({4.76+\i*1.44+1.36},-2.66);
    \foreach \d in {0.34,0.68,1.02}
      \draw[cellsep] ({4.76+\i*1.44+\d},-2.97)
        -- ({4.76+\i*1.44+\d},-2.69);}
  \node[wl] at (-0.35,-2.83) {$\tfrac38$};
  \node[wr] at (12.15,-2.83) {$\tfrac58$};
\end{tikzpicture}
\end{center}

Each cell is one question and each rounded group is one block, so
the counts $K_{\kappa,n}$ can be read off:
$(K_1, K_2) = (1,0),\ (0,1),\ (2,1),\ (2,3),\ (6,5)$. The pair share
$w_{1,n} = 1,\ 0,\ \tfrac12,\ \tfrac14,\ \tfrac38$ closes in on its
target by halving steps, and at $n = 5$ the tiling is exact:
$6 \times 2 + 5 \times 4 = 32 = Q$.
\end{example}

\subsection{Block laws and their map priors}

The profile machinery never displayed the prior itself. It is
constructive: each block law prices the restriction of a map to its
block, and independence across blocks multiplies the prices. A map
$\phi_j$ restricts to the answer tuple $\phi_j(\mathcal V)$ on block
$\mathcal V$, and
\begin{equation}
  p_j \;=\; \prod_{\mathcal V\in\mathcal P_n}
  \Pr\big(\psi(\mathcal V) = \phi_j(\mathcal V)\big).
  \label{eq:blockproduct}
\end{equation}
The classical block laws, with the factor each contributes to
\eqref{eq:blockproduct} and the curve it produces:

\begin{table}[H]
\begin{center}
\begin{tabular}{l p{3.6cm} p{4.0cm} p{3.1cm}}
\toprule
block law & a size-$r$ block is sampled by &
  factor $\Pr(\psi(\mathcal V) = y)$ & leverage \\
\midrule
clique & one value $a_{\mathcal V}\sim\nu_{\mathcal V}$ copied to
  every question &
  $\nu_{\mathcal V}(a)$ if $y = (a,\ldots,a)$, else $0$ &
  $L \equiv r$, exact at every $n$ \\
parity ($\sigma$) & uniform on the coset
  $\bigoplus_q y_q = \sigma$ &
  $2^{-(r-1)}$ on the coset, else $0$ &
  rises $1 \to r/(r-1)$, late \\
tilted parity & i.i.d.\ Bernoulli($\theta$) bits conditioned on the
  coset &
  $\theta^{|y|}(1-\theta)^{r-|y|}/\mathcal Z_{r,\sigma}$ on the coset,
  else $0$ &
  same shape, bent by $\varepsilon = 1-2\theta$ \\
$[r,k]_A$ MDS & uniform on an MDS code ($A \ge r$) &
  $A^{-k}$ on codewords, else $0$ &
  $L(1) = r/k$; interior peak when mixed with fresh sites \\
lapse ($\delta$) & any law above, replaced by the uniform block with
  probability $\delta$ &
  $(1-\delta)P_{\mathcal V}(y) + \delta A^{-r}$ &
  full support; $\gamma(1) > 0$ \\
\bottomrule
\end{tabular}
\end{center}
\caption{The classical block laws: how a block of size $r$ is sampled, the factor it contributes to \eqref{eq:blockproduct}, and the leverage that follows.}
\label{tab:blocklaws}
\end{table}

The supports multiply as well: cliques carry the $A^{Q/r}$ maps
constant on every block, parity carries the $2^{(r-1)Q/r}$ maps
landing in every coset, a code carries $A^{kQ/r}$ maps, and the
lapse restores $p_j > 0$ everywhere, playing the role the top class
played in Chapter~\ref{sec:shapeprior}.

\begin{example}
Cliques at $(n,m) = (2,1)$: partition the four questions by their
high bit, $\mathcal V_1 = \{00, 01\}$ and
$\mathcal V_2 = \{10, 11\}$, with $\nu_{\mathcal V_1} = (0.8, 0.2)$ and
$\nu_{\mathcal V_2} = (0.55, 0.45)$ on the answers $(0,1)$. A supported map must
be constant on both blocks, so it can depend only on $b_1$: of the
sixteen maps of Table~\ref{tab:sixteen}, only FALSE ($j = 0$),
$\lnot b_1$ ($j = 3$), $b_1$ ($j = 12$) and TRUE ($j = 15$) survive,
with product weights
$p_0 = 0.8 \times 0.55 = 0.44$,
$p_3 = 0.11$, $p_{12} = 0.36$, $p_{15} = 0.09$, and $p_j = 0$ for
the other twelve. One answer from either block settles that block's
remaining question for free: $L \equiv 2$.
\end{example}

From any row of the table the route to leverage is the one already
walked in Chapter~\ref{sec:shapeprior}. The block law fixes the
subset entropies \eqref{eq:blockprofile}, their differences give the
increments $\zeta$ \eqref{eq:zetakappa}, and the hypergeometric average
\eqref{eq:blockfinitegamma} is exact at every finite $n$; its
binomial limit is the Bernstein profile \eqref{eq:bernsteinprofile}.
No capacity theorem is needed: blocks are bounded, so elementary
hypergeometric-to-binomial convergence replaces it. The finite law
\eqref{eq:finitelev} then gives $\mathcal L_\ell$ exactly at
every $n$, and the master law \eqref{eq:master} gives the curve, with
$\eta_0 = \sum_\kappa w_\kappa \zeta_\kappa(0)$.

\subsection{Bernstein polynomials as a profile language}

For a single block type, suppressing the type label,
equation~\eqref{eq:bernsteinprofile} reads in the Bernstein basis
\cite{lorentz}
\begin{equation}
  \gamma(t) = \zeta(0)(1-t)^2 + 2\zeta(1)t(1-t) + \zeta(2)t^2
  \label{eq:bernstein3}
\end{equation}
at $r = 3$, and
$\zeta(0)(1-t)^3 + 3\zeta(1)t(1-t)^2 + 3\zeta(2)t^2(1-t) + \zeta(3)t^3$
at $r = 4$. Some useful coefficient vectors, with $h_{\rm fresh}$ the
entropy of one independent fresh answer and $h_{\rm clique}$ the
entropy of the value copied through a clique:

\begin{table}[H]
\begin{center}
\begin{tabular}{lll}
\toprule
block & $(\zeta(0),\ldots,\zeta(r-1))$ & profile \\
\midrule
independent fresh answers
  & $(h_{\rm fresh},\ldots,h_{\rm fresh})$ & $h_{\rm fresh}$ \\
clique & $(h_{\rm clique},0,\ldots,0)$
  & $h_{\rm clique}(1-t)^{r-1}$ \\
parity & $(m,\ldots,m,0)$ & $m(1-t^{r-1})$ \\
$(r,k)$ MDS & $k$ entries $m$, then zeros
  & $m\Pr[\operatorname{Bin}(r-1,t)\le k-1]$ \\
\bottomrule
\end{tabular}
\end{center}
\caption{Coefficient vectors $\zeta$ for the block laws above, and the profile each one generates.}
\label{tab:zeta}
\end{table}

Entropy submodularity gives
$m \ge \zeta_\kappa(0) \ge \cdots \ge \zeta_\kappa(r_\kappa-1)\ge0$,
and the derivative identity for Bernstein polynomials makes the
macroscopic monotonicity explicit,
\begin{equation}
  \gamma'(t)
  = \sum_\kappa w_\kappa(r_\kappa-1)\sum_{i=0}^{r_\kappa-2}
  \big(\zeta_\kappa(i+1)-\zeta_\kappa(i)\big)\beta_{r_\kappa-2,i}(t)
  \;\le\; 0 ,
  \label{eq:bernsteinderiv}
\end{equation}
so $\gamma$ is nonincreasing as the general entropy calculus requires.
Leverage itself need not be monotone.

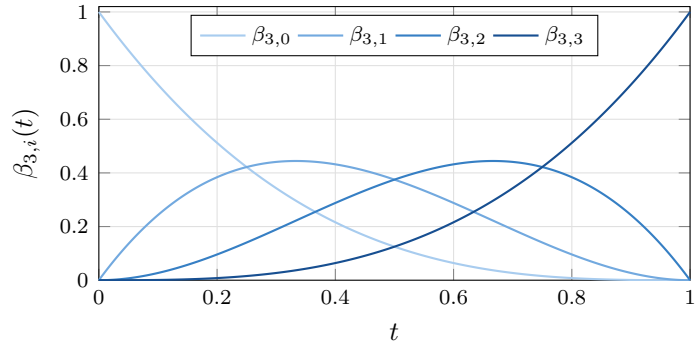
\begin{figure}[H]
\centering
\begin{tikzpicture}[baseline=(current bounding box.north)]
\begin{axis}[width=9.4cm, height=5.2cm,
  xlabel={$t$}, ylabel={$\beta_{3,i}(t)$},
  xmin=0, xmax=1, ymin=0, ymax=1.02,
  tick label style={font=\scriptsize}, label style={font=\small},
  grid=major, grid style={black!12},
  legend style={font=\scriptsize, at={(0.5,0.97)}, anchor=north,
    legend columns=4},
  legend cell align=left]
\dataplot{cn1, thick}{x}{b0}{data/bernstein3.dat}
\dataplot{cn2, thick}{x}{b1}{data/bernstein3.dat}
\dataplot{cn3, thick}{x}{b2}{data/bernstein3.dat}
\dataplot{cn4, thick}{x}{b3}{data/bernstein3.dat}
\datalegend{$\beta_{3,0}$, $\beta_{3,1}$, $\beta_{3,2}$, $\beta_{3,3}$}
\end{axis}
\end{tikzpicture}
\caption{The degree-three Bernstein basis: bump functions scanning
left to right, the $i$-th peaked at $i/3$. A block profile is a
positive combination of such bumps weighted by the microscopic
conditional entropies $\zeta(i)$, so the shape of $L$ becomes a
question about \emph{where the coefficients drop}.}
\label{fig:bernstein}
\end{figure}

Equation~\eqref{eq:bernsteinprofile} describes every prior in the
stated class of independent blocks drawn from a fixed finite list of
bounded types, and mixing types mixes their profiles linearly. There
is also a useful approximation result when the answer alphabet is
allowed to grow: mixtures of common-length MDS blocks realize any
nonincreasing Bernstein coefficient sequence, and Bernstein
polynomials approximate every continuous nonincreasing profile. A
Reed--Solomon block of length $r$ needs $A \ge r$, however, so
sending $r\to\infty$ in that approximation also requires
$m\to\infty$, or a different family of realizable entropy vectors.
It is not a completeness theorem at fixed answer width.

\subsection{Leverage curves at finite \texorpdfstring{$n$}{n}}

Because \eqref{eq:blockfinitegamma} is exact, the finite-$n$ curves
need no simulation: the hypergeometric increments feed the finite law
\eqref{eq:finitelev} directly, and the limits below are the master
law \eqref{eq:master} applied to the Bernstein profiles of the table
above. Three representative members, at $Q = 16$ and $Q = 64$
against their limits:

\begin{figure}[H]
\centering
\begin{tikzpicture}[baseline=(current bounding box.north)]
\begin{axis}[width=5.1cm, height=5.2cm,
  title={clique, $r=4$}, title style={font=\small},
  xlabel={$t = \ell/Q$}, ylabel={$\mathcal L_\ell$},
  xmin=0, xmax=1, ymin=0.9, ymax=4.6,
  tick label style={font=\scriptsize}, label style={font=\small},
  grid=major, grid style={black!12},
  legend style={font=\tiny, at={(0.97,0.03)}, anchor=south east},
  legend cell align=left]
\addplot[black, dashed, domain=0:1] {4};
\dataplot{cf1, only marks, mark=*, mark size=1.3}{t}{lev}
  {data/blockfinite_clique4_q16.dat}
\dataplot{cf3, only marks, mark=o, mark size=1.5}{t}{lev}
  {data/blockfinite_clique4_q64.dat}
\datalegend{{limit $L\equiv4$}, {$Q=16$}, {$Q=64$}}
\end{axis}
\end{tikzpicture}\hfill
\begin{tikzpicture}[baseline=(current bounding box.north)]
\begin{axis}[width=5.1cm, height=5.2cm,
  title={parity, $r=4$}, title style={font=\small},
  xlabel={$t = \ell/Q$},
  xmin=0, xmax=1, ymin=0.95, ymax=1.45,
  tick label style={font=\scriptsize}, label style={font=\small},
  grid=major, grid style={black!12}]
\addplot[black, dashed, domain=0.02:1, samples=120]
  {(x+(1-x)*x^3)/(x-x^4/4)};
\dataplot{cf1, only marks, mark=*, mark size=1.3}{t}{lev}
  {data/blockfinite_parity4_q16.dat}
\dataplot{cf3, only marks, mark=o, mark size=1.5}{t}{lev}
  {data/blockfinite_parity4_q64.dat}
\end{axis}
\end{tikzpicture}\hfill
\begin{tikzpicture}[baseline=(current bounding box.north)]
\begin{axis}[width=5.1cm, height=5.2cm,
  title={MDS $(4,2)$ + fresh}, title style={font=\small},
  xlabel={$t = \ell/Q$},
  xmin=0, xmax=1, ymin=0.95, ymax=1.75,
  tick label style={font=\scriptsize}, label style={font=\small},
  grid=major, grid style={black!12}]
\addplot[black, dashed, domain=0.02:1, samples=120]
  {(0.75*(1-(1-x)*(1-3*x^2+2*x^3))+0.25*x)
   /(0.75*(x-x^3+x^4/2)+0.25*x)};
\dataplot{cf1, only marks, mark=*, mark size=1.3}{t}{lev}
  {data/blockfinite_mdsfresh_q16.dat}
\dataplot{cf3, only marks, mark=o, mark size=1.5}{t}{lev}
  {data/blockfinite_mdsfresh_q64.dat}
\end{axis}
\end{tikzpicture}
\caption{Exact finite-$n$ leverage (dots, $Q = 16$ and $64$) against
the macroscopic limits (dashed), from
\eqref{eq:blockfinitegamma} and \eqref{eq:finitelev} with no
simulation. Left: cliques of four with a fair shared bit; the dots
sit exactly on the constant $L \equiv 4$ at every finite $n$.
Middle: parity blocks of four; deductions concentrate late and the
finite curves approach the rise to $r/(r-1) = \tfrac43$ from below.
Right: a $3{:}1$ mixture of $[4,2]$ MDS code blocks ($m = 2$) with
fresh questions; the code releases its deductions around the
threshold and the fresh sites keep receiving afterwards, an interior
maximum of $1.645$ at $t \simeq 0.72$ against the completion value
$1.6$.}
\label{fig:blockfinite}
\end{figure}
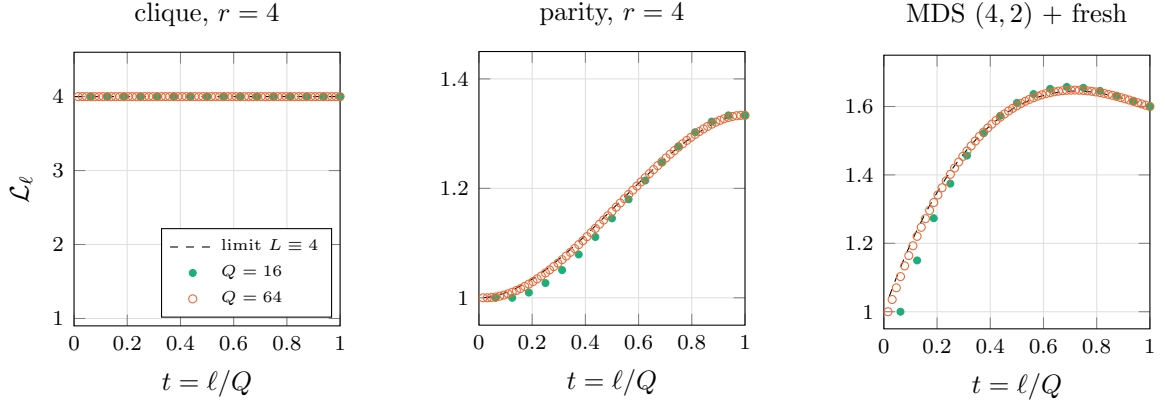

\section{More characteristics of the spike prior}\label{app:spike}

This appendix continues the toy example of Section~\ref{sec:spike}:
the anatomy of a single question, the fraction of the dependency
stock it extracts, and the sharp-prior limit with its
order-of-limits structure.

\subsection{One question, branch by branch}

For a single question abbreviate $P := P_1$ and $U := U_1$, so that
$P + (A-1)U = 1$: the zero answer has probability $P$, while each
particular nonzero answer has probability $U$. Here \emph{correct}
means that the observed answer agrees with the spike map at this
question; it does not assert that the entire spike map is the truth.
The surprisals are
\begin{equation}
  s_{\rm correct} = -\log_2 P,
  \qquad
  s_{\rm wrong} = -\log_2 U,
  \qquad
  \langle\!\langle s\rangle\!\rangle = P\,s_{\rm correct} + (A-1)U\,s_{\rm wrong} = G_1 ,
  \label{eq:spikesurprisal}
\end{equation}
a particular wrong answer having probability $U$, not $1 - P$.
Before asking, every one of the $Q$ columns carries the same answer
distribution $(P, U, \ldots, U)$ and hence entropy $G_1$, so
$H_{\rm before} = H(M^{(0)}) = Q\,G_1$. Conditional on confirmation the
next unasked answer has probabilities $P_2/P$ for zero and $U_2/P$
for each nonzero symbol, with entropy
\begin{equation}
  H_{\rm c}
  \;:=\; -\frac{P_2}{P}\log_2\frac{P_2}{P}
  - (A-1)\frac{U_2}{P}\log_2\frac{U_2}{P},
  \qquad
  H_{\rm after, correct} = (Q-1)H_{\rm c},
  \label{eq:spikeHc}
\end{equation}
the asked column being fixed while all $Q-1$ unasked columns share
that conditional marginal. Refutation instead leaves equal posterior
weights on all compatible maps, so every unasked answer is uniform on
$A$ symbols and $H_{\rm after, wrong} = (Q-1)m$. Weighting the two
branches,
\begin{equation}
  \big\langle\!\big\langle H\big(M^{(1)}\big)\big\rangle\!\big\rangle
  = P\,H_{\rm after, correct} + (A-1)U\,H_{\rm after, wrong}
  = (Q-1)(G_2 - G_1),
  \label{eq:spikebranchavg}
\end{equation}
the first equality ordinary branch averaging, the second the
$\ell=1$ case of \eqref{eq:law-remaining}.

\begin{figure}[H]
\centering
\begin{tikzpicture}[baseline=(current bounding box.north)]
\begin{axis}[width=7.4cm, height=5.6cm,
  title={$(n,m) = (4,1)$}, title style={font=\small},
  xlabel={$p$}, ylabel={bits}, xmin=0, xmax=1, ymin=0, ymax=18,
  tick label style={font=\scriptsize}, label style={font=\small},
  grid=major, grid style={black!12},
  legend style={font=\tiny, at={(0.5,-0.22)}, anchor=north,
    legend columns=4, /tikz/every even column/.append style={column sep=4pt}},
  legend cell align=left, restrict y to domain=0:20]
\dataplot{cpre, dotted, very thick}{p}{hbefore}{data/spike_one_4to1.dat}
\dataplot{cok}{p}{scorrect}{data/spike_one_4to1.dat}
\dataplot{cbad}{p}{swrong}{data/spike_one_4to1.dat}
\dataplot{cexp}{p}{sexpected}{data/spike_one_4to1.dat}
\dataplot{cok, dashed}{p}{hcorrect}{data/spike_one_4to1.dat}
\dataplot{cbad, dashed}{p}{hwrong}{data/spike_one_4to1.dat}
\dataplot{cexp, dashed}{p}{hexpected}{data/spike_one_4to1.dat}
\datalegend{{before}, {$s_{\rm correct}$}, {$s_{\rm wrong}$},
  {$\langle\!\langle s\rangle\!\rangle$}, {after (correct)},
  {after (wrong)}, {$\langle\!\langle H(M^{(1)})\rangle\!\rangle$}}
\end{axis}
\end{tikzpicture}\hfill
\begin{tikzpicture}[baseline=(current bounding box.north)]
\begin{axis}[width=7.4cm, height=5.6cm,
  title={$(n,m) = (3,2)$}, title style={font=\small},
  xlabel={$p$}, xmin=0, xmax=1, ymin=0, ymax=18,
  tick label style={font=\scriptsize}, label style={font=\small},
  grid=major, grid style={black!12}, restrict y to domain=0:20]
\dataplot{cpre, dotted, very thick}{p}{hbefore}{data/spike_one_3to2.dat}
\dataplot{cok}{p}{scorrect}{data/spike_one_3to2.dat}
\dataplot{cbad}{p}{swrong}{data/spike_one_3to2.dat}
\dataplot{cexp}{p}{sexpected}{data/spike_one_3to2.dat}
\dataplot{cok, dashed}{p}{hcorrect}{data/spike_one_3to2.dat}
\dataplot{cbad, dashed}{p}{hwrong}{data/spike_one_3to2.dat}
\dataplot{cexp, dashed}{p}{hexpected}{data/spike_one_3to2.dat}
\end{axis}
\end{tikzpicture}
\caption{One-question anatomy of the spike prior against the special
map's mass $p$: the branch surprisals \eqref{eq:spikesurprisal}
(solid), the answer-matrix entropy before the question (dotted) and
on each branch after it \eqref{eq:spikeHc} (dashed). Both panels
carry $Qm = 16$ bits of table entropy at the uniform point.}
\label{fig:spikeone}
\end{figure}
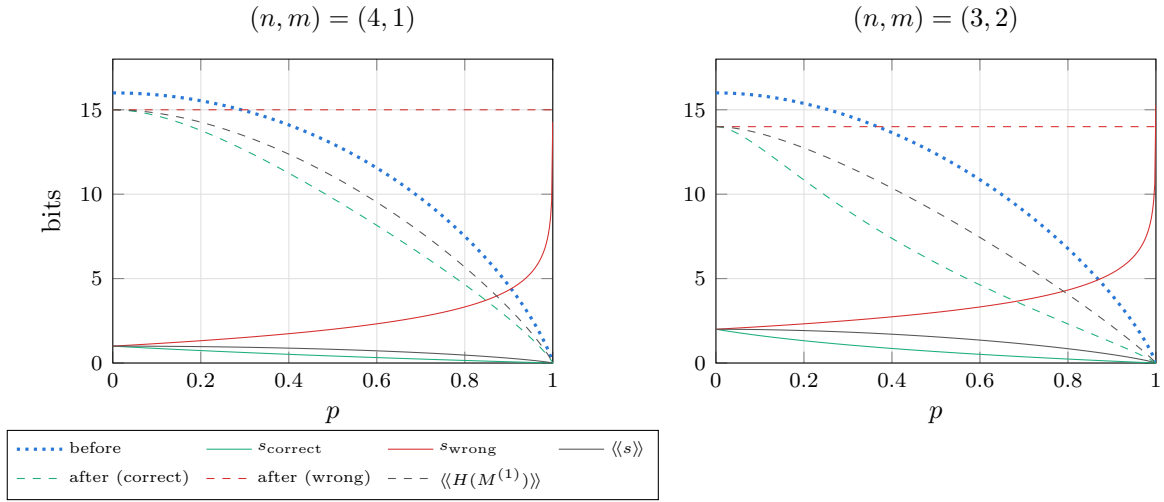

For $p < 1$ the branch leverages divide the vertical distance between
the pre-question curve and each branch curve by that branch's own
surprisal,
\begin{equation}
  L_{\rm correct}
  = \frac{Q G_1 - (Q-1)H_{\rm c}}{-\log_2 P},
  \qquad
  L_{\rm wrong}
  = \frac{Q G_1 - (Q-1)m}{-\log_2 U},
  \label{eq:spikebranchlev}
\end{equation}
whereas the aggregate leverage is the ratio of the two expected
totals,
\begin{equation}
  \mathcal L_1
  = \frac{P\big[Q G_1 - (Q-1)H_{\rm c}\big]
    + (A-1)U\big[Q G_1 - (Q-1)m\big]}{G_1}
  = \frac{Q G_1 - (Q-1)(G_2-G_1)}{G_1}. 
  \label{eq:spikeL1}
\end{equation}
The branch formulas use realized surprisals; the expected formula
divides the expected entropy drop by the expected surprisal $G_1$ and
is therefore \emph{not} the probability-weighted average of the two
branch ratios.

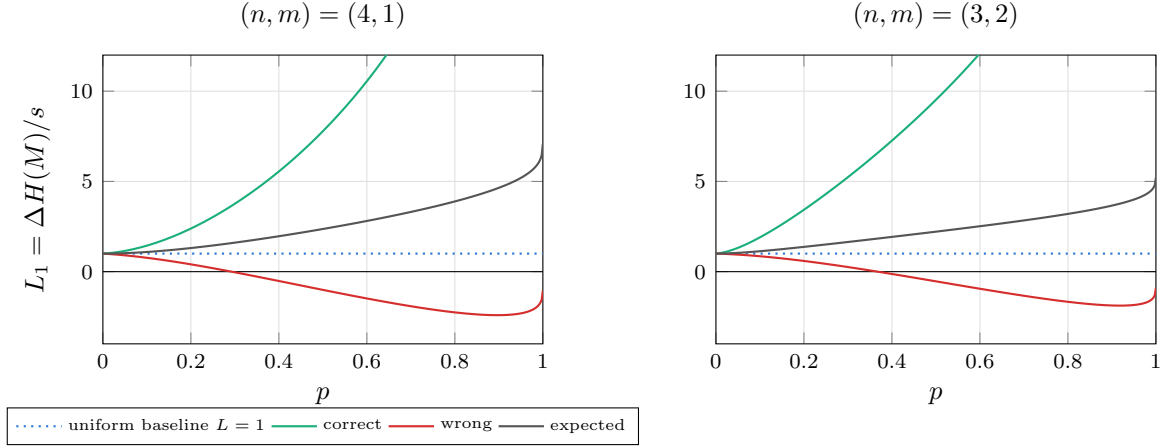
\begin{figure}[H]
\centering
\begin{tikzpicture}[baseline=(current bounding box.north)]
\begin{axis}[width=7.4cm, height=5.4cm,
  title={$(n,m) = (4,1)$}, title style={font=\small},
  xlabel={$p$}, ylabel={$L_1 = \Delta H(M)/s$},
  xmin=0, xmax=1, ymin=-4, ymax=12,
  tick label style={font=\scriptsize}, label style={font=\small},
  grid=major, grid style={black!12}, restrict y to domain=-6:14,
  legend style={font=\tiny, at={(0.5,-0.22)}, anchor=north, legend columns=5},
  legend cell align=left]
\addplot[black, thin, forget plot, domain=0:1] {0};
\addplot[cpre, dotted, thick, domain=0:1] {1};
\dataplot{cok, thick}{p}{lcorrect}{data/spike_lev_4to1.dat}
\dataplot{cbad, thick}{p}{lwrong}{data/spike_lev_4to1.dat}
\dataplot{cexp, thick}{p}{lexpected}{data/spike_lev_4to1.dat}
\datalegend{{uniform baseline $L=1$}, {correct}, {wrong}, {expected}}
\end{axis}
\end{tikzpicture}\hfill
\begin{tikzpicture}[baseline=(current bounding box.north)]
\begin{axis}[width=7.4cm, height=5.4cm,
  title={$(n,m) = (3,2)$}, title style={font=\small},
  xlabel={$p$}, xmin=0, xmax=1, ymin=-4, ymax=12,
  tick label style={font=\scriptsize}, label style={font=\small},
  grid=major, grid style={black!12}, restrict y to domain=-6:14]
\addplot[black, thin, domain=0:1] {0};
\addplot[cpre, dotted, thick, domain=0:1] {1};
\dataplot{cok, thick}{p}{lcorrect}{data/spike_lev_3to2.dat}
\dataplot{cbad, thick}{p}{lwrong}{data/spike_lev_3to2.dat}
\dataplot{cexp, thick}{p}{lexpected}{data/spike_lev_3to2.dat}
\end{axis}
\end{tikzpicture}
\caption{Branchwise \eqref{eq:spikebranchlev} and expected
\eqref{eq:spikeL1} leverage of the first question. A refuting answer
carries so much surprisal that its own ratio falls below the uniform
baseline and eventually below zero, yet the expectation stays well
above $1$ because the confirming branch is both likely and cheap.}
\label{fig:spikelev}
\end{figure}

\subsection{The extraction ratio}

The total correlation $C := H(M^{(0)}) - G_Q = Q G_1 - G_Q$ is the
information stored in dependencies among answers rather than visible
in their separate marginals. Only the part of the expected entropy
drop beyond the received entropy $G_1$ comes out of that store, so
when $C > 0$ the share of it freed by a single question is
\begin{equation}
  R_1 \;:=\; \frac{\langle\!\langle \Delta H(M)\rangle\!\rangle - G_1}{C}
  \;=\; \frac{(Q-1)(2G_1 - G_2)}{Q G_1 - G_Q}. 
  \label{eq:extraction}
\end{equation}
The two ratios differ only in their denominator:
$R_1$ measures against the whole dependency stock, leverage against
the received surprisal, and
\begin{equation}
  R_1 = \frac{G_1}{C}\big(\mathcal L_1 - 1\big),
  \qquad
  \mathcal L_1 = 1 + \frac{C}{G_1}R_1 .
  \label{eq:extractionvslev}
\end{equation}
Thus $R_1$ is a stock-extraction fraction while
$\mathcal L_1 - 1$ is released dependency information per
received bit. At $p = 1$ both the numerator of $R_1$ and $C$ vanish;
expanding \eqref{eq:spikeGk} with $\varepsilon = 1 - p \to 0^+$ gives
the exact finite-$(n,m)$ ratio limit
\begin{equation}
  \lim_{p\to1^-}R_1(p)
  = \frac{(Q-1)N(1-u)^2}{Q N(1-u) - (N-1)}, 
  \qquad
  \frac{(Q-1)(1-u)^2}{Q(1-u)-1}
  \label{eq:extractionsharp}
\end{equation}
the second expression being the additional large-$N$ approximation.
At $(n,m) = (2,1)$ the exact value is $12/17 \simeq 0.706$ against
the approximation's $0.75$. Both share the thermodynamic limit
\begin{equation}
  \lim_{n\to\infty}\lim_{p\to1^-}R_1(p) \;=\; 1 - u \;=\; 1 - 2^{-m}. 
  \label{eq:extractiontdl}
\end{equation}

\begin{figure}[H]
\centering
\begin{tikzpicture}[baseline=(current bounding box.north)]
\begin{axis}[width=7.4cm, height=5.4cm,
  title={$m = 1$, varying $n$}, title style={font=\small},
  xlabel={special-map mass $p$},
  ylabel={$R_1$ (fraction of $C$ extracted)},
  xmin=0, xmax=1, ymin=0, ymax=1.02,
  tick label style={font=\scriptsize}, label style={font=\small},
  grid=major, grid style={black!12},
  legend style={font=\tiny, at={(0.5,-0.22)}, anchor=north, legend columns=5},
  legend cell align=left]
\dataplot{cf1, thick}{p}{ratio}{data/extraction_n2.dat}
\dataplot{cf2, thick}{p}{ratio}{data/extraction_n3.dat}
\dataplot{cf3, thick}{p}{ratio}{data/extraction_n4.dat}
\dataplot{cf4, thick}{p}{ratio}{data/extraction_n5.dat}
\dataplot{cf5, thick}{p}{ratio}{data/extraction_n6.dat}
\datalegend{{$n=2$}, {$n=3$}, {$n=4$}, {$n=5$}, {$n=6$}}
\addplot[cf1, dashed, thin, forget plot, domain=0:1] {0.7059};
\addplot[cf2, dashed, thin, forget plot, domain=0:1] {0.5826};
\addplot[cf3, dashed, thin, forget plot, domain=0:1] {0.5357};
\addplot[cf4, dashed, thin, forget plot, domain=0:1] {0.5167};
\addplot[cf5, dashed, thin, forget plot, domain=0:1] {0.5081};
\end{axis}
\end{tikzpicture}\hfill
\begin{tikzpicture}[baseline=(current bounding box.north)]
\begin{axis}[width=7.4cm, height=5.4cm,
  title={$n = 3$, varying $m$}, title style={font=\small},
  xlabel={special-map mass $p$},
  xmin=0, xmax=1, ymin=0, ymax=1.02,
  tick label style={font=\scriptsize}, label style={font=\small},
  grid=major, grid style={black!12},
  legend style={font=\tiny, at={(0.5,-0.22)}, anchor=north, legend columns=5},
  legend cell align=left]
\dataplot{cf1, thick}{p}{ratio}{data/extraction_m1.dat}
\dataplot{cf2, thick}{p}{ratio}{data/extraction_m2.dat}
\dataplot{cf3, thick}{p}{ratio}{data/extraction_m3.dat}
\dataplot{cf4, thick}{p}{ratio}{data/extraction_m4.dat}
\datalegend{{$m=1$}, {$m=2$}, {$m=3$}, {$m=4$}}
\addplot[cf1, dashed, thin, forget plot, domain=0:1] {0.5826};
\addplot[cf2, dashed, thin, forget plot, domain=0:1] {0.7875};
\addplot[cf3, dashed, thin, forget plot, domain=0:1] {0.8932};
\addplot[cf4, dashed, thin, forget plot, domain=0:1] {0.9465};
\end{axis}
\end{tikzpicture}
\caption{The exact one-question extraction ratio \eqref{eq:extraction}
of the spike prior, plotted only where it is defined
($1/N < p < 1$). Dashed lines are the exact finite-$N$ limits
\eqref{eq:extractionsharp} as $p \to 1^-$. Growing $n$ drives them
down to $1-u = 1/2$; growing $m$ lifts them toward one.}
\label{fig:extraction}
\end{figure}
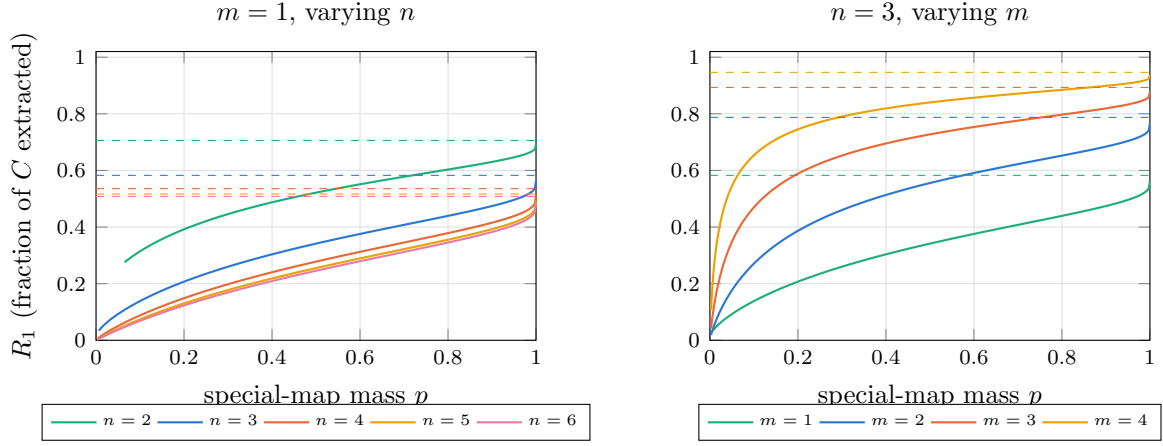

For the spike-plus-uniform family \eqref{eq:extractiontdl} is a
theorem. Extending it to a broader class of uninformative hedges
would require a precise definition and a separate extremal argument:
full support alone is not sufficient.

\subsection{The sharp-prior limit and the order of limits}

The sharp-prior calculation takes $\varepsilon = 1 - p \to 0^+$ with
$n$ and $m$ fixed. It is a limit toward the delta prior, not
evaluation at $p = 1$, where leverage is $0/0$. With the entropy
coefficient $\iota_k := \tfrac{N}{N-1}(1 - u^k)$ we have
$P_k = 1 - \varepsilon\iota_k$, the other $u^{-k}-1$ outcomes share
mass $\varepsilon\iota_k$, and
\begin{equation}
  G_k = h_2(\varepsilon\iota_k)
  + \varepsilon\iota_k\log_2(u^{-k}-1)
  \;\sim\; \varepsilon\,\iota_k\log_2\frac1\varepsilon
  \qquad (\varepsilon \downarrow 0)
  \label{eq:sharpGk}
\end{equation}
at fixed finite $Q$ and $k$. Substituting into \eqref{eq:Lk-G}
cancels the common factor $\varepsilon\log_2(1/\varepsilon)$ and
leaves, for $1 \le \ell \le Q$,
\begin{equation}
  \lim_{p\to1^-}\mathcal L_\ell
  = (1-u)\,\frac{Q - (Q-\ell)u^\ell}{1-u^\ell}
  = \underbrace{Q(1-u)}_{\text{extensive plateau}}
  + \underbrace{(1-u)\frac{\ell\,u^\ell}{1-u^\ell}}_{O(1)\ \text{boundary layer}} .
  \label{eq:sharpLk}
\end{equation}
The second term decays geometrically in $\ell$ and does not depend on
$Q$ at all. In particular
$\lim_{p\to1^-}\mathcal L_1 = Q - (Q-1)2^{-m}$, so
\begin{equation}
  \lim_{n\to\infty}\lim_{p\to1^-}\frac{\mathcal L_1}{Q}
  \;=\; 1 - 2^{-m}. 
  \label{eq:culling}
\end{equation}
The coefficient $1-2^{-m}$ is the fraction of the hypothesis count
eliminated by one answer. Its equality with the normalized leverage
follows from the entropy asymptotics and should not be read as
literally settling that fraction of every individual column. The
limiting operation is necessary: at $p = 1$ the learner already
assigns probability one to the truth, so received and destroyed
information both vanish and leverage is $0/0$. The finite value is
the ratio at which the two vanish as $p \to 1^-$. Taking
$\ell = \lfloor tQ\rfloor$ in \eqref{eq:sharpLk} gives the same constant
at every $t > 0$, with the uniform bound
\begin{equation}
  \sup_{1\le \ell\le Q}
  \left|\frac1Q\lim_{p\to1^-}\mathcal L_\ell - (1-u)\right|
  = O(Q^{-1}),
  \label{eq:plateaubound}
\end{equation}
so the microscopic boundary term survives only as an $O(1)$
correction to the unnormalized leverage.

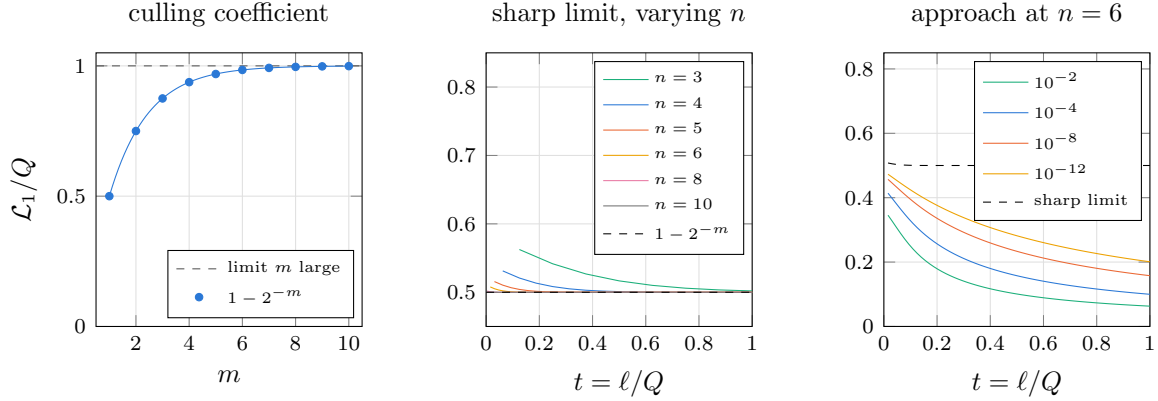
\begin{figure}[H]
\centering
\begin{tikzpicture}[baseline=(current bounding box.north)]
\begin{axis}[width=5.1cm, height=5.2cm,
  title={culling coefficient}, title style={font=\small},
  xlabel={$m$}, ylabel={$\mathcal L_1/Q$},
  xmin=0.5, xmax=10.5, ymin=0, ymax=1.05, xtick={2,4,6,8,10},
  tick label style={font=\scriptsize}, label style={font=\small},
  grid=major, grid style={black!12},
  legend style={font=\tiny, at={(0.97,0.03)}, anchor=south east},
  legend cell align=left]
\addplot[cexp, dashed, domain=0.5:10.5] {1};
\addplot[cpre, thin, forget plot, domain=1:10, samples=80] {1-2^(-x)};
\addplot[cpre, only marks, mark=*, mark size=1.4]
  coordinates {(1,0.5)(2,0.75)(3,0.875)(4,0.9375)(5,0.96875)
    (6,0.984375)(7,0.9921875)(8,0.99609)(9,0.998047)(10,0.999023)};
\legend{{limit $m$ large}, {$1-2^{-m}$}}
\end{axis}
\end{tikzpicture}\hfill
\begin{tikzpicture}[baseline=(current bounding box.north)]
\begin{axis}[width=5.1cm, height=5.2cm,
  title={sharp limit, varying $n$}, title style={font=\small},
  xlabel={$t = \ell/Q$}, xmin=0, xmax=1, ymin=0.45, ymax=0.85,
  tick label style={font=\scriptsize}, label style={font=\small},
  grid=major, grid style={black!12},
  legend style={font=\tiny, at={(0.97,0.97)}, anchor=north east},
  legend cell align=left]
\dataplot{cf1}{x}{lev}{data/spikeflow_sharp_n3.dat}
\dataplot{cf2}{x}{lev}{data/spikeflow_sharp_n4.dat}
\dataplot{cf3}{x}{lev}{data/spikeflow_sharp_n5.dat}
\dataplot{cf4}{x}{lev}{data/spikeflow_sharp_n6.dat}
\dataplot{cf5}{x}{lev}{data/spikeflow_sharp_n8.dat}
\dataplot{black!55}{x}{lev}{data/spikeflow_sharp_n10.dat}
\addplot[black, dashed, domain=0:1] {0.5};
\datalegend{{$n=3$}, {$n=4$}, {$n=5$}, {$n=6$}, {$n=8$}, {$n=10$},
  {$1-2^{-m}$}}
\end{axis}
\end{tikzpicture}\hfill
\begin{tikzpicture}[baseline=(current bounding box.north)]
\begin{axis}[width=5.1cm, height=5.2cm,
  title={approach at $n=6$}, title style={font=\small},
  xlabel={$t = \ell/Q$}, xmin=0, xmax=1, ymin=0, ymax=0.85,
  tick label style={font=\scriptsize}, label style={font=\small},
  grid=major, grid style={black!12},
  legend style={font=\tiny, at={(0.97,0.97)}, anchor=north east},
  legend cell align=left]
\dataplot{cf1}{x}{e2}{data/spikeflow_eps.dat}
\dataplot{cf2}{x}{e4}{data/spikeflow_eps.dat}
\dataplot{cf3}{x}{e8}{data/spikeflow_eps.dat}
\dataplot{cf4}{x}{e12}{data/spikeflow_eps.dat}
\dataplot{black, dashed}{x}{sharp}{data/spikeflow_eps.dat}
\datalegend{{$10^{-2}$}, {$10^{-4}$}, {$10^{-8}$}, {$10^{-12}$},
  {sharp limit}}
\end{axis}
\end{tikzpicture}
\caption{Left: the culling coefficient \eqref{eq:culling} against the
answer width. Middle: the sharp-prior curve \eqref{eq:sharpLk}
normalized by $Q$, whose $O(1)$ boundary layer is squeezed into the
origin as $n$ grows, leaving the plateau at $1-u$. Right: the
approach at fixed $n = 6$, $m = 1$, labeled by $1-p$. Tightening the
prior lifts the curve only logarithmically, so the plateau is
approached pointwise in $t$ but never uniformly.}
\label{fig:sharpflow}
\end{figure}

As $p \to 1^-$ the fixed-$p$ coefficient of
\eqref{eq:spikehyperbola} diverges,
$\big[\eta_0(p,m)-(1-p)m\big]/\big[(1-p)m\big]
\sim \tfrac{1-2^{-m}}{m}\log_2\tfrac1{1-p}$, which is the
noncommutation in explicit form. Writing
$L_{p,n}(t) := \mathcal L_{\lfloor tQ\rfloor}$ and comparing
under the same normalization,
\begin{equation}
  \lim_{n\to\infty}\lim_{p\to1^-}\frac{L_{p,n}(t)}{Q} = 1-2^{-m},
  \qquad
  \lim_{p\to1^-}\lim_{n\to\infty}\frac{L_{p,n}(t)}{Q} = 0 .
  \label{eq:noncommute}
\end{equation}
Before dividing by $Q$ the first ordering is extensive while the
second diverges only logarithmically in $1/(1-p)$. The two orders are
the endpoints of a family of joint scalings. Choose $p_n \to 1$ with
prescribed exponential sharpness
\begin{equation}
  \frac{\log_2\big(1/(1-p_n)\big)}{Q}\longrightarrow
  \lambda\in(0,\infty),
  \label{eq:lambdascaling}
\end{equation}
so that $\lambda = 0$ recovers the scaling behind the second limit and
the formal endpoint $\lambda = \infty$ the first. Putting
$\varepsilon_n = 1-p_n$, the leading entropies for $0 < t < 1$ and
$\ell = \lfloor tQ\rfloor$ are
$G_1 \sim (1-u)\varepsilon_n\lambda Q$,
$G_\ell \sim \varepsilon_n Q(\lambda + mt)$ and
$G_{\ell+1}-G_\ell \sim \varepsilon_n m$, and substitution into
\eqref{eq:Lk-G} gives the crossover
\begin{equation}
  \frac{L_{p_n,n}(t)}{Q}
  \longrightarrow (1-u)\,\frac{\lambda}{\lambda + mt}. 
  \label{eq:crossover}
\end{equation}
At $t = 1$ the increment $G_{\ell+1}-G_\ell$ is undefined and the same value
follows instead from the completion formula. The two familiar orders
are thus endpoints of a larger family.

\section{Choosing the questions}\label{app:strategies}

The uniform question order of Chapter~\ref{sec:expectations} was a symmetry
choice, not a strategy.
Choosing the next question is Bayesian experimental design
\cite{chalonerverdinelli95}, and Werner can do better along two independent axes: how far he looks ahead,
and whether he commits to an order beforehand or pivots contingent on the answers so far. A \textbf{Greedy} Werner asks, at
every step, the single question with the largest expected drop of
$H(M)$: one step of lookahead. \textbf{$\ell$-optimal} maximizes the
expected knowledge after exactly $\ell$ questions, looking the
whole horizon ahead. All four combinations are straightforward
with the machinery at hand. $1$-optimal coincides with Greedy at the start.

\begin{table}[H]
\begin{center}
\begin{tabular}{|c|>{\raggedright\arraybackslash}m{5.35cm}|>{\raggedright\arraybackslash}m{5.35cm}|}
\hline
 & \textbf{order fixed beforehand} & \textbf{contingent on answers} \\
\hline
\rotatebox{90}{~\textbf{greedy}~} &
Add planned questions to the docket, each maximizing the immediate
expected drop of $H(M)$, averaged over the still-unseen answers.
All of this before any answer is seen, so the plan cannot react to luck. &
Recompute the forecasts after every
answer and ask the current argmax. Question order is a branching tree, each branch
contingent on last answer. \\
\hline
\rotatebox{90}{~\textbf{$\ell$-optimal}~} &
Updating commutes, so a fixed plan is just a set $|\mathcal{S}|=\ell$: search the
$\binom{Q}{\ell}$ subsets for the smallest expected remainder
$\sum_{q' \notin \mathcal S} H(\psi(q') \mid \psi(\mathcal S))$. The best non-adaptive
knowledge; sees correlations that the one-step view misses. &
Backward induction over posteriors. The
Bayes-optimal policy: recompute optimal further set for remainder at $\ell$, after 
every answer. \\
\hline
\end{tabular}
\end{center}
\caption{The four strategies open to Werner: one step of lookahead or $\ell$, with the question order fixed in advance or chosen contingent on the answers already heard.}
\label{tab:strategies}
\end{table}

Expected knowledge is non-decreasing left to right and top to bottom. The
dynamic program in the last cell is exact but grows quickly: its
states are the reachable posteriors, one per pair of asked set and
answer pattern, $(1 + A)^{Q}$ in all. That is $81$ for the
guiding example and astronomical beyond; the underlying problem, an
optimal decision tree over the answers, is NP-complete in general
\cite{hyafilrivest76}. Its policy is also
horizon-dependent, since the best first question for $\ell = k$
need not open the best plan for $\ell = k+1$. For
the guiding prior of Table~\ref{tab:sixteen}, contingent greedy outperforms the uniform order, 
and already attains the contingent $\ell$-optimum at every horizon. 
That is a quirk of this prior, not a law: a prior that
stores its knowledge in deep shells can defeat Greedy in either
column: a question worth little now may unlock a cascade
later.

Both $\ell$-optimal protocols produced the same trajectory in the
examples tested above, but this is not a law. Adaptive questioning can
strictly improve on every fixed set of the same size, as Example~\ref{ex:adaptive} demonstrates.
The distinction is that an answer can tell Werner which question
will be useful next.
\begin{example}[label=ex:adaptive]
Take $n=2$, $m=1$, and put equal prior weight $1/4$ on the four
maps below, with zero weight on the others. The columns retain the
descending question order used throughout.
\begin{center}
\begin{tabular}{@{}c cccc c@{}}
\toprule
$j$ & $\phi_j(11)$ & $\phi_j(10)$ & $\phi_j(01)$ & $\phi_j(00)$ & $p_j$\\
\midrule
0 & 0 & 0 & 0 & 0 & $1/4$\\
2 & 0 & 0 & 1 & 0 & $1/4$\\
1 & 0 & 0 & 0 & 1 & $1/4$\\
5 & 0 & 1 & 0 & 1 & $1/4$\\
\bottomrule
\end{tabular}
\end{center}
Ask $q=00$ first. If $a=\psi(00)=0$, the survivors are maps
$0$ and $2$, distinguished by asking $q'=01$. If $a=1$, they are
maps $1$ and $5$, distinguished by asking $q'=10$. Thus two
adaptive questions identify the map completely and leave $H(M)=0$.

The tree below compares protocols: the second question is whichever
splits the pair that survived. No single question splits both. The
other examples are isomorphic to this one. The colored digit is the
one the question at that node reads.

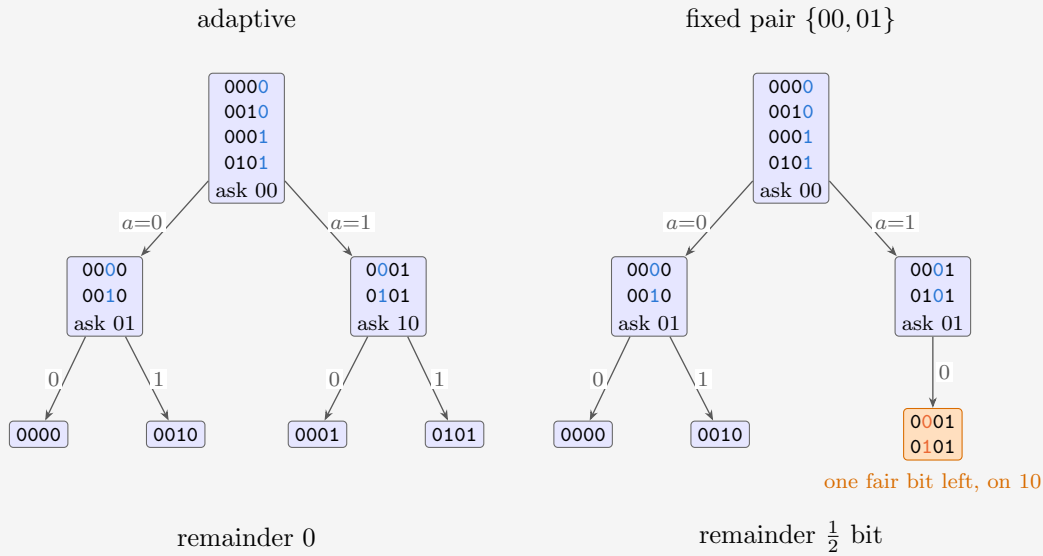
\begin{center}
\begin{tikzpicture}[x=1.25cm, y=1.35cm,
  hyp/.style={draw=black!60, rounded corners=1.5pt, inner sep=2.5pt,
              fill=blue!10, font=\scriptsize, align=center},
  ans/.style={font=\scriptsize, inner sep=1pt, fill=white},
  ed/.style={-{Stealth[length=4pt]}, black!65, line width=0.5pt}]
  \node[font=\small] at (0,4.55) {adaptive};
  \node[hyp] (ar) at (0,3.4) {\texttt{000\textcolor{cpre}{0}}\\\texttt{001\textcolor{cpre}{0}}\\\texttt{000\textcolor{cpre}{1}}\\\texttt{010\textcolor{cpre}{1}}\\[1pt]ask $00$};
  \node[hyp] (al) at (-1.5,1.85) {\texttt{00\textcolor{cpre}{0}0}\\\texttt{00\textcolor{cpre}{1}0}\\[1pt]ask $01$};
  \node[hyp] (arr) at (1.5,1.85) {\texttt{0\textcolor{cpre}{0}01}\\\texttt{0\textcolor{cpre}{1}01}\\[1pt]ask $10$};
  \node[hyp] (al0) at (-2.2,0.5) {\texttt{0000}};
  \node[hyp] (al1) at (-0.75,0.5) {\texttt{0010}};
  \node[hyp] (ar0) at (0.75,0.5) {\texttt{0001}};
  \node[hyp] (ar1) at (2.2,0.5) {\texttt{0101}};
  \draw[ed] (ar) -- node[ans, pos=0.55, left=2pt] {$a{=}0$} (al);
  \draw[ed] (ar) -- node[ans, pos=0.55, right=2pt] {$a{=}1$} (arr);
  \draw[ed] (al) -- node[ans, pos=0.5, left=1pt] {$0$} (al0);
  \draw[ed] (al) -- node[ans, pos=0.5, right=1pt] {$1$} (al1);
  \draw[ed] (arr) -- node[ans, pos=0.5, left=1pt] {$0$} (ar0);
  \draw[ed] (arr) -- node[ans, pos=0.5, right=1pt] {$1$} (ar1);
  \node[font=\small] at (0,-0.5) {remainder $0$};
  \begin{scope}[xshift=7.2cm]
  \node[font=\small] at (0,4.55) {fixed pair $\{00,01\}$};
  \node[hyp] (fr) at (0,3.4) {\texttt{000\textcolor{cpre}{0}}\\\texttt{001\textcolor{cpre}{0}}\\\texttt{000\textcolor{cpre}{1}}\\\texttt{010\textcolor{cpre}{1}}\\[1pt]ask $00$};
  \node[hyp] (fl) at (-1.5,1.85) {\texttt{00\textcolor{cpre}{0}0}\\\texttt{00\textcolor{cpre}{1}0}\\[1pt]ask $01$};
  \node[hyp] (frr) at (1.5,1.85) {\texttt{00\textcolor{cpre}{0}1}\\\texttt{01\textcolor{cpre}{0}1}\\[1pt]ask $01$};
  \node[hyp] (fl0) at (-2.2,0.5) {\texttt{0000}};
  \node[hyp] (fl1) at (-0.75,0.5) {\texttt{0010}};
  \node[hyp, fill=orange!25, draw=orange!85!black] (fu) at (1.5,0.5)
    {\texttt{0\textcolor{clim}{0}01}\\\texttt{0\textcolor{clim}{1}01}};
  \draw[ed] (fr) -- node[ans, pos=0.55, left=2pt] {$a{=}0$} (fl);
  \draw[ed] (fr) -- node[ans, pos=0.55, right=2pt] {$a{=}1$} (frr);
  \draw[ed] (fl) -- node[ans, pos=0.5, left=1pt] {$0$} (fl0);
  \draw[ed] (fl) -- node[ans, pos=0.5, right=1pt] {$1$} (fl1);
  \draw[ed] (frr) -- node[ans, pos=0.5, right=1pt] {$0$} (fu);
  \node[font=\scriptsize, orange!85!black, below=1pt] at (fu.south)
    {one fair bit left, on $10$};
  \node[font=\small] at (0,-0.5) {remainder $\tfrac12$ bit};
  \end{scope}
\end{tikzpicture}
\end{center}

No fixed pair does so. The pairs $\{00,01\}$ and $\{00,10\}$ each
leave one fair unknown answer with probability $1/2$, hence an
expected remainder of $1/2$ bit. The pair $\{01,10\}$ leaves maps
$0$ and $1$ indistinguishable when both answers are zero, again an
event of probability $1/2$ with one fair unknown answer. These are
the three best fixed pairs. Pairs containing the constant question
$11$ leave at least $1$ bit. The optimal fixed remainder is therefore
$1/2$ bit, strictly above the adaptive remainder of zero.
\end{example}

\section{The Gibbs complexity prior}\label{app:gibbs}

Every prior so far was an arbitrary set of instructively chosen $\{p_j\}$. 
This appendix builds the prior the other way around, from a
single philosophical principle: \emph{simple explanations are likelier}.
The principle has formal standing, as Solomonoff's universal prior
\cite{solomonoff64,livitanyi08} and as a learning guarantee for hypotheses
with short descriptions \cite{blumer87}. 
How do we define simplicity? With circuit
complexity as the measure, made quantitative as a Gibbs
ensemble over the hypothesis space. It is an illustration; the body
does not depend on it.

\subsection{AIG complexity}
\label{sec:aigcomplexity}

A compact circuit description supplies a concrete notion of
simplicity. We use an \emph{and-inverter graph} (AIG): a feed-forward
circuit with two-input AND nodes and complemented edges, so inversion
on any wire is free. Let $X_j$ be the fewest AND nodes in any such
graph computing $\phi_j$, with constants and input wires available
at zero cost. This is the standard AIG-size metric of logic
synthesis~\cite{aigrewrite}. It differs from counting ordinary
NAND gates, where inversion itself has a cost. Reference values in
the AIG convention:
\begin{itemize}
  \item AND, OR, NAND and NOR all cost $1$;
  \item XOR costs $3$;
  \item constants, projections and negations cost $0$.
\end{itemize}   
For $m > 1$ outputs, $X_j$ is the size of one \emph{joint} circuit computing all
output bits together, gates shared, which is practical. It is the minimum
cost in this AIG convention, generally smaller than the sum of per-output circuits.

The $(n,m) = (4,1)$ and $(3,2)$ are the largest tractable setups, at $65{,}536$ maps apiece. 
The tabulation is feasible because $X_j$ is invariant under
relabeling: permuting or negating inputs, negating or (for $m \ge 2$)
swapping outputs are all free operations, so the $65{,}536$ maps collapse to $222$ and $308$
equivalence classes respectively. The $(4,1)$ class values are the published
optimal AIG sizes of \cite{krinkin}, cross-checked against our own
SAT-based synthesizer on random classes with zero mismatches; the $(3,2)$
values come from that synthesizer: circuit sizes $X = 0, 1, 2,
\ldots$ are encoded as satisfiability instances, and the first
satisfiable $X$ is provably minimal, every smaller size having been
refuted. Appendix~\ref{app:aig} gives the encoding, the symmetry
reduction that makes it tractable, and the provenance of every value.

Alongside the interior cost $X_j$, which takes a SAT solver to see,
two descriptors are visible in one pass over the truth table:
\begin{itemize}
\item the \textbf{footprint} $F_j = n \cdot \text{[Image]} + \text{[Support]}$, where the support
 counts the input bits the map actually depends on and the
  image size counts the distinct answers it attains (from $1$
  for constants up to $A$). The combination is bijective to the pair, as Support $\le n$.
  Compare to base-$n$ digits, quotient Image, remainder Support. The footprint can be viewed as the
  map's \emph{interface}: does it react to all its inputs, does it
  use all its values?
\item the \textbf{bias} $B_j$: the number of $1$s minus the number of
  $0$s among the $mQ$ truth-table bits. This is the
  symmetry breaker: $X_j$ and $F_j$ are even under negating all
  outputs, $B_j$ is odd: an energy term coupling to it acts like a
  magnetic field, favoring 0s or 1s off the bat. In the language of Appendix~\ref{app:correlators}: 
  switching on the odd correlator shells that a pure
  complexity prior forbids.
\end{itemize}

\begin{example}
Take again the $(n,m)=(2,1)$ space. Each map's data is listed for illustration.
We graph two of the Gibbs priors built from these inputs, defined in Section~\ref{sec:gibbsensemble}. 
Occam in blue growing left, the Pragmatist in orange growing
right. The Pragmatist bars are logarithmic, five decades to the full
width, since a linear scale on the same axis as Occam's would show
nothing but the two constants.

\begin{center}
\begin{tabular}{clccccc r@{\hspace{2pt}}|@{\hspace{2pt}}l}
\toprule
$j$ & map & $X_j$ & Support & Image & $F_j$ & $B_j$
  & \multicolumn{2}{c}{prior mass} \\
 & & & & & & & {\scriptsize\color{blue!55!black} Occam}
  & {\scriptsize\color{orange!85!black} Pragmatist} \\
\midrule
0  & FALSE                          & 0 & 0 & 1 & 2 & $-4$ & \pbarO[3]{0.1106} & \pbarP{0.2799} \\
1  & $\lnot(b_1 \vee b_0)$ (NOR)    & 1 & 2 & 2 & 6 & $-2$ & \pbarO[3]{0.0407} & \pbarP{0.0014} \\
2  & $\lnot b_1 \wedge b_0$         & 1 & 2 & 2 & 6 & $-2$ & \pbarO[3]{0.0407} & \pbarP{0.0014} \\
3  & $\lnot b_1$                    & 0 & 1 & 2 & 5 & $0$  & \pbarO[3]{0.1106} & \pbarP{0.0208} \\
4  & $b_1 \wedge \lnot b_0$         & 1 & 2 & 2 & 6 & $-2$ & \pbarO[3]{0.0407} & \pbarP{0.0014} \\
5  & $\lnot b_0$                    & 0 & 1 & 2 & 5 & $0$  & \pbarO[3]{0.1106} & \pbarP{0.0208} \\
6  & $b_1 \oplus b_0$ (XOR)         & 3 & 2 & 2 & 6 & $0$  & \pbarO[3]{0.0055} & \pbarP{0.0001} \\
7  & $\lnot(b_1 \wedge b_0)$ (NAND) & 1 & 2 & 2 & 6 & $+2$ & \pbarO[3]{0.0407} & \pbarP{0.0021} \\
8  & $b_1 \wedge b_0$ (AND)         & 1 & 2 & 2 & 6 & $-2$ & \pbarO[3]{0.0407} & \pbarP{0.0014} \\
9  & $b_1 = b_0$ (XNOR)             & 3 & 2 & 2 & 6 & $0$  & \pbarO[3]{0.0055} & \pbarP{0.0001} \\
10 & $b_0$ (projection)             & 0 & 1 & 2 & 5 & $0$  & \pbarO[3]{0.1106} & \pbarP{0.0208} \\
11 & $b_1 \to b_0$                  & 1 & 2 & 2 & 6 & $+2$ & \pbarO[3]{0.0407} & \pbarP{0.0021} \\
12 & $b_1$ (projection)             & 0 & 1 & 2 & 5 & $0$  & \pbarO[3]{0.1106} & \pbarP{0.0208} \\
13 & $b_0 \to b_1$                  & 1 & 2 & 2 & 6 & $+2$ & \pbarO[3]{0.0407} & \pbarP{0.0021} \\
14 & $b_1 \vee b_0$ (OR)            & 1 & 2 & 2 & 6 & $+2$ & \pbarO[3]{0.0407} & \pbarP{0.0021} \\
15 & TRUE                           & 0 & 0 & 1 & 2 & $+4$ & \pbarO[3]{0.1106} & \pbarP{0.3114} \\
\bottomrule
\end{tabular}
\end{center}

Occam spreads its belief mostly across the six gate-free maps ($0.1106$
apiece) and thins gradually with cost; the Pragmatist, punishing
footprint and rewarding bits with value 1, piles $0.90$ of its mass onto the two
constants ($0.62$ on TRUE against $0.28$ on FALSE), the bias
breaking the tie. It all but abandons the interior of the table. If the map turns out non-constant, renormalization scales that interior up, and it becomes relevant.
Grouped by the full tuple $(X, F, B)$, the sixteen maps fall into
six energy classes:
\begin{center}
\begin{tabular}{cc@{\hskip 2em}cc}
\toprule
$(X, F, B)$ & members & $(X, F, B)$ & members \\
\midrule
$(0, 2, -4)$ & 1 & $(1, 6, -2)$ & 4 \\
$(0, 2, +4)$ & 1 & $(1, 6, +2)$ & 4 \\
$(0, 5, 0)$  & 4 & $(3, 6, 0)$  & 2 \\
\bottomrule
\end{tabular}
\end{center}

Note the table skips $X = 2$ entirely: with two inputs,
nothing a two-gate circuit computes is out of reach of one or
zero gates.
\end{example}

At the larger sizes the tuple $(X, F, B)$ breakdown runs to hundreds of rows
($102$ distinct tuples at $(n,m)=(4,1)$, $196$ at $(3,2)$). It is still interesting to tabulate
the complexity $X$ shells alone, with each shell's \emph{aggregate} prior
mass under the Occam setting alongside its head count:

\begin{example}
These tables illustrate the compromise between exponentially suppressed weight per map, and combinatorics
putting most maps at middling complexity.
\begin{center}
\begin{tabular}{crr@{\hspace{4pt}}l}
\toprule
\multicolumn{4}{c}{$(4,1)$} \\
$X$ & maps & \multicolumn{2}{c}{Occam mass} \\
\midrule
0  & 10       & 0.041 & \pbarB[2]{0.0407} \\
1  & 48       & 0.072 & \pbarB[2]{0.0718} \\
2  & 256      & 0.141 & \pbarB[2]{0.1408} \\
3  & 940      & 0.190 & \pbarB[2]{0.1903} \\
4  & 2{,}336  & 0.174 & \pbarB[2]{0.1739} \\
5  & 6{,}464  & 0.177 & \pbarB[2]{0.1771} \\
6  & 10{,}616 & 0.107 & \pbarB[2]{0.1070} \\
7  & 18{,}984 & 0.070 & \pbarB[2]{0.0704} \\
8  & 17{,}680 & 0.024 & \pbarB[2]{0.0241} \\
9  & 7{,}882  & 0.004 & \pbarB[2]{0.0040} \\
10 & 320      & 0.000 & \pbarB[2]{0.0001} \\
\midrule
total & 65{,}536 & 1 & \\
\bottomrule
\end{tabular}
\hspace{1.6em}
\begin{tabular}{crr@{\hspace{4pt}}l}
\toprule
\multicolumn{4}{c}{$(3,2)$} \\
$X$ & maps & \multicolumn{2}{c}{Occam mass} \\
\midrule
0  & 64       & 0.058 & \pbarB[2]{0.0579} \\
1  & 432      & 0.144 & \pbarB[2]{0.1438} \\
2  & 2{,}064  & 0.253 & \pbarB[2]{0.2528} \\
3  & 4{,}860  & 0.219 & \pbarB[2]{0.2190} \\
4  & 10{,}832 & 0.180 & \pbarB[2]{0.1795} \\
5  & 16{,}208 & 0.099 & \pbarB[2]{0.0988} \\
6  & 17{,}060 & 0.038 & \pbarB[2]{0.0383} \\
7  & 10{,}672 & 0.009 & \pbarB[2]{0.0088} \\
8  & 3{,}312  & 0.001 & \pbarB[2]{0.0010} \\
9  & 32       & 0.000 & \pbarB[2]{0.0000} \\
\midrule
total & 65{,}536 & 1 & \\
\bottomrule
\end{tabular}
\end{center}
\end{example}

The head counts are thin at both ends: ten gate-free maps at $(4,1)$
(two constants, eight literals) against only $320$ at the ceiling
$X = 10$. The aggregate mass shows the thermodynamic compromise this
forces on Occam: the count grows faster than $e^{-X}$ decays until
$X = 3$ at $(4,1)$ ($X = 2$ at $(3,2)$), so even under a pure
simplicity prior Werner has most faith that the map has $X_j \in \{2,3\}$.
Shell degeneracy beats shell energy for the simplest maps.

\subsection{The ensemble}\label{sec:gibbsensemble}

Take the energy linear in the three descriptors and the prior as its
Boltzmann weight,
\begin{equation}
E_j = \beta\, X_j + \alpha\, F_j + \mu\, B_j,
\qquad
p_j = \frac{e^{-E_j}}{\sum_k e^{-E_k}},
\label{eq:gibbs}
\end{equation}
with no overall temperature: it's absorbed in the coupling fields. Positive
$\beta$ suppresses complex maps, positive $\alpha$ suppresses large
footprints, and $\mu$ couples to the signed bias like a magnet. $\mu < 0$ tilts
toward $1$-heavy tables and thereby breaks the output-negation
symmetry. We explore three settings below, chosen out of principle, as well as to exhibit
interesting learning behavior.

\begin{table}[H]
\begin{center}
\begin{tabular}{lccc}
\toprule
 & $\beta$ & $\alpha$ & $\mu$ \\
\midrule
Occam      & $1$   & $0$ & $0$    \\
Occamer    & $2$   & $0$ & $0$    \\
Pragmatist & $1.5$ & $1$ & $-0.1$ \\
\bottomrule
\end{tabular}
\end{center}
\caption{The three Gibbs settings of \eqref{eq:gibbs}, as couplings to complexity, footprint and bias.}
\label{tab:gibbssettings}
\end{table}

\textbf{Occam} suppresses circuit complexity and nothing else.
\textbf{Occamer} doubles the same field: an even stronger simplicity
bias stores more knowledge in the prior, and stores it deeper in the
correlation shells, where only the later questions of a run can
release it. The \textbf{Pragmatist} judges by the obvious heuristics
instead: a footprint reads the visible interface rather
than the interior circuit, a moderate complexity field seconds it,
and the tilt $\mu = -0.1$ leans toward $1$-heavy tables, breaking
the output-negation symmetry that the other two settings keep. This last choice is to illustrate
that Werner isn't bound to start with a uniform distribution $M$ before the first question.

\subsection{The layered entropy books}\label{sec:gibbsbooks}

Let us visualize the uniform question-order average for the three ensembles,
at each system size. The following graphs illustrate the flow of bits. 
The three quantities of \eqref{eq:Lk-avg} partition
the starting table entropy at every $\ell$: what is still in the table,
what was paid for in surprisal, and what was inferred. The third has no
standing name, so give it one,
\begin{equation}
  \big\langle\!\big\langle \text{deduced}_\ell \big\rangle\!\big\rangle
  \;:=\; C_0 - \langle\!\langle C_\ell\rangle\!\rangle
  \;=\; H\big(M^{(0)}\big)
        - \big\langle\!\big\langle H\big(M^{(\ell)}\big)\big\rangle\!\big\rangle
        - G_\ell ,
  \label{eq:deduced}
\end{equation}
the expected withdrawal from the correlation store. The second equality
is \eqref{eq:C} at both ends together with $\langle\!\langle \Delta H(p)\rangle\!\rangle =
G_\ell$. In expectation, the belief entropy falls by exactly the
surprisal received, so whatever else the table lost came out of $C$.
The three therefore sum to $H(M^{(0)})$ at every $\ell$, and each
panel below is a full-height bar whose composition shifts from all
remaining uncertainty to received plus deduced. One row per system size, sharing a
vertical scale in bits; one column per prior.
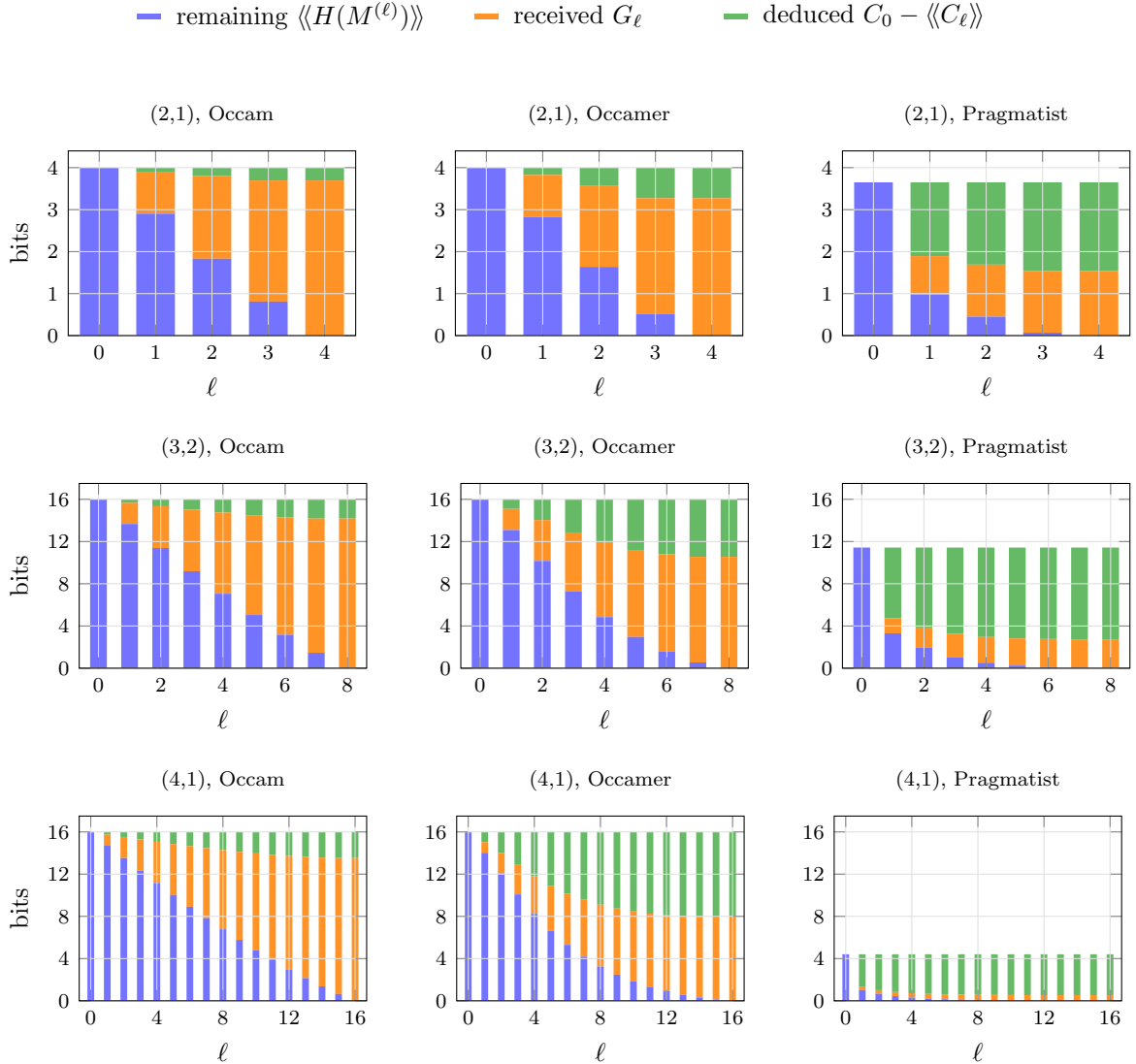
\begin{center}
\begin{tikzpicture}[y=0.32cm]
  \fill[blue!55] (0,0) rectangle (0.32,0.24);
  \node[right, font=\small] at (0.40,0.12)
    {remaining $\langle\!\langle H(M^{(\ell)})\rangle\!\rangle$};
  \fill[orange!85] (4.6,0) rectangle (4.92,0.24);
  \node[right, font=\small] at (5.0,0.12) {received $G_\ell$};
  \fill[green!55!black!60] (8.0,0) rectangle (8.32,0.24);
  \node[right, font=\small] at (8.4,0.12)
    {deduced $C_0 - \langle\!\langle C_\ell\rangle\!\rangle$};
\end{tikzpicture}
\end{center}

\begin{center}
\begin{tikzpicture}[baseline=(current bounding box.north)]
\begin{axis}[width=5.5cm, height=4.1cm,
  ybar stacked, bar width=15pt, xmin=-0.55, xmax=4.55,
  ymin=0, ymax=4.4, ytick={0,1,2,3,4}, xtick={0,1,2,3,4},
  title={(2,1), Occam}, title style={font=\scriptsize},
  xlabel={$\ell$}, ylabel={bits},
  tick label style={font=\scriptsize}, label style={font=\small},
  grid=major, grid style={black!12}, axis on top]
  \addplot[fill=blue!55, draw=none] coordinates {(0,4.00000) (1,2.90370) (2,1.83001) (3,0.81104) (4,0.00000)};
  \addplot[fill=orange!85, draw=none] coordinates {(0,-0.00000) (1,1.00000) (2,1.96790) (3,2.88290) (4,3.69394)};
  \addplot[fill=green!55!black!60, draw=none] coordinates {(0,0.00000) (1,0.09630) (2,0.20209) (3,0.30606) (4,0.30606)};
\end{axis}
\end{tikzpicture}\hfill
\begin{tikzpicture}[baseline=(current bounding box.north)]
\begin{axis}[width=5.5cm, height=4.1cm,
  ybar stacked, bar width=15pt, xmin=-0.55, xmax=4.55,
  ymin=0, ymax=4.4, ytick={0,1,2,3,4}, xtick={0,1,2,3,4},
  title={(2,1), Occamer}, title style={font=\scriptsize},
  xlabel={$\ell$},
  tick label style={font=\scriptsize}, label style={font=\small},
  grid=major, grid style={black!12}, axis on top]
  \addplot[fill=blue!55, draw=none] coordinates {(0,4.00000) (1,2.82620) (2,1.63686) (3,0.51162) (4,0.00000)};
  \addplot[fill=orange!85, draw=none] coordinates {(0,0.00000) (1,1.00000) (2,1.94207) (3,2.76050) (4,3.27212)};
  \addplot[fill=green!55!black!60, draw=none] coordinates {(0,0.00000) (1,0.17380) (2,0.42107) (3,0.72788) (4,0.72788)};
\end{axis}
\end{tikzpicture}\hfill
\begin{tikzpicture}[baseline=(current bounding box.north)]
\begin{axis}[width=5.5cm, height=4.1cm,
  ybar stacked, bar width=15pt, xmin=-0.55, xmax=4.55,
  ymin=0, ymax=4.4, ytick={0,1,2,3,4}, xtick={0,1,2,3,4},
  title={(2,1), Pragmatist}, title style={font=\scriptsize},
  xlabel={$\ell$},
  tick label style={font=\scriptsize}, label style={font=\small},
  grid=major, grid style={black!12}, axis on top]
  \addplot[fill=blue!55, draw=none] coordinates {(0,3.65070) (1,0.98017) (2,0.45071) (3,0.06902) (4,0.00000)};
  \addplot[fill=orange!85, draw=none] coordinates {(0,0.00000) (1,0.91268) (2,1.23940) (3,1.46475) (4,1.53378)};
  \addplot[fill=green!55!black!60, draw=none] coordinates {(0,0.00000) (1,1.75786) (2,1.96059) (3,2.11692) (4,2.11692)};
\end{axis}
\end{tikzpicture}
\end{center}

\begin{center}
\begin{tikzpicture}[baseline=(current bounding box.north)]
\begin{axis}[width=5.5cm, height=4.1cm,
  ybar stacked, bar width=6.5pt, xmin=-0.62, xmax=8.62,
  ymin=0, ymax=17.5, ytick={0,4,8,12,16}, xtick={0,2,4,6,8},
  title={(3,2), Occam}, title style={font=\scriptsize},
  xlabel={$\ell$}, ylabel={bits},
  tick label style={font=\scriptsize}, label style={font=\small},
  grid=major, grid style={black!12}, axis on top]
  \addplot[fill=blue!55, draw=none] coordinates {(0,16.00000) (1,13.66381) (2,11.38982) (3,9.18077) (4,7.05614) (5,5.04253) (6,3.16734) (7,1.46045) (8,0.00000)};
  \addplot[fill=orange!85, draw=none] coordinates {(0,0.00000) (1,2.00000) (2,3.95197) (3,5.85028) (4,7.68643) (5,9.45046) (6,11.13131) (7,12.71498) (8,14.17543)};
  \addplot[fill=green!55!black!60, draw=none] coordinates {(0,0.00000) (1,0.33619) (2,0.65821) (3,0.96896) (4,1.25743) (5,1.50700) (6,1.70136) (7,1.82457) (8,1.82457)};
\end{axis}
\end{tikzpicture}\hfill
\begin{tikzpicture}[baseline=(current bounding box.north)]
\begin{axis}[width=5.5cm, height=4.1cm,
  ybar stacked, bar width=6.5pt, xmin=-0.62, xmax=8.62,
  ymin=0, ymax=17.5, ytick={0,4,8,12,16}, xtick={0,2,4,6,8},
  title={(3,2), Occamer}, title style={font=\scriptsize},
  xlabel={$\ell$},
  tick label style={font=\scriptsize}, label style={font=\small},
  grid=major, grid style={black!12}, axis on top]
  \addplot[fill=blue!55, draw=none] coordinates {(0,16.00000) (1,13.09000) (2,10.16649) (3,7.29628) (4,4.84181) (5,2.96269) (6,1.56885) (7,0.54633) (8,0.00000)};
  \addplot[fill=orange!85, draw=none] coordinates {(0,0.00000) (1,2.00000) (2,3.87000) (3,5.56442) (4,7.02367) (5,8.23412) (6,9.22169) (7,10.00612) (8,10.55244)};
  \addplot[fill=green!55!black!60, draw=none] coordinates {(0,0.00000) (1,0.91000) (2,1.96351) (3,3.13930) (4,4.13452) (5,4.80318) (6,5.20946) (7,5.44756) (8,5.44756)};
\end{axis}
\end{tikzpicture}\hfill
\begin{tikzpicture}[baseline=(current bounding box.north)]
\begin{axis}[width=5.5cm, height=4.1cm,
  ybar stacked, bar width=6.5pt, xmin=-0.62, xmax=8.62,
  ymin=0, ymax=17.5, ytick={0,4,8,12,16}, xtick={0,2,4,6,8},
  title={(3,2), Pragmatist}, title style={font=\scriptsize},
  xlabel={$\ell$},
  tick label style={font=\scriptsize}, label style={font=\small},
  grid=major, grid style={black!12}, axis on top]
  \addplot[fill=blue!55, draw=none] coordinates {(0,11.43030) (1,3.30547) (2,1.95822) (3,1.05175) (4,0.51932) (5,0.25230) (6,0.11491) (7,0.02763) (8,0.00000)};
  \addplot[fill=orange!85, draw=none] coordinates {(0,0.00000) (1,1.42879) (2,1.90100) (3,2.22737) (4,2.43772) (5,2.56755) (6,2.65165) (7,2.70911) (8,2.73673)};
  \addplot[fill=green!55!black!60, draw=none] coordinates {(0,0.00000) (1,6.69605) (2,7.57109) (3,8.15118) (4,8.47326) (5,8.61046) (6,8.66374) (7,8.69357) (8,8.69357)};
\end{axis}
\end{tikzpicture}
\end{center}

\begin{center}
\begin{tikzpicture}[baseline=(current bounding box.north)]
\begin{axis}[width=5.5cm, height=4.1cm,
  ybar stacked, bar width=2.6pt, xmin=-0.70, xmax=16.70,
  ymin=0, ymax=17.5, ytick={0,4,8,12,16}, xtick={0,4,8,12,16},
  title={(4,1), Occam}, title style={font=\scriptsize},
  xlabel={$\ell$}, ylabel={bits},
  tick label style={font=\scriptsize}, label style={font=\small},
  grid=major, grid style={black!12}, axis on top]
  \addplot[fill=blue!55, draw=none] coordinates {(0,16.00000) (1,14.74710) (2,13.52461) (3,12.32975) (4,11.16169) (5,10.02089) (6,8.90878) (7,7.82738) (8,6.77913) (9,5.76681) (10,4.79345) (11,3.86229) (12,2.97690) (13,2.14140) (14,1.36113) (15,0.64341) (16,0.00000)};
  \addplot[fill=orange!85, draw=none] coordinates {(0,0.00000) (1,1.00000) (2,1.98314) (3,2.94918) (4,3.89763) (5,4.82777) (6,5.73876) (7,6.62963) (8,7.49934) (9,8.34673) (10,9.17056) (11,9.96947) (12,10.74193) (13,11.48616) (14,12.19996) (15,12.88052) (16,13.52393)};
  \addplot[fill=green!55!black!60, draw=none] coordinates {(0,0.00000) (1,0.25290) (2,0.49225) (3,0.72106) (4,0.94069) (5,1.15134) (6,1.35246) (7,1.54299) (8,1.72153) (9,1.88646) (10,2.03599) (11,2.16823) (12,2.28117) (13,2.37244) (14,2.43891) (15,2.47607) (16,2.47607)};
\end{axis}
\end{tikzpicture}\hfill
\begin{tikzpicture}[baseline=(current bounding box.north)]
\begin{axis}[width=5.5cm, height=4.1cm,
  ybar stacked, bar width=2.6pt, xmin=-0.70, xmax=16.70,
  ymin=0, ymax=17.5, ytick={0,4,8,12,16}, xtick={0,4,8,12,16},
  title={(4,1), Occamer}, title style={font=\scriptsize},
  xlabel={$\ell$},
  tick label style={font=\scriptsize}, label style={font=\small},
  grid=major, grid style={black!12}, axis on top]
  \addplot[fill=blue!55, draw=none] coordinates {(0,16.00000) (1,14.02313) (2,12.05565) (3,10.07789) (4,8.24016) (5,6.64167) (6,5.29847) (7,4.18012) (8,3.24929) (9,2.47724) (10,1.84273) (11,1.32708) (12,0.91199) (13,0.58080) (14,0.32107) (15,0.12665) (16,0.00000)};
  \addplot[fill=orange!85, draw=none] coordinates {(0,0.00000) (1,1.00000) (2,1.93488) (3,2.79599) (4,3.57122) (5,4.25790) (6,4.86168) (7,5.39153) (8,5.85599) (9,6.26215) (10,6.61604) (11,6.92316) (12,7.18858) (13,7.41658) (14,7.61018) (15,7.77071) (16,7.89736)};
  \addplot[fill=green!55!black!60, draw=none] coordinates {(0,-0.00000) (1,0.97687) (2,2.00947) (3,3.12612) (4,4.18863) (5,5.10044) (6,5.83985) (7,6.42835) (8,6.89472) (9,7.26062) (10,7.54123) (11,7.74976) (12,7.89943) (13,8.00262) (14,8.06875) (15,8.10264) (16,8.10264)};
\end{axis}
\end{tikzpicture}\hfill
\begin{tikzpicture}[baseline=(current bounding box.north)]
\begin{axis}[width=5.5cm, height=4.1cm,
  ybar stacked, bar width=2.6pt, xmin=-0.70, xmax=16.70,
  ymin=0, ymax=17.5, ytick={0,4,8,12,16}, xtick={0,4,8,12,16},
  title={(4,1), Pragmatist}, title style={font=\scriptsize},
  xlabel={$\ell$},
  tick label style={font=\scriptsize}, label style={font=\small},
  grid=major, grid style={black!12}, axis on top]
  \addplot[fill=blue!55, draw=none] coordinates {(0,4.40913) (1,1.02252) (2,0.66648) (3,0.44062) (4,0.29637) (5,0.20423) (6,0.14404) (7,0.10281) (8,0.07319) (9,0.05137) (10,0.03532) (11,0.02367) (12,0.01528) (13,0.00922) (14,0.00480) (15,0.00170) (16,0.00000)};
  \addplot[fill=orange!85, draw=none] coordinates {(0,-0.00000) (1,0.27557) (2,0.34374) (3,0.39134) (4,0.42524) (5,0.44994) (6,0.46850) (7,0.48291) (8,0.49433) (9,0.50348) (10,0.51082) (11,0.51670) (12,0.52144) (13,0.52526) (14,0.52833) (15,0.53073) (16,0.53243)};
  \addplot[fill=green!55!black!60, draw=none] coordinates {(0,0.00000) (1,3.11104) (2,3.39891) (3,3.57716) (4,3.68752) (5,3.75497) (6,3.79659) (7,3.82341) (8,3.84161) (9,3.85428) (10,3.86299) (11,3.86876) (12,3.87241) (13,3.87466) (14,3.87600) (15,3.87670) (16,3.87670)};
\end{axis}
\end{tikzpicture}
\end{center}

\subsection{Trajectories}\label{sec:gibbstraj}

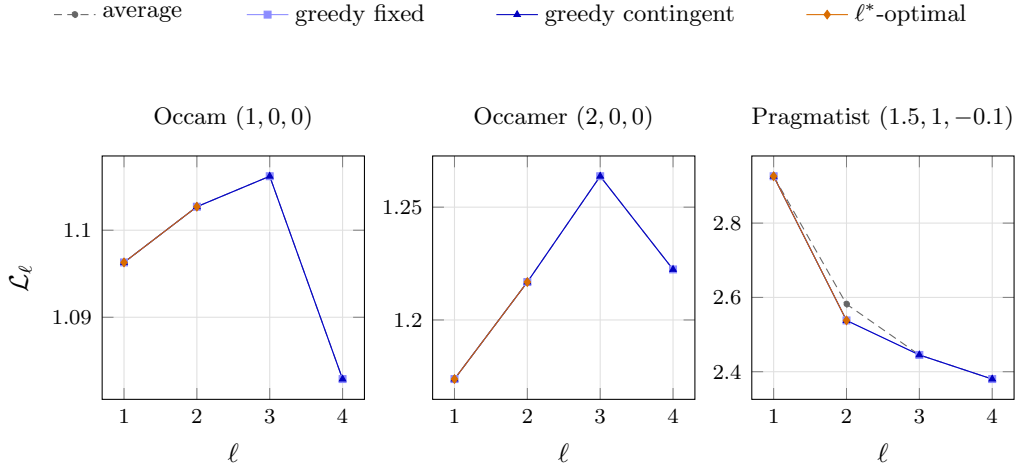
\begin{figure}[H]
\centering
\begin{tikzpicture}
\begin{axis}[width=5.05cm, height=4.8cm,
  title={Occam $(1,0,0)$}, title style={font=\footnotesize},
  xlabel={$\ell$}, ylabel={$\mathcal L_\ell$},
  xtick={1,2,3,4}, grid=major, grid style={black!12},
  tick label style={font=\scriptsize}, label style={font=\small}]
\addplot[black!60, densely dashed, mark=*, mark size=1.1]
  coordinates {(1,1.0963)(2,1.1027)(3,1.1062)(4,1.0829)};
\addplot[blue!45, mark=square*, mark size=1.3]
  coordinates {(1,1.0963)(2,1.1027)(3,1.1062)(4,1.0829)};
\addplot[blue!75!black, mark=triangle*, mark size=1.4]
  coordinates {(1,1.0963)(2,1.1027)(3,1.1062)(4,1.0829)};
\addplot[orange!85!black, mark=diamond*, mark size=1.6]
  coordinates {(1,1.0963)(2,1.1027)};
\end{axis}
\end{tikzpicture}\hspace{1mm}%
\begin{tikzpicture}
\begin{axis}[width=5.05cm, height=4.8cm,
  title={Occamer $(2,0,0)$}, title style={font=\footnotesize},
  xlabel={$\ell$}, xtick={1,2,3,4}, grid=major,
  grid style={black!12},
  tick label style={font=\scriptsize}, label style={font=\small}]
\addplot[black!60, densely dashed, mark=*, mark size=1.1]
  coordinates {(1,1.1738)(2,1.2168)(3,1.2637)(4,1.2224)};
\addplot[blue!45, mark=square*, mark size=1.3]
  coordinates {(1,1.1738)(2,1.2168)(3,1.2637)(4,1.2224)};
\addplot[blue!75!black, mark=triangle*, mark size=1.4]
  coordinates {(1,1.1738)(2,1.2168)(3,1.2637)(4,1.2224)};
\addplot[orange!85!black, mark=diamond*, mark size=1.6]
  coordinates {(1,1.1738)(2,1.2168)};
\end{axis}
\end{tikzpicture}\hspace{1mm}%
\begin{tikzpicture}
\begin{axis}[width=5.05cm, height=4.8cm,
  title={Pragmatist $(1.5,1,-0.1)$}, title style={font=\footnotesize},
  xlabel={$\ell$}, xtick={1,2,3,4}, grid=major,
  grid style={black!12},
  tick label style={font=\scriptsize}, label style={font=\small}]
\addplot[black!60, densely dashed, mark=*, mark size=1.1]
  coordinates {(1,2.9261)(2,2.5819)(3,2.4452)(4,2.3802)};
\addplot[blue!45, mark=square*, mark size=1.3]
  coordinates {(1,2.9261)(2,2.5380)(3,2.4452)(4,2.3802)};
\addplot[blue!75!black, mark=triangle*, mark size=1.4]
  coordinates {(1,2.9261)(2,2.5380)(3,2.4452)(4,2.3802)};
\addplot[orange!85!black, mark=diamond*, mark size=1.6]
  coordinates {(1,2.9261)(2,2.5380)};
\end{axis}
\end{tikzpicture}
\caption{Aggregate leverage under the Gibbs priors at $(2,1)$,
$\ell^* = 2$. In both complexity-only settings every schedule
coincides exactly; the Pragmatist's fields finally separate them.}
\label{fig:gibbs21}
\end{figure}

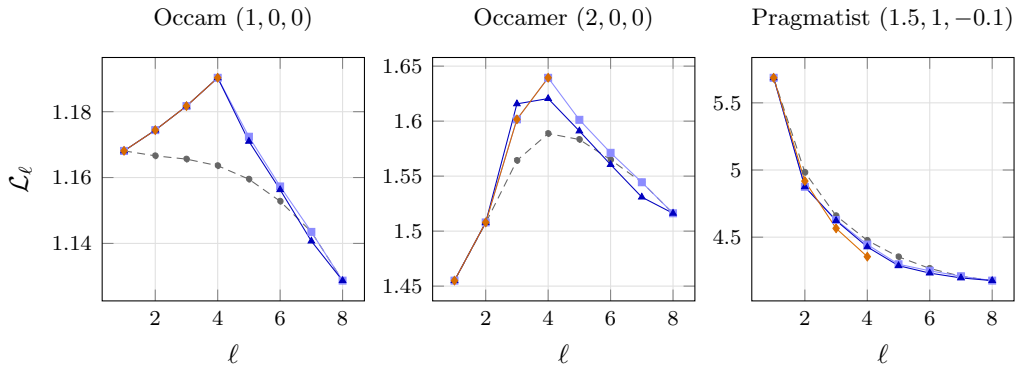
\begin{figure}[H]
\centering
\begin{tikzpicture}
\begin{axis}[width=5.05cm, height=4.8cm,
  title={Occam $(1,0,0)$}, title style={font=\footnotesize},
  xlabel={$\ell$}, ylabel={$\mathcal L_\ell$},
  xtick={2,4,6,8}, grid=major, grid style={black!12},
  tick label style={font=\scriptsize}, label style={font=\small}]
\addplot[black!60, densely dashed, mark=*, mark size=1.1]
  coordinates {(1,1.1681)(2,1.1666)(3,1.1656)(4,1.1636)(5,1.1595)
    (6,1.1528)(7,1.1435)(8,1.1287)};
\addplot[blue!45, mark=square*, mark size=1.3]
  coordinates {(1,1.1681)(2,1.1744)(3,1.1817)(4,1.1903)(5,1.1724)
    (6,1.1573)(7,1.1435)(8,1.1287)};
\addplot[blue!75!black, mark=triangle*, mark size=1.4]
  coordinates {(1,1.1681)(2,1.1744)(3,1.1817)(4,1.1903)(5,1.1710)
    (6,1.1563)(7,1.1407)(8,1.1287)};
\addplot[orange!85!black, mark=diamond*, mark size=1.6]
  coordinates {(1,1.1681)(2,1.1744)(3,1.1817)(4,1.1903)};
\end{axis}
\end{tikzpicture}\hspace{1mm}%
\begin{tikzpicture}
\begin{axis}[width=5.05cm, height=4.8cm,
  title={Occamer $(2,0,0)$}, title style={font=\footnotesize},
  xlabel={$\ell$}, xtick={2,4,6,8}, grid=major,
  grid style={black!12},
  tick label style={font=\scriptsize}, label style={font=\small}]
\addplot[black!60, densely dashed, mark=*, mark size=1.1]
  coordinates {(1,1.4550)(2,1.5074)(3,1.5642)(4,1.5887)(5,1.5833)
    (6,1.5649)(7,1.5444)(8,1.5162)};
\addplot[blue!45, mark=square*, mark size=1.3]
  coordinates {(1,1.4550)(2,1.5079)(3,1.6016)(4,1.6394)(5,1.6012)
    (6,1.5712)(7,1.5444)(8,1.5162)};
\addplot[blue!75!black, mark=triangle*, mark size=1.4]
  coordinates {(1,1.4550)(2,1.5079)(3,1.6157)(4,1.6205)(5,1.5910)
    (6,1.5605)(7,1.5309)(8,1.5162)};
\addplot[orange!85!black, mark=diamond*, mark size=1.6]
  coordinates {(1,1.4550)(2,1.5079)(3,1.6016)(4,1.6394)};
\end{axis}
\end{tikzpicture}\hspace{1mm}%
\begin{tikzpicture}
\begin{axis}[width=5.05cm, height=4.8cm,
  title={Pragmatist $(1.5,1,-0.1)$}, title style={font=\footnotesize},
  xlabel={$\ell$}, xtick={2,4,6,8}, grid=major,
  grid style={black!12},
  tick label style={font=\scriptsize}, label style={font=\small}]
\addplot[black!60, densely dashed, mark=*, mark size=1.1]
  coordinates {(1,5.6865)(2,4.9827)(3,4.6596)(4,4.4759)(5,4.3536)(6,4.2673)(7,4.2090)(8,4.1766)};
\addplot[blue!45, mark=square*, mark size=1.3]
  coordinates {(1,5.6865)(2,4.8751)(3,4.6281)(4,4.4450)(5,4.2984)(6,4.2497)(7,4.2090)(8,4.1766)};
\addplot[blue!75!black, mark=triangle*, mark size=1.4]
  coordinates {(1,5.6865)(2,4.8751)(3,4.6236)(4,4.4275)(5,4.2877)(6,4.2325)(7,4.1966)(8,4.1766)};
\addplot[orange!85!black, mark=diamond*, mark size=1.6]
  coordinates {(1,5.6865)(2,4.9176)(3,4.5657)(4,4.3548)};
\end{axis}
\end{tikzpicture}
\caption{Aggregate leverage under the Gibbs priors at $(3,2)$,
$\ell^* = 4$. Greedy Werner's fixed set is the optimal set under Occam and,
up to an input negation, under Occamer. For the Pragmatist, the $\ell^*$-optimal has the 
worst leverage at the horizon, a reminder that we optimize for information received, while
accepting that may cost even more surprisal.}
\label{fig:gibbs32}
\end{figure}

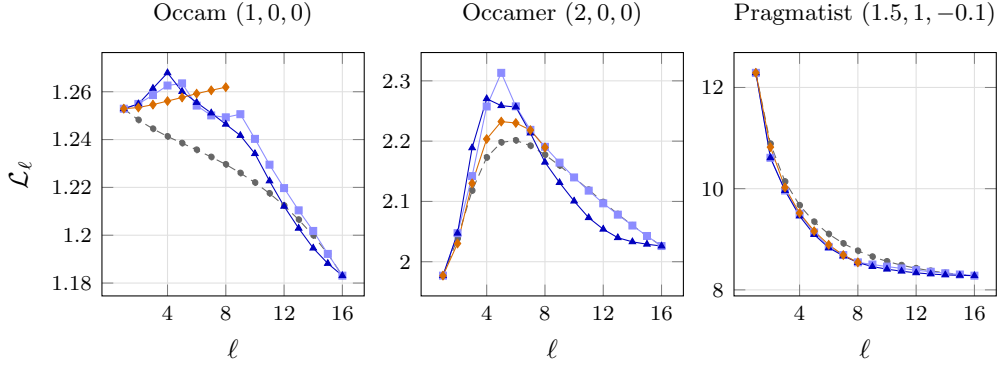
\begin{figure}[H]
\centering
\begin{tikzpicture}
\begin{axis}[width=5.05cm, height=4.8cm,
  title={Occam $(1,0,0)$}, title style={font=\footnotesize},
  xlabel={$\ell$}, ylabel={$\mathcal L_\ell$},
  xtick={4,8,12,16}, grid=major, grid style={black!12},
  tick label style={font=\scriptsize}, label style={font=\small}]
\addplot[black!60, densely dashed, mark=*, mark size=1.1]
  coordinates {(1,1.2529)(2,1.2482)(3,1.2445)(4,1.2413)(5,1.2385)
    (6,1.2357)(7,1.2327)(8,1.2296)(9,1.2260)(10,1.2220)(11,1.2175)
    (12,1.2124)(13,1.2065)(14,1.1999)(15,1.1922)(16,1.1831)};
\addplot[blue!45, mark=square*, mark size=1.3]
  coordinates {(1,1.2529)(2,1.2549)(3,1.2586)(4,1.2626)(5,1.2635)
    (6,1.2542)(7,1.2501)(8,1.2493)(9,1.2506)(10,1.2403)(11,1.2295)
    (12,1.2197)(13,1.2104)(14,1.2018)(15,1.1922)(16,1.1831)};
\addplot[blue!75!black, mark=triangle*, mark size=1.4]
  coordinates {(1,1.2529)(2,1.2549)(3,1.2614)(4,1.2679)(5,1.2601)
    (6,1.2555)(7,1.2511)(8,1.2464)(9,1.2417)(10,1.2341)(11,1.2227)
    (12,1.2121)(13,1.2029)(14,1.1946)(15,1.1882)(16,1.1831)};
\addplot[orange!85!black, mark=diamond*, mark size=1.6]
  coordinates {(1,1.2529)(2,1.2534)(3,1.2546)(4,1.2561)(5,1.2576)
    (6,1.2592)(7,1.2606)(8,1.2619)};
\end{axis}
\end{tikzpicture}\hspace{1mm}%
\begin{tikzpicture}
\begin{axis}[width=5.05cm, height=4.8cm,
  title={Occamer $(2,0,0)$}, title style={font=\footnotesize},
  xlabel={$\ell$}, xtick={4,8,12,16}, grid=major,
  grid style={black!12},
  tick label style={font=\scriptsize}, label style={font=\small}]
\addplot[black!60, densely dashed, mark=*, mark size=1.1]
  coordinates {(1,1.9769)(2,2.0386)(3,2.1181)(4,2.1729)(5,2.1979)
    (6,2.2012)(7,2.1923)(8,2.1774)(9,2.1594)(10,2.1398)(11,2.1194)
    (12,2.0989)(13,2.0790)(14,2.0603)(15,2.0427)(16,2.0260)};
\addplot[blue!45, mark=square*, mark size=1.3]
  coordinates {(1,1.9769)(2,2.0473)(3,2.1420)(4,2.2575)(5,2.3133)
    (6,2.2577)(7,2.2187)(8,2.1906)(9,2.1644)(10,2.1398)(11,2.1179)
    (12,2.0968)(13,2.0781)(14,2.0600)(15,2.0427)(16,2.0260)};
\addplot[blue!75!black, mark=triangle*, mark size=1.4]
  coordinates {(1,1.9769)(2,2.0473)(3,2.1889)(4,2.2703)(5,2.2588)
    (6,2.2566)(7,2.2134)(8,2.1650)(9,2.1312)(10,2.1005)(11,2.0730)
    (12,2.0539)(13,2.0398)(14,2.0330)(15,2.0290)(16,2.0260)};
\addplot[orange!85!black, mark=diamond*, mark size=1.6]
  coordinates {(1,1.9769)(2,2.0302)(3,2.1300)(4,2.2035)(5,2.2324)
    (6,2.2301)(7,2.2183)(8,2.1896)};
\end{axis}
\end{tikzpicture}\hspace{1mm}%
\begin{tikzpicture}
\begin{axis}[width=5.05cm, height=4.8cm,
  title={Pragmatist $(1.5,1,-0.1)$}, title style={font=\footnotesize},
  xlabel={$\ell$}, xtick={4,8,12,16}, grid=major,
  grid style={black!12},
  tick label style={font=\scriptsize}, label style={font=\small}]
\addplot[black!60, densely dashed, mark=*, mark size=1.1]
  coordinates {(1,12.2894)(2,10.8881)(3,10.1407)(4,9.6717)(5,9.3456)(6,9.1037)(7,8.9175)(8,8.7714)(9,8.6553)(10,8.5624)(11,8.4874)(12,8.4264)(13,8.3767)(14,8.3363)(15,8.3045)(16,8.2812)};
\addplot[blue!45, mark=square*, mark size=1.3]
  coordinates {(1,12.2894)(2,10.6106)(3,9.9628)(4,9.5034)(5,9.1348)(6,8.8509)(7,8.6881)(8,8.5439)(9,8.5015)(10,8.4603)(11,8.4241)(12,8.3888)(13,8.3582)(14,8.3282)(15,8.3045)(16,8.2812)};
\addplot[blue!75!black, mark=triangle*, mark size=1.4]
  coordinates {(1,12.2894)(2,10.6106)(3,9.9628)(4,9.4586)(5,9.0926)(6,8.8324)(7,8.6717)(8,8.5393)(9,8.4604)(10,8.4089)(11,8.3691)(12,8.3367)(13,8.3137)(14,8.2962)(15,8.2873)(16,8.2812)};
\addplot[orange!85!black, mark=diamond*, mark size=1.6]
  coordinates {(1,12.2894)(2,10.8208)(3,10.0230)(4,9.5178)(5,9.1604)(6,8.8912)(7,8.6881)(8,8.5439)};
\end{axis}
\end{tikzpicture}
\caption{Aggregate leverage under the Gibbs priors at $(4,1)$,
$\ell^* = 8$. Occamer's deep correlations $C$ bend even the uniform
average curve into an interior hump; the Pragmatist releases all its instinct at the start: whether it's constant.}
\label{fig:gibbs41}
\end{figure}

Figures~\ref{fig:gibbs21}--\ref{fig:gibbs41} show the expected
cumulative leverage $\mathcal L_\ell$ under the three
prior hypotheses, for the uniform question-order average and the strategy grid of
Appendix~\ref{app:strategies}. The $\ell$-optimal schedule targets
the halfway horizon $\ell^* = Q/2$ and is plotted only up to it:
past its own horizon the optimizer's continuation is not defined.
Inside the optimal set the questions are asked in random
order (the graph averages that order). 
We omit its contingent twin everywhere: wherever we can
afford to compute it, it reproduces the fixed optimum's curve to
machine precision, and one for which no symmetry argument yet is known. 
Two consistency checks hold in every panel: all schedules agree at $\ell = 1$, because
input relabeling acts transitively on the questions and the Gibbs
energies are relabeling-invariant, so single questions are
exchangeable; and every full-length curve ends at the same
$\mathcal L_{Q} = H(M^{(0)})/H(p)$, since any policy that asks everything
receives exactly $H(p)$ on average, and clears the table $M$. The
same invariance acts on whole sets, so the $\ell$-optimal objective
is a function of a set's relabeling orbit and not of the set:
minimizers arrive in orbits.

\begin{center}
\begin{tikzpicture}[font=\footnotesize]
  \draw[black!60, densely dashed] (0,0) -- (0.55,0)
    plot[mark=*, mark size=1.1] coordinates {(0.275,0)};
  \node[anchor=west, inner sep=1.5pt] at (0.58,0) {average};
  \draw[blue!45] (2.55,0) -- (3.10,0)
    plot[mark=square*, mark size=1.3] coordinates {(2.825,0)};
  \node[anchor=west, inner sep=1.5pt] at (3.13,0) {greedy fixed};
  \draw[blue!75!black] (5.85,0) -- (6.40,0)
    plot[mark=triangle*, mark size=1.4] coordinates {(6.125,0)};
  \node[anchor=west, inner sep=1.5pt] at (6.43,0) {greedy contingent};
  \draw[orange!85!black] (9.95,0) -- (10.50,0)
    plot[mark=diamond*, mark size=1.6] coordinates {(10.225,0)};
  \node[anchor=west, inner sep=1.5pt] at (10.53,0) {$\ell^*$-optimal};
\end{tikzpicture}
\end{center}

Three discussions, one per system size.

At $(2,1)$, Occam and Occamer don't have enough structure to tell
subsets apart, and all subsets of equal size share a single joint
entropy.
Occamer shows what doubling the field buys within those constraints.
The whole curve lifts (endpoint $1.2224$ against Occam's
$1.0829$, store $C_0 = 0.728$ against $0.306$ bits) without creating
a single strategic choice to make. The Pragmatist's fields split the
pair orbits at last, and the strategy yields
a \emph{lower} leverage: greedy's pair $\{00, 11\}$ knows more at
$\ell = 2$ than the random order but pays so much more surprisal that it is less leveraged
($2.538$ against $2.582$).

At $(3,2)$ there are exactly two optimal quadruples under \emph{all
three} settings: the even-weight inputs $\{000, 011, 101, 110\}$, the
maximally spread design, pairwise at Hamming distance two, and its
odd-weight image under negating one input bit. Relabeling makes them
exactly degenerate. Under the Pragmatist greedy
finally falls short: its set $\{000, 011, 101, 111\}$ leaves twice the
ignorance of the optimal four ($0.405$ against $0.198$ bits
remaining) while receiving $0.099$ bits less and it posts the
\emph{higher} leverage, $4.445$ against $4.355$, with the
strategyless average higher still at $4.476$. The $\ell$-optimal
schedule maximizes knowledge, not leverage, and a ratio is buoyed by 
cheaper ignorance.

$(4,1)$ is the richest, with the most questions, the choice becomes
impactful. Doubling $\beta$
triples the store ($C_0 = 8.10$ bits against Occam's $2.48$) and
bends even the average curve into an interior hump, rising to
$2.20$ at $\ell = 6$ before decaying. This is the deep-shell signature
promised in Section~\ref{sec:gibbsensemble}, knowledge that only
becomes spendable once enough answers are in hand to cash out the
higher correlations. The optimal eight questions are a parity class
of the input hypercube in every setting, and as at $(3,2)$ both
classes minimize exactly; greedy misses the class
under both complexity-only settings ($\mathcal L = 1.249$ against the optimum's
$1.262$ at the Occam $\ell^*$ horizon) but finds it exactly under the
Pragmatist whose prior, $0.90$ of its mass on the two constants,
is a two-sided spike in disguise: $H(p) = 0.53$ bits, leverage
running from $12.29$ down to $8.28$, a great deal of inference
about a game that is mostly over after two questions.

\section{Computing the circuit complexities}\label{app:aig}

We read the footprint and the bias of
Section~\ref{sec:aigcomplexity} off a truth table in one pass. The
complexity $X_j$ we cannot: it is a minimum over all circuits, and we
have to compute it. This appendix records how we obtained the
$65{,}536$ values in each of the $(4,1)$ and $(3,2)$
tables. We describe the pipeline in
\texttt{gibbs\_complexity.md}, and implement it in
\texttt{aig\_synth.py}, \texttt{groups.py},
\texttt{run\_synth\_3to2.py} and the
\texttt{build\_table} drivers.

Reduction by symmetry: Relabeling wires is free in an
and-inverter circuit: permuting inputs is rewiring, negating an input
or an output is an edge attribute, and for $m = 2$ swapping the two
outputs renames which wire is which. So $X_j$ is constant on the orbits
of the relabeling groups
\begin{equation}
  |G_{4,1}| = 4!\cdot 2^4\cdot 2 = 768,
  \qquad
  |G_{3,2}| = 3!\cdot 2^3\cdot 2^2\cdot 2 = 384,
\end{equation}
the first being the standard NPN group of logic synthesis \cite{knuth4a}.
In \texttt{groups.py} we apply every group element to every truth table
and keep the numerically smallest image as the class representative,
which collapses $65{,}536$ maps to $222$ classes at $(4,1)$ and $308$ at
$(3,2)$. We then run exact synthesis a few hundred times instead of
$65{,}536$.

Exact synthesis as satisfiability: For a fixed gate budget $r$, in
\texttt{aig\_synth.py} we build a formula that is satisfiable exactly when
some $r$-gate and-inverter circuit computes the target. Three families
of variables carry it. First $t_{s,\rho}$, the bit that signal $s$
holds on truth-table row $\rho$, with signal $0$ the constant and
signals $1,\ldots,n$ the inputs, all pinned by unit clauses. Then a
one-hot selector $\sigma_{g,a,b,\iota_a,\iota_b}$ saying that gate $g$
draws its two fanins from \emph{earlier} signals $a \le b$, inverted
according to $\iota$; restricting to earlier signals is what makes the
circuit acyclic by construction, so we need no ordering constraints.
Last a one-hot selector $o_{k,s,\iota}$ naming which signal, at which
polarity, carries output $k$. The AND semantics costs three clauses per
gate, candidate and row, and the output constraint pins the chosen
signal to the target column on every row. We add two structural clauses
that tighten the search without excluding any minimum circuit: every
gate must feed a later gate or an output, and no two gates may carry
identical truth tables. Then
\begin{equation}
  X_j \;=\; \min\{\, r \ge 0 \;:\; \text{the $r$-gate formula is
  satisfiable} \,\},
  \label{eq:aigmin}
\end{equation}
which we search by trying $r = 0, 1, 2, \ldots$ upward. We get a proven
minimum rather than a best effort: the solver refuted every smaller
budget, and did not merely fail to find one. We run CaDiCaL \cite{cadical}
as the back end, through PySAT \cite{pysat}. Run as a script, the synthesizer
reproduces the textbook two-input values, AND, OR and NAND at $1$ and
XOR at $3$.

Provenance: At $(4,1)$ we adopt the class values from the
published reproducibility dataset \cite{krinkin} of the $222$ NPN
classes, and in \texttt{crosscheck.py} we re-derive ten of them
at random with our own synthesizer, finding no mismatch. At $(3,2)$
there is no reference table, so we synthesized all $308$ classes from
scratch with \texttt{run\_synth\_3to2.py} over a process pool. The
resulting ranges are $X \le 10$ at $(4,1)$ and $X \le 9$ at $(3,2)$,
and we tabulate the head counts in Section~\ref{sec:aigcomplexity}.

Why we adopt the $(4,1)$ values rather than recompute them: All
$222$ entries of \cite{krinkin} are proven optimal, the two hardest
classes included: $\mathtt{0x1669}$ and $\mathtt{0x166b}$, at $10$
gates each, which is the ceiling of the range. Our own synthesizer
agrees wherever it can reach, but it cannot reach that far. For
$\mathtt{0x1669}$ it refutes $r = 5$ in under a second and $r = 7$ in
half a minute, and does not settle $r = 9$ within fifteen: the instance
grows steeply with the budget, since the selector family for gate $g$
already carries $O\big((n+g)^2\big)$ candidates and its at-most-one
encoding is quadratic in that. Hence our division of labor. We take the
$(4,1)$ column from the reference dataset and sample it against
our own synthesizer, while the $(3,2)$ classes, being one input bit
smaller and so a much smaller formula, we synthesize here from
scratch.

\section{Symbols}\label{app:symbols}

\begin{table}[H]
\begin{center}
\begin{tabular}{@{}r@{\qquad}l@{}}
\toprule
$n$ & number of bits in a question \\
$m$ & number of bits in an answer \\
$q \in \mathcal{Q}$ & set of all questions \\ 
$a \in \mathcal {A}$ & set of all answers \\
$Q = 2^n$ & how many distinct questions exist, the width of $M$ \\
$A = 2^m$ & how many answers any one question admits \\
$N = A^Q$ & how many candidate maps exist, the size of the hypothesis space \\
\addlinespace[5pt]
$\psi$ & the truth: the single map nature uses to answer, unknown to Werner \\
$\phi_j$ & candidate map $j$, indexed by its truth table read as a numeral \\
$p_j$ & Werner's prior belief that $\psi = \phi_j$ \\
$M_{a,q}$ & the answer matrix: his belief that question $q$ is answered $a$ \\
$H(M)$ & table entropy, the column entropies of $M$ added \\
$H(p)$ & entropy of the belief itself, over the $N$ maps \\
$C$ & correlation store $H(M) - H(p)$, knowledge no single column sees \\
$\chi_{\mathcal S}$ & correlator: the mean parity of the answers on a set of sites \\
\addlinespace[5pt]
$\mathcal V \subseteq \mathcal S$ & subset and set of questions asked so far \\
$\ell$ & how many have been asked, $|\mathcal S|$ \\
$y$ & the pattern of answers received on $\mathcal S$ \\
$s_k$ & surprisal of round $k$. Negative log of the probability of the answer \\
$\langle\,\cdot\,\rangle$ & average over the truth $\psi \sim p_j$, at a question we have named \\
$\langle\!\langle\,\cdot\,\rangle\!\rangle$ & the same, averaged over every set of $\ell$ questions as well \\
$L_\ell$ & leverage: table entropy cleared per bit of surprisal received \\
$\mathcal L_\ell$ & ratio of expected table-entropy reduction to expected surprisal \\
$\eta_n(v)$ & fresh-answer entropy at independent reveal probability $v$ \\
$G_k$ & mean entropy of the answers to $k$ questions, averaged over which $k$ \\
$\gamma_{n,k}$ & increment $G_{n,k+1} - G_{n,k}$, what one fresh answer is worth \\
\addlinespace[5pt]
$t = \ell/Q$ & macroscopic time, the fraction of the table already asked \\
$\gamma(t)$ & the profile: the increment's limit, entropy of a fresh answer at $t$ \\
$g(t)$ & $\int_0^t\gamma$, the surprisal received per question by time $t$ \\
$\eta_0$ & marginal entropy of one answer before anything has been asked \\
$\nu$ & an answer law: a discrete distribution on the alphabet $\mathcal A$ \\
$\Lambda$ & the directing measure: a distribution over answer laws $\nu$ \\
\addlinespace[5pt]
$F$, $R$ & the statistic: a CDF probability law on $[0,1]$ and the value drawn from it \\
$D$ & the complexity class that draw selects \\
$w_d$ & prior weight of class $d$, the $F$-measure of its rate gap \\
$K_d$ & dimension of class $d$, the number of monomials of degree $\le d$ \\
$R_{n,d}$ & rate of class $d$, its dimension per question $K_d/Q$ \\
$\rho_{\mathcal B}$ & coefficient of the monomial $\mathcal B$ in a map's polynomial \\
$\Omega$ & generator matrix: a class's monomials evaluated at every question \\
$\xi$ & string of random bits, from a coin ($m=1$) or sampled from $\mathcal A$. \\
\bottomrule
\end{tabular}
\end{center}
\caption{Most important symbols used in this work, in order of appearance.}
\label{tab:symbols}
\end{table}

\section{Scripts}\label{app:scripts}

Numbers, curves, and combinatoric sets in this work are produced by the scripts
below. They are hosted with the source, at
\url{https://github.com/DMChernowitz/boolean-maps-learning}.

\begin{table}[H]
\begin{center}
\small
\begin{tabular}{@{}l p{10.3cm}@{}}
\toprule
script & what it does \\
\midrule
\texttt{tikz\_data.py} & Tabulates the spike-prior curves of
  Appendix~\ref{app:spike}, the one-question anatomy, the branch leverages,
  the extraction ratio and the sharp-prior flow, and the Bernstein basis of
  Appendix~\ref{app:blockpriors}. \\
\texttt{spike\_finite\_data.py} & Exact finite-$n$ leverage of the spike
  prior from its closed block entropies. \\
\texttt{rm\_gamma\_data.py} & Mean rank increment
  $\gamma^{(d)}_{n,\ell}$ of one Reed-Muller class along a random question
  order. \\
\texttt{rm\_finite\_data.py} & Exact finite-$n$ leverage of the
  polynomial-degree prior, for the three statistics of
  Figure~\ref{fig:Fexamples}. \\
\texttt{block\_finite\_data.py} & Exact finite-$n$ leverage of the clique,
  parity and MDS block priors, through the hypergeometric increment
  \eqref{eq:blockfinitegamma}. \\
\addlinespace[5pt]
\texttt{rm\_label\_data.py} & Splits the received entropy into the class
  average and the label term $I_{n,\ell}$ of \eqref{eq:Gmix}. \\
\texttt{rm\_carpet\_fig.py} & Draws the layered increments that
  illustrate \eqref{eq:ungroup} in the limit, from those tables. \\
\texttt{gibbs\_curves.py} & Aggregate leverage trajectories under the three
  Gibbs settings, and the descriptor tables they are read from. \\
\texttt{gibbs\_stack\_data.py} & The three-layer split of the table entropy
  at every $\ell$, for the panels of Section~\ref{sec:gibbsbooks}. \\
\addlinespace[5pt]
\texttt{aig\_synth.py} & Exact minimum-size and-inverter synthesis as
  satisfiability, one formula per gate budget. \\
\texttt{groups.py} & Canonical representatives of the relabeling classes,
  of order $768$ at $(4,1)$ and $384$ at $(3,2)$. \\
\texttt{run\_synth\_3to2.py} & Runs the synthesizer over all $308$ classes
  of the $(3,2)$ space on a process pool. \\
\texttt{crosscheck.py} & Re-derives random classes of the published $(4,1)$
  dataset with our own synthesizer. \\
\texttt{build\_table\_4to1.py} & Assembles the $65{,}536$-row $(4,1)$ table
  of complexity, footprint and bias. \\
\texttt{build\_table\_3to2.py} & The same at $(3,2)$, where the complexity
  is one joint circuit for both output bits. \\
\texttt{build\_table\_3to1.py} & The same at $(3,1)$, which the checks below
  use as a small test space. \\
\addlinespace[5pt]
\texttt{expected\_trajectory.py} & Checks the two exact laws, and the
  expected run of the guiding example, by enumerating every question set. \\
\texttt{walsh\_update.py} & Checks the correlator ledger and the two-term
  update rule \eqref{eq:corr-update} at $m = 1$. \\
\texttt{correlator\_shells.py} & The same at general $m$, on the $(3,2)$
  space, against \eqref{eq:corr-update-m}. \\
\texttt{spike\_flow.py} & Checks the two-state renormalization flow
  \eqref{eq:spikeflow} and the invariant $P_k r_k = p$. \\
\bottomrule
\end{tabular}
\end{center}
\caption{The scripts behind this work, grouped by role: the tables read at
compile time, the figures generated and pasted in, the circuit-complexity
pipeline of Appendix~\ref{app:aig}, and the checks.}
\label{tab:scripts}
\end{table}

\clearpage
\phantomsection
\stdsection*{\refname}
\addcontentsline{toc}{section}{\refname}
\begingroup
\renewcommand{\section}[2]{}
\small
\begin{multicols}{2}

\end{multicols}
\endgroup

\end{document}